\documentstyle[12pt]{article}

\newcommand \beq{\begin{eqnarray}}
\newcommand \eeq{\end{eqnarray}}
\begin{document}
\input epsf

\def\bfgamma{\mbox{\boldmath$\gamma$}}
\def\bfalpha{\mbox{\boldmath$\alpha$}}
\def\bftau{\mbox{\boldmath$\tau$}}
\def\bfnabla{\mbox{\boldmath$\nabla$}}
\def\bfsigma{\mbox{\boldmath$\sigma$}}
\def\bfxi{\mbox{\boldmath$\xi$}}
\def\BN{\hbox{Bloch-Nordsiek}}
\def\vp{\mbox{$\bf v\cdot p$}}
\def\vq{\mbox{$\bf v\cdot q$}}
\def\vpq{\mbox{$\bf v\cdot(p+ q)$}}
\def\tilA{\mbox{$v\cdot A$}}
\def\tilQ{\mbox{v\cdot q}}
\def\tilQ1{\mbox{$v\cdot q_1$}}
\def\tilQ2{\mbox{$v\cdot q_2$}}
\def\bfp{\mbox{\boldmath$p$}}
%

\newenvironment{petitchar}{\begin{list}{}
{\leftmargin1.5em\rightmargin0.0cm}%
\item\small}{\end{list}}


\def\nbfepsilon{\mbox{\boldmath$\epsilon$}}
\def\nbfgrad{\mbox{\boldmath$\grad$}}
\def\bfgrad{\mbox{\boldmath$\grad$}}
\def\bfgamma{\mbox{\boldmath$\gamma$}}
\def\bfxi{\mbox{\boldmath$\xi$}}
\def\bfcalA{\mbox{\boldmath${\cal A}$}}
\def\bfcalS{\mbox{\boldmath${\cal S}$}}
\def\bfp{\mbox{\boldmath$p$}}
\def\bfv{\mbox{\boldmath$v$}}
\def\bfj{\mbox{\boldmath$j$}}
\def\bfhp{\mbox{\boldmath$\hat p$}}
\def\bfei{\mbox{\boldmath$e_i$}}
\def\bfe{\mbox{\boldmath$e$}}
\def\bfej{\mbox{\boldmath$e_j$}}
\def\bfk{\mbox{\boldmath$k$}}
\def\bfq{\mbox{\boldmath$q$}}
\def\bfR{\mbox{\boldmath$R$}}
\def\bfC{\mbox{\boldmath$C$}}
\def\bfR{\mbox{\boldmath$R$}}
\def\bfX{\mbox{\boldmath$X$}}
\def\bfx{\mbox{\boldmath$x$}}
\def\bfE{\mbox{\boldmath$E$}}
\def\bfB{\mbox{\boldmath$B$}}
\def\bfy{\mbox{\boldmath$y$}}
\def\bfr{\mbox{\boldmath$r$}}
\def\rmRe{\mbox{\rm$Re$}}
\def\rmIm{\mbox{\rm$ImR$}}

\def\bfgamma{\mbox{\boldmath$\gamma$}}
\def\bfalpha{\mbox{\boldmath$\alpha$}}
\def\bfsigma{\mbox{\boldmath$\sigma$}}
\def\bfalpha{\mbox{\boldmath$\alpha$}}
\def\bfsigma{\mbox{\boldmath$\sigma$}}
\def\bfSigma{\mbox{\boldmath$\Sigma$}}
\def\bfepsilon{\mbox{\boldmath$\epsilon$}}


\def\T{\hbox{temperature}}
\def\bk{\hbox{background}}
\def\P{\mbox{\psi}}
\def\BP{\mbox{\bar\Psi}}
\def\E{\hbox{equation}}
\def\Es{\hbox{equations}}
\def\QGP{\hbox{quark-gluon plasma}}
\def\HTL{\hbox{hard thermal loops}}
\def\htl{\hbox{hard thermal loop}}
\def\se{\hbox{self-energy}}
\def\pt{\hbox{polarization tensor}}
\def\pov{\hbox{point of view}}
\def\pth{\hbox{perturbation theory}}
\def\wr{\hbox{with respect to}}
\def\fn{\hbox{function}}
\def\FN{\hbox{functions}}
\def\BN{\hbox{Bloch-Nordsieck}}
\def\vp{\mbox{$\bf v\cdot p$}}
\def\vq{\mbox{$\bf v\cdot q$}}
\def\vpq{\mbox{$\bf v\cdot(p+ q)$}}
\def\tilA{\mbox{$v\cdot A$}}
\def\tilQ{\mbox{v\cdot q}}
\def\tilQ1{\mbox{$v\cdot q_1$}}
\def\tilQ2{\mbox{$v\cdot q_2$}}
\def\bfp{\mbox{\boldmath$p$}}

\hyphenation{approxima-tions}
\hyphenation{par-ti-cu-le}
\hyphenation{par-ti-cu-les}

\hyphenation{ac-com-pa-gnees}
\hyphenation{cons-tan-te}
\hyphenation{e-lec-tro-ma-gne-ti-que}
\hyphenation{e-lec-tro-ma-gne-ti-ques}
\hyphenation{im-pe-ra-tif}

\newcommand{\theo}{th\'{e}orie\,\,}
\newcommand{\mod}{mod\`ele\,\,}
\newcommand{\mods}{mod\`eles\,\,}
\newcommand{\theos}{th\'{e}ories\,\,}

\def\square{\hbox{{$\sqcup$}\llap{$\sqcap$}}}   
\def\grad{\nabla}                               
\def\del{\partial}                              

\def\frac#1#2{{#1 \over #2}}
\def\smallfrac#1#2{{\scriptstyle {#1 \over #2}}}
\def\half{\ifinner {\scriptstyle {1 \over 2}}
   \else {1 \over 2} \fi}

\def\bra#1{\langle#1\vert}              
\def\ket#1{\vert#1\rangle}              

\def\simge{\mathrel{%
   \rlap{\raise 0.511ex \hbox{$>$}}{\lower 0.511ex \hbox{$\sim$}}}}
\def\simle{\mathrel{
   \rlap{\raise 0.511ex \hbox{$<$}}{\lower 0.511ex \hbox{$\sim$}}}}


\def\parenbar#1{{\null\!                        
   \mathop#1\limits^{\hbox{\fiverm (--)}}       
   \!\null}}                                    
\def\nunubar{\parenbar{\nu}}
\def\ppbar{\parenbar{p}}


\def\buildchar#1#2#3{{\null\!                   
   \mathop#1\limits^{#2}_{#3}                   
   \!\null}}                                    
\def\overcirc#1{\buildchar{#1}{\circ}{}}


\def\slashchar#1{\setbox0=\hbox{$#1$}           
   \dimen0=\wd0                                 
   \setbox1=\hbox{/} \dimen1=\wd1               
   \ifdim\dimen0>\dimen1                        
      \rlap{\hbox to \dimen0{\hfil/\hfil}}      
      #1                                        
   \else                                        
      \rlap{\hbox to \dimen1{\hfil$#1$\hfil}}   
      /                                         
   \fi}                                         %


\def\subrightarrow#1{
  \setbox0=\hbox{
    $\displaystyle\mathop{}
    \limits_{#1}$}
  \dimen0=\wd0
  \advance \dimen0 by .5em
  \mathrel{
    \mathop{\hbox to \dimen0{\rightarrowfill}}
       \limits_{#1}}}                           

\def\real{\mathop{\rm Re}\nolimits}     
\def\imag{\mathop{\rm Im}\nolimits}     

\def\tr{\mathop{\rm tr}\nolimits}       
\def\Tr{\mathop{\rm Tr}\nolimits}       
\def\Det{\mathop{\rm Det}\nolimits}     

\def\mod{\mathop{\rm mod}\nolimits}     
\def\wrt{\mathop{\rm wrt}\nolimits}     


\def\TeV{{\rm TeV}}                     
\def\GeV{{\rm GeV}}                     
\def\MeV{{\rm MeV}}                     
\def\KeV{{\rm KeV}}                     
\def\eV{{\rm eV}}                       

\def\mb{{\rm mb}}                       
\def\mub{\hbox{$\mu$b}}                 
\def\nb{{\rm nb}}                       
\def\pb{{\rm pb}}                       

%
\def\journal#1#2#3#4{\ {#1}{\bf #2} ({#3})\  {#4}}

\def\AdvPhys{\journal{Adv.\ Phys.}}
\def\AnnPhys{\journal{Ann.\ Phys.}}
\def\EurophysLett{\journal{Europhys.\ Lett.}}
\def\JApplPhys{\journal{J.\ Appl.\ Phys.}}
\def\JMathPhys{\journal{J.\ Math.\ Phys.}}
\def\LettNuovoCimento{\journal{Lett.\ Nuovo Cimento}}
\def\Nature{\journal{Nature}}
\def\NPA{\journal{Nucl.\ Phys.\ {\bf A}}}
\def\NPB{\journal{Nucl.\ Phys.\ {\bf B}}}
\def\NuovoCimento{\journal{Nuovo Cimento}}
\def\Physica{\journal{Physica}}
\def\PLA{\journal{Phys.\ Lett.\ {\bf A}}}
\def\PLB{\journal{Phys.\ Lett.\ {\bf B}}}
\def\PhysRev{\journal{Phys.\ Rev.}}
\def\PR{\journal{Phys.\ Rev.}}
\def\PRC{\journal{Phys.\ Rev.\ {\bf C}}}
\def\PRD{\journal{Phys.\ Rev.\ {\bf D}}}
\def\PRE{\journal{Phys.\ Rev.\ {\bf E}}}
\def\PRB{\journal{Phys.\ Rev.\ {\bf B}}}
\def\PRL{\journal{Phys.\ Rev.\ Lett.}}
\def\PhysRept{\journal{Phys.\ Repts.}}
\def\ProcNatlAcadSci{\journal{Proc.\ Natl.\ Acad.\ Sci.}}
\def\ProcRoySoc{\journal{Proc.\ Roy.\ Soc.\ London Ser.\ A}}
\def\RevModPhys{\journal{Rev.\ Mod.\ Phys.}}
\def\Science{\journal{Science}}
\def\SovPhysJETP{\journal{Sov.\ Phys.\ JETP}}
\def\SovPhysJETPLett{\journal{Sov.\ Phys.\ JETP Lett.}}
\def\SovJNuclPhys{\journal{Sov.\ J.\ Nucl.\ Phys.}}
\def\SovPhysDoklady{\journal{Sov.\ Phys.\ Doklady}}
\def\ZPhys{\journal{Z.\ Phys.}}
\def\ZPhysA{\journal{Z.\ Phys.\ A}}
\def\ZPhysB{\journal{Z.\ Phys.\ B}}
\def\ZPhysC{\journal{Z.\ Phys.\ C}}


\begin{titlepage}
\begin{flushright}
SACLAY--T01/005\\ CERN--TH/2000--272\\hep-ph/0101103
\end{flushright}
\vspace*{1.cm}
\begin{center}
\baselineskip=13pt
{\Large{\bf  The Quark-Gluon Plasma:\\
Collective Dynamics and Hard Thermal Loops\\}}
\vskip0.5cm
Jean-Paul BLAIZOT\footnote{Member of CNRS.
E-mail: blaizot@spht.saclay.cea.fr}  and 
Edmond IANCU\footnote{Member of CNRS. E-mail: iancu@spht.saclay.cea.fr}\\
{\small\it Service de Physique Th\'eorique\footnote{Laboratoire de la Direction
des
Sciences de la Mati\`ere du Commissariat \`a l'Energie
Atomique}, CE-Saclay \\ 91191 Gif-sur-Yvette, France}\\
\end{center}

\date{\today}

\vskip 1cm
\begin{abstract} 
We present a unified description of the high temperature phase of QCD, the
so-called quark-gluon plasma, in a regime where the effective gauge coupling
$g$ is sufficiently small to allow for weak coupling calculations.  The main
focuss is the construction of the effective theory for the collective
excitations which develop at a typical scale $gT$, 
which is well separated from the typical energy of single particle
excitations which is  the temperature $T$. We show that the short
wavelength thermal fluctuations, i.e., the plasma particles, provide a
source for long wavelength oscillations of average fields which carry the
quantum numbers of the plasma constituents, the quarks and the
gluons. To leading order in
$g$, the plasma particles obey simple  gauge-covariant kinetic
equations, whose derivation from the general Dyson-Schwinger equations
is outlined. By solving  these equations, 
we effectively integrate out the hard
degrees of freedom, and are left with an  
 effective theory for the soft collective excitations. As a by-product, 
 the ``hard thermal loops'' emerge naturally in a physically transparent
framework.  We show that the collective 
excitations can be described in terms of
classical fields, and develop for these a Hamiltonian formalism. This can be
used to estimate the effect of the soft thermal fluctuations on the
correlation functions. The effect of 
collisions among the hard particles is also studied. In
particular we discuss how the collisions affect 
the lifetimes of quasiparticle
excitations in a regime where the mean free
 path is comparable with the range of
the relevant interactions. Collisions play also a decisive role 
in the construction
of the effective theory for ultrasoft excitations, with momenta $\sim g^2T$,
a topic which is briefly addressed at the end of this paper.
 \end{abstract}
\vskip .5cm

\centerline{Submitted to Physics Reports}
\end{titlepage}
\tableofcontents
\newpage
\setcounter{equation}{0}
\section{Introduction}

It is currently believed that matter at high density
(several  times ordinary nuclear matter
density) or high temperature (beyond a few hundred MeV)
becomes {\it simple}: all known
hadrons are expected to dissolve into a plasma of their elementary 
constituents,
the quarks
and the gluons, forming  a new state of matter: the
{\it quark-gluon plasma} \cite{Shuryak,Hwa}. 

  The transition from the quark-gluon plasma to hadronic matter is 
one of several  transitions occurring in the early universe \cite{Linde}.
It is supposed to
 take  place during the first few microseconds after the big bang,
when the temperature is of the order of 200 MeV. At a higher
temperature, of the order of 250 GeV, another transition takes place, the
electroweak transition above which all  particles become massless and  form
another ultrarelativistic plasma.
The study of this phase transition and of the corresponding plasma is an
interesting and active field of research (see e.g. \cite{Shapo,Rumm98}). The
electroweak plasma has many features in common 
 with the quark-gluon plasma, and
we shall allude to some of them in the course of this paper. However  we shall
concentrate  here mainly  on the quark-gluon plasma.

Indeed, much of the present interest in
the quark-gluon plasma is coming from the  hope to observe it
  in laboratory experiments, by
colliding heavy nuclei  at high energies. An important experimental
program is underway, both in the USA (RHIC at Brookhaven), and in
Europe at CERN. (For general references on the field, see
\cite{Hwa,QM,Blaizot96}.)
It is therefore of the utmost importance to try and specify theoretically the
expected properties of such a plasma. Part of our motivations in writing
this report is to contribute to this effort. 
 
The existence of weakly interacting quark matter was
  anticipated on the
basis of asymptotic freedom of QCD \cite{CP75}. But the most
compelling theoretical evidences for the existence of  the
quark-gluon plasma are coming from lattice gauge calculations (for recent
reviews see e.g. 
\cite{DeTar,Karsch96,Karsch99}). These are at present the
unique tools allowing  a detailed study of the transition region where
various interesting phenomena are taking place, such as colour deconfinement or
chiral symmetry restoration. In this report, we shall not consider this
transition region, but focus rather on the high temperature phase, exploiting
the fact that at sufficiently high temperature the effective gauge coupling
constant becomes small enough to allow for weak coupling  calculations
\cite{Kapusta89,Blaizot92,MLB96,Smilga96}.

The picture of the quark gluon plasma which emerges from these
weak coupling calculations is a simple one, and in many respect the quark-gluon
plasma is very much like  an ordinary  electromagnetic
plasma in the ultrarelativistic regime \cite{Silin60,Fradkin65,qed}, 
with however specific 
effects related to the non Abelian gauge symmetry 
\cite{BP90,FT90,BIO95}. To zeroth order in an expansion
in powers of the coupling $g$, 
the quark gluon plasma is a gas of noninteracting
quarks and gluons. The interactions appear to alter only slightly this simple
picture: they  turn  those plasma particles which have  momenta of the order of
the temperature into massive quasiparticles, and generate collective modes at
small momenta which can be described accurately in terms
of classical fields. One thus see emerging a hierarchy of scales and degrees of
freedom which invites us to  construct effective
theories  for these various degrees of freedom. Weak
coupling techniques can be used to this aim 
\cite{BP90,FT90,TW90,qcd,Nair93,Bodeker,BE}; 
once the effective theories are known they can be used to also describe non
perturbative phenomena \cite{BMR99,Moore98}. 

It is indeed important to keep in mind that weak coupling 
approximations are not
 to be  identified with strictly perturbative calculations. A celebrated
counter example is that of the presently much discussed phenomenon of color
superconductivity
\cite{Wilczek}. Staying in the realm of high temperature QCD, we note that weak
coupling expansions generate terms which are odd in $g$, and these can only be
obtained through infinite resummations. Such resummations appear naturally in
the construction of effective theories alluded to earlier.  The possibility
to identify and perform such resummations 
 offers a chance to extrapolate weak coupling results down
to temperature where the coupling is not really small (recall that the
dependence of the coupling on the temperature is only logarithmic, and it is
only for $T\gg T_c$, where $T_c$ is the deconfinement temperature, that the
coupling is truly small). 
Recent works indicate that  this strategy may indeed be
successful \cite{ABS99,BIR99,BIR00}.

As well known, severe infrared divergences occur in high order
perturbative calculations. These divergences, usually
associated with those of an effective three dimensional theory,
are not easily overcome by analytic tools. Lattice calculations
indicate that the 
 strong longwavelength
fluctuations responsible for such divergences survive at high temperature and
 give significant contributions to the parameters characterizing the long
distance behaviour of the correlation functions (e.g. the so-called screening
masses \cite{Kajantie97,LP98}). While those results may suggest the existence
of new, nonperturbative, degrees of freedom, there is no evidence that these
degrees of freedom contribute significantly to thermodynamical quantities. On
the contrary, both recent lattice results \cite{Laine00}, and the
analytical resummations mentioned above, support the conclusion that this
contribution is small. 

A  final motivation for pushing these analytical techniques is the
possibility they offer to study dynamical quantities. These are difficult to
obtain on  the lattice, but are essential in any attempt to study real
phenomena. Indeed much of this report will be devoted to dynamical features of
the quark gluon plasma, emphasizing in particular its kinetic and transport
properties. In fact, as we shall discover, kinetic theory appears to be a
powerful tool for integrating out degrees of freedom when constructing effective
theories. Finally, 
it may be added that dynamical information, in particular that
on the plasma quasiparticles and its collective modes, can be
relevant also for the 
calculation of thermodynamical quantities \cite{ABS99,BIR99,BIR00}.

The  goal of this review is twofold. On the one hand, we wish to offer a
consistent description of the quark-gluon plasma in 
the weak coupling regime,
 summarizing recent progress and pointing out some open problems. On
the other hand, we shall give a pedagogical introduction to some of the
techniques that we have found useful in dealing with this problem. We
emphasize that most of the discussion will concern a plasma in
equilibrium or close to equilibrium, and the present work is but a
little step towards the ultimate goal of treating more realistic
situations such as met in nuclear collisions for instance. We hope
nevertheless that some of the techniques introduced here can be 
extended to treat these more complex situations,  and indeed some have
already been used to this aim.

A more precise view of the content of this paper is detailed in the 
rest of this section, where we shall introduce, in an elementary fashion,
most of the important concepts to be used.
An explicit outline is given in Sect. 1.7.

\subsection{Scales and degrees of freedom in ultrarelativistic plasmas}

In the absence of interactions, the plasma particles are distributed 
in momentum space
according to the Bose-Einstein  or  Fermi-Dirac distributions:
\beq\label{BFINTRO}
N_k\,=\,\frac{1}{{\rm e}^{\beta \varepsilon_k}\,-\,1},\qquad\qquad
n_k\,=\,\frac{1}{{\rm e}^{\beta \varepsilon_k}\,+\,1},\eeq
where $\varepsilon_k=k\equiv |{\bf k}|$ (massless particles),
$\beta\equiv 1/T$, and chemical potentials are assumed to vanish. In such an
ultrarelativistic system, the particle density
$n$ is not an independent parameter, but is
  determined by the temperature: $n\propto T^3$. Accordingly,
the mean interparticle distance $n^{-1/3}\sim 1/T$ is of the same
order as the thermal wavelength $\lambda_T=1/k$ of  a
typical particle in the thermal bath for which $k\sim T$. Thus the particles of
an ultrarelativistic plasma are  quantum degrees of freedom for which in
particular  the Pauli principle can never be ignored.

In the weak coupling regime ($g\ll 1$), the interactions do not alter
significantly the picture. The {\it hard} degrees of freedom, i.e. the plasma
particles with momenta $k\sim T$,
remain the dominant degrees of freedom and since the coupling to
gauge fields occurs typically through covariant derivatives,
$D_x=\del_x+igA(x)$,  the effect of interactions on particle motion is a small
perturbation unless the fields are very large, i.e., unless
$A\sim T/g$, where $g$ is the gauge coupling: only then do we have
$\del_x\sim T\sim gA$, where $\del_x\sim k $ is a hard space-time gradient. 
We should note here that  often in this report we
shall rely on considerations, such as the one just presented, which are  based
on the magnitude of the gauge fields. Obviously, such considerations 
depend on the
choice of a gauge. What we mean is  that  there exists  a large
class of gauge choices for which they are valid. And we
shall verify a posteriori that within such a class, our final results are
gauge invariant. Note however that  thermal
fluctuations could make it difficult to give a gauge independent
meaning to colour inhomogeneities on scales much larger than $1/g^2T$
\cite{Moore00}.

Considering now more generally the effects of the interactions, we note that
these depend both on the strength of the gauge fields and on the wavelength of
the modes under study. A measure of the strength of the gauge fields in
typical situations is obtained from the magnitude of their thermal
fluctuations, that is
$\bar A\equiv \sqrt{\langle A^2(t,{\bf x})\rangle}$. In equilibrium $\langle
A^2(t,{\bf
x})\rangle$ is independent of $t$ and ${\bf x}$ and given by
$\langle A^2\rangle= G(t=0,{\bf x=0})$
where $G(t,{\bf x})$ is the gauge field propagator.
In the non interacting case  we have (with $\varepsilon_k=k$):
\beq\label{fluctuationsA}
\langle A^2\rangle=
\int \frac{{\rm d}^3
k}{(2\pi)^3}\frac{1}{2\varepsilon_k}(1+2N_k).
\eeq
Here we shall use this formula also in the interacting case,
assuming that  the  effects of the interactions can be accounted for  simply
by a  change of $\varepsilon_k$ (a more complete calculation is presented in
Appendix B).
  We shall also ignore
  the (divergent) contribution of
  the vacuum fluctuations (the term independent of the temperature
  in eq.~(\ref{fluctuationsA})).

  For
the plasma particles   $\varepsilon_k=k\sim T$ and  $\langle A^2\rangle_T\sim T^2$.
The associated electric (or magnetic) field fluctuations are
$\langle E^2 \rangle_T\sim
\langle (\del A)^2\rangle_T \sim k^2 \langle A^2\rangle_T\sim
T^4$ and give a dominant contribution to the plasma energy density.
 As already mentioned, these short wavelength,
or {\it hard}, gauge field
fluctuations produce a small perturbation on the motion of a plasma particle.
However, this is not so for an excitation at the momentum scale  $k\sim gT$,
since then  the two terms in the covariant derivative
$\del_x$ and $g\bar A_T$ become comparable. That is, the properties of an
excitation with momentum $gT$ are expected to be nonperturbatively renormalized
by the hard thermal fluctuations. And indeed, the scale
$gT$ is that  at which collective phenomena develop, the study of which is one
of the main topic of this report. The emergence of the Debye screening mass
$m_D\sim gT$ is one of the simplest examples of such phenomena. 

Let us now consider the thermal fluctuations at this scale
$gT\ll T$, to be referred to as the {\it soft} scale.
 We shall see that these fluctuations can be accurately described by classical
fields. In fact, since $\varepsilon_k\sim gT \ll T$, one can replace
$N_k$ by $ T/\varepsilon_k$ in eq.~(\ref{fluctuationsA}); thus,
the associated occupation numbers are large, $N_k\gg 1$.
Introducing  an upper cut-off $gT$ in the momentum integral, one  then
gets:
\beq
\langle A^2\rangle_{gT} \sim \int^{gT}{\rm d}^3k \, \frac{T}{k^2}\sim
gT^2.
\eeq
Thus $\bar A_{gT}\sim \sqrt{g} T$ so that $g\bar A_{gT}
\sim g^{3/2}T$ is still of higher order than the kinetic term $\del_x\sim gT$.
In that sense the soft modes with $k\sim gT$ are still perturbative, i.e. their
self-interactions can be ignored in a first approximation. Note however that
they generate contributions to physical observables which are not analytic in
$g^2$, as shown by the example of the order
$g^3$ contribution to the
energy density of the plasma:
\beq
\epsilon^{(3)}\sim  \int_0^{\omega_{pl}}
{\rm d}^3 k \,\,\omega_{pl}\,\frac{1}{{\rm e}^{\omega_{pl}/T}-1}\sim
\omega_{pl}^3\,\omega_{pl}\,\frac{T}{\omega_{pl}}\,\sim\, g^3T^4,
\eeq
where $\omega_{pl}\sim gT$ is the typical frequency of a collective mode.

Moving down to lower momenta, one meets the contribution of the
unscreened magnetic fluctuations which play a dominant role for
$k\sim g^2T$.  At that scale, to be referred to as the {\it ultrasoft }
scale,  it becomes necessary to distinguish  the electric and the magnetic
sectors (which provide comparable contributions at the scale
$gT$). The  electric 
fluctuations are damped by the Debye screening mass
($N_k/\varepsilon_k\simeq T/(k^2+m_D^2)\approx T/m_D^2$ when $k\sim g^2T$)
and their contribution, of order  $ g^4 T^2$, is negligible in
comparison with that of the magnetic fluctuations. Indeed, because of
the absence of static screening in the magnetic sector, we have there
$\varepsilon_k\sim k$ and
\beq\label{fluctg2t}
\langle A^2\rangle_{g^2T}\sim T\int_0^{g^2T}{\rm d}^3k \frac{1}{k^2}\,
\sim\,g^2 T^2,\eeq
so that $g\bar A_{g^2T}\sim g^2T$ is now of the same order as the 
 ultrasoft 
derivative $\del_x\sim g^2T$: the fluctuations are no longer perturbative. This
is the origin of the breakdown of perturbation theory in high temperature QCD.

\begin{figure}
\protect \epsfxsize=11.cm{\centerline{\epsfbox{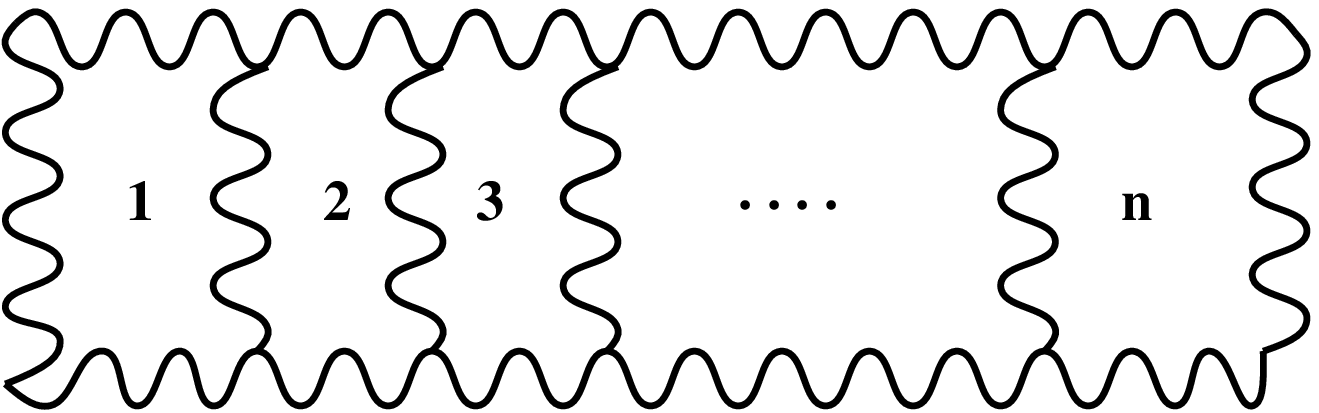}}}
          \caption{}
\label{ladder}
\end{figure}

To appreciate the difficulty from another perspective, 
let us first observe that the dominant contribution to
the fluctuations at scale $g^2T$ comes from the zero Matsubara frequency:
\beq
\langle A^2\rangle_{g^2T}= T\sum_n \int_0^{g^2T}{\rm d}^3k\,\,
\frac{1}{\omega_n^2+k^2}\sim T\int_0^{g^2T}{\rm d}^3k \,\,\frac{1}{k^2}.
\eeq
Thus the fluctuations that we are discussing are those of a three
dimensional theory of static fields. Following
Linde \cite{Linde79,Linde80}  consider then the higher order
corrections to the pressure in hot  Yang-Mills theory. Because of the strong
static fluctuations most of the diagrams of perturbation theory are infrared (IR)
divergent. By power counting, the strongest IR divergences arise from ladder
diagrams, like the one depicted in Fig.~\ref{ladder}, in which  all the
propagators are static, and the loop integrations are
three-dimensional. Such
$n$-loop diagrams can be estimated as ($\mu$ is an IR cutoff):
\beq\label{Linde}
g^{2(n-1)}\left(T\int {\rm d}^3k\right)^n\frac{k^{2(n-1)}}
{(k^2+\mu^2)^{3(n-1)}}\,,\eeq
which is of the order $g^6 T^4 \ln(T/\mu)$ if $n=4$ and of the order
$g^6T^4\left(g^2T/\mu\right)^{n-4}$  if $n>4$. (The various factors in
eq.~(\ref{Linde}) arise, respectively, from the $2(n-1)$ three-gluon vertices,
the $n$ loop integrations, and the $3(n-1)$ propagators.) According 
to this equation, if $\mu \sim g^2T$,
{\it all} the diagrams with
$n\ge 4$ loops contribute to the same order, namely to ${\cal O}(g^6)$.
In other words, the correction of ${\cal O}(g^6)$ to the pressure cannot be
computed in perturbation theory.

Having identified the main scales and degrees of freedom, our  task will
be to construct appropriate effective theories at the various scales, obtained
by eliminating the degrees of freedom at higher scales. This will be done in
steps. In fact the main part of this work will be devoted to the construction
of the effective theory at the scale
$g$T obtained by eliminating the hard degrees of freedom with momenta $k\sim
T$. We shall consider  some aspects of the effective theory at the scale
$g^2T$ only in section 7.

The soft
excitations at the scale $gT$ can be described in terms of
{\it average  fields}. Such average fields develop for example when the system
is exposed to an external perturbation, such as an external electromagnetic
current. Staying with QED, we can summarize the  effective theory for the
soft modes by the equations of motion:
\beq\label{maxwell00}
\del_\mu F^{\mu\nu}\,=\,j_{ind}^\nu+j_{ext}^\nu
\eeq
that is, Maxwell equations with a source term composed of the external
perturbation $j_{ext}^\nu$, and an extra contribution $j_{ind}^\nu$ which we
shall refer to as the {\it induced current}. The induced current is generated
by  the collective motion of the charged particles, i.e. the hard fermions. 
In the absence of the external current,
eq.~(\ref{maxwell00})  describes the longwavelength
collective modes which carry the quantum numbers of the photon, 
i.e., the soft plasma waves. Similarly, we shall see that the Dirac
equation with an appropriate induced source $\eta^{ind}(x)$ describes
collective   longwavelength excitations with fermionic quantum numbers
\cite{qed} :
\beq\label{avpsi00}
i\slashchar{D} \,\Psi(x)\,=\,\eta^{ind}(x).
\eeq
The induced sources $j_{ind}$ and $\eta_{ind}$ 
 may be regarded as a functionals of the average gauge  fields
$A_\mu(x)$ and fermion field $\Psi(x)$.  Once these functionals are
known,  the equations above constitute a closed  system of equations for
the soft fields.

The main problem is to calculate the induced sources $j_{ind}$
and $\eta^{ind}$. This is done by considering the dynamics of the hard 
particles in the
background of the soft fields $A^\mu$ and $\Psi$. Let us restrict
ourselves here to the induced current. This  
can be obtained using linear response theory.
  To be more specific, consider as an
example a system of charged particles on which is acting a perturbation
of the form  $\int
{\rm d}x \,j_\mu (x)
A^\mu (x)$, where $j_\mu(x)$ is the current tensor and $A^\mu (x)$ some
applied gauge  potential. Linear response theory leads to the 
following relation
for the induced current:
\beq\label{responsej}
j_\mu^{ind}=\int {\rm d}^4 y\, \Pi^R_{\mu\nu}(x-y) A^\nu (y), \qquad
\Pi^R_{\mu\nu}(x-y)=-i\theta(x_0-y_0)\langle
[j_\mu(x),j_\nu (y)]\rangle_{eq.},
\eeq
where the (retarded) response function $\Pi^R_{\mu\nu}(x-y)$ is also referred to
as the polarization operator. Note that in eq.~(\ref{responsej}), the
expectation value is  taken in the equilibrium state.  Thus, within
linear response, the task of calculating the basic  ingredients of the effective
theory for soft modes  reduces to  that of calculating appropriate
equilibrium correlation functions. This can be done by a variety of techniques
which will be reviewed in Section 2. In fact we shall need the 
response function only in the
weak coupling regime, and for particular kinematical conditions
which allow for
important simplifications. In leading order in weak coupling, the
polarization tensor is given by the one-loop approximation. In the kinematical
regime of interest, where the incoming momentum is soft while the loop momentum
is hard, we can write
$\Pi(\omega,p)=g^2T^2f(\omega/p,p/T)$ with $f$ a dimensionless function, and in
leading order in
$p/T\sim g$,
$\Pi$ is of the form $g^2T^2f(\omega/p)$. This particular contribution of the
one-loop polarization tensor is an example of what has been called a ``hard
thermal loop''
\cite{Klimov81,Weldon82a,Weldon82b,Pisarski89,BP90,FT90}; 
for photons in QED, 
this is the only one. It turns out that this  hard thermal loop can be
obtained from simple {\it kinetic theory}, and the corresponding calculation is
done in the next subsection.

In non Abelian theory,
linear response is not sufficient:
constraints due to gauge symmetry force us to take into account
specific non linear effects and a more complicated formalism needs to
be worked out. Still,  simple kinetic equations can  be
obtained in this case also, but in contrast to QED, the resulting induced
current is a non linear  functional of the gauge fields. As a result, it
generates  an infinite number of  ``hard thermal loops''. Actually, we
shall see that even in QED, gauge invariance forces the
fermionic induced source $\eta_{ind}$ to depend non linearly upon the
gauge fields, which entails the occurence of an infinite number of hard
thermal loops with two external fermion lines and an arbitrary number of
photon external lines.

\subsection{One-loop polarization tensor from kinetic theory}

As indicated in the previous subsection, in the kinematical regime
considered, the one loop polarization tensor can be obtained using elementary
kinetic theory. Since this approach will be at the heart of the forthcoming
developments in this paper, we present here this elementary
calculation.
  We consider an electromagnetic plasma and  momentarily assume that we
can describe its charged particles in terms of  classical distribution
functions
$f_q({\bf p},x)$ giving the density of particles of charge  $q$ ($q=\pm e$)
and
momentum ${\bf p}$ at the space-time point $x=(t,{\bf r})$
\cite{PhysKin}.
We consider then the case where collisions among the charged particles can be
neglected and where the only relevant interactions are those of particles with
average electric (${\bf E}$) and magnetic (${\bf B}$) fields. Then the
distribution functions obey the following simple kinetic equation, known as
the Vlasov equation \cite{Vlasov38,PhysKin} :
\beq\label{vlasov}
\frac{\del f_q}{\del t}+{\bf v}\cdot \frac{\del f_q}{\del {\bf r}}+
{\bf F}\cdot \frac{\del f_q}{\del {\bf p}}=0,
\eeq
where ${\bf v}={\rm d}\varepsilon_p/{\rm d}{\bf p}$ is the velocity of a
particle with momentum ${\bf p}$ and energy $\varepsilon_p$ (for massless
particles
${\bf v}=\hat{\bf p}$), and
${\bf F}=q({\bf E}+{\bf v}\wedge{\bf B})$ is the Lorentz force.
  The average fields ${\bf E}$ and ${\bf
B}$ depend themselves on the distribution functions $f_q$. Indeed, the induced
current
\beq\label{indcurrent1}
j^\mu_{ind}(x)=e\int \frac{{\rm d}^3p}{(2\pi)^3}\,v^\mu\,\left(f_+({\bf
p},x)-f_-({\bf p},x)\right),
\eeq
where  $v^\mu\equiv (1,{\bf v})$, is the source term in
the Maxwell equations (\ref{maxwell00}) for the mean fields.

  When the
plasma is in equilibrium, the distribution functions, denoted  as
$f_q^0(p)\equiv f^0(\varepsilon_p)$, are
isotropic in momentum space and independent of the
space-time coordinates; the
induced current vanishes, and so do the average fields
${\bf E}$ and
${\bf B}$. When the plasma is weakly perturbed, the distribution functions
deviate slightly from their equilibrium values, and we can write:
$f_q({\bf p},x)=f^0(\varepsilon_p)+\delta f_q({\bf p},x)$.  
In the linear approximation,  $\delta f$ obeys
\beq\label{vlasovlinear}
(v\cdot\del_x)\delta f_q({\bf p},x)=-
q{\bf v}\cdot {\bf E}\frac{{\rm d}f^0}{{\rm
d}\varepsilon_p},
\eeq
where $v\cdot\del_x\equiv \del_t+{\bf v}\cdot {\bf \grad}$. The magnetic field
does not contribute because of the isotropy of the
equilibrium distribution function. 

It is convenient here to set
\beq\label{DEFW}
\delta f_q({\bf p},x)\,\equiv\,-qW(x,{\bf v})\,\frac{{\rm d}f^0}{{\rm
d}\varepsilon_p},
\eeq
thereby introducing a notation which will be used in various forms
throughout this report. Since
\beq
f_q({\bf p},x)\,=\,f^0(\varepsilon_p)
-qW(x,{\bf v})\,\frac{{\rm d}f^0}{{\rm
d}\varepsilon_p}\,\simeq\, f^0(\varepsilon_p-qW(x,{\bf v})),\eeq
$W(x,{\bf v})$ may be 
viewed as a local distortion of the momentum distribution of the 
plasma particles. The equation for $W$ is simply:
\beq\label{EQW}
(v\cdot\del_x)W(x,{\bf v})\,=\,{\bf v}\cdot {\bf E}(x).\eeq

Contrary to eq.~(\ref{vlasov}), 
the linearized eqs.~(\ref{vlasovlinear}) or (\ref{EQW})
do not involve the derivative of $f$ with respect to
${\bf p}$, and can be solved by the method of characteristics: 
$v\cdot\del_x$
is the time derivative of
$\delta f({\bf p},x)$ along the characteristic defined by ${\rm d }{\bf x}/{\rm
d}t={\bf v}$. Assuming then that
the perturbation is introduced adiabatically so that the fields and the
fluctuations vanish as ${\rm e}^{\eta t_0}$ ($\eta\to 0^+$) when
$t_0\to -\infty$, we obtain the retarded solution:
\beq\label{fluctuation}
W(x,{\bf v})\,=\,
\int_{-\infty}^t{\rm d}t'\, {\rm
e}^{-\eta(t-t')}\,{\bf v}\cdot {\bf E}({\bf
x-v}(t-t'),t'),
\eeq
and the corresponding  induced current:
\beq\label{JMU0}
j^\mu_{ind}(x)= -2e^2\int \frac{{\rm d}^3p}{(2\pi)^3}\,v^\mu\frac{{\rm
d}f^0}{{\rm d}\varepsilon_p}\int_0^\infty{\rm d}\tau\, {\rm
e}^{-\eta\tau}\,{\bf v}\cdot{\bf E}(x-v\tau).
\eeq
Since ${\bf E}=-{\bfgrad}A^0-\del {\bf
A}/\del t$, the induced current is a linear functional of
$A^\mu$.

At this point we assume explicitly that the particles are
massless. In this case, ${\bf v}$ is a unit vector, and the angular
integral over the direction of ${\bf v}$ factorizes in eq.~(\ref{JMU0}).
Then, using  eq.~(\ref{responsej}) as definition for the polarization tensor
$\Pi^{\mu\nu}(x-y)$, and the
fact that the Fourier transform of $\int_0^\infty {\rm d}\tau\, {\rm
e}^{-\eta\tau}f(x-v\tau)$ is $i\,f(Q)/(v\cdot Q+i\eta)$, with
$Q^\mu=(\omega,{\bf q})$ and $f(Q)$ the Fourier transform of $f(x)$, one
gets, after a simple calculation \cite{Silin60} :
\beq\label{polarisation2}
\Pi_{\mu\nu}(\omega,{\bf q})=m_D^2\left\{
-\delta_{\mu 0}\delta_{\nu 0}+\omega\int\frac{{\rm d}\Omega}{4\pi} \frac{ v_\mu
v_\nu}{\omega-{\bf v}\cdot{\bf q}+i\eta}\right\},
\eeq
where the angular integral $\int {\rm d}\Omega$
  runs over all the orientations of ${\bf v}$, and 
$m_D$ is the Debye screening mass:
\beq\label{plasmafrequency}
m_D^2 = -\frac{2e^2}{\pi^2}\int_0^\infty{\rm d}p\, p^2
\frac{{\rm d}f^0}{{\rm d}\varepsilon_p}\,.
\eeq
As we
shall see, eq.~(\ref{polarisation2})  is  the dominant contribution at
high temperature to the one-loop polarization tensor in QED \cite{Fradkin65},
provided one substitutes
for $f^0$ the actual quantum equilibrium distribution function,
that is, $f^0 (\varepsilon_p)=n_p$, with $n_p$ given in eq.~(\ref{BFINTRO}).
After insertion in eq.~(\ref{plasmafrequency}), this yields $m_D^2=e^2T^2/3$. 

In
the next subsection, we shall address the question of how simple kinetic
equations emerge in the description of systems of quantum particles, and under
which conditions such systems can be described by seemingly classical
distribution functions where both positions and momenta are simultaneously
specified.

We shall later find that the expression obtained for the polarization tensor
using simple kinetic theory generalizes to the non Abelian case.
This is so in
particular because  the kinematical regime  remains that of the linear Vlasov
equation, with  straight line characteristics.

\subsection{Kinetic equations for quantum particles}

In order to discuss in a simple setting how  kinetic equations emerge in the
description of collective motions of quantum particles, we consider in this
subsection a system of non relativistic fermions coupled to classical gauge
fields. Since we are dealing with a system of independent particles in
an external field,  all the information on the quantum many-body state  is
encoded in the one-body density matrix \cite{FW71,BR86,NO} :
\beq\label{spdm0}
\rho({\bf r},{\bf r'},t)\equiv\langle \Psi^\dagger({\bf r'},t)\Psi({\bf
r},t)\rangle\,,
\eeq
where $\Psi$ and $\Psi^\dagger$ are the annihilation and creation 
operators, and
the average is over the initial equilibrium state. It is on this object that we
shall  later implement the relevant kinematical approximations.  To 
this aim, we introduce the {\it Wigner transform} of
$\rho({\bf r}, {\bf r}',t)$ \cite{Wigner1,Wigner2} :
\beq
f({\bf p}, {\bf R},t)=\int {\rm d}^3 s \, {\rm e}^{-i {\bf p}\cdot{\bf s}}\,
\rho\left({\bf R}+\frac{{\bf s}}{2}, {\bf R}-\frac{{\bf
s}}{2},t\right).
\eeq
The Wigner function has many properties that one expects of a classical
phase space distribution function as may be seen by calculating the
expectation values of simple one-body observables. For instance the
average density of particles $n({\bf R})$ is given by:
\beq
n({\bf R}, t)=\rho({\bf R}, {\bf R}, t)=\int\frac{{\rm d}^3 p}{(2\pi)^3}
\,f({\bf p}, {\bf R}, t).
\eeq
Similarly, the current operator:
$({1}/{2mi})\left( \psi^\dagger\bfnabla\psi- (\bfnabla
\psi^\dagger)\psi\right)$
has for expectation value:
\beq\label{jnonrel}
{\bf j}({\bf R}, t)=\frac{1}{2mi}\left(\bfnabla_y-\bfnabla_x\right)\rho({\bf
y},{\bf x}, t)|_{|{\bf y}-{\bf x}|\to 0}=
  \int\frac{{\rm d}^3 p}{(2\pi)^3}\,\frac{{\bf p}}{m}\,
f({\bf p}, {\bf R}, t).
\eeq
These results are indeed those one would obtain in a classical description
with $f({\bf p},{\bf R},t)$ the probability density to find a particle with
momentum ${\bf p}$ at point ${\bf R}$ and time $t$.  Note however that while
$f$ is real, due to the hermiticity of
$\rho$,  it is not always  positive as a truly classical distribution
function would be. Of course $f$ contains the same
quantum information as
$\rho$, and it does  not make sense to specify quantum
mechanically both the position and the momentum. However,  $f$ behaves as
  a classical distribution function in the  calculation of one-body
observables  for which the typical momenta $p$ that are involved in the
integration are large in comparison with the scale $1/\lambda$ characterizing
the range of spatial variations of $f$, i.e. $p\lambda\gg 1$.

    By using the equations of motion for the
field operators,
$i\dot\Psi({\bf r},t)=[H,\Psi]$, where $H$ is the single particle Hamiltonian,
one obtains easily the following
  equation of
motion for the density matrix
\beq \label{hrho}
i \del_t
\rho=[H,\rho].
\eeq
  In fact we shall need the Wigner transform of this equation  in
cases where  the gradients with respect to
$R$ are small compared to the typical values of $p$. Under such conditions, the
  equation of motion
  reduces to
\beq\label{kineteq0}
\frac{\del}{\del t}f+\bfnabla_p \,H\cdot \bfnabla_R \,f-\bfnabla_R \,H\cdot
\bfnabla_p \,f=0.
\eeq
where we have kept only the leading terms in an expansion in $\bfnabla_R $. For
particles interacting with gauge potentials
$A^\mu(X)$, the Wigner transform of the single particle Hamiltonian in
eq.~(\ref{kineteq0}) takes  the form:
\beq
H({\bf R},{\bf p},t)=\frac{{\bf p}^2}{2m}-\frac{e}{m}{\bf A}\cdot{\bf p}
+\frac{e^2}{m} {\bf A}^2({\bf R},t)+eA_0({\bf R},t).
\eeq
Assuming that the field is weak and neglecting the term in $A^2$,
one can write eq.(\ref{kineteq0}) in the form:
\beq\label{vlasovnoco}
\del_t f+{\bf v}\cdot\bfnabla_R f+e ({\bf E}+{\bf v}\wedge{\bf
B})\cdot\bfnabla_p f +\frac{e}{m}(p_j\del_j
A^i)\nabla^i_p f=0,
\eeq
where we have set ${\bf v}=({\bf p-eA})/{m}$. This equation is almost
  the Vlasov equation (\ref{vlasov}): it differs
from it by the last term which is not gauge
invariant. The presence of such a term, and the related gauge dependence of
the  Wigner function, obscure the physical interpretation. It is then convenient
to  define a gauge invariant density matrix:
\beq\label{rhocov}
\acute\rho({\bf r},{\bf r'},t)\equiv
\langle\psi^\dagger({\bf r'},t)\psi({\bf
r},t)\rangle U({\bf r},{\bf r'}, t),
\eeq
where (${\bf s}={\bf r}-{\bf r}'$)
\beq\label{approxU}
U({\bf r},{\bf r'})=\exp\left(-ie\int_{\bf r'}^{\bf r} {\rm d}{\bf z}
\cdot {\bf A}({\bf z}, t))\right)\approx
\exp\left(-ie{\bf s}\cdot {\bf A}({\bf R})\right)
\eeq
and the integral is along an arbitrary path
 going from ${\bf r}'$ to ${\bf r}$. Actually, in the
last step we have used an approximation which amounts to chose for this path
the straight line between ${\bf r}'$ to ${\bf r}$;
 furthermore, we have assumed
that the gauge potential does not vary 
significantly between ${\bf r}'$ to ${\bf
r}$.  (Typically, $\rho({\bf r},{\bf r}')$ is peaked at $s=0$ and drops to zero
when $s\simge \lambda_T$ where $\lambda_T$ is the thermal wavelength of the
particles. What we assume is that over a 
distance of order $\lambda_T$ the gauge
potential remains approximately constant.)  
Note that in the calculation of
the current (\ref{jnonrel}), only the limit 
$s\to 0$ is required, and that is given correctly by eq.~(\ref{approxU}) (see
also eq.~(\ref{jcov00}) below). With the approximate expression (\ref{approxU})
the Wigner transform of eq.~(\ref{rhocov}) is simply 
$\acute f({\bf R},{\bf k})=f({\bf R},{\bf k} +e{\bf A})$. By making the
substitution
$f({\bf R},{\bf p})=\acute f({\bf R},{\bf p}-e{\bf A})$ in
eq.~(\ref{vlasovnoco}), one verifies that the  non covariant
term cancels out and that the covariant Wigner function 
$\acute f$ obeys indeed Vlasov's equation.

In the presence of a gauge field, the previous definition (\ref{jnonrel})
of the current suffers
from  the lack of gauge covariance. It is however easy to construct  a gauge
invariant expression for the current operator, 
\beq \label{jcov}{\bf j}=\frac{1}{2m}\left(
\psi^\dagger(\frac{1}{i}\bfnabla-e{\bf A})
\psi-\left((\frac{1}{i}\bfnabla+e{\bf A})\psi^\dagger\right)\,\psi\right),\eeq
whose expectation value in terms of the Wigner transforms reads:
\beq\label{jcov00}
{\bf j}({\bf R}, t)=\int \frac{{\rm d}^3 p}{(2\pi)^3}\,
 \left(\frac{{\bf p}-e{\bf 
A}}{m}\right) f({\bf R},{\bf p}, t)=\int\frac{{\rm d}^3 k}{(2\pi)^3}\,
\left(\frac{{\bf k}}{m}\right) \acute f({\bf R},{\bf k}, t).
\eeq
The last expression involving the covariant Wigner function  makes it clear
that ${\bf j}({\bf R}, t)$ is gauge invariant, as it should. The momentum
variable of the gauge covariant Wigner transform is often referred to as the
{\it kinetic}
 momentum. It
is directly related to the velocity of the particles: ${\bf k}=m{\bf v}={\bf
p}-e{\bf A}$.  As for ${\bf p}$, the argument  of the non-covariant Wigner
function, it is related to the gradient operator  and is often referred to as the
{\it canonical} momentum.

In order to understand the structure of the equations that we shall obtain for
the QCD plasma, it is  finally instructive to consider the case where the
particles possess internal degrees of freedom (such spin, isospin, 
or colour).
The density matrix is then a matrix in internal space. As a specific example,
consider a system of spin
$1/2$ fermions. The Wigner distribution reads \cite{BP91}:
\beq
f({\bf p},{\bf R})=f_0({\bf p},{\bf R})+f_a({\bf p},{\bf R})\, \sigma_a,
\eeq
where the $\sigma_a$ are the Pauli matrices, 
and the  $f_a$ are three independent
distributions which  describe the  excitations of the system in the
various spin channels; together they form a vector that we can interpret as
a local spin density, ${\bf f}= (1/2) {\rm Tr} (f \bfsigma)$. When 
the system is
in a   magnetic field with Hamiltonian
$H=-\mu_0\,\bfsigma\cdot {\bf B}$ the equation of motion for ${\bf f}$ acquires
a new component:
\beq\label{spinprecession}
\frac{{\rm d}{\bf f}}{{\rm d}t}=- 2\mu_0 {\bf B}\wedge {\bf f},
\eeq
which accounts for the  spin precession in the magnetic field (in writing
eq.~(\ref{spinprecession}), we have ignored the gradients).  In
the linear approximation this precession may be viewed as an extra time
dependence of the distribution function along the characteristics:
\beq
\frac{{\rm d}}{{\rm d}t}=\frac{\del}{\del t}+{\bf v}\cdot\bfnabla-2\mu_0
{\bf B}\wedge\,.
\eeq

It is important to realize that all the differential operators above and in
the Vlasov equation apply to the arguments of the distribution functions, and not
to the coordinates of the actual particles. Note however that  equations similar
to the ones presented here  can be obtained for classical  spinning particles.
When the  angular momentum of such particles is large, it can indeed be treated
as a classical degree of freedom, and the corresponding  equations of motion have
been obtained by Wong
\cite{Wong}. After replacing spin by colour, these equations
have been used by Heinz
\cite{Heinz83,EHPRept} in order to write down transport equations for
classical coloured particles. By implementing the relevant kinematical
approximations  one then recovers \cite{Kelly94}
the non-Abelian Vlasov equations to be derived below,
i.e., eqs.~(\ref{n0INTRO}) and (\ref{N0INTRO}). 
(See also Refs.~\cite{Gyulassy93,Markov95,Hu,MHM98,Manuel99,Pisarski97} for
related work.)
We shall not pursue this line of reasoning however, since we
do not find it technically useful (it does not bring any simplifications)
and it is physically misleading. Besides, the kinetic equation
describing soft fermionic excitations (like eq.~(\ref{L0INTRO})
below) are not readily obtained in this way. 

\subsection{QCD Kinetic equations and hard thermal loops }

We are now ready to present the equations that we shall obtain for the QCD
plasma. These equations are for generalized one-body 
density matrices describing the long wavelength collective motions of 
the hard particles. They look formally as Vlasov equations, 
the main ones being  \cite{qcd,qed} :
\beq\label{n0INTRO}
\left[ v\cdot D_x,\,\delta n_\pm({{\bf k}},x)\right]=\mp\, g\,{\bf v}
\cdot{\bf E}(x)\,\frac{{\rm d}n_k}{{\rm d}k},\eeq
\beq\label{N0INTRO}
\left[ v\cdot D_x,\,\delta N({{\bf k}},x)\right]=-\, g\,
{\bf v}\cdot{\bf E}(x)\frac{{\rm d}N_k}{{\rm d}k},\eeq
\beq\label{L0INTRO}
(v\cdot D_x){\slashchar\Lambda}({\bf k}, x)=
-igC_f\,(N_k+n_k)\,{\slashchar v}\,\Psi(x),\eeq
In these equations, $v^\mu=(1, {\bf v})$, ${\bf v}={\bf k}/k$, $A^\mu_a(x)$
and $\Psi(x)$ are average gauge and fermionic fields, and
$\delta n_\pm$, $\delta N$ and ${\slashchar\Lambda}$ are
gauge-covariant Wigner functions for the hard particles. The first two Wigner
functions ar density matrices describing the colour oscillations of the
quarks and the gluons, respectively: $\delta n_\pm=\delta n_\pm^a t^a$
and $\delta N=\delta N_aT^a$. The last one (${\slashchar\Lambda}$)
is that of a more exotic density matrix which mixes
bosons and fermions degrees of freedom, ${\slashchar\Lambda}
\sim \langle \psi A\rangle$; it determines the induced fermionic source
$\eta^{ind}$ in eq.~(\ref{avpsi00}).
The right hand sides of the equations specify the quantum numbers of the
excitations that they are describing: 
soft gluon for the first two, and soft quark for the last one.  

One of the major difference between the QCD equations above and the
linear Vlasov equation for QED is the presence of covariant derivatives in the
left hand sides of eqs.~(\ref{n0INTRO})--(\ref{L0INTRO}). 
These play a role similar to that of the
magnetic field in eq.~(\ref{spinprecession}) 
for the distribution functions of particles with spin. 
(Note that the equation for ${\slashchar\Lambda}$ holds also in
QED, with a covariant derivative there as well.)

Eqs.~(\ref{n0INTRO})--(\ref{L0INTRO}) 
have a number of interesting properties which will be discussed
in section 3. In particular, they
  are covariant under local gauge transformations
of the classical fields, and independent of the gauge-fixing
in the underlying quantum theory.

By solving these equations, one can express the induced sources as
functionals of the background fields. To be specific,
  consider the {\it induced colour current}:
\beq\label{jb1INTRO}
j_a^\mu(x) \equiv 2g\int\frac{{\rm d}^3k}{(2\pi)^3}\,v^\mu
\,{\rm Tr}\,\Bigl(T^a\delta N({\bf k},x)\Bigr),\eeq
where $\delta N$ is the gluon density matrix (the quark contribution
reads similarly).
Quite generally, the induced colour current may be expanded in powers 
of $A_\mu$,
thus generating the one-particle irreducible amplitudes  
of the soft gauge fields \cite{qcd}:
\beq\label{exp0}
j^{a}_\mu \,=\,\Pi_{\mu\nu}^{ab}A_b^\nu
+\frac{1}{2}\, \Gamma_{\mu\nu\rho}^{abc} A_b^\nu A_c^\rho+\,...
\eeq
Here, $\Pi_{\mu\nu}^{ab}=\delta^{ab}\Pi_{\mu\nu}$
is the polarization tensor, and the other terms represent
vertex corrections. These  amplitudes are the ``hard thermal loops'' (HTL)
\cite{Pisarski89,BP90,FT90,TW90}  which define the
{\it effective theory} for the soft gauge fields 
at the  scale $gT$. Similar HTL's for the soft fermionic fields
are generated by expanding $\eta^{ind}$.
Diagrammatically, the HTL's are obtained by isolating the leading
order contributions to one-loop diagrams with soft external lines
(see Appendix B for some explicit such calculations).
It is worth noticing  that the kinetic equations
isolate  directly these hard thermal loops, in
a gauge invariant manner, without further approximations.

The  gluon density matrix
can be parameterized as in eq.~(\ref{DEFW}) :
\beq\label{dn0INTRO}
\delta N({\bf k}, x)\,=\,
  - gW(x,{\bf v})\,({\rm d}N_k/{\rm d}k),\eeq
where $N_k\equiv 1/({\rm e}^{\beta k}-1)$
is the Bose-Einstein thermal distribution,
and $W(x,{\bf v})\equiv W_a(x,{\bf v}) T^a$
is a colour matrix in the adjoint representation
which depends upon the velocity ${\bf v}={\bf k}/k$
(a unit vector), but not upon the magnitude $k
=|{\bf k}|$ of the  momentum. A similar representation holds for the
quark density matrices $\delta n_\pm({{\bf k}},x)$. 
Then the colour current takes the form:
\beq\label{jindINTRO}
j_{ind}^{\mu\,a}(x)=m_D^2\int \frac{{\rm d}\Omega}{4\pi}v^\mu W^a(x,{\bf v})
\eeq
with $m_D^2\sim g^2 T^2$. 
The kinetic equations for 
$\delta N$ and $\delta n_\pm$ can then be  written
as an equation  for
$W_a(x,{\bf v})$:
\beq\label{VLASINTRO}
(v\cdot D_x)^{ab}W_b(x,{\bf v})&=&{\bf v}\cdot{\bf E}^a(x).\eeq
This differs from the corresponding Abelian equation (\ref{EQW})
merely by the replacement of the ordinary   derivative
$\del_x \sim gT$ by the covariant one $D_x=\del_x+igA$.
Accordingly, the soft gluon polarization tensor derived from
eqs.~(\ref{jindINTRO})--(\ref{VLASINTRO}),
i.e., the ``hard thermal loop'' $\Pi_{\mu\nu}$,
is formally identical to the photon
polarization tensor obtained from eq.~(\ref{EQW}) and 
given by eq.~(\ref{polarisation2}) \cite{Klimov81,Weldon82a}.
The reason for the existence of an infinite number of hard thermal loops in QCD
is the presence of the covariant derivative in the left hand side of
eq.~(\ref{VLASINTRO}). A similar observation can be made by  writing the
induced electromagnetic current in  the form:
\beq
j^\mu_{ind}(x)=m_D^2\int \frac{{\rm d}\Omega}{4\pi} v^\mu\int {\rm d}^4y
\,\bra{x}\frac{1}{v\cdot\del}\ket{y}\,{\bf v}\cdot{\bf E}(y)=\int{\rm d}^4y
\,\sigma^{\mu j}(x,y) E^j(y).
\eeq
This expression, which is easily obtained from the expression
(\ref{JMU0}) of $\delta f$, defines the conductivity tensor
$\sigma^{\mu\nu}$. As we shall see, the generalization of this expression to
QCD amounts essentially to
  replacing the ordinary derivative by a covariant one.

\subsection{Effect of collisions}

Until now, we have been discussing independent particles moving in average
self-consistent fields. It can be argued that in weak coupling and for long
wavelength excitations, this is the dominant picture. There are situations
however where collisions among the plasma particles cannot be ignored. We shall
consider in this report two such cases. One concerns the lifetime of the single
particle excitations to be discussed in section 6. The other refers 
to the study
of ultrasoft excitations at the scale $g^2T$ which will be presented in
section 7.

The determination of the lifetimes of single particle excitations played
an essential role in the development of the subject and led in particular
to the identification of the hard thermal loops  
\cite{Pisarski89,BP90a,Pisarski91}. Physically,
the lifetime of a quasiparticle excitation is 
limited by its collisions with the
other particles in the plasma. The collision rate can be  estimated directly in
the form $\gamma=n\sigma v$, where
$n\sim T^3$ is the density of plasma particles,  $\sigma$ the collision
cross section, and $v$ the velocity equal to the speed of light. Restricting
ourselves first to the Coulomb interaction, we can write
$\sigma=\int{\rm d}q^2({\rm d}\sigma/{\rm d}q^2)$, with $ {\rm d}\sigma/{\rm
d}q^2\sim g^4/q^4$. Thus,
\beq
\gamma\sim g^4\,T^3\,\int {\rm d}q^2\frac{1}{q^4},
\eeq
which is badly infrared divergent. One knows, however, that in  the plasma the
Coulomb interaction is screened, so that the effective electric
photon propagator is not $1/q^2$ but $1/(q^2 +m_D^2)$, where $m_D\sim 
gT$ is the Debye
screening mass. With this correction taken into account, the collision rate
becomes
\beq\label{electric}
\gamma\sim g^4T^3\frac{1}{m_D^2}\sim g^2T,
\eeq
which is now finite, and of order $g^2T$.

However, screening corrections at the scale $gT$ \cite{Baym90}, 
as encoded in the
hard thermal loops,  are not sufficient to eliminate all the
divergences due to the magnetic interactions
\cite{Pisarski89,Rebhan93,Gyulassy93,Heisel94,debye};
 they leave an estimate for the lifetime
\beq\label{gammadiv}
\gamma\sim g^2T\int^{m_D}\frac{{\rm d}q}{q}
\eeq
which is logarithmically divergent \cite{Pisarski89}.
This infrared problem occurs  both in  Abelian and non-Abelian
gauge theories. In QCD, it is commonly bypassed
  by advocating the  infrared cut-off provided by a 
``magnetic  mass'' $\sim g^2T$, so that $\gamma\sim g^2T\ln{(1/g)}$.
 But such a solution cannot apply for
QED where one does not expect  any magnetic screening \cite{Fradkin65,sqed}.

In section 6,
we shall analyze the origin of these infrared divergences and show that the
dominant ones can be resummed in closed form for the retarded propagator of the
quasiparticle excitation. This will be achieved by considering as an
intermediate step the  propagation of a test particle in a background of
random ultrasoft (and mostly static) thermal fluctuations.  The retarded
propagator is obtained by averaging over these
  fluctuations. Remarkably, the
resulting damping is non exponential, the retarded propagator being of the
form
$S_R(t)\sim
\exp\left(-g^2 T \, t\ln (t\, m_D)\right)$ \cite{lifetime}. 
We shall see that such a particular
behaviour also emerges in a treatment of the collisions using a
generalized Boltzmann equation in a regime where the  mean free path
is comparable with the  range of the relevant interactions \cite{BDV98}.

The second case where the collisions become
important is in the study of ultrasoft  perturbations at the scale
$g^2T$ or smaller.  To give a crude estimate of these collisional effects,
one may use the relaxation time approximation, and write the kinetic equation
as
\beq\label{RTA}
(v\cdot D_x)^{ab}W_b(x,{\bf v})&=&{\bf v}\cdot{\bf E}^a(x)
\,-\,\frac{W^a(x,{\bf v})}{\tau_{col}},\eeq
where $\tau_{col}$ is a typical relaxation time. It is important here to
distinguish between colour and colourless excitations. The relaxation of colour
is dominated by the singular forward scattering  which yields
$\tau_{col}  \sim 1/(g^2T\ln(1/g))$
\cite{Gyulassy93,Heisel94,Bodeker}. Then, eq.~(\ref{RTA}) shows that
the effect of the collisions become a leading order effect
for inhomogeneities at the scale $\del_x \sim g^2 T$, or less.
Colourless fluctuations, such as fluctuations
in the momentum or the electric charge distributions, involve a colour
independent fluctuation $W$. The corresponding kinetic equation reduces to a
simple drift term $v\cdot\del_x$ in the left hand side (no colour mean
field) and a collision term in
the right hand side. This collision term involves now large angle scatterings,
and the resulting  relaxation time is much larger,
$\tau_{el} \sim 1/(g^4T\ln(1/g))$ \cite{Baym90,Heisel94a,Baym97}. In that
case, collisions become important only for space-time inhomogeneities at scale
$\sim 1/g^4T$.

Of course, the relaxation time approximation is only a crude approximation.
(For coloured excitations, this is not even a gauge-invariant
approximation \cite{USA}.) A
complete Boltzmann equation \cite{BE}
 will be derived in the last section of this
report, by extending the techniques used to derive the collisionless kinetic
equations in section 3. 
In the same way as the induced current calculated
from the solution of the Vlasov equation (\ref{VLASINTRO})
generates directly the hard thermal loops, we
shall see that the induced current calculated with the solution of the
Boltzmann equation isolates the leading-order contributions
to an infinite set of multi-loop diagrams where 
the external momenta are ultrasoft \cite{USA}. These
amplitudes share many properties with the hard thermal loops, although they
correspond typically to multiloop diagrams. These amplitudes are 
logarithmically
infrared divergent, so are best understood as ingredients of the effective
theory  for ultrasoft excitations at a scale $\Lambda\ll gT$,
with $\Lambda$ playing in their calculation the role of an IR cutoff
\cite{Bodeker}.

\subsection{Effective theory for soft and ultrasoft excitations}

We have concentrated so far on the dynamics of hard degrees of freedom in
external background fields, possibly taking into account the effect of
collisions when considering very long wavelength excitations. But it is also of
interest to consider the effective theory for the soft degrees of freedom
obtained by ``eliminating'' the hard ones. As mentioned earlier, for 
soft photons in QED this
effective theory reduces to the Maxwell equations with an induced current, and
the same holds for gluons in QCD, 
with the Maxwell equations replaced by the Yang
Mills equations and with the colour current (\ref{jb1INTRO}). 
Similarly, the soft fermionic excitations are described, in both
QED and QCD, by the Dirac equation (\ref{avpsi00}) with the
induced source $\eta^{ind}$ built out of ${\slashchar\Lambda}({\bf k}, x)$,
cf. eq.~(\ref{L0INTRO}).
If we want to study for instance the collective excitations of
the plasma these equations of motion are all what is needed.

There are cases however where one needs to take into account the effect of such
collective modes on correlation functions (an example is actually provided by
the calculation of the damping rate of quasiparticle excitations). To 
do so, one
needs to go one step further and determine the Boltzmann weight associated with
such modes. The problem is made easier by the fact that soft 
bosonic excitations
can be described by classical fields \cite{GR,McLerran}
which may be identified with the average
fields introduced before. For excitations at the scale
$gT$, one can construct a Hamiltonian description of the dynamics of
these classical fields. In terms of the fields $W^a$ introduced earlier, the
Hamiltonian is remarkably simple \cite{Nair93,emt,gauge} :
\beq\label{Hintro}
H\,=\,\frac{1}{2}\int {\rm d}^3 x\left\{{\bf E}^a\cdot{\bf E}^a\,+\,
{\bf B}^a\cdot{\bf B}^a\,+\,m_D^2
\int\frac{{\rm d}\Omega}{4\pi}\,W^a(x, {\bf v})\,W^a(x, {\bf v})\right\}.\eeq
As we shall see in Sect. 4, when appropriate Poisson brakets are introduced,
the Hamiltonian (\ref{Hintro}) generates indeed the correct dynamics
\cite{Nair93,baryo}.
It will also be shown in Sect. 4 that this Hamiltonian provides the correct
Boltzmann weight to integrate over soft fluctuations \cite{baryo}. 
The calculation of real
time correlation functions reduces then to the calculation of a functional
integral where the integration variables are the gauge fields and the auxiliary
fields $W$, and the functional integration amounts to an average over the
initial conditions for the classical field equations of motion. 
This allows in particular for numerical
calculations of the real time correlation functions on a three-dimensional
lattice. An important application, which has received much attention in
recent years \cite{Shapo}, \cite{ASY97}--\cite{Moore98}, \cite{BMR99}
is the evaluation of the anomalous baryon number violation
rate at high temperature. This is defined as \cite{Shapo}
\beq\label{Baryorate}
\Gamma\,\equiv\,\int_{-\infty}^\infty {\rm d}t\int {\rm d}^3 x
\left(\frac{g^2}{8\pi^2}\right)^2\left\langle\Bigl[
E^i_aB^i_a(t,{\bf x})\Bigr]\,\Bigl[
E^i_aB^i_a(0,{\bf 0})\Bigr]\,\right\rangle,\eeq
and receives contributions typically from the non-perturbative 
magnetic modes with momenta $k \sim g^2T$ and energies
$\omega \simle g^4T$ \cite{ASY97}. Recently, this has been computed
via lattice simulations of the classical effective theory with
Hamiltonian (\ref{Hintro}) \cite{BMR99}.
(See also Refs. \cite{Hu,MHM98} for a different
lattice implementation of the HTL effects, and Refs. 
\cite{AmbK,TS96,MT97,MR99} for numerical calculations within 
purely Yang-Mills classical theory, without HTL's.)

The effective theory that we have just outlined leads to
ultraviolet divergences. However, it is defined with an ultraviolet cutoff
$gT\ll\Lambda\ll T$. The  coefficients of the
effective theory, which are the hard thermal loops, must also be calculated
with an infrared cutoff $\Lambda$, so that the cutoff dependence of the
parameters in the effective theory (here the Debye mass) cancels against the
cutoff dependence of the classical thermal loops. Without such a matching,
which turns out to be difficult to implement in QCD, the calculation of
correlation functions within the classical effective theory remains linearly
sensitive to the ultraviolet cutoff
\cite{McLerran,ASY97,baryo,Arnold97,Nauta99,ANW99}. 
This is clearly exhibited by the numerical results for $\Gamma$,
eq.~(\ref{Baryorate}), obtained in \cite{MR99}.

In order to reduce the sensitivity to the scale $\Lambda$ it has been
suggested to go one step further and eliminate also the soft degrees of
freedom down to a scale $g^2T\ll
\Lambda\ll gT$. This can be done starting from the classical effective
theory for soft field and  integrating explicitly over the soft degrees of
freedom. This is the approach followed by B\"odeker, who showed that the
resulting theory at the scale $g^2T$ takes the form of a Boltzmann-Langevin
equation \cite{Bodeker}. 
Results of numerical simulations based on (a simplified
version of) this equation have been given in \cite{Moore98} (cf. Sect.
7 below). The collision term obtained by B\"odeker
is identical to that appearing in the Boltzmann equation that we have obtained
following a completely different route \cite{BE}.
The reason for this will be detailed in
section 7, where we also show that the noise term in the Langevin equation is
simply related to the collision integral through the fluctuation-dissipation
theorem.  The building blocks of the new effective theory are the
ultrasoft amplitudes mentioned above. As already mentioned, these amplitudes
depend logarithmically on the separation scale $\Lambda$,
but this dependence will eventually cancel against the cutoff 
dependence of the loop corrections in the effective theory.

\subsection{Outline of the paper}

We now  present the outline of this paper.

Section 2 is a
pedagogical introduction to most of the techniques that we shall be using. This
includes a short review of equilibrium thermal field theory in the imaginary
time formalism, a description of near equilibrium longwavelength excitations,
the use of Wigner transform to obtain kinetic equations. To keep things as
simple as possible, the formalism is developed for the case of a real scalar
field.

In section 3, we begin to implement these techniques in the case of QCD.
In particular, we present the main steps in the derivation of the kinetic
equations for the hard particles.

These kinetic equations are solved explicitly in
section 4. This leads to effective equations of motion  for the soft modes of
the plasma. These soft modes could be excitations of the plasma driven by
external disturbances. They also appear as long-wavelength fluctuations in the
plasma in equilibrium. The issue of calculating the effect of such fluctuations
on real time correlation functions is addressed. We show that this can be
formulated conveniently in terms of an effective theory for classical fields.
The construction of this effective theory is explicitly given.

The induced
current which  provides the source for the soft mode propagation is a non
linear functional of the gauge fields. It may be viewed as a generating
functional for an infinite set of one loop amplitudes, the so-called hard
thermal loops. Some of these hard thermal loops are explicitly constructed in
section 5 and their properties analyzed. A few applications are mentioned.

Section 6 addresses the issue of the damping of the plasma 
excitations. This is
a problem which has triggered much of the work on the hard thermal loops, but
whose general
solution requires going beyond the hard thermal loop approximation. It
provides an interesting illustration of the effects of collisions in a regime
where the range of the relevant interactions is comparable with the mean free
path of the particles.

In section 7 we consider some of the physics taking place at the scale $g^2
T$. For modes with such momenta, collision  terms in the Boltzmann equation
become relevant. We show that there exists  an infinite set of
amplitudes, which we called ultrasoft amplitudes, which become of the same
order of magnitude as the hard thermal loops, and which are generated by the
Boltzmann equation. This equation is an essential ingredient in the effective
theory for ultrasoft excitations which is briefly presented.

Finally section 8 summarizes the conclusions.

Appendix A contains a summary of the notation used throughout. Appendix B
presents detailed calculations, in the hard thermal loop
approximation, of one loop diagrams that are referred to in the main text.

\setcounter{equation}{0}
\setcounter{equation}{0}
\section{Quantum fields near thermal equilibrium}

In most of this paper, we shall
study generically how a system initially
in thermal  equilibrium responds to a weak and slowly
varying disturbance. This section summarizes the main tools that will 
be needed in such a study. 
It starts with a short review of equilibrium thermal field theory
using the imaginary time formalism. Then we turn to off-equilibrium situations
and derive the equations of motion for the appropriate Green's 
functions. The last
subsection is devoted to 
the implementation of the longwavelength approximation
using gradient expansions. 
This allows us to transform the general equations of
motion into simpler kinetic equations. Much of the material of this section is
fairly standard, and many results will be mentioned without proof. 
More complete
presentations can be found for instance in Refs.
\cite{Bernard74,GPY81,Landsman87,Kapusta89,Blaizot92,MLB96}
for equilibrium situations, and in Refs. 
\cite{KB62,PhysKin,Serene83,chine,Rammer86,Daniel83,Botermans90}
for the non-equilibrium ones.

In order to bring out the essential aspects of the formalism
while avoiding the complications specific to gauge theories,
we shall consider in this section only a scalar field theory,
   with Lagrangian
\beq\label{Lagran}
{\cal L}&=& {1\over2}\partial_\mu\phi\partial^\mu\phi-
{m^2\over2}\phi^2-V(\phi)\nonumber\\
&=& {1\over2}(\partial_0\phi)^2-{1\over2}(\nbfgrad\phi)^2
-{m^2\over2}\phi^2-V(\phi),
\eeq
where $V(\phi)$ is a local potential.

The initial equilibrium state is described by the canonical density operator:
\beq\label{canonicalD}
{\cal D}=\frac{{\rm e}^{-\beta H}}{Z},
\eeq
where $H$ is the hamiltonian of the system and $Z$ the partition
function. For the scalar field,
\beq
H= \int{\rm d}^3x\,\left(\frac{1}{2}\pi^2+
\frac{1}{2}(\nbfgrad\phi)^2
+\frac{m^2}{2}\phi^2+V(\phi)\right),\eeq
where $\pi(x)$ is the field canonically conjugate to $\phi(x)$.
We may express ${\cal D}$ in terms of the
eigenstates $\ket{n}$ of $H$ ($H\ket{n}=E_n\ket{n}$) and probabilities
$p_n$ ($p_n={\rm e}^{-\beta E_n}/Z$):
\beq
{\cal D}=\sum_n \ket{n}\, p_n \,\bra{n}.
\eeq

We consider a time-dependent perturbation of the form:
\beq\label{Hj1}
H_j(t)-H\,=\,\int {\rm d}^3{x}\,j(t,{\bf x}) \phi({\bf x}).
\eeq
Under the action of such a perturbation, the system evolves away
from the equilibrium  state. The density operator at time $t$ is given
by  the  equation of motion:
\beq\label{evolutionD}
i\dot {\cal D}_j=[H_j(t),{\cal D}(t)],
\eeq
where $\dot{\cal D}_j\equiv \del_t{\cal D}_j$. It can be written as:
\beq
{\cal D}_j(t)=\sum_n \ket{n;t}\, p_n \,\bra{n;t},
\eeq
with time-independent $p_n$'s (the same as in equilibrium); 
the state $\ket{n;t}$ is  the
solution of the  Schr\"odinger equation which coincides initially
with the eigenstate
$\ket{n}$.
Note that the evolution described by eq.~(\ref{evolutionD})
conserves the entropy $S=-k_B {\rm Tr}{\cal D}\ln {\cal D}=-
k_B\sum_n\,p_n\,\ln p_n$.  All the approximations  that we shall
consider here fulfill this property.

\subsection{Equilibrium thermal field theory}

Before embarking into the discussion of the non equilibrium dynamics, it is
useful to review briefly  the formalism of thermal field
theory in equilibrium. We shall in particular recall how  perturbation theory
can be used  to calculate the partition function:
\beq\label{Z1}
Z\equiv {\rm Tr} \exp\left\{ -\beta H\right\}
=\sum_n \exp\left\{-\beta E_n
\right\},
\eeq
from which all the thermodynamical functions can be obtained.

The simplest formulation of the perturbation theory
   for thermodynamical quantities
is based on the formal analogy between the partition function
(\ref{Z1}) and the evolution operator
   $U(t,t_0)=\exp\{-i(t-t_0)H\}$, where the time variable $t$
is allowed to take complex values. Specifically, we can write
$Z={\rm Tr}\,U(t_0-i\beta,t_0)$, with arbitrary (real) $t_0$.
   More generally, we
shall define an operator $U(\tau)\equiv \exp (-\tau H)$,
where $\tau$  is real,
but often referred to as the {\it imaginary time} ($\tau=i(t-t_0)$ with $t-t_0$
purely imaginary). The evaluation of the partition function
(\ref{Z1}) by a  perturbative expansion  involves the splitting of
the hamiltonian into
$ H=H_0+H_1 $,
with $H_1\ll H_0$. For instance, for the scalar field
theory in eq.~(\ref{Lagran}), it is convenient to take:
\beq
H_0= \int{\rm d}^3x\,\frac{1}{2}\left( \pi^2+
\frac{1}{2}(\nbfgrad\phi)^2
+\frac{m^2}{2}\phi^2\right) ,\qquad \,\,\,\,H_1= \int{\rm d}^3x\,
V(\phi).\eeq
We then set:
\beq\label{IMTU}
U(\tau)&=&\exp\left( -\tau H\right)  \nonumber \\
   &=&\exp\left( -\tau H_0\right) \exp\left(
\tau H_0 \right)\exp\left( -\tau H \right)  \nonumber \\
   &=& U_0(\tau) \,U_I(\tau) ,
\eeq
where $U_0(\tau)\equiv\exp(-H_0\tau)$. The operator $U_I(\tau)$ is called
the {\it interaction representation} of $U$. We also define the interaction
representation of the perturbation $H_1$:
\beq\label{interactionH}
   H_1(\tau)=e^{\tau H_0}H_1 e^{-\tau
H_0},
\eeq
   and similarly for other operators.
   It is easy to
verify that $U_I(\tau)$ satisfies the following differential
equation:
\beq
   {d\over d\tau}U_I(\tau)+H_1(\tau)U_I(\tau)=0,
\eeq
   with the boundary condition
\beq\label{bdcond}
   U_I(0)=1. \eeq
The solution of the above differential equation, with the
   boundary condition (\ref{bdcond}), can be written formally in
terms of the time ordered exponential:
\beq\label{e_betaH}
e^{-\beta H} = e^{-\beta H_0} \,{\rm
T}_\tau \exp\left\{-\int_0^\beta d\tau H_1(\tau)\right\}.
\eeq
The symbol ${\rm T}_\tau$ implies an ordering of the operators on its
right,  from
left to right in decreasing order of their imaginary time 
arguments:
\beq
\lefteqn{
{\rm
T}_\tau\exp\left\{-\int_0^\beta d\tau H_1(\tau)\right\} }\nonumber\\
& & =1-\int_0^\beta d\tau H_1(\tau)+\frac{1}{2}\int_0^\beta d\tau_1d\tau_2\,
{\rm T}[\,H_1(\tau_1)H_1(\tau_2)]+\cdots \nonumber\\
& & =1-\int_0^\beta d\tau H_1(\tau)+\int_0^\beta d\tau_1\int_0^{\tau_1}
d\tau_2\,H_1(\tau_1)H_1(\tau_2)+\cdots\nonumber\\
   & &
\eeq
Using eq.~(\ref{e_betaH}), one can
rewrite $Z$ in the form:
\beq\label{Z_pert}
Z=Z_0\,\langle {\rm T}\exp\left\{-\int_0^\beta d\tau
H_1(\tau)\right\}\rangle_0,
\eeq
where, for any operator $\cal O$,
\beq\label{operator_ev}
\langle {\cal O} \rangle_0\equiv{\rm Tr}\left({e^{-\beta
H_0}\over {Z}_0}{\cal O}\right)
.\eeq

Alternatively, one may write the partition function
as the following path integral:
\beq\label{ZCE}
Z\,=\,{\cal N}
\int_{\phi(0)=\phi(\beta)}{\cal D}(\phi)\exp\left\{
- \int_0^\beta {\rm d}\tau \int{\rm d}^3x \,{\cal L}_E(x)\right\}\,,\eeq
where ${\cal N}$ is a normalization constant which cancels out in the 
calculation
of expectation values, but which needs to be treated with care in the 
evaluation of
thermodynamical functions.  In the equation (\ref{ZCE}),
$\phi({\tau}, {\bf x})\equiv
\phi(t=t_0-i\tau, {\bf x})$ and:
\beq\label{LE}
{\cal L}_E\,=\,{1\over2}(\partial_\tau\phi)^2+{1\over2}(\nbfgrad\phi)^2+
{m^2\over2}\phi^2 + V(\phi)\,.\eeq
The functional integration
runs over field configurations which are periodic in imaginary
time, ${\phi(\tau=0)=\phi(\tau=\beta)}$.

In the rest of this report, we shall refer to both the path integral and the
operator formalisms, the choice of either one depending on which is
the most convenient for the question under study. In both formalisms,
imaginary time Green's functions or
propagators appear. These have special properties which are recalled
in the next subsections.

\subsubsection{The imaginary time Green's functions}

By adding to ${\cal L}_E$ in eq.~(\ref{ZCE}) a source term $-j(x)\phi(x)$,
where $j(x)$ is an arbitrary external current, one transforms $Z$ into
the generating functional $Z[j]$ of the imaginary-time
   Green's functions:
\beq\label{THG}
\langle {\rm T}_\tau
\phi(x_1) \phi(x_2)...\phi(x_n)\rangle=
{\rm Tr}\left\{{\cal D} \,{\rm T}_\tau
\phi(x_1)  \phi(x_2) ...\phi(x_n)\right\} =\,\frac{1}{Z}\,
\frac{\delta^n
Z[j]}{\delta j(x_1)\delta j(x_2)...\delta j(x_n)}\biggl |_{j=0}.\nonumber\\
\eeq
In this formula, $\phi(x)$ (with $x^\mu=(t,{\bf x})$, 
$t=t_0-i\tau$, $0\le\tau\le\beta$) is a field operator
in the Heisenberg representation,
\beq\label{phiH}
\phi(t,{\bf x})={\rm e}^{iH(t-t_0)}\phi({\bf x})\, {\rm e}^{-iH(t-t_0)}
={\rm e}^{H\tau}\phi({\bf x})\, {\rm e}^{-H\tau}.\eeq

The {\it connected}  Green's functions, for which we reserve
throughout the notation $G^{(n)}(x_1,x_2, ... ,x_n)$, are given by:
\beq\label{GEdef}
G^{(n)}(x_1,x_2, ... ,x_n)\,=\, \frac{\,\delta^n \ln
Z[j]}{\delta j(x_1)\delta j(x_2)...\delta j(x_n)}\biggl|_{j=0}.\eeq
For space-time translational invariant systems, they
depend effectively only on $n-1$ relative coordinates.
In particular, for the 2-point function,
we shall write $G^{(2)}(x_1,x_2)=G^{(2)}(x_1-x_2)\equiv G(x)$.

The imaginary time Green's functions obey periodicity conditions.
For instance, for the 2-point function, we have:
   \beq\label{KMSE}
G(\tau-\beta)&=&G(\tau)\qquad {\rm for}\,\,\,
0\le \tau \le \beta,\nonumber\\
G(\tau+\beta)&=&G(\tau)\qquad {\rm for}\,\,\,
-\beta \le \tau \le 0,
\eeq
where $\tau\equiv \tau_1-\tau_2$.
(Here, and often below, when focusing on temporal properties  we do 
not mention
the   spatial coordinates for simplicity.) To prove these relations, 
note that, for
$0\le \tau
\le
\beta$,
\beq\label{G>tau}
   G(\tau)={\rm Tr}\left\{{\cal D}\,\phi(\tau)\phi(0)\right\}
=\frac{1}{Z}\,{\rm Tr}\left\{{\rm e}^{-H(\beta-\tau)}\,
\phi\,{\rm e}^{-\tau H}\phi\right\},\eeq
where eq.~(\ref{phiH}) has been used.
On the other hand, $-\beta\le \tau-\beta \le 0$, so that:
\beq\label{G<tau}
   G(\tau-\beta)&=&{\rm Tr}\left\{{\cal D}\,
\phi(0)\phi(\tau-\beta)\right\}\nonumber\\
&=&\frac{1}{Z}\,{\rm Tr}\left\{{\rm e}^{-\beta H}\,\phi
\,{\rm e}^{-H(\beta-\tau)}\,\phi\,{\rm e}^{(\beta-\tau) H}\right\},\eeq
which coincides indeed with $G(\tau)$, eq.~(\ref{G>tau}),
because of the cyclic invariance of the trace.
The periodicity conditions (\ref{KMSE}) allow us to represent $G(\tau)$ by
a Fourier series:
\beq\label{SUM}
G(\tau)=\frac{1}{\beta}\sum_n
{\rm e}^{-i\omega_n\tau}G(i\omega_n),
\eeq
where the frequencies
$ \omega_n = 2\pi n T$, with integer $n$,
   are called {\it Matsubara frequencies}.

The  free propagator $G_0(x-y)$
is defined as (see eq.~(\ref{operator_ev})):
\beq\label{DeltaDEF}
G_0(x-y)\equiv \langle {\rm T}_\tau
\phi_I(x) \phi_I(y)\rangle_0\,,\eeq
where $\phi_I(x)$ is the interaction
representation of the field operator (cf. eq.(\ref{interactionH})):
\beq\label{phiI}
\phi_I(t,{\bf x})={\rm e}^{H_0\tau}\phi({\bf x})\,
{\rm e}^{-H_0\tau}.\eeq
It satisfies the equation of motion:
\beq\label{eqD0}
\left(-\del_{\tau}^2-\nbfgrad_{\bf x}^2+m^2\right)
G_0(\tau, {\bf x})&=&\delta(\tau)\,
\delta^{(3)}({\bf x}),\eeq
   with the periodic boundary conditions (\ref{KMSE}).
This equation is easily solved using the Fourier
representation (\ref{SUM}). One gets:
\beq\label{D0F}
G_0(i\omega_n,{\bf k})={1\over\varepsilon_k^2+\omega_n^2},
\eeq
where $\varepsilon_k=\sqrt{{\bf k}^2+m^2}$.
The imaginary-time propagator $G_0(\tau)$ can be recovered
from its Fourier coefficients (\ref{D0F}) by performing
the frequency sum in eq.~(\ref{SUM}).
After a simple calculation (see  Appendix B),
   one obtains the following expression
for $G_0(\tau, {\bf k})$, valid for $-\beta\le \tau \le \beta$:
\beq\label{Dtau}
G_0(\tau, {\bf k})&=&\int_{-\infty}^{+\infty}\frac{{\rm d}k_0}{2\pi}\,
{\rm e}^{-k_0\tau}\rho_0(k)\,\Bigl(\theta(\tau)+N(k_
0)\Bigr)\,,\eeq
where $N(k_0)=1/({\rm e}^{\beta k_0}-1)$
   is the Bose-Einstein occupation factor, and:
   \beq\label{rho0}\rho_0(k)=2\pi\epsilon(k_0)\delta(k^2 -m^2)
=\frac{\pi}{\varepsilon_k}\,\Bigl(\delta(k_0-\varepsilon_k)\,-\,
\delta(k_0+\varepsilon_k)\Bigr), \eeq
(with  $\epsilon(k_0)=k_0/|k_0|$) is the spectral function
of a free relativistic scalar particle
of mass $m$.

In the imaginary-time
formalism,  the thermal perturbation theory has the same structure as the
perturbation theory in the vacuum, the only difference being that the integrals
over the loop energies are replaced by sums over the Matsubara frequencies
\cite{Kapusta89,MLB96}.
Appendix  B  provides examples of explicit computations
in this formalism. Note that because the heat bath provides a preferred frame,
explicit Lorentz invariance is lost, which makes some calculations 
more complicated
than the equivalent ones at zero temperature.

\subsubsection{Analyticity properties and real-time propagators}

The imaginary-time propagator
$G(\tau)$ may be written quite generally as:
\beq\label{GT><}
G(\tau)=\theta(\tau)\,G^>(\tau)+\theta(-\tau)\,G^<(\tau),\eeq
where the functions $G^>$ and $G^<$ are defined by
\beq\label{G>G<}
G^>(x,y)&\equiv&{\rm Tr}\left\{{\cal D}\,
\phi(x)\phi(y)\right\},\nonumber\\
G^<(x,y)&\equiv&{\rm Tr}\left\{{\cal D}\,
\phi(y)\phi(x)\right\}\,=\,G^>(y,x),\eeq
with the fields $\phi(x)$ in the Heisenberg representation (\ref{phiH}).
In these equations, all the time variables are
complex variables of the form $t=t_0-i\tau$ to start with.
However, as we shall see, the functions $G^>$ and $G^<$ are analytic functions
or their time arguments, with certain domains of analyticity to be specified
below. They can be used to construct  {\it real-time} Green's functions, such as 
the {\it time-ordered}, or {\it Feynman}, propagator:
\beq\label{D><}
G(x,y)\,=\,\theta(x_0-y_0)\, G^>(x,y)
+\theta(y_0-x_0) \,G^<(x,y),\eeq
as well as 
{\it retarded} and {\it advanced} propagators:
\beq\label{A}
G_R(x,y)&\equiv &
i\theta(x_0-y_0)\Bigl[G^>(x,y)\,-\,G^<(x,y)\Bigr]\,,\nonumber\\
G_A(x,y)&\equiv & - 
i\theta(y_0-x_0)\Bigl[G^>(x,y)\,-\,G^<(x,y)\Bigr]\,,\eeq
where $x_0$ and $y_0$ are  both  real time variables. 
These functions enter the description of the response of the system
to small external perturbations (cf. Sect. 2.2.1).

To see the origin of the analyticity, note that, by definition,
\beq\label{G>z}
   G^>(t)={\rm Tr}\left\{{\cal D}\,\phi(t)\phi(0)\right\}
=\frac{1}{Z}\,{\rm Tr}\left\{{\rm e}^{-H(\beta-it)}\,
\phi\,{\rm e}^{-i Ht}\phi\right\}.\eeq
To evaluate the trace in eq.~(\ref{G>z}), we may introduce a complete
set $|n\rangle$  of energy eigenstates, $H|n\rangle=E_n|n\rangle$,
and thus obtain:
\beq\label{G>S}
G^>(t)\,=\,\frac{1}{Z}\,\sum_{m,n}{\rm e}^{-\beta E_n}\,
|\langle n|\phi|m\rangle |^2 \,{\rm e}^{it(E_n-E_m)}\,.\eeq
If we assume that the exponentials control the convergence of this sum,
we expect the trace to be convergent as long as  $ -\beta < {\rm Im}\, t < 0$.
   Similarly, we expect $G^<(t)$ to exist  for all $t$ in the region 
$0<{\rm Im}\,t
<\beta$. In these respective domains,
   $G^>(t)$ and $G^<(t)$ are both analytic functions.
They also exist, as generalized functions, 
for $t$ approaching the boundaries
of their respective analyticity domains, and, in particular,
for real values of $t$ \cite{Baym61,KB62,Landsman87}.

For complex time variables, the periodicity conditions (\ref{KMSE})
translate into the following condition on the analytic functions
$G^>$ and $G^<$ ($0\le {\rm Im} \, t \le \beta$):
\beq\label{KMSG}
G^<(t)&=&G^>(t-i\beta),\eeq
also known as the Kubo-Martin-Schwinger (KMS) 
condition \cite{Kubo57,Martin}.

The  real-time 
functions $G^>$ and $G^<$ satisfy hermiticity properties:
$(G^>(y,x))^* =G^>(x,y)$ and $(G^<(y,x))^*=G^<(x,y)$. This is easily verified.
For instance
\beq\label{HERM}
(G^>(y,x))^*&=&
\Bigl({\rm Tr}\left\{{\cal D}\,\phi(y)\phi(x)\right\}\Bigr)^*
\nonumber\\&=&{\rm Tr}\Bigl\{\Bigl({\cal D}\,
\phi(y)\phi(x)\Bigr)^\dagger\Bigr\}\,=\,
{\rm Tr}\left\{\phi(x)\phi(y){\cal D}\right\}\,=\,G^>(x,y),\eeq
where in writing the last two equalities we have used the hermiticity
of $\phi(x)$ ($x_0$ real) and of ${\cal D}$, and the cyclic
invariance of the trace.

The hermiticity property (\ref{HERM}), together with the definitions
(\ref{A}), imply $(G_R(x,y))^*=G_A(y,x)$. For the real scalar
field we have
the additional symmetry condition $G^>(x,y)=G^<(y,x)$ (cf.
eq.~(\ref{G>G<})), which ensures that $G_R(x,y)$ and $G_A(x,y)$ are real
functions, with $G_A(x,y)=G_R(y,x)$.

 The  analytic functions $G_0^>$ 
and $G_0^<$ corresponding to the free
scalar field can be read off eq.~(\ref{Dtau}):
\beq\label{Delta>}
G_0^>(\tau, {\bf k})
&=&\frac{1}{2\varepsilon_k}\left\{
(1+N_k){\rm e}^{-\varepsilon_k\tau}
+N_k{\rm e}^{\varepsilon_k\tau}\right\},
\nonumber\\
G_0^<(\tau, {\bf k})&=&
\frac{1}{2\varepsilon_k}\left\{
N_k {\rm e}^{-\varepsilon_k\tau}+(1+N_k)
 {\rm e}^{\varepsilon_k\tau}\right\}.
\eeq
where $N_k\equiv N(\varepsilon_k)$.
By replacing $\tau \,\to it$ (with real $t$) in these equations,
and using the definitions (\ref{D><}) and (\ref{A}), we get:
\beq
G_0(t,{\bf k})&=&\frac{1}{2\varepsilon_k}\left\{\theta(t)\,
{\rm e}^{-i\varepsilon_k t}\,+\,\theta(-t)\,
{\rm e}^{i\varepsilon_k t}\,+\,2N_k \,\cos \varepsilon_kt\right\},\eeq
\beq
G_0^R(t,{\bf k})\,=\,\theta(t)\,\frac{\sin\varepsilon_k t}{\varepsilon_k}
\,,\qquad\,\,
G_0^A(t,{\bf k})\,=\,-\theta(-t)\,\frac{\sin\varepsilon_k t}{\varepsilon_k}\,.
\eeq

\subsubsection{Spectral representations for the propagator and the self-energy}

The analytic properties of the Green's functions, considered as
functions of complex times, entail corresponding properties of their
Fourier transforms, which we shall now summarize.

Let $G^>(k)$  (with $k^\mu=(k^0, {\bf k})$, $k_0$ real) be the
Fourier transform of the real-time function $G^>(x)$:
\beq\label{G>k}
G^>(k)&=&\int {{\rm d}^4x}\,{\rm e}^{ik\cdot x}\,G^>(x),\eeq
and similarly for $G^<(k)$. The hermiticity of $G^>(x)$
(cf. eq.~(\ref{HERM})) implies that $G^>(k)$ is a real function,
and similarly for $G^<(k)$. Furthermore, the KMS condition (\ref{KMSG})
implies the following relation:
\beq\label{KMSk}
G^>(k_0,{\bf k})&=&{\rm e}^{\beta k_0}\,
G^<(k_0,{\bf k}).\eeq

Consider then the {\it spectral density} $\rho(k)$.
   This is related to the functions
$G^>(k)$ and $G^<(k)$ by: \beq\label{defrho}
\rho(k)&\equiv& G^>(k)\,-\,G^<(k)
\,=\,\int {{\rm d}^4x}\,{\rm e}^{ik\cdot x}\,
\left\langle [\phi(x),\,\phi(0)]\right\rangle \nonumber\\&=&
\frac{2\pi}{Z}\,\sum_{m,n}{\rm e}^{-\beta E_n}\,|\langle n|\phi|m\rangle |^2
\Bigl(\delta(k_0+E_n-E_m)-\delta(k_0-E_n+E_m)\Bigr).\eeq
In writing the second line, all reference to the
spatial momenta has been omitted, for simplicity.
For the non-interacting system, eq.~(\ref{defrho}) reduces to the
free spectral density (\ref{rho0}). By rotational symmetry,
$\rho(k)$ is a function of $k_0$ and $|{\bf k}|$. The 
dependence on $k_0$ it is such that $\rho(-k_0)=-\rho(k_0)$ and
$k_0\rho(k_0)\ge 0$, as can be deduced from  eq.~(\ref{defrho}). 
Furthermore, the equal-time commutation relation $[i\del_t
\phi(t,{\bf x}),\,\phi(0,{\bf x'})]_{t=0}
=\delta({\bf x-x'})$ can be used to obtain the sum rule:
\beq\label{SRrho}
\int\frac{{\rm d}k_0}{2\pi}\,k_0
\rho(k)\,=\,\int {{\rm d}^3x}\,{\rm e}^{-i{\bf k}\cdot{\bf x}}\,
\Bigl\langle [i\del_t\phi(x),\,\phi(0)]_{t=0}\Bigr\rangle
\,=\,1.\eeq

By combining eqs.~(\ref{KMSk})
and (\ref{defrho}), we obtain
\beq\label{G><k}
G^>(k)=\rho(k)\Bigl[1+N(k_0)\Bigr],\,\,\qquad\,\,
G^<(k)=\rho(k)\,N(k_0)\,.\eeq
The formulae (\ref{defrho}) and (\ref{G><k}) show that
the functions $G^>(k)$ and $G^<(k)$ are positive definite,
and suggest the following interpretation for them
\cite{KB62}: For positive $k_0$,
$G^<(k)$ is proportional to the average density of particles with
momentum ${\bf k}$ and energy $k_0$, while
$G^>(k)$ measures the density of states available for the addition
of an extra particle with four-momentum $k^\mu$.
A similar interpretation may be given for negative $k_0$,
by exchanging the roles of $G^>$ and $G^<$ (recall
the identity $1+N(-k_0)=-N(k_0)$, so that $G^>(-k_0)=G^<(k_0)$).

By inverting eq.~(\ref{G>k}), and using eq.~(\ref{G><k}),
one obtains, for   $-\beta \le {\rm Im}\, x_0 \le 0 $,
\beq\label{G>t}
G^>(x)\,=\,\int\frac{{\rm d}^4k}{(2\pi)^4}\,{\rm e}^{-ik\cdot x}\,
\rho(k)\Bigl[1+N(k_0)\Bigr]\,,\eeq
which generalizes eq.~(\ref{Dtau}).
This expression, when continued to imaginary time $t\to -i\tau$,
$0\le \tau \le \beta$, gives the function $G^>(\tau, {\bf k})$ and,
by inversion of eq.~(\ref{SUM}), the Matsubara propagator:
\beq\label{invsum}
   G(i\omega_n,{\bf k})
&=&\int_0^\beta {\rm d}\tau \,{\rm e}^{i\omega_n\tau}\,
G^>(\tau,{\bf k})\nonumber\\
&=&\int_{-\infty}^\infty \frac{{\rm d}k_0}{2\pi}\,
\frac{\rho(k_0,k)}{k_0-i\omega_n}\,.\eeq
In going from the first to the second line,  we used
   ${\rm e}^{i\beta\omega_n}=1$ and
$[1+N(k_0)]({\rm e}^{-\beta k_0}-1)=-1$.
According to eq.~(\ref{invsum}),
   the Fourier coefficient $G(i\omega_n,{\bf k})$
can be regarded as the value of the function:
\beq\label{Gana}
   G(\omega,{\bf k})&=&
\int_{-\infty}^\infty \frac{{\rm d}k_0}{2\pi}\,
\frac{\rho(k_0,k)}{k_0-\omega}\,,\eeq
for $\omega=i\omega_n$. This function is often referred to as
the {\it analytic propagator}. It is the unique continuation
of the Matsubara propagator $G(i\omega_n,{\bf k})$ which is
analytic off the real axis and does not grow as fast as an 
exponential as $|\omega| \to \infty$ \cite{Baym61}. 
Note that eq.~(\ref{Gana}) relates the spectral density to
   the discontinuity of $G(\omega)$ across the real axis:
\beq\label{disc}
i\rho(k_0,k)\,=\,G(k_0+i\eta, {\bf k})\,-\,
G(k_0-i\eta, {\bf  k})\,,\eeq
with $\eta\to 0_+$.

The causal Green's functions are also simply related
   to the analytic propagator.
For instance, the Fourier transform of the  retarded 2-point function
(\ref{A}):
\beq\label{DRk}
G_R(k)&=&\int {{\rm d}^4x}\,{\rm e}^{ik\cdot x-\eta
x_0}\,G_R(x),\eeq may be obtained as the limit of the analytic
propagator $G(\omega,{\bf k})$, eq.~(\ref{Gana}), as $\omega$
approaches the real
energy axis from above:
\beq\label{DGR}
G_R(k_0, {\bf k})\,=\,
\int_{-\infty}^\infty \frac{{\rm d}k_0^\prime}{2\pi}\,
\frac{\rho(k_0^\prime,k)}{k_0^\prime-k_0-i\eta}\,,\eeq
that is,
\beq\label{DGA}
G_R(k_0,{\bf k})\,=\,G(\omega=k_0+i\eta,{\bf k}).\eeq
Similarly, for the advanced 2-point function (see eq.~(\ref{A}))
we have:
\beq
G_A(k_0,{\bf k})\,=\,G(\omega=k_0-i\eta,{\bf
k})\,=\,G_R^{\,*}(k_0,{\bf k}).\eeq
By using the spectral representation (\ref{DGR}), we can
extend the definition of the retarded propagator to any complex
energy $\omega$ such that Im $\omega >0$: it then follows that
$G_R(\omega)$ is an analytic function in the upper half-plane,
   where it coincides with the analytic propagator (\ref{Gana}).
In  the lower half plane,  on the other hand, $G_R(\omega)$ is
defined by continuation across the real axis, and it may have there
singularities. Similarly, the advanced propagator $G_A(\omega)$
can be defined as an analytic function in the lower half plane.

The analyticity properties that we have discussed have an important 
consequence,
known as the {\it fluctuation-dissipation theorem}, which relates the
  dissipation properties of a system to its
 various correlations.  To exhibit such
a relation, let us first observe that by combining eqs.~(\ref{DGR}), 
(\ref{DGA}) and
(\ref{disc}), we can write
\beq\label{spectraldef}
\rho(k) = 2 \,{\rm Im}\,G_R(k).
\eeq
  Thus the
  spectral function $\rho(k)$  may be obtained from the imaginary part of the
retarded propagator which  describes the dissipation
of the single particle excitations (see the end of this subsection). 
But once the
spectral density is known, the various correlations can be calculated 
according to
Eqs.~(\ref{G><k}).

The previous discussion can be readily extended to
the self-energy $\Sigma$,  defined by the
Dyson-Schwinger equation:
\beq \label{DEFSIGMA}
G^{-1}\,=\,G_0^{-1} + \Sigma\,.\eeq
Up to a possible singular part at $\tau=0$ (see
eqs.~(\ref{Sig<>})--(\ref{SDEL}) below for an example), we can write:
\beq\label{ST><}
\Sigma(\tau)=\theta(\tau)\,\Sigma^>(\tau)+\theta(-\tau)\,\Sigma^<(\tau),\eeq
where the self-energies $\Sigma^>$ and $\Sigma^<$
share the analytic properties of the 2-point functions
$G^>$ and $G^<$, respectively. After continuation to
complex values of time,   they satisfy the KMS condition
   $ \Sigma^<(t)\,=\, \Sigma^> (t-i\beta)$
(for $ 0 \le{\rm Im}\,t \le \beta$), and can be given the following
representations in momentum space:
\beq\label{SIG><k}
\Sigma^>(k)=-\Gamma(k)\Bigl[1+N(k_0)\Bigr],\qquad\,\,
\Sigma^<(k)=-\Gamma(k)\,N(k_0),\eeq
where:
\beq\label{defGamma}
-\Gamma(k)\,\equiv\,\Sigma^>(k) \,-\,\Sigma^<(k)\,.\eeq
One can also define an analytic self-energy
(analytic continuation of $\Sigma(i\omega_n)$) with
   the spectral representation (up to the possible subtraction
of a singular part):
\beq\label{Sana}
\Sigma(\omega,{\bf k})&=&
\int_{-\infty}^\infty \frac{{\rm d}k_0}{2\pi}\,
\frac{-\,\Gamma(k_0,{\bf k})}{k_0-\omega}\,,\eeq
with $\Gamma(k_0,{\bf k})$ defined in eq.~(\ref{defGamma}).

The Dyson-Schwinger equation
can be used to relate the retarded propagator to the retarded self-energy:
\beq\label{Greteq}
G_R(k_0,{\bf k})\,=\,\frac{-1}{
(k_0+i\eta)^2-\varepsilon_k^2-\,\Sigma_R(k_0,{\bf k})}\,,\eeq
where $\Sigma_R(k_0,{\bf k})\equiv
\Sigma(k_0+i\eta,{\bf k})$.
   Note that, with the present conventions,
$\Gamma(k_0,{\bf k})=- 2{\rm Im}\,\Sigma_R(k_0,{\bf k})$.

By using eq.~(\ref{disc}), one obtains the spectral density as
\beq\label{rholor}
\rho (k_0,k)\,=\,\frac{\Gamma(k_0,{k})}{\Bigl(k_0^2-\varepsilon_k^2-{\rm Re}
\,\Sigma_R(k_0,{k})\Bigr)^2\,+\,
\Bigl(\Gamma(k_0,{k})/2\Bigr)^2}\,.\eeq
The sign properties of  $\rho (k_0,k)$, discussed after
eq.~(\ref{defrho}), require ${\rm Re}
\,\Sigma_R(k_0)$ to be even and $\Gamma (k_0)$ to be odd functions
of $k_0$, with $k_0\,\Gamma (k_0)\ge 0$.
In particular,  $\Sigma^>(k)$ and $\Sigma^<(k)$ are {\it negative}
definite in our present conventions (see eqs.~(\ref{SIG><k})).

For a free particle, $\Sigma=0$, and the spectral function is a sum of delta
functions (see eq.~(\ref{rho0})). When $\Gamma$ is small and not too strongly
dependent on $k_0$,  the spectral density (\ref{rholor}) does not differ too
much from the free particle one. In such cases,  the associated single-particle
excitations are often referred to as {\it quasiparticles}. To be more
specific, let
$E_k$ be   the positive-energy solution (whenever it exists) of the equation
$k_0^2=\varepsilon_k^2+{\rm Re}\,\Sigma_R(k_0,k)$.
If $\Gamma$ is a slowly varying function of $k_0$
in the vicinity of $E_k$, then, for $k_0$ close to $E_k$,
the spectral density (\ref{rholor}) has a Lorentzian shape:
\beq\label{rhoLOR}
\rho (k_0\simeq E_k,k)\,\simeq\,\frac{z_k}{2E_k}\,
\frac{2\gamma_k}{(k_0- E_k)^2\,+\,
\gamma_k^2}\,,\eeq
while the  retarded propagator (\ref{Greteq}) develops a simple pole
at $k_0=E_k-i\gamma_k$:
\beq\label{GRLOR}
G_R (k_0\simeq E_k,k)\,\simeq\,\frac{z_k}{2E_k}\,
\frac{-1}{k_0\,-\, E_k\,+\,i\gamma_k}\,.\eeq
In writing these equations, we have denoted:
\beq\label{defzg}
z_k^{-1}&\equiv&1\,-\,\frac{1}{2E_k}\,\frac{\del
{\rm Re}\,\Sigma_R}{\del k_0}\biggl |_{k_0=E_k}\,,\nonumber\\
\gamma_k&\equiv& \frac{z_k}{4E_k}\,\Gamma(k_0=E_k,{k})\,,\eeq
and we have assumed that $\gamma_k\ll E_k$.
Eqs.~(\ref{rhoLOR}) and (\ref{GRLOR}) describe quasiparticles
with  energy $E_k$ and width $\gamma_k$.
For negative energy, there is another pole in $G_R$,
at $k_0=-E_k-i\gamma_k$.
Note that both poles lie in the lower  half plane, in agreement
with the analytic structure of the retarded propagator discussed before.

The quantity $\gamma_k$ controls the {\it lifetime}
   of  the corresponding quasiparticle
excitation, as measured by the behaviour of
the retarded propagator at large times.
The retarded propagator is given by (cf. eq.~(\ref{DGR})) :
\beq\label{GRTrho}
G_R(t, {\bf k})\,=\,
\int_{-\infty}^\infty {{\rm d}k_0 \over 2\pi}\,{\rm e}^{-ik_0 t}
G_R(k_0, {\bf k})\,=\,i\theta(t)
\int_{-\infty}^\infty {{\rm d}k_0 \over 2\pi}\,{\rm e}^{-ik_0 t}
\rho(k_0,{\bf k}).\eeq
Whenever eqs.~(\ref{rhoLOR}) and (\ref{GRLOR}) are valid,
$G_R(t,{\bf k})\sim {\rm e}^{-iE_k t}\,{\rm e}^{-\gamma_k t}$ at large
times (for a positive energy state),
so that $|G_R(t,{\bf k})|^2 \propto {\exp}\{-2\gamma_k t\}$.
We shall refer to the quantity  $\tau(k)=1/2\gamma_k$ as the lifetime of the
excitation, and to $\gamma_k$  as the quasiparticle
{\it damping rate}.

Note that, even if it is generic, the exponential decay  is by no
means universal. A more complicated behaviour can occur whenever
some of the aforementioned assumptions are not satisfied. In
section 6, we shall encounter an example of such a non-trivial
evolution in time \cite{lifetime,damping,TBN,BVHS98}.

\subsubsection{Classical field approximation and dimensional reduction}

In the high temperature limit, $\beta\to 0$, the imaginary-time dependence
of the fields frequently becomes unimportant and can be ignored in a first 
approximation. The
integration over imaginary time becomes then trivial  and the
partition function (\ref{ZCE})  reduces to:
\beq\label{ZCEcl}
Z\,\approx\,{\cal N}
\int{\cal D}(\phi)\exp\left\{
- \beta \int{\rm d}^3x \,{\cal H}({\bf x})\right\}\,,\eeq
where { $\phi\equiv\phi({\bf x})$ is now a three-dimensional
field, and}
\beq\label{LEcl}
{\cal H}\,=\,{1\over2}\,(\nbfgrad\phi)^2+
{m^2\over2}\,\phi^2 + V(\phi)\,.\eeq
The functional integral { in eq.~(\ref{ZCEcl})}
is recognized as the partition function for static  three-dimensional
field configurations with energy $ \int{\rm d}^3x \,{\cal H}(x)$.
We shall refer to this limit as the {\it classical field approximation}.

Ignoring the time dependence of the fields is equivalent to retaining only
the zero Matsubara frequency in their Fourier decomposition. Then the Fourier
transform of the free propagator is simply:
\beq
G_0({\bf k})\,=\,\frac{T}{\varepsilon_k^2}\,.
\eeq
This may be obtained directly from eq.~(\ref{SUM}) keeping only the term
with $\omega_n=0$, or from eq.~(\ref{Dtau}) by ignoring
the time dependence and using the approximation
\beq\label{BESOFT}
N(\varepsilon_k)\,=\,\frac{1}{{\rm e}^{\beta\varepsilon_k} -1}\approx
\frac{T}{\varepsilon_k}\,.
\eeq
Both approximations make sense only for $\varepsilon_k\ll T$, implying
$N(\varepsilon_k)\gg 1$. In this limit, the energy density per mode
$\varepsilon_k N(\varepsilon_k)\approx T$ is as expected from the
classical equipartition theorem. Also, because $N(\varepsilon_k)
\approx 1+N(\varepsilon_k) \approx T/\varepsilon_k$, 
the two propagators $G_0^>$ and $G_0^<$ in
eq.~(\ref{Delta>}) become equal, and the analytic properties
discussed in Sect. 2.1.2 are lost. 
That $G^>\approx G^<$ in the classical limit
is in agreement with the fact that the field operator 
$\phi(x)$ becomes a commuting $c$-number in this limit.
We shall discuss later, in 
Sect. 2.2.5, how to construct real time propagators 
in the classical field approximation.

The classical field approximation may be viewed as the leading
term  in a systematic expansion. To see that, let us expand the field
variables in the path integral (\ref{ZCE}) in terms of their Fourier
components:
\beq
\phi(\tau)=\frac{1}{\beta} \sum_n {\rm e}^{-i\omega_n \tau}
\phi(i\omega_n),
\eeq
where the $\omega_n$'s are the Matsubara frequencies.
This takes care automatically of the periodic boundary conditions.
The path integral (\ref{ZCE}) can then be written as:
\beq\label{ZCE0}
Z\,=\,{\cal N}_1
\int{\cal D}(\phi_0)\exp\left\{-S[\phi_0]\right\}\,,\eeq
where $\phi_0\equiv \phi(\omega_n=0)$ and
\beq\label{ZCE1}
\exp\left\{-S[\phi_0]\right\}=\,{\cal N}_2
\int{\cal D}(\phi_{n\ne 0})\exp\left\{
- \int_0^\beta {\rm d}\tau \int{\rm d}^3x \,{\cal L}_E(x)\right\}\,.\eeq
The quantity $S[\phi_0]$ may be called the effective action for the ``zero
mode'' $\phi_0$. Aside from the direct classical field contribution 
that we have
already considered, this effective action  receives also contributions from
integrating out the non-vanishing Matsubara frequencies.
  Diagrammatically, $S[\phi_0]$ is the sum of all the connected
diagrams with external lines associated to
$\phi_0$, and in which the internal lines are the
propagators of the non-static
modes $\phi_{n\ne 0}$. Thus, a priori,  $S[\phi_0]$ contains
operators of arbitrarily high order in $\phi_0$, which are
also non-local.
In practice, however, one wishes to expand  $S[\phi_0]$ in terms  of 
{\it local} operators, i.e., operators with the schematic structure 
$a_{m,\,n}\grad^m \phi_0^n$ with coefficients $a_{m,\,n}$
to be computed in perturbation theory.

To implement this strategy, it is useful to introduce an intermediate scale
$\Lambda$ ($\Lambda\ll  T$) which  separates {\it hard} 
($k\simge \Lambda$)
and {\it soft} ($k\simle \Lambda$) momenta. All the non-static modes, as well
as the static ones with $k \simge \Lambda$ are {\it hard}
(since $K^2\equiv \omega_n^2 +k^2 \simge \Lambda^2$
for these modes), while the static ($\omega_n=0$) modes
with $k\simle \Lambda$ are {\it soft}.
Thus, strictly speaking, in the construction of the effective
theory along the lines indicated above, one has to integrate out also 
the static
modes with $k \simge\Lambda$. The benefits of this separation of
scales are that ({\it a}) the resulting effective action for the
soft fields can be made
  {\it local} (since the initially non-local amplitudes can be
expanded out in powers of $p/K$, where $p \ll \Lambda$
is a typical external momentum, and $K\simge
\Lambda$ is a hard momentum on an
internal line), and  ({\it b}) the effective theory is now
used exclusively at soft momenta, where  classical approximations such as
(\ref{BESOFT}) are expected to be valid.
This strategy, which consists in integrating out the non-static modes
  in perturbation theory in order to obtain an effective
three-dimensional  theory for the soft static modes (with
$\omega_n=0$ and $k\equiv |{\bf k}|\simle \Lambda$), is generally 
referred  to as ``dimensional reduction''
\cite{Gins80,Appel81,Nadkarni83,Landsman89,Braaten94,Kajantie94}.
It is especially useful in view of non-perturbative lattice calculations,
which are easier to perform in lower dimensions
\cite{Kajantie94,Kajantie96} (see also Sect. 5.4.3 below).

  As an illustration
let us consider a massless scalar theory with quartic interactions;
that is, $m=0$ and $V(\phi)=({g^2}/{4!})\phi^4$ in
eq.~(\ref{Lagran}). The ensuing effective action for the soft fields
(which we shall still denote as $\phi_0$) reads
\beq\label{EFFLagran}
S[\phi_0] =\beta {\cal F}(\Lambda)+\int{\rm d}^3x \,\left\{
{1\over 2}\,(\nbfgrad\phi_0)^2+{1\over 2}\,M^2(\Lambda)\phi^2_0+
\,{g^2_3(\Lambda)\over 4!}\,\phi^4_0+{h(\Lambda)\over 6!}\,\phi^6_0
+\Delta {\cal L}\right\},
\eeq
where ${\cal F}(\Lambda)$ is the contribution of the hard modes
to the free-energy, and
$\Delta {\cal L}$ contains all the other local operators which are
invariant under rotations and under the  symmetry $\phi\to -\phi$, 
i.e., all the
local operators which are consistent with the symmetries of the 
original Lagrangian.
We have changed the normalization of the field ($\phi_0\rightarrow
\sqrt{T}\phi_0$) with respect to  eqs.~(\ref{ZCEcl})--(\ref{LEcl}),
so as to
absorb the factor $\beta$ in front of the effective action. The 
effective ``coupling
constants'' in eq.~(\ref{EFFLagran}), i.e.
$M^2(\Lambda)$,
$g^2_3(\Lambda)$,
$h(\Lambda)$ and the infinitely many parameters in $\Delta {\cal L}$, 
are computed
in perturbation theory, and depend upon the
separation scale $\Lambda$, the temperature $T$ and the original
coupling $g^2$. To lowest order in $g$, $g^2_3\approx g^2T$,
$h\approx 0$ (the first contribution to $h$
arises at order $g^6$, via one-loop diagrams), and $M\sim gT$,
as we shall see shortly.
Note that eq.~(\ref{EFFLagran}) involves in general non-renormalizable
operators, via $\Delta {\cal L}$. This is not a difficulty, however,
since this is only an { effective} theory, in which the scale
$\Lambda$ acts as an explicit ultraviolet (UV) cutoff for
the loop integrals. Since the scale $\Lambda$ is
arbitrary, the dependence on $\Lambda$ coming from such soft loops
must cancel against the dependence on $\Lambda$ of the parameters
in the effective action.

\begin{figure}
\protect \epsfysize=4.cm{\centerline{\epsfbox{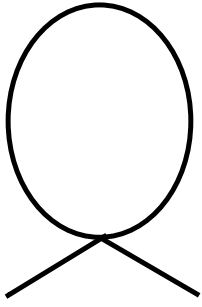}}}
	 \caption{One-loop tadpole diagram for the self-energy of the scalar field.}
\label{tadpole}
\end{figure}
Let us verify this cancellation explicitly
in the case of the thermal mass $M$ of the scalar field,
and to lowest order in perturbation theory. To this order,
the scalar self-energy is given by the tadpole diagram in 
fig.~\ref{tadpole}. The
mass parameter $M^2(\Lambda)$ in the effective action is obtained by 
integrating
over hard momenta within the loop in
  fig.~\ref{tadpole} (cf. eq.~(\ref{TP})) :
\beq\label{MLAM}
M^2(\Lambda)&=&{g^2\over 2}\,T\sum_n
\int\frac{{\rm d}^3k}{(2\pi)^3}\, \frac{(1-\delta_{n 0})+
\theta(k-\Lambda)\delta_{n 0}}{\omega_n^2+k^2}\nonumber\\
&=&{g^2\over 2}\int\frac{{\rm d}^3k}{(2\pi)^3}\,
\left\{\frac{N(k)}{k}\,+\,\frac{1}{2k}\,-\,\theta(\Lambda-k)
\frac{T}{k^2}\right\},\eeq
where the $\theta$-function in the second line has been
generated by writing  $\theta(k-\Lambda) = 1 -\theta(\Lambda-k)$.
The first term, involving the thermal distribution, gives
the contribution
\beq\label{MSHTL}
\hat M^2\,\equiv\,
\frac{g^2}{2}\int\frac{{\rm d}^3k}{(2\pi)^3}\,\frac{N(k)}{k}
\,=\,\frac{g^2}{24}\,T^2\,.\eeq
As it will turn out, this is the leading-order (LO)
scalar thermal mass, and also the simplest example of what will be called
 ``hard thermal loops''
(HTL). The second term, involving $1/2k$, in eq.~(\ref{MLAM}) is
quadratically UV divergent, but independent of the temperature;
  the standard
renormalization procedure at $T=0$
amounts to simply removing this term
(see Sect. 2.3.3).
The third term, involving the $\theta$-function, is easily
evaluated. One finally gets:
\beq\label{MLAM1}
M^2(\Lambda)\,=\,\hat M^2\,-\,\frac{g^2}{4\pi^2}\,\Lambda T\,
\equiv\,
\frac{g^2T^2}{24}\left(1-\frac{6}{\pi^2}\,\frac
{\Lambda}{T}\right).\eeq
The $\Lambda$-dependent term above is subleading, by a
factor $\Lambda/T\ll 1$.

The one-loop correction to the thermal mass within the effective
theory is given by the same diagram in fig.~\ref{tadpole}, but
where the internal field is static and soft, with the massive
propagator $1/(k^2+M^2(\Lambda))$,
and coupling constant $g^2_3\approx g^2T$. Since the typical momenta in the
integral will be $k\simge M$, and $M\sim\hat M\sim gT$, we choose $\Lambda\gg
gT$. We then obtain
\beq\label{MDCL1}
\delta M^2(\Lambda) &=&
{g^2\over 2}\int\frac{{\rm d}^3k}{(2\pi)^3}\,\Theta(\Lambda-k)\,
\frac{T}{k^2+M^2(\Lambda)}\nonumber\\
&=&\frac{g^2T\Lambda}{4\pi^2}\left(
1-{\pi M\over 2\Lambda}\,\arctan\,{M\over \Lambda}\right)
\,\simeq\,
\frac{g^2T\Lambda}{4\pi^2}
\,-\,\frac{g^2}{8\pi}\hat MT\,,\eeq
where the terms neglected in the last step
are of higher order in $\hat M/\Lambda$
or $\Lambda/T$.

As anticipated, the $\Lambda$-dependent terms cancel in the
sum $M^2\equiv M^2(\Lambda)+\delta M^2(\Lambda)$, which then provides
the physical thermal mass within the present accuracy:
  \beq \label{MTOTS}
M^2\,=\,M^2(\Lambda)+\delta M^2(\Lambda)
\,=\,\frac{g^2T^2}{24}
\,-\,\frac{g^2}{8\pi}\hat MT\,.\eeq
The LO term, of order $g^2T^2$, is the HTL $\hat M$.
The next-to-leading order (NLO) term, which involves
the resummation of the thermal mass $M(\Lambda)$ in the
soft propagator, is of order $g^2\hat MT\sim g^3T^2$, 
and therefore non-analytic in $g^2$.
This non-analyticity is related to the fact that the integrand
in eq.~(\ref{MDCL1}) cannot be expanded in powers of $M^2/k^2$
without running into infrared divergences.

In the Sect. 2.2.5, we shall see how effective theories based on a classical
field approximation can be used to compute
{\it time-dependent} correlations.
Then, in Sect. 4.4 we shall extend this strategy to gauge theories. 
In that case
however, the problem of matching the coefficients of the 
effective theory with
those of the original one can be a delicate one.

\subsection{Non-equilibrium evolution of the quantum fields}

We consider now situations where the
   system, initially in equilibrium, is perturbed  by an external
source which starts acting at some time $t_0$.
We take the external source to be a current $j(x)$
linearly coupled to the scalar field. The evolution of the system
is then described by the Hamiltonian $H_j(t)$ of
eq.~(\ref{Hj1}). The density operator at time $t$ is given by
(cf. eq.~(\ref{evolutionD})):
\beq\label{rhoj}
{\cal D}_j(t)\,=\, U_j(t,t_0)\,{\cal D}  \,U_j^{-1}(t,t_0),
\eeq
where ${\cal D}$ is the density operator at time $t_0$ and
   $U_j(t,t_0)$, the evolution operator, satisfies:
\beq\label{eqUj}
i\del_t \,U_j(t,t_0)\,=\,H_j(t)U_j(t,t_0),\qquad\qquad
U_j(t_0,t_0)\,=\,1.\eeq

An operator $U_j(t_2, t_1)$ can be defined similarly for arbitrary $t_1$. Such
   operators, which may be viewed as ``time translation operators",
   satisfy the group property:
\beq\label{group}  U_j(t_2,t_1)\,=\,U_j(t_2,t_3)\,U_j(t_3,t_1)\,.\eeq
In particular since for any $t_1$, $U_j(t_1,t_1)=1$, we have
$U_j^{-1}(t_2,t_1)=U_j(t_1,t_2)$. Eq.~(\ref{eqUj}) can be formally
integrated to
yield the following expression for $U_j(t_2,t_1)$:
\beq\label{Uj}
U_j(t_2,t_1)
&=&\tilde{\rm T}\exp\left\{-i\int_{t_1}^{t_2} H_j(t){\rm d}t\right\},
\eeq
where the symbol $\tilde{\rm T}$ orders  the operators from
right to left, in {\it increasing or  decreasing order of their time
arguments depending respectively on whether} $t_2>t_1$ or $t_2<t_1$
(i.e., we use the same symbol for what are usually distinguished as
chronological or antichronological ordering operators; the reason for this will
become more evident when we discuss contour propagators). In other words,
$\tilde T [H_j(t)\,H_j(t')]=H_j(t')\,H_j(t)$ if,
   in going from $t_1$ to $t_2$ along
the time axis, one first meets
$t$ and then $t'$; in the opposite case, $\tilde
T[H_j(t)\,H_j(t')]=H_j(t)\,H_j(t')$.

\subsubsection{Retarded response functions}

The expectation value of any operator  ${\cal O}$ at time $t$
can be calculated from the density operator solution of the equation of motion
(\ref{eqUj}). We assume here that
${\cal O}$ does
not depend explicitly on time. Then,
\beq\label{avOj} {\rm Tr} {\cal D}_j(t){\cal O}\,=\,{\rm Tr}
{\cal D}\,{\cal O}_j(t)
\,=\,\langle {\cal O}_j(t)\rangle\,,\eeq
where
\beq\label{Oj}
{\cal O}_j(t)\equiv
   U_j^{-1}(t,t_0)\,{\cal O}\,U_j(t,t_0)\,=\,
   U_j(t_0,t)\,{\cal O}\,U_j(t,t_0),\eeq
and $U_j(t,t_0)$ is the evolution operator defined in the previous subsection.

If $j=0$, ${\rm Tr} {\cal D}\,{\cal O} =\langle {\cal O}\rangle$
is time-independent and corresponds to the
equilibrium expectation value. The difference $\delta
\langle {\cal O}_j(t)\rangle\equiv \langle {\cal O}_j(t)\rangle
-\langle {\cal O}\rangle$ is
a measure  of the response of the system to the external perturbation. If the
departure from equilibrium is small,
   we may attempt to calculate  $\langle {\cal
O}_j(t)\rangle$ as an expansion in powers of $j$. To do so, the
following identities are useful ($t,\,t'>t_0$):
\beq
i\,\frac{\delta U_j(t,t_0)} {\delta j(t')}=
\theta(t-t')\,U_j(t,t_0)\phi_j(t'),\qquad
i\,\frac{\delta U_j(t_0,t)} {\delta j(t')}=
-\theta(t-t')\,\phi_j(t')U_j(t_0,t).\eeq
(The second identity follows from the first one by noticing that
$U_j(t_0,t)=U_j^{-1}(t,t_0)$, and that $\delta U_j^{-1}=- U_j^{-1}\, \delta
U_j\,U_j^{-1}$.) From these identities, we get easily:
\beq
i\,\frac{\delta } {\delta j(t')}{\cal O}_j(t)=\theta (t-t')\left[{\cal
O}_j(t),\phi_j(t')
\right],
\eeq
and, more generally:
\beq\label{delO}
\delta\langle {\cal O}_j(t)\rangle&=&
\sum_{n=1}^\infty
(-i)^n\int {\rm d}^4y_1 {\rm d}^4y_2 ... {\rm d}^4y_n\,
\theta(t-y^0_1)\theta(y^0_1-y^0_2) ... \theta(y^0_{n-1}-y^0_n)
\nonumber\\
&&\qquad \left\langle \Bigl[\,...\,\Bigl[
\Bigl[{\cal O}(t), \phi(y_1)\Bigr],\,\phi(y_2)\Bigr]
\,...\,\phi(y_n)\Bigr]\right\rangle\,j(y_1)j(y_2)...j(y_n)\,.\eeq
In this equation,  the symbol $[\,...\,[[\, , \,]]\,...\,]$ denotes
nested commutators, and $\phi(t)$ and ${\cal O}(t)$ are operators
in the Heisenberg representation {\it without} the source, that is:
\beq\label{heish}
{\cal O}(t)&=&{\rm e}^{iH(t-t_0)}\,{\cal O}\, {\rm e}^{-iH(t-t_0)}\,,
\eeq
and similarly for $\phi(t)$.
The expectation values in the r.h.s. of
eq.~(\ref{delO}) are {\it equilibrium} expectation values,
   computed in the absence of external sources.

In particular, the average field induced in the system by
the external current can be  expanded as:
\beq\label{phiL}
\Phi(x)= -\sum_{n=1}^\infty\frac{1}{n!}
\int {\rm d}^4y_1 {\rm d}^4y_2 ... {\rm d}^4y_n
\,G_R^{(1+n)}(x;y_1,y_2, ... ,y_n)\,
j(y_1)j(y_2)...j(y_n),\eeq
where: \beq\label{GRn}
G_R^{(1+n)}(x;y_1,y_2, ... ,y_n)&\equiv&
(-i)^{n+2} \sum_{P}
\theta(t-y^0_1)\theta(y^0_1-y^0_2) ... \theta(y^0_{n-1}-y^0_n)
\nonumber\\ &&\qquad\,\,\,\,
   \left\langle \bigl[\,...\,\bigl[
\bigl[\phi(x), \phi(y_1)\bigr],\,\phi(y_2)\bigr]
\,...\,\phi(y_n)\bigr]\right\rangle\,\eeq
is a retarded Green's function with $n+1$ external legs. The sum
in eq.~(\ref{GRn}) runs over all the $n!$ permutations of the
labels $y_1$, $y_2$, ... , $y_n$, so that the function
$G_R^{(1+n)}$ is symmetric with respect to its $y$-arguments.
On the other hand, this is a causal function with respect to
$x$,  since it vanishes for $x^0<y^0_i$ ($i=1,\,2,...,n$),
that is, prior to the action of the perturbation.

As already noted the statistical averages in the formulae above are
taken over the initial {\it equilibrium} thermodynamical ensemble,
with the canonical density operator ${\cal D}={\cal D}_j(t_0)={\rm
e}^{-\beta H}/Z$. Thus, in principle, it is possible  to
study the response of the system to external perturbations
by computing only {\it equilibrium} Green's functions.
This is especially convenient for weak perturbations,
when the expansion in eqs.~(\ref{delO}) and (\ref{phiL}) can be limited
to its first term: this is the {\it linear response }approximation.
In this case,
eq.~(\ref{phiL}) reduces to
\beq\label{phiL1}
\Phi(x)= - \int {\rm d}^4y\,\,G_R(x-y)\,j(y),\eeq
where $G_R(x-y)= i\theta(x_0-y_0)\langle [\phi(x),\phi(y)]\rangle$
is the retarded propagator (\ref{A}), studied in the previous section.
If one could limit oneself to the study of linear response,
the imaginary time formalism presented in the previous section
could therefore be sufficient.

However, as we shall see later,
in non Abelian gauge theories, Green's functions
with different numbers of external legs are related by Ward identities.
In other words, non Abelian gauge symmetry forces us to go beyond
linear response,
even when studying the response to weak external perturbations.
This means that we shall need to consider
$n$-point functions such as (\ref{GRn}), whose  calculation  is
generally difficult. At this stage, some extra formalism is needed,
and this will be developed in the next section.

\subsubsection{Contour Green's functions}

The main technical feature of the formalism to be described now,
and which allows one to exploit the full power of field theoretical
techniques in the calculation of non
equilibrium $n$-point functions, is the use of a
complex time path surrounding the real-time axis. This has been originally
introduced  by Schwinger
\cite{Schwinger61} and Keldysh
\cite{Keldysh64} (see also Refs. \cite{Bakshi63}; for a recent
  presentation of this formalism see \cite{Niemi84,Landsman87,MLB96}).
We shall also refer to the formalism of Kadanoff and Baym
\cite{KB62}, which exploits the
analytic properties of the Green's functions in order
to derive real time equations
of motion for the Green's functions from the corresponding
equations in imaginary-time.

Consider then the time-ordered 2-point function in the presence of $j$ :
\beq\label{EX}
G(t_1,t_2)&=&\langle {\rm T}\,\phi_j(t_1)\phi_j(t_2)\rangle\,
\equiv\, {\rm Tr}\left({\cal D}\,
{\rm T}\,\phi_j(t_1)\phi_j(t_2)\right)\nonumber\\
&\equiv&\theta(t_1-t_2)G^>(t_1,t_2)+\theta(t_2-t_1)G^<(t_1,t_2),\eeq
where $\phi_j(t)$ is the field operator in the Heisenberg representation in the
presence of the sources, given by  eq.~(\ref{Oj}).
By making explicit the various
evolution operators in eq.~(\ref{EX}), we can write, e.g.,
\beq\label{GOE}
G^>(t_1,t_2)\,=\,\frac{1}{Z}\,{\rm Tr}\left\{{\cal D}U_j(t_0,t_1)
\,\phi \,U_j(t_1,t_2)\, \phi\, U_j(t_2,t_0)
\right\},
\mbox{ }
\eeq
where we have used the
group property (\ref{group}).
Imagine now writing all the
evolution operators in terms of  ordered exponentials, as in eq.~(\ref{Uj}),
thus generating a chain of time dependent Hamiltonians.
These operators follow, along the
chain, different ordering prescriptions, depending from which $U_j$ they
originate. Assume, for instance, that $t_0<t_2<t_1$, in which case the
2-point function $G^>(t_1,t_2)$ coincides with the time-ordered
propagator in eq.~(\ref{EX}); then, the evolution operators are
chronologically ordered from $t_0$ to $t_2$ and from $t_2$ to $t_1$,
and anti-chronologically ordered from $t_1$ to $t_0$. This is a source of
complications which, however, can be bypassed by allowing all time
variables to run  on an appropriate contour in the complex time plane.

\begin{figure}
\protect \epsfxsize=16.cm{\centerline{\epsfbox{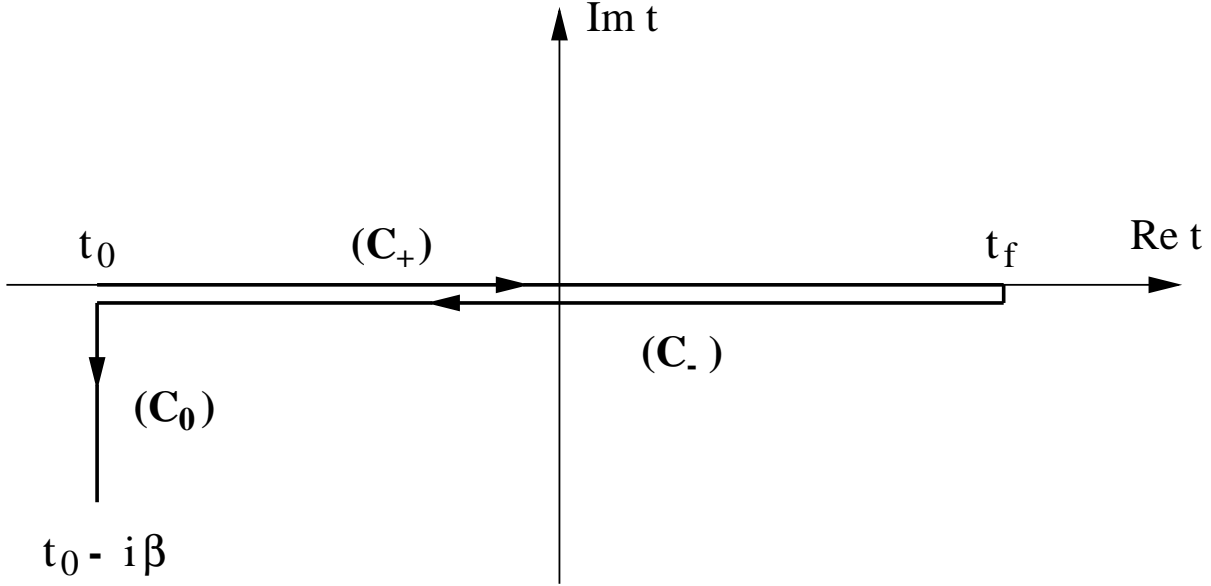}}}
           \caption{Complex-time contour for the evaluation of
the thermal expectation values: $C=C_+\cup C_-\cup C_0$.}
\label{CONT}
\end{figure}

We then extend the definition of the evolution operator to complex time
variables, i.e., we define $U_j(z_2,z_1)$ as the solution of
eq.~(\ref{eqUj}) with $t$ replaced by a complex variable $z$.
The evolution operator becomes then a translation operator
in the complex time plane and the equation can be formally
integrated along any given contour $C$. Such a contour can be specified
by a function $z(u)$, where the real parameter
$u$ is continuously increasing along $C$.
The contour evolution operator can then be written as (compare with
eq.~(\ref{Uj})):
\beq\label{Vz}
   U_j(C)\,=\,{\rm T}_C\exp\left\{
-i\int_C H_j(z){\rm d}z\right\}\,,\eeq
where the operator ${\rm T}_C$
orders the operators $H_j(z_i)$ from right to left in increasing
order of the parameters $u_i$ ($z_i=z(u_i)$).  Note that eq.~(\ref{Vz})
involves
\beq\label{Hjz}
H_j(z)\,\equiv\,H\,+\,\int {\rm d}^3{x}\,j(z,{\bf x}) \,\phi({\bf x})\,;
\eeq
this requires the extension of the external source $j$ to complex time
arguments, which we leave arbitrary at this stage.

We define a contour $\theta$-function $\theta_C$:
$\theta_C(z_1,z_2)=1$ if $z_1$ is further than $z_2$
along the contour (we write then $z_1 \succ z_2$),
while $\theta_C(z_1,z_2)=0$ if the opposite situation holds
($z_1 \prec z_2$). In terms of the coordinate $u$ along the contour,
   $\theta_C(z_1,z_2)=
\theta(u_1-u_2)$, with $z_1=z(u_1)$ and $z_2=z(u_2)$. We shall need also
a contour delta function, which we  define by:
\beq\label{contdelta}
\delta_C(z_1,z_2)\equiv \left(\frac {\del z}{\del u}\right)^{-1}
\delta(u_1-u_2).\eeq

Consider now the specific contour   depicted in fig.~\ref{CONT}. This
may be seen as the juxtaposition of three pieces:
$C=C_+\cup C_-\cup C_0$. On $C_+$, $z=t$ takes all the
real values between $t_0$ to $t_f$, with $t_f$ larger than all the
times of interest. On  $C_-$, we set $z=t-i\eta$
($\eta\to 0_+$) and $t$ runs backward from
$t_f$ to $t_0$. Finally, on $C_0$, $z=t_0-i\tau$, with $0<\tau\le \beta$.
This particular contour allows us to replace the various orderings of
operators that we have met by a single ordering along the
contour. Thus, the ordering along the contour coincides
with the chronological ordering on $C_+$, with
antichronological ordering on $C_-$, and with
ordering according to the imaginary time on $C_0$.

   We then generalize the
Heisenberg representation (\ref{Oj}) to fields defined on the contour:
\beq\label{phiIz}
\phi_j(z)\,=\, U_j^{-1}(z,t_0)\,\phi\, U_j(z,t_0)\,,\eeq
and define the ordered  product of two such operators by:
\beq\label{TimeC}
{\rm T}_C(\phi_j(z_1)\phi_j(z_2))\,=\,\theta_C(z_1,z_2)
\phi_j(z_1)\phi_j(z_2) + \theta_C(z_2,z_1)\phi_j(z_2)\phi_j(z_1),\eeq
This allows us to extend the definition (\ref{EX})
of the off-equilibrium propagator as follows:
\beq\label{EXCONT}
G(z_1,z_2)&=&
{\rm Tr}\left\{ {\cal D}\,
{\rm T}_C\,\phi_j(z_1)\phi_j(z_2)\right\}\nonumber\\
   &\equiv&\theta_C(z_1,z_2) G^>(z_1,z_2)\,+\,
\theta_C(z_2,z_1) G^<(z_1,z_2).
\eeq

The physical non-equilibrium Green's functions in real-time
(cf. eq.~(\ref{EX})) are  obtained from the corresponding
contour functions  by
choosing appropriately the time arguments on $C_+$ and $C_-$ and
identifying the external source $j(z,{\bf x})$ with the
physical perturbation in eq.~(\ref{Hj1}), i.e.,
$j(z=t) = j(z=t-i\eta)\equiv j(t)$.
Thus, one easily verifies that for the choice $z_1=t_1-i\eta \in C_-$
and $z_2=t_2
\in C_+$, the contour two-point function $G^>(z_1,z_2)$ obtained from
eq.~(\ref{EXCONT}) reduces to the
physical Green's function $G^>(t_1,t_2)$ in eq.~(\ref{GOE}).
Similarly, by choosing  $z_1=t_1\in C_+$ and $z_2=t_2-i\eta\in C_-$
in eq.~(\ref{EXCONT}), one obtains $G^<(t_1,t_2)$, while for
both $z_1$ and $z_2$ on $C_+$ one gets the time-ordered,
or Feynman, propagator $G(t_1,t_2)$ of  eq.~(\ref{EX}).

So far we have not used the part $C_0$ of the contour. It becomes
useful whenever the initial density operator is the canonical density
operator (\ref{canonicalD}). Indeed, as already noticed in Sect.
2.1, this can be
represented as an evolution operator along the contour $C_0$.
Such a representation allows us to treat the statistical average over
the initial state on the same footing as the time evolution, and in
particular to perform approximations on the initial state  which are
consistent with those made on the evolution equation.

With this in mind, we then define the following generating functional:
\beq\label{ZCC}
Z[j]\,\equiv\,{\rm Tr}\, U_j(t_0-i\beta,t_0)\,,\eeq
where $j(z)$
is an arbitrary function along the contour to start with. This
generating functional may be written as
   as the following
path integral:
\beq\label{ZC0}
Z[j]&=&{\rm Tr}\left\{{\rm e}^{-\beta H}
\,{\rm T}_C\exp\left[-i\int_C{\rm d}^4x \,j(x)\phi(x)\right]\right\}
\nonumber\\&=&
{\cal N}
\int_{\phi(t_0)=\phi(t_0-i\beta)}{\cal D}\phi\,\exp\left\{
i\int_C {\rm d}^4x \Bigl({\cal L}(x)-j(x)\phi(x)
\Bigr)\right\},\eeq
where ${\rm d}^4x = {\rm d}z \,{\rm d}^3x$,
and the integral $\int_C {\rm d}z$ runs
from $t_0$ to $t_0-i\beta$ along the contour; the
periodicity conditions at $t_0$ and $t_0-i\beta$ are the same as in
eq.~(\ref{ZCE}).
General (connected) $n$-point contour Green's functions are obtained as
\beq\label{ZGNE}
G^{(n)}(z_1,z_2, ... ,z_n)\,=\,
\frac{i^n\,\delta^n \ln Z[j]}{\delta j(z_1)\delta j(z_2)...\delta j(z_n)}
\,.\eeq
All the manipulations which lead to the perturbative
expansion can now be simply extended to complex time arguments
lying on the contour. To do perturbative calculations,
   we  need the free contour propagator:
\beq\label{GC0}
G_0(z_1,z_2)&\equiv&
\theta_C(z_1,z_2) G_0^>(z_1,z_2)\,+\,\theta_C(z_2,z_1) G_0^<(z_1,z_2)
\nonumber\\ &=&
\int\frac{{\rm d}^4k}{(2\pi)^4}\,{\rm e}^{-i(k_0 z-
{\bf k\cdot x})}\,\rho_0(k)\Bigl[\theta_C(z)+N(k_0)\Bigr]\,,\eeq
where $z=z_1-z_2$, ${\bf x}={\bf x}_1-{\bf x}_2$,
$\rho_0(k)$ is the free  spectral density, eq.~(\ref{rho0}),
and the second line follows from eq.~(\ref{G>t}) for $G^>$
together with a similar equation for $G^<$ (cf. eq.~(\ref{G><k})).
A similar representation, but with $\rho_0(k)$ replaced by
$\rho(k)$ (the full spectral density), holds also for
the exact contour propagator in thermal equilibrium.

Note finally that we can  exploit the analytic properties
of the thermal Green's functions to deform the contour
$C$ in the complex time plane. This is clear at least
in thermal equilibrium, where the analyticity properties
of the functions $G^>(z)$ and $G^<(z)$ discussed in Sect. 2.1.2
imply that the contour 2-point function (i.e., eq.~(\ref{EXCONT})
with $j=0$) is well defined for any contour $C$ such that ${\rm Im}\,z$ is
non-increasing along the contour.  The choice of a specific contour
is a matter of convenience, and different contours may lead
to slightly different formalisms
(see, e.g., \cite{Landsman87,Evans93,MLB96}).
In fact, any such a contour may be viewed as a particular deformation
of the Matsubara contour used in  Sect. 2.1.
For any contour $C$, the Green's functions  satisfy
boundary conditions which generalize the KMS conditions
in equilibrium (cf. eq.~(\ref{KMSG})). For instance:
\beq\label{KMS}
G(t_0,z)&=&G(t_0-i\beta,z)\nonumber\\
G^{(3)}(t_0,z',z'')&=&G^{(3)}(t_0-i\beta,z',z''),\qquad {\rm etc.}\eeq
Under suitable conditions these analytic properties may hold  also for
the {\it non-equilibrium} ($j\ne 0$)
Green's functions \cite{KB62}, although no rigorous
proof can be given in general.

\subsubsection{Equations of motion for Green's functions}

After all these preparations, we are now ready to write down the  equations of
motion satisfied by the contour Green's  functions. The mean field
equation is most easily obtained by taking the expectation value with
the statistical operator of  the equations of motion obeyed by the field
operators in the Heisenberg representation:
\beq\label{av1}
(-\del^2-m^2)\Phi(x)\,-\,
   \left \langle \frac{dV}{d\phi}(x)\right\rangle\,=\,j(x),\eeq
where  $\Phi(x)\equiv \langle \phi(x)\rangle$,
   $\del^2=\del_z^2-\nbfgrad^2$, and the angular brackets
denote expectation values.  Eq.~(\ref{av1}) is conveniently rewritten
as:
\beq\label{av1'}
(-\del^2-m^2)   \Phi(x)
\,=\,j(x)\,+\,j^{ind}(x),\eeq
where:
\beq\label{jind0}
j^{ind}(x)\equiv \left \langle
   \frac{dV}{d\phi}(x)\right\rangle
\eeq
will be called  the {\it induced current} because it
plays, in the r.h.s. of eq.~(\ref{av1}), the same role as
the external current, namely the role of a source for the
average field $\Phi$. For a $\phi^4$-theory, where
$V(\phi)\,=\,({g^2}/{4!})\phi^4$, we have explicitly:
\beq\label{jind4}
j^{ind}(x)\,=\,\frac{g^2}{3!}\left \langle\phi^3(x)
\right\rangle
\,=\,\frac{g^2}{3!}\left(\Phi^3(x)+ G^{(3)}(x,x,x)\right)
+\frac{g^2}{2!}\,\Phi(x) G(x,x)\,.\eeq

By differentiating eq.~(\ref{av1'})
   with respect to $j(y)$, and using
$i({\delta \Phi(x)}/{\delta j(y)})=G(x,y)$,
one obtains an equation for $G(x,y)$:
\beq\label{D1}
(-\del^2_x-m^2)G(x,y)\,-\,i\int_C {\rm d}^4z\,\Sigma(x,z) \,G(z,y)\,=
\,i\delta_C(x,y),\eeq
where $\delta_C(x_0,y_0)$ is the contour delta function, 
eq.~(\ref{contdelta}), and the self-energy $\Sigma$ is given by:
\beq\label{defSig}
   \Sigma(x,y)\equiv -i\,\frac{\delta j^{ind}(x)}{\delta \Phi(y)}
\,=\,-i\,\frac{\delta}{\delta \Phi(y)}
   \left \langle \frac{dV}{d\phi}(x)\right\rangle\,.\eeq
In writing eq.~(\ref{D1}), the
following chain of identities has been used:
\beq
i\,\frac{\delta j^{ind}(x)}{\delta j(y)}\,=\,
i\int_C {\rm d}^4z\,\frac{\delta j^{ind}(x)}{\delta \Phi(z)}\,
\frac{\delta \Phi(z)}{\delta j(y)}\,=\,
i\int_C {\rm d}^4z\,\Sigma(x,z) \,G(z,y)\,.\eeq
For a free theory ($\Sigma=0$), the solution to eq.~(\ref{D1}) with the
KMS boundary conditions (i.e., the free contour propagator) is given by
eq.~(\ref{GC0}).

The self-energy (\ref{defSig}) admits  the following
decomposition,  similar to that of $G$, eq.~(\ref{D><}):
\beq\label{Sig<>}
\Sigma(x,y)\,=\,-i\Sigma^\delta(x)\delta_C(x,y)\,+\,
\theta_C(x_0,y_0) \Sigma^>(x,y)
+\theta_C(y_0,x_0) \Sigma^<(x,y).\eeq
We have isolated here a possible singular piece
$\Sigma^\delta$. For instance, for the $\phi^4$ theory,
\beq\label{SDEL}
\Sigma^\delta(x)\,=\,
\frac{g^2}{2}\left(\Phi^2(x) \,+\,G(x,x)\right).\eeq
The non-singular components obey the KMS condition:
\beq\label{KMSSIG} \Sigma^<(t_0,z)\,=\, \Sigma^> (t_0-i\beta,z),\eeq
a consequence of the definition (\ref{defSig}), and
of the conditions (\ref{KMS})
which are  satisfied by the $n$-point Green's functions.

The equations of motion in real-time for the mean field and the
2-point functions are obtained by choosing the contour in fig.~\ref{CONT}
and letting the external time variables $x_0$ and $y_0$
   take values on the real-time pieces of this contour.
For $x_0\in C_+$, eq.~(\ref{av1'}) goes formally unmodified:
\beq\label{avt}
-\left(\del^2+m^2\right)
    \Phi(x) \,=\,j(x)\,+\,j^{ind}(x).\eeq

Consider now eq.~(\ref{D1}):
by choosing $x_0\in C_+$ and $y_0\in C_-$,
and by using the decompositions (\ref{D><})
and (\ref{Sig<>}),
we obtain an equation for $G^<(x,y)$:
\beq\label{eqG<}
\Bigl(\del^2_x+m^2+\Sigma^{\delta}(x)\Bigr)
G^<(x,y)&=&-i\int_{t_0}^{x_0} {\rm d}^4z\,
\Bigl[\Sigma^>(x,z)-\Sigma^<(x,z)\Bigr] \,G^<(z,y)
\nonumber \\
&+&i\int_{t_0}^{y_0} {\rm d}^4z\,\Sigma^<(x,z) \,\Bigl[G^>(z,y)
-G^<(z,y)\Bigr]\nonumber \\
&-&i\int_{t_0}^{t_0-i\beta} {\rm d}^4z\,\Sigma^<(x,z)G^>(z,y),\eeq
where the first two integrals in the r.h.s. run along
the real axis and we have isolated in the third integral
the  contribution from
the imaginary-time piece of the contour. A similar equation for $G^>$
follows similarly if, starting from  eq.~(\ref{D1}), one chooses
$x_0\in C_-$ and $y_0\in C_+$. 

It is instructive to consider the restriction of the equation above to the case
of equilibrium, and in particular to verify that it is then independent of the
inital time
$t_0$, as it should. In equilibrium,  the various
propagators and self-energies are expected to be functions only of time
differences and to admit spectral representations like eq.~(\ref{G>t}) for
$G^>_{eq}(x-z)$ and, similarly (cf. eq.~(\ref{SIG><k})),
\beq\label{S>t}
\Sigma^>_{eq}(x-z)\,=\,-
\int\frac{{\rm d}^4k}{(2\pi)^4}\,{\rm e}^{-ik\cdot (x-z)}\,
\Gamma(k)\Bigl[1+N(k_0)\Bigr]\,.\eeq
It is easy to verify that such representations are indeed
consistent with eq.~(\ref{eqG<}). In fact, by
inserting these representations in eq.~(\ref{eqG<}), and using 
properties like ${\rm e}^{-\beta k_0}[1+N(k_0)]=N(k_0)$, one finds
that, in thermal equilibrium, this equation is
independent of the initial time $t_0$, and it can be rewritten in
momentum space as:
\beq\label{eqG<eq}
\Bigl(-k^2+m^2+\Sigma^{\delta}\Bigr)
G^<(k)\,=\,-\Sigma_R(k) \,G^<(k) - \Sigma^<(k)\,G_A(k)\,
\eeq
where we have used the  definitions (\ref{A}) for the retarded and advanced
Green's functions, together with similar definitions for $\Sigma_R$ and
$\Sigma_A$. This equation is solved indeed by $G^<(k)=\rho(k)N(k_0)$
with $\rho(k)$ given by eq.~(\ref{rholor}).

The  equation   (\ref{eqG<}) and the corresponding one for $G^>$
could have been obtained also by  analytic continuation
of the {\it imaginary-time} Dyson-Schwinger equations \cite{KB62}.
Specifically, one may start with eq.~(\ref{D1})  written along
the Matsubara contour, that is
(with $x_0=t_0-i\tau_x$ and $y_0=t_0-i\tau_y$):
\beq\label{D1M}
\left(-\del_{\tau_x}^2-\nbfgrad_{\bf x}^2+m^2+
\Sigma^\delta(x)\right)G(x,y)\,+\,
\int_0^\beta {\rm d}\tau_z\int {\rm d}^3z\,\Sigma(x,z) \,G(z,y)
\nonumber\\=\,\delta(\tau_x-\tau_y)\delta^{(3)}({\bf x-y}),
\eeq
and then deform the contour in the complex time plane,
by exploiting the analytic properties
of the non-equilibrium Green's functions (see previous subsection).
This simple technique will be used in connection with
gauge theories, in sections 3 and 7 below \cite{qed,qcd,BE}.

In what follows it will often be convenient to let $t_0\to -\infty$
and to assume that the external sources are switched off adiabatically
in the remote past. Then, for fixed values of the real-time arguments
$x_0$ and $y_0$, and for any $z_0$ on the vertical piece of the contour,
the real parts of the time differences $x_0-z_0$ and  $y_0-z_0$
go to infinity. In this limit,
the 2-point correlations $G^>(z,y)$ and $G^>(z,x)$ are expected
to die away sufficiently fast, for the contributions of
the imaginary-time integrals in eq.~(\ref{eqG<})
to become negligible \cite{KB62}. In fact, for the kind of non-equilibrium
situations to be considered below, and which involve only
longwavelength perturbations, the correlation function
$G^>(z,y)$ is dominated by hard degrees of freedom ($k\sim T$),
and decays over a characteristic range $|x_0-y_0|\sim 1/T$
(cf. eq.~(\ref{largex})).
Thus, neglecting the imaginary-time integrals
in eq.~(\ref{eqG<})  is justified as soon as  $x_0\gg 1/T$ or $y_0\gg 1/T$.
We are thus led to the following set of equations:
\beq\label{eqsG}
\Bigl(\del^2_x+m^2+\Sigma^{\delta}(x)\Bigr)
G^<(x,y)&=&-\int_{-\infty}^{\infty} {\rm d}^4z\,
\Bigl[\Sigma_R(x,z) \,G^<(z,y) \,+\,\Sigma^<(x,z)\,G_A(z,y)\Bigr]\nonumber \\
\Bigl(\del^2_x+m^2+\Sigma^{\delta}(x)\Bigr)
G^>(x,y)&=&-\int_{-\infty}^{\infty} {\rm d}^4z\,
\Bigl[\Sigma_R(x,z) \,G^>(z,y) \,+\,\Sigma^>(x,z)\,G_A(z,y)\Bigr]\,,
\nonumber\\\eeq
where we have extended the  definitions of the 
retarded and advanced Green's functions and self-energies
to non-equilibrium situations.
The presence of these functions has allowed us to extend the upper
bound of the $z_0$ integral to $+\infty$.
(In equilibrium, the Fourier transform of the first equation
(\ref{eqsG}) coincides with eq.~(\ref{eqG<eq}), as it should.)

By taking the difference of the two equations above, one obtains an
equation satisfied by the
retarded propagator $G_R(x,y)$ (cf. eq.~(\ref{A})):
\beq\label{eqGR}
\Bigl(\del^2_x+m^2+\Sigma^{\delta}(x)\Bigr)
G_R(x,y)\,+\,\int_{-\infty}^{\infty} {\rm d}^4z\,\Sigma_R(x,z) \,G_R(z,y) \,
=\,\delta^{(4)}(x-y)\,,\eeq
together with an independent equation where the differential
operator in the l.h.s. is acting on $y$ rather than on $x$.
Note that, while the correlation functions $G^>$ and  $G^<$ and the
corresponding self-energies are coupled by eqs.~(\ref{eqsG}),
the retarded Green's
function $G_R$ is determined by the retarded self-energy $\Sigma_R$ alone.
A similar observation applies to the advanced functions
$G_A$ and $\Sigma_A$.

Eqs.~(\ref{eqsG}) and (\ref{eqGR}) are
   the equations obtained by Kadanoff and Baym \cite{KB62},
   in the framework
of non-relativistic many-body theory. In these equations,
any explicit reference to the initial conditions has disappeared.
Thus the KMS conditions only enter as boundary conditions to be satisfied
by the various Green's functions in the remote past.
The same set of equations has been shown
by Keldysh  \cite{Keldysh64} to describe the general non-equilibrium
evolution of a quantum system, with the density matrix  of the initial state
determining the appropriate boundary conditions 
(see also \cite{PhysKin,Daniel83}).

To make progress, the above equations must be supplemented
with some approximation allowing us to express the
   self-energy $\Sigma$ in terms of
the propagator $G$. This generally results in complicated, non-linear and
integro-differential, equations for $G$. Moreover, in off-equilibrium
situations, we generally loose translational invariance, so
we cannot analyze these equations with the help of
Fourier transforms. However, for  slowly varying (or soft) off-equilibrium
perturbations, these equations can be transformed into kinetic equations
\cite{KB62,Daniel83,Calzetta88,MD90,Heinz94},
as it will be  explained in Sect. 2.3 below.

\subsubsection{Correlation functions in the classical field
approximation}

There are situations where one wishes to evaluate the real time correlation
functions in the classical field approximation (cf. Sect. 2.1.4). 
Although the techniques developed
above could in principle be used, it is more efficient to proceed differently.
Consider for instance the  calculation of the  correlation
function
\beq\label{GCLDEF}
G_{cl}(x,y)\,\equiv\,\langle \Phi_{cl}(x) \Phi_{cl}(y)\rangle\,,\eeq
where the brackets denote the classical thermal averaging (cf.
eq.~(\ref{GCL}) below), and 
$\Phi_{cl}(t,{\bf x})$ and $\Pi_{cl}(t,{\bf x})$ are classical fields,
whose time dependence is obtained by solving the classical
equations of motion,
\beq\label{GEQ}
\del_0 \Pi \,=\,-\frac{\del {\cal H}}{\del \Phi}\,,\qquad
\del_0 \Phi \,=\,\frac{\del {\cal H}}{\del \Pi}\,=\,\Pi,\eeq
with  $\Pi$  the field  canonically conjugate to $\Phi$, and $H$ the classical
Hamiltonian
\beq\label{HAM}
H\,=\,\int{\rm d}^3x\,\left\{{1\over2}\,\Pi^2\,+\,{1\over2}(\nbfgrad\Phi)^2\,
+\,\frac{M^2}{2}\,\Phi^2+\,V(\Phi)\right\}\,\equiv\,\int{\rm d}^3x\,{\cal H}
({\bf x})\, .\eeq
Note that $V$ may contain  a perturbation which drives the system out of
equilibrium. Initially however the perturbation vanishes and the system is in
thermal equilibrium. To be specific we shall denote by $H_{eq}$ the 
corresponding Hamiltonian. The  initial conditions are
\beq\label{GIN}
\Phi_{cl}(t_0,{\bf x})=\Phi({\bf x}),\qquad
\Pi_{cl}(t_0,{\bf x})=\Pi({\bf x})\,, \eeq
and the classical field configurations $\Phi({\bf x})$ and $\Pi({\bf x})$ are
statistically distributed according to the Boltzmann weight ${\rm e}^{-\beta
H_{eq}}$. The correlator (\ref{GCLDEF}) is then obtained by averaging over the
initial conditions according to
\beq\label{GCL}
G_{cl}(t,{\bf x}, t',{\bf y})=Z_{cl}^{-1}
\,\int\,{\cal D}\Pi({\bf x})\,{\cal D}\Phi({\bf x})\,\,
\Phi_{cl}(t,{\bf x})\,\Phi_{cl}(t',{\bf y})\,
{\rm e}^{-\beta H_{eq}(\Pi, \Phi)}
\eeq
where $Z_{cl}$ is
  the classical partition
function:
\beq\label{ZCLS}
Z_{cl}\,=\,\int\,{\cal D}\Pi({\bf x})\,{\cal D}\Phi({\bf x})\,
\,{\rm e}^{-\beta H_{eq}(\Pi, \Phi)}.\eeq

As an application of eq.~(\ref{GCL}), let us compute
$G_{cl}(x,y)$ for a free scalar field ($V=0$).
The solution $\Phi_{cl}(x)$ to the free equations of motion,
\beq
(\del_0^2 - \nbfgrad^2+M^2)\Phi(x)\,=\,0,\eeq
with the initial conditions (\ref{GIN}) reads
\beq
\Phi_{cl}(t,{\bf x})=\int\frac{{\rm d}^3k}{(2\pi)^3}\,{\rm e}^{i{\bf k\cdot x}}
\,\biggl\{\Phi({\bf k})\cos \varepsilon_k t + \Pi({\bf k})\,\frac{\sin
   \varepsilon_k t}{\varepsilon_k}\biggr\},\eeq
with $\varepsilon_k = \sqrt{k^2+M^2}$, and
$\Pi({\bf k})$ the Fourier transform of $\Pi({\bf x})$, etc.
In this case, the functional integral in eq.~(\ref{GCL})
can be exactly computed since Gaussian, and yields:
\beq\label{GCL0}
G_{cl}^{\,0}(x,y)\,\equiv\,
T\int\frac{{\rm d}^3k}{(2\pi)^3}\,{\rm e}^{i{\bf k\cdot
(x-y)}}\,\frac{\cos \varepsilon_k (x_0-y_0)}{\varepsilon_k^2}
\,=\, \int\frac{{\rm d}^4k}{(2\pi)^4}\,\,{\rm e}^{-i k\cdot(x-y)}\,\,
{T\over k_0}\,\rho_0(k),\eeq
where $\rho_0(k)$ is the free spectral density (\ref{rho0}).
One recognizes in Eq.~(\ref{GCL0}) the classical limit of the
  correlators in eq.~(\ref{G><k}). Indeed, at soft momenta, 
$N(k_0)\simeq T/k_0 \gg
1$, and therefore
\beq\label{G><CL}
G^>_0(k)\,\simeq\,G^<_0(k)\,\simeq\,{T\over k_0}\,\rho_0(k)
\,=\,G_{cl}^{\,0}(k)\,.\eeq
Note that in the classical field approximation the spectral function is still
related to the imaginary part of the retarded propagator as in
eq.~(\ref{spectraldef}), but is no longer given by the difference of 
the functions $G^>(k)$ and $G^<(k)$ (see eq.~(\ref{defrho})).

In the presence of interactions, 
the averaging over initial conditions using the
functional integral (\ref{GCL})
will develop ultraviolet divergences, so these
make sense only if supplemented with an UV cutoff $\Lambda$.
This situation is quite similar to that discussed in Sect. 2.1.4  for the
static case, and will not be discussed further here 
(see Refs. \cite{McLerran,AM97} for more details).

\subsection{Mean field and kinetic equations}
We are now ready to implement in the scalar case the general 
approximation scheme
that will be used in the rest of this paper for gauge theories. Starting from
the general equations for the Green's functions that we have derived
in the previous subsection, we specialize to longwavelength perturbations and
use a gradient expansion to reduce the general equations of motion to simpler
kinetic equations.
The self-energies in these equations are obtained through weak
coupling expansion  combined with  mean field approximations. 
Assuming furthermore
weak deviations from the equilibrium, we then arrive at a closed 
system describing
the dynamics of the longwavelentgth excitations of the system in the high
temperature limit. Then, we analyze the role of the collisions in the 
damping of
single particle excitations. In the last subsection we reconsider the kinetic
equations  in the time representation; this will be useful later when 
dealing with
problems where
  coherence effects are important.

\subsubsection{Wigner functions}

In thermal equilibrium, the system is homogeneous, the average
field vanishes (we do not consider here the possibility of spontaneous
symmetry breaking), and the two-point functions
depend only on the relative coordinates $s^\mu=x^\mu-y^\mu$.
In the high temperature limit $T\gg m$, the thermal particles
have typical momenta $k\sim T$ and typical energies $\varepsilon_k \sim T$;
the 2-point functions are peaked around
$s^\mu=0$, their range of variation being determined by the thermal
wavelength $\lambda_T=1/k\sim 1/T$ (cf. eq.~(\ref{largex})).

In what follows, we  are interested in off-equilibrium deviations
   which are slowly varying in space and time. That is, we assume that
the system  acquires space-time inhomogeneities over
a typical scale $\lambda \gg \lambda_T$. The  field $\phi$
   develops then a  non-vanishing average value $\Phi(x)$,
   and the 2-point functions
depend on both coordinates $x$ and $y$. It is then convenient
to introduce relative and central coordinates:
   \beq\label{Rel}
   s^\mu\equiv x^\mu-y^\mu,\qquad\qquad
X^\mu\equiv{x^\mu+y^\mu\over2}\,,\eeq
and use the {\it Wigner transforms} of the 2-point functions. These are
defined as Fourier transforms with respect to the
relative coordinates $s^\mu$. For instance, the Wigner transform
of $G^<(x,y)$ is:
\beq\label{G<WIG}
   { G}^<(k,X)\equiv\int {\rm d}^4s
\,{\rm e}^{ik\cdot s} \,G^<\left(X+{s\over 2},X-{s\over 2}\right), \eeq
with similar definitions for the other 2-point functions like $G^>$,
$G_R$, $G_A$ and the various self-energies.
Note that in order to avoid the proliferation of symbols, we use the 
same symbols
for the 2-point functions and their Wigner transforms, considering that the
different functions can be recognized from their arguments.

The hermiticity properties of the 2-point functions, as
discussed in Sect. 2.1.2  (cf. eqs.~(\ref{HERM}) and (\ref{A})),
imply similar properties for the corresponding Wigner functions.
For instance, from $(G^>(x,y))^*=G^>(y,x)$ we deduce that
$G^<(k,X)$ is a real function, $(G^<(k,X))^*=G^<(k,X)$, as
in thermal equilibrium, and similarly for $G^>(k,X)$.
Also, $(G_A(k,X))^*=G_R(k,X)$. Similar properties
hold for the various self-energies. Moreover, for a {\it real} scalar
field, we have the additional relations $G^>(k,X)=G^<(-k,X)$ and
$G_A(k,X)=G_R(-k,X)$, which follow since
   $G^>(x,y)=G^<(y,x)$ and $G_A(x,y)=G_R(y,x)$
(cf. eqs.~(\ref{G>G<}) and (\ref{A})).

For slowly varying disturbances,  taking place over a scale
$\lambda \gg \lambda_T$, we expect the $s^\mu$ dependence of
the 2-point functions to be close to that in equilibrium.
Thus, typically, $k\sim\del_s \sim T$, while $\del_X \sim 1/\lambda \ll T$.
   The general equations of motion written down
in Sect. 2.2.3 can then be simplified with the help of
a {\it gradient expansion}, using $k$ and $X$ as most
convenient variables.

\subsubsection{Kinetic equations}

We shall construct below the equation satisfied by $ { G}^<(k,X)$
to leading order in the gradient expansion.
The starting point is eq.~(\ref{eqsG}) for $G^<(x,y)$,
namely,
\beq\label{eq1}
\Bigl(\del^2_x+m^2+\Sigma^{\delta}(x)\Bigr)
G^<(x,y) \,=\,-\int {\rm d}^4 z\,
\Bigl[\Sigma_R(x,z) \,G^<(z,y)
\,+\,\Sigma^<(x,z)\,G_A(z,y)\Bigr],\nonumber\\\eeq
together with an analogous equation where the differential
operator is acting on $y$:
\beq\label{eq2}
\Bigl(\del^2_y+m^2+\Sigma^{\delta}(y)\Bigr)
G^<(x,y)\,=\,-\int  {\rm d}^4 z\,
\Bigl[G^<(x,z)\,\Sigma_A(z,y)
\,+\,G_R(x,z)\,\Sigma^<(z,y)\Bigr].\,\nonumber\\\eeq
(To obtain eq.~(\ref{eq2}), start with the second
eq.~(\ref{eqsG}) for $G^>(x,y)$, interchange the space-time
variables $x^\mu$ and $y^\mu$, and use symmetry properties
like $G^>(y,x)=G^<(x,y)$, $G_A(z,x)=G_R(x,z)$, etc.)
When the system is inhomogeneous,  the 2-point functions like
$G^<(x,y)$ depend separately on the two arguments $x$ and  $y$,
so that the two equations
(\ref{eq1}) and (\ref{eq2}) are independent.

In order to carry out  the gradient expansion, we consider the difference of
eqs.~(\ref{eq1}) and (\ref{eq2}), to be briefly referred to as
{\it the difference equation} in what follows.
After replacing  $x$ and $y$ by the
coordinates $s$ and $X$ (see eq.~(\ref{Rel})), we rewrite the derivatives as
\beq\label{DERIV}
\del_x=\del_s+\half\del_X,\qquad\,\, \del_y=-\del_s+\half\del_X
\qquad\,\, \del_x^2-\del_y^2=2\del_s\cdot\del_X,\eeq  and
perform an  expansion in powers of $\del_X$, keeping only the terms
involving at most one soft derivative $\del_X$. For instance,
\beq
\Sigma^{\delta}(x)-\Sigma^{\delta}(y)=
\Sigma^{\delta}\left(X+{s\over 2}\right)-
\Sigma^{\delta}\left(X-{s\over 2}\right)\simeq (s\cdot \del_X)
\Sigma^{\delta}(X)\,.\eeq
We then perform a Fourier transform $s^\mu \to k^\mu$ and get
   an equation involving Wigner functions.
By Fourier transform,
\beq
(s\cdot \del_X)\Sigma^{\delta}(X) \,\longrightarrow\,-i\left(\del^\mu_X
\Sigma^\delta\right)\del_\mu^k\,.\eeq
Furthermore, it is easily verified that
the convolutions in the r.h.s. of eqs.~(\ref{eq1})---(\ref{eq2}) transform as:
\beq\label{AOB}
\int  {\rm d}^4 z\,A(x,z)\,B (z,y)\,
\longrightarrow\,A(k,X) B(k,X)\,+\,
\frac{i}{2}\,\Bigl\{A,\, B\Bigr\}_{P.B.}\,+\,...\,,\eeq
where $\{A,B\}_{P.B.}$ denotes a  Poisson bracket:
\beq\label{poisson}
\Bigl\{A,\,{ B}\Bigr\}_{P.B.}\equiv \del_k A\cdot\del_X{ B}
\,-\,\del_X A\cdot\del_k{ B}\,,\eeq
and the dots  stand for terms which
   involve at least two powers of the soft derivative. Thus, 
the difference equation involves, for instance,
\beq
\int {\rm d}^4 z
\Bigl[\Sigma_R(x,z)G^<(z,y)-
G^<(x,z)\Sigma_A(z,y)\Bigr] \longrightarrow
(\Sigma_R-\Sigma_A)G^<+\,\frac{i}{2}\Bigl\{\Sigma_R+
\Sigma_A,G^<\Bigr\}_{P.B.},\nonumber\\\eeq
where all the functions  in the r.h.s. are Wigner transforms
(i.e., they are functions of $k$ and $X$).

At this stage, it is convenient to introduce
   the following Wigner functions:
\beq\label{NEQAG}
\rho(k,X)&\equiv& { G}^>(k,X) - { G}^<(k,X)\nonumber\\
\Gamma(k,X)&\equiv& \Sigma^<(k,X) - \Sigma^>(k,X)
\,,\eeq
which provide non-equilibrium generalizations
of the spectral densities $\rho(k)$, eq.~(\ref{defrho}), and
$\Gamma(k)$, eq.~(\ref{defGamma}).
In terms of them, the Wigner transforms $G_R(k,X)$ and
$\Sigma_R(k,X)$ admit the following representations:
\beq\label{RETWIG}
   G_R(k,X)=\int_{-\infty}^\infty \frac{{\rm d}k_0^\prime}{2\pi}\,
\frac{\rho(k_0^\prime,{\bf k}, X)}{k_0^\prime-k_0-i\eta}\,,\qquad
\Sigma_R(k,X)=-\int_{-\infty}^\infty \frac{{\rm d}k_0^\prime}{2\pi}\,
\frac{{\Gamma}(k_0^\prime,{\bf k}, X)}{k_0^\prime-k_0-i\eta}\,.\eeq
Similar relations (with $-i\eta \to i\eta$) hold
for the corresponding advanced  functions.
Note also the relations:
\beq
   G_R(k,X)- G_A(k,X)&=&i\rho(k,X),\nonumber\\
G_R(k,X)+G_A(k,X)&=&2\,{\rm Re}\, G_R(k,X).\eeq
Similar relations hold for the self-energies $\Sigma_R$, $\Sigma_A$,
and $\Gamma$. By using these relations, and the manipulations indicated above,
the difference equation reduces to the following equation for $G^<(k,X)$:
\beq\label{kinG<}
2(k\cdot \del_X)\,{ G}^<\,+\,\left(\del^\mu_X
\Sigma^\delta\right)\del_\mu^k { G}^<\,=\,
- \Gamma { G}^< - \rho \Sigma^<
+\Bigl\{{\Sigma}^<,\,{\rm Re}\,G_R\Bigr\}_{P.B.}
+\Bigl\{{\rm Re}\,\Sigma_R,\,{ G}^<\Bigr\}_{P.B.}\,.\nonumber\\\eeq
By using the definition (\ref{poisson}) of the Poisson bracket,
together with the identity $\Gamma G^< + \rho \Sigma^< = G^>\Sigma^<
- \Sigma^>{ G}^<$ (cf. eq.~(\ref{NEQAG})), one finally rewrites
this equation as\footnote{The first two terms in the l.h.s.
of eq.~(\ref{kinG1}) can be recognized as the Poisson bracket
$-\,\bigl\{{\rm Re}\,G_R^{-1},\,G^<\bigr\}_{P.B.}$, where
$G_R^{-1}(k,X)\equiv -k^2+m^2+\Sigma_R(k,X)$; see eq.~(\ref{Gretoff})
below.} :
\beq\label{kinG1}
\left(2k^\mu - \frac{\del {\rm Re}\Sigma}{\del k_\mu}\right)
\frac{\del G^<}{\del X^\mu}&+&\frac{\del {\rm Re}\Sigma}
{\del X_\mu}\,\frac{\del G^<}{\del k^\mu} \,-\,
\Bigl\{{\Sigma}^<,\,{\rm Re}\,G_R\Bigr\}_{P.B.}=\,
-\Bigl(G^>\Sigma^<- \Sigma^>{ G}^<\Bigr),
\nonumber\\&{}&\eeq
where ${\rm Re}\Sigma \equiv {\rm Re}\Sigma_R+\Sigma^\delta$.
Eq.~(\ref{kinG1})  holds to leading order in the gradient expansion
(that is, up to
terms involving at least two powers of the soft derivative),
   and to all orders in the interaction strength.

In equilibrium, both sides of eq.~(\ref{kinG1}) are identically zero.
This is obvious for the terms in the l.h.s.,
which involve the soft derivative $\del_X$, and can be easily
verified for the terms in the r.h.s. by using the KMS conditions
for $G$ and $\Sigma$ (cf. eq.~(\ref{KMSk})).
Thus eq.~(\ref{kinG<}) describes the off-equilibrium
inhomogeneity in ${ G}^<(k,X)$, and can be seen
as a quantum generalization of the Boltzmann equation (see below).
The Wigner function
$G^<(k,X)$ plays here the role of the phase-space
distribution function $f({\bf k}, X)$. The drift term on the
l.h.s. of eq.~(\ref{kinG1}) generalizes the usual kinetic drift term
$\del_t+{\bf v}\cdot\bfgrad_X$ by including self-energy corrections:
The real part
of the self-energy acts as an effective potential whose space-time derivative
provides a  ``force'' term $(\del^\mu_X {\rm Re}\Sigma)
(\del_\mu^k { G}^<)$. The momentum dependence of $\Sigma$ modifies
the ``velocity"
of the particles: $v^\mu\to v^\mu-(1/2k_0) \del_k^\mu {\rm Re}\Sigma$.
The terms on the
r.h.s. describe collisions.
We shall see below that, for on-shell excitations, these collision 
terms acquire
the standard Boltzmann form. Finally,  the Poisson bracket
$\{{\Sigma}^<,\,{\rm Re}\,G_R\}_{P.B.}$
has a less transparent physical interpretation, which should be
clarified, however, by the following discussion of the spectral density.

The off-equilibrium spectral density $\rho(k,X)$ is most easily
obtained from the retarded propagator $G_R(k,X)$
(cf. eq.~(\ref{RETWIG})), for which we can get  kinetic equations. These are
derived in the same way as above,
   by performing a gradient expansion in eq.~(\ref{eqGR})
and an analogous  equation involving $\del_y^2$. Unlike the previous
calculation, however,
the gradient expansion is performed here on the {\it sum} of the two
equations for $G_R(k,X)$. One gets then:
\beq\label{Gretoff}
\Bigl(k^2-m^2-\Sigma^\delta(X) - \Sigma_R(k,X)\Bigr)\,G_R(k,X)\,=\,
-1\,.\eeq
This equation contains {\it no} soft derivative $\del_X$ (the first corrections
   involve at least {\it two} powers of the soft gradients).
Accordingly, the retarded off-equilibrium Green's function
   $G_R(k,X)$
is related to the corresponding self-energy $\Sigma_R(k,X)$
in the same way as  the respective functions in equilibrium (recall
eq.~(\ref{Greteq})). In particular, the associated spectral density
is the straightforward generalization of eq.~(\ref{rholor}),  namely:
\beq\label{AOE}
\rho (k,X)\,=\,\frac{\Gamma(k,X)}{\Bigl(k^2
-m^2-\Sigma^\delta(X) - {\rm Re}\,\Sigma_R(k,X)\Bigr)^2\,+\,
\Bigl(\Gamma(k,X)/2\Bigr)^2}\,.\eeq
The off-equilibrium inhomogeneity  enters eqs.~(\ref{Gretoff})
and (\ref{AOE}) only via their parametric dependence on $X$
\cite{KB62,Daniel83}.

It is also useful to note that
$\rho(k,X)= { G}^>(k,X) - { G}^<(k,X)$ satisfies a kinetic
equation which follows
from  eq.~(\ref{kinG<}) for ${ G}^<(k,X)$ together with a
corresponding equation for $G^>(k,X)$. This reads:
\beq\label{kinrho}
\left(2k^\mu - \frac{\del {\rm Re}\Sigma}{\del k_\mu}\right)
\frac{\del \rho}{\del X^\mu}&+&\frac{\del {\rm Re}\Sigma}
{\del X_\mu}\,\frac{\del \rho}{\del k^\mu}
\,=\,-\,
\Bigl\{\Gamma,\,{\rm Re}\,G_R\Bigr\}_{P.B.}\,.\eeq
(It is straightforward to verify that eq.~(\ref{AOE}) satisfies 
indeed this kinetic
equation.)
Remarkably, the collision terms have mutually canceled in the difference
of the two equations for $G^<$ and $G^>$.
The terms in the l.h.s. of eq.~(\ref{kinrho}) describe drift and mean
field effects, as discussed in connection with eq.~(\ref{kinG1}).
The Poisson bracket in the r.h.s. (the difference
of the corresponding P.B.'s in the equations for $G^<$ and $G^>$)
accounts for the off-equilibrium inhomogeneity in the width
$\Gamma(k,X)$. In fact, if this term is neglected,
then the corresponding solution of eq.~(\ref{kinrho}) is simply
   \beq\label{rhoPB}
\rho(k,X) \,=\, 2\pi\epsilon(k_0)\delta\Bigl(k^2-m^2-{\rm Re}\Sigma(k,X)
\Bigr).\eeq
This defines a {\it quasiparticle approximation}, to be further
discussed in Sects. 2.3.3 and 2.3.4 below. Conversely, whenever one needs
to go beyond such an approximation and include finite width effects,
one has to also take into account the Poisson brackets, for consistency
\cite{Bornath,Lipavsky}. For instance, the role of the P.B.'s for insuring
conservation laws in systems with broad resonances is discussed in Ref.
\cite{Knoll98}.

\subsubsection{Mean field approximation}

A mean field approximation is obtained if we neglect the interactions
among the particles beyond their interactions with the average 
fields, that is, in
particular, if we neglect the collision terms.
  From the point of view of the Dyson-Schwinger equations, this corresponds
to a truncation of the hierarchy at the level of the 2-point functions:
all the connected $n$-point functions with $n\ge 3$
are set to zero. Thus, in  the kinetic
equations (\ref{kinG<})--(\ref{kinG1}), we shall neglect
all self-energy terms except for the tadpole
$\Sigma^\delta$, eq.~(\ref{SDEL}).

We thus get the following closed set of equations
for the mean field $\Phi$ and the Wigner function ${ G}^<(k,X)$:
\beq\label{mfa1}
-\left(\del_X^2+ m^2 \right)  \Phi(X) \,=\,j(X)\,+\,
j^{ind}(X)\,,\eeq
   \beq\label{mfa2}
\left[k\cdot \del_X +\,\frac{1}{2}(\del^\mu_X
\Sigma^\delta)\del_\mu^k\,\right] { G}^<(k,X)\,=\,0\,,\eeq
where the induced current is (cf. eq.~(\ref{jind4})):
\beq\label{jMFA}
j^{ind}(X)\,=\,\frac{g^2}{6}\,\Phi(X)\Bigl(\Phi^2(X)\,+
\,3 G^<(X,X)\Bigr)\,.\eeq
In this equation, $G^<(X,X)$ denotes the function
$G^<(x,y)$ for $x=y=X$, that is, the integral over $k$ of the Wigner
transform $G^<(k,X)$.

The spectral density of the hard quasiparticles ($k\sim T$)
in the mean field approximation follows from eq.~(\ref{Gretoff})
with $\Sigma_R(k,X)=0$.  We obtain (cf.
eq.~(\ref{AOE})):
   \beq\label{MFAA}
\rho(k,X) \,=\, 2\pi\epsilon(k_0)\delta\Bigl(k_0^2-E_{\bf k}^2(X)\Bigr),\eeq
with $E_{\bf k}^2(X)\equiv {\bf k}^2 + M^2(X)$ and:
\beq\label{M20}
M^2(X)&\equiv& m^2+\Sigma^\delta(X)\,=\,m^2+
\frac{g^2}{2}\left(\Phi^2(X) \,+\,G^<(X,X)\right)\,.\eeq
Thus, the mean field approximation automatically leads
to a quasiparticle approximation.

The spectral density (\ref{MFAA})  satisfies (cf.
eq.~(\ref{kinrho})):
\beq\label{kinrhoMFA}
\left[k\cdot \del_X +\,\frac{1}{2}(\del^\mu_X
\Sigma^\delta)\del_\mu^k\right]\rho(k,X)\,=\,0\,.\eeq
By using this equation, it is easy to see that the
solution to eq.~(\ref{mfa2}) can be written as
\beq\label{solG<}
{ G}^<(k,X)&=&\rho(k,X)\,N(k,X)\\
&=&2\pi\delta(k_0^2-E_{\bf k}^2(X))\left\{
\theta(k_0)N({\bf k},X)+\theta(-k_0)[N(-{\bf k},X)+1]\right\},
\nonumber\eeq
where we have separated, in the second line, the positive and negative
energy components of  the on-shell Wigner function $N(k,X)$.
The structure of the second line follows by using
$G^>(k,X)=G^<(k,X)+\rho(k,X)$, together with the symmetry property
$G^>(k,X)=G^<(-k,X)$.
   The density matrix  $N({\bf k},X)$
satisfies a kinetic equation analogous to the Vlasov equation:
\beq\label{scVL}
\Bigl (\del_t\,+\,{\bf v}_k\cdot\nbfgrad_{\bf x}\,-\,
\nbfgrad_{\bf x} E_k\cdot \nbfgrad_{\bf k}\Bigr) N({\bf k},t,{\bf x})\,
=\,0\,,\eeq
and can be interpreted as a phase-space distribution function for
quasiparticles with momentum ${\bf k}$ and energy $E_k(X)$.
In eq.~(\ref{scVL}), ${\bf v}_k(X)={\bf k}/E_k(X)=\nbfgrad_{\bf k} E_k(X)$
is the quasiparticle velocity, and the spatial gradient of the
quasiparticle energy $E_k(X)$ acts like a force on the quasiparticle.

Since, for a given field configuration $\Phi(X)$,
   the quasiparticle mass squared  $M^2(X)$ depends
on the distribution functions, via eq.~(\ref{M20}), the kinetic equations
should in principle be solved simultaneously with the ``gap equation'':
\beq\label{GAP}
M^2(X)&=&m^2+
\frac{g^2}{2}\Phi^2(X) \,+\frac{g^2}{2}\int\frac{{\rm d}^4k}{(2\pi)^4}\,
{ G}^<(k,X)\nonumber\\
&=&m^2+
\frac{g^2}{2}\Phi^2(X) \,+\frac{g^2}{2}\int\frac{{\rm d}^3k}{(2\pi)^3}\,
\frac{2N({\bf k},X)+1}{2E_k(X)}\,.\eeq
That would  correspond  to a self-consistent one-loop approximation,
which is similar to the large-$N$ limit for the O$(N)$ scalar model
\cite{Mottola94}. However,
we shall not pursue here the analysis of these equations in full 
generality, but
rather restrict ourselves to the case of {\it  small
field oscillations}, $\Phi \to 0$.
In this case, the mean field and the kinetic equations decouple
since, to leading order in $\Phi$, $j^{ind}(X)\simeq M^2 \Phi(X)$,
where $M^2$ is now a constant, solution of the  gap equation:
\beq\label{GAP1}
M^2\,=\,m^2\,+\,\frac{g^2}{2}\int\frac{{\rm d}^3k}{(2\pi)^3}\,
\frac{2N(E_k)+1}{2E_k}\,,\eeq
and $E_k^2= k^2+M^2$.
The linearized  mean field equation reads then:
\beq\label{linphi}
-\left(\del_X^2+M^2 \right)  \Phi(X) \,=\,0\,.\eeq
Thus, in this weak field approximation, the same mass $M$ characterizes the
longwavelength oscillations of the mean field $\Phi$, which we can regard as
collective excitations of the system, and the short wavelength excitations
associated rather to single particle excitations. Eq.~(\ref{linphi}) 
shows that
this  mass
$M$  sets the scale of the soft space-time variations:
$\lambda^{-1}\sim \del_X \sim M$. Furthermore, from  eq.~(\ref{jMFA}) 
one deduces
that ``small fields'' means
$g\Phi \ll M$: the contribution to the mass (or to the induced current) is then
dominated by the short wavelength, or hard, thermal fluctuations.

In order to compute $M$,
one should first eliminate the UV divergences from the gap equation
(\ref{GAP1}). Although such questions will play a minor role in our 
discussions,
it is nevertheless instructive to see  how this can be
achieved in this simple example. Divergences occur in the following integral,
which we compute with an upper cut-off
$\Lambda\,$:
\beq\label{LAMBDA} \frac{1}{2} \int\frac{{\rm d}^3k}{(2\pi)^3}\,
\frac{1}{2E_k}&=& I_1(\Lambda)\,-\,M^2 I_2(\Lambda) \,+\,\frac{M^2}{2(4\pi)^2}
\,\ln \frac{ M^2}{\mu^2}\,+\,...\,,\eeq
where $\mu$ is an arbitrary subtraction scale,
\beq
I_1(\Lambda)\equiv \frac{\Lambda^2}{2(4\pi)^2}\,,\qquad\,\,
I_2(\Lambda)\equiv \frac{1}{2(4\pi)^2}\,\ln \frac{\Lambda^2}{\mu^2}\,,\eeq
and the dots stand for terms which vanish as $\Lambda \to \infty$.
If we define the renormalized mass and coupling constant via:
\beq\label{renMG}
\frac{m_r^2}{g^2_r}\,\equiv\, \frac{m^2}{g^2}\,+\,I_1(\Lambda),\qquad\,\,
\,\,\frac{1}{g^2_r}\,\equiv\, \frac{1}{g^2}\,+\,I_2(\Lambda),\eeq
then we obtain a gap equation which is free of UV divergences:
\beq\label{GAP2}
M^2\,=\,m_r^2\,+\,\frac{g^2_r M^2}{2(4\pi)^2}
\,\ln \frac{ M^2}{\mu^2}
\,+\,\frac{g^2_r}{2}\int\frac{{\rm d}^3k}{(2\pi)^3}\,
\frac{N(E_k)}{E_k}\,.\eeq
It is easy to verify that the above renormalization procedure
renders finite also the inhomogeneous gap eq.~(\ref{GAP}).
Furthermore, the relations (\ref{renMG})
between the bare and renormalized parameters
do {\it not}  involve the temperature, that is, they are the same
as in the vacuum. In particular, eq.~(\ref{renMG}) implies
that the renormalized coupling constant satisfies
\beq
\frac{{\rm d}\,g_r^2}{{\rm d}\ln \mu}\,=\,\frac{g_r^4}{(4\pi)^2}\,,
\eeq
which ensures that the solution $M^2$ of eq.~(\ref{GAP2}) is
independent of $\mu$.

The gap equation (\ref{GAP2}) can be solved numerically;
the corresponding result is discussed, e.g., 
in Refs. \cite{DHLR98,BIR00}.
Alternatively, in the high temperature limit $T\gg m$, 
and in the weak coupling regime $g^2\ll 1$,
we have $T\gg M$ as well, and
the solution to eq.~(\ref{GAP2}) can be obtained in a high
temperature expansion:
\beq\label{M21}
M^2(T)\,=\,m^2 + \frac{g^2}{24}\,T^2 - \frac{g^2}{8\pi}MT
+{ O}(g^2 M^2 \ln (T/\mu))\,.\eeq
(Here and below,
we denote the renormalized parameters simply as $m^2$ and $g^2$.)
The leading contribution of the thermal fluctuations
(the ``hard thermal loop'')
is of the order of $g^2 T^2$, and comes from hard momenta $k\sim T \gg M$
within the integral of eq.~(\ref{GAP2}), for which one can neglect
$M$ as compared to $k$:
\beq\label{MS0}
\frac{g^2}{2}\int\frac{{\rm d}^3k}{(2\pi)^3}\,\frac{N(k)}{k}
\,=\,\frac{g^2}{24}\,T^2=\hat M^2\,.\eeq
The subleading thermal effect in eq.~(\ref{M21}) comes from the
contribution of {\it soft} momenta ($k\sim M \ll T$) to
the integral of eq.~(\ref{GAP2}), for which one can approximate
$N(E_k)\approx T/E_k\,$:
\beq\label{MTSL}
\frac{g^2}{2}\int\frac{{\rm d}^3k}{(2\pi)^3}\left(
\frac{N(E_k)}{E_k}-\frac{N(k)}{k}\right)&\approx&\frac{g^2}{4\pi^2}\int
{\rm d}k\,k^2\left(\frac{T}{E_k^2}\,-\,\frac{T}{k^2}\right)
\nonumber\\&=&
\frac{g^2M^2T}{4\pi^2}\int\frac{{\rm d}k}{k^2+M^2}
\,=\,-\,\frac{g^2}{8\pi}MT\,.\eeq
In particular, for $m=0$ and to lowest order in $g$ one can replace
$M$ by $\hat M$ in eq.~(\ref{MTSL}), and thus recover the NLO result
for the thermal mass given in eq.~(\ref{MTOTS}).

\subsubsection{Damping rates from kinetic equations}

The approximations developed in the previous subsection are 
sufficient to give a
consistent description of the dynamics of the longwavelenth excitations of an
ultrarelativistic plasma of scalar particles. In the case of gauge 
theories we shall show explicitly, in sections 3, 4 and 5,
that this approximation scheme isolates the dominant
contributions in a systematic expansion in powers of the gauge 
coupling. Now, there
are many interesting physical phenomena whose description requires 
going beyond this mean field approximation. This is the case in particular of 
transport phenomena, or of the damping of various excitations. 
In both cases, the
collisions play an essential role.

We shall then consider the kinetic
equation with the collision terms included,
that is, the quantum Boltzmann equation (\ref{kinG1}).
For a weakly interacting system,
which has long-lived single-particle excitations,
it is a good approximation to work in a
quasiparticle approximation (see Sect. 2.3.2). 
We shall then look for solutions of the form:
\beq\label{MFAGG}
{ G}^<(k,X)\,=\,\rho(k,X)\,N(k,X),\qquad
{ G}^>(k,X)\,=\,\rho(k,X)\,[1+N(k,X)],\eeq
where $\rho(k,X)= 2\pi\epsilon(k_0)\delta(k_0^2-E_k^2(X))$ is the
spectral density in the mean-field approximation, eq.~(\ref{MFAA}).
With this Ansatz for $\rho$, the l.h.s. of eq.~(\ref{kinrhoMFA}) vanishes,
which suggests to neglect, for consistency,
the Poisson brackets in the r.h.s. of 
eq.~(\ref{kinrho}) and those in
the r.h.s. of eq.~(\ref{kinG<}) \cite{KB62,Daniel83}.
With these approximations,  eq.~(\ref{kinG1}) becomes:
\beq\label{Boltz}
(k\cdot \del_X)\,{ G}^<\,+\,\frac{1}{2}\left(\del^\mu_X
\Sigma^\delta\right)\del_\mu^k { G}^<\,=\,
-\,\frac{1}{2}\Bigl({ G}^>\Sigma^<
- \Sigma^>{ G}^<\Bigr)\,.\eeq
This equation has to be complemented with approximations for
$\Sigma^>$ and $\Sigma^<$ consistent with the previous 
approximations. Then, as we shall see, it becomes
the Boltzmann equation.

As in the previous subsection, we can decompose the Wigner
functions into positive and negative energy components
(cf. eq.~(\ref{solG<})). Then, by isolating the positive-energy component
of eq.~(\ref{Boltz}), we obtain the following equation for the
distribution function $N({\bf k}, X)\,$:
\beq\label{BMFA}
\Bigl (\del_t+{\bf v}_k\cdot\nbfgrad_{\bf x} &-&
\nbfgrad_{\bf x} E_k\cdot \nbfgrad_{\bf k}\Bigr) N({\bf k},X)
\\&=&-\,\frac{1}{2E_k}\left\{\Bigl[1+N({\bf k},X)\Bigr]
\Sigma^<({\bf k}, X) -
N({\bf k},X)\Sigma^>({\bf k},X)\right\},\nonumber\eeq
where $\Sigma({\bf k}, X)\equiv \Sigma(k_0=E_k, {\bf k},X)$
is the on-shell self-energy, and the other notations are as
in eq.~(\ref{scVL}).

As an application, let us now consider the
single particle excitation which is obtained by adding, at $t_0=0$,
a particle with momentum ${\bf p}$ (with $p\sim T$) to a system
initially in equilibrium. We want to compute the relaxation rate
for this elementary excitation.
Since, for a large system, this is a small perturbation,
we can neglect all mean field effects (so that, e.g.,
$\nbfgrad_{\bf x} E_p=0$), and assume $N({\bf p},t)$
to be only a function of time. From eq.~(\ref{BMFA}) we get:
\beq\label{b1}
2E_p\frac{\del}{\del t}N({\bf p},t)\,=\,-\Bigl[1+N({\bf p},t)\Bigr]
\Sigma^<({\bf p},t) +
N({\bf p},t)\Sigma^>({\bf p},t)\,.\eeq
Here, $E_p=({\bf p}^2 + \hat M^2)^{1/2}$, with $\hat M^2=g^2T^2/24$
(the zero-temperature mass $m$ is set to zero).
Since the self-energies in the r.h.s. depend a priori on
$N({\bf p},t)$ itself, this equation is generally non-linear.
However, for momenta ${\bf k}\ne {\bf p}$, the distribution
function does not change appreciably from the equilibrium value
$N(E_k)$, so that, to leading order in the perturbation, we can
use the equilibrium self-energies (\ref{SIG><k}). These read
$\Sigma^>({\bf p})=-\Gamma({\bf p})[1+N(E_p)]$
and $\Sigma^<({\bf p})=-\Gamma({\bf p})\,N(E_p)$, where
$\Gamma({\bf p})\equiv \Gamma(p^0=E_p,{\bf p})$ is
the discontinuity of the (equilibrium) self-energy on the mass shell.
With this approximation, we get a {linear} equation:
\beq\label{b2}
2E_p\frac{\del}{\del t}N({\bf p},t)
\,=\,-\Bigl[N({\bf p},t)- N(E_p)\Bigr]
\Gamma({\bf p}),\eeq
whose solution is of the form
($\delta N({\bf p},t)\equiv N({\bf p},t) - N(E_p)$):
\beq\label{expdamp}
\delta N({\bf p},t)\,=\,\delta N({\bf p},0)\,{\rm e}^{-2\gamma(p)t}\,,\eeq
with $\gamma(p)\equiv \Gamma(p)/4E_p$.
We thus recover the relation between the lifetime of the single particle
excitation and the imaginary part of the self-energy on the mass shell,
already mentioned at the end of Sect. 2.1.3.

\begin{figure}
\protect \epsfxsize=6.cm{\centerline{\epsfbox{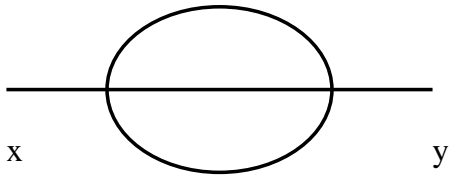}}}
           \caption{Leading-order contribution to the
collisional self-energy in $\phi^4$ theory.}
\label{sun-set}
\end{figure}

The
leading order contribution to $\Gamma$ comes from the two loop
diagram in fig.~\ref{sun-set}. Thus, $\Gamma \sim g^4T^2$, and therefore
$\gamma \sim (\Gamma/E_p) \sim g^4 T$
for a hard excitation ($E_p\sim T$), while $\gamma \sim g^3 T$ for a
soft one ($E_p  \sim M\sim g T$). In both cases we have
$\tau\sim 1/\gamma \gg 1/E_p$, which corresponds to long-lived excitations,
as required for the validity of the quasiparticle approximation. Although the
latter is not a self-consistent
approximation (the collision
term generates a width which is not included in the spectral
densities which are used to estimate it),  the neglected
terms are of higher order than those we have kept.

To compute $\Gamma$, one can directly evaluate the on-shell
imaginary part of the self-energy in fig.~\ref{sun-set},
using equilibrium perturbation theory \cite{RP92,Jeon95,Heinz95}.
Alternatively,  one can first
construct the collision term associated to this self-energy,
and then extract $\Gamma$ as the coefficient of $\delta N({\bf p},t)$
in eq.~(\ref{b2}). Since
the resulting collision term is interesting for other
applications \cite{JY96} 
than the one discussed here, and since it clarifies the
physical interpretation of the damping in terms of
collisions, this is the method we shall follow here.

The self-energy in fig.~\ref{sun-set} can be easily evaluated
in the $x$ representation:
\beq\label{SSS1}
\Sigma(x,y)\,=\,-\frac{g^4}{6}\,\Bigl(G(x,y)\Bigr)^3,\eeq
where the time variables $x_0$ and $y_0$ take values along the
contour of fig.~\ref{CONT}. By taking $x_0$ and $y_0$ real,
with $x_0$ later (respectively, earlier) than $y_0$,
we get:
\beq\label{SSS2}
\Sigma^>(x,y)\,=\,-\frac{g^4}{6}\,\Bigl(G^>(x,y)\Bigr)^3,\qquad
\Sigma^<(x,y)\,=\,-\frac{g^4}{6}\,\Bigl(G^<(x,y)\Bigr)^3,\eeq
or, after a Wigner transform,
\beq\label{SSS3}
\Sigma^>(p,X)\,=\,-\frac{g^4}{6}\,\int {\rm d}[k_1,k_2,k_3]\,{ G}^>(k_1,X)
{ G}^>(k_2,X){ G}^>(k_3,X),\eeq
with a similar expression for $\Sigma^<(p,X)$.
Here, we have set:
\beq
{\rm d}[k_1,k_2,k_3]\equiv \frac{{\rm d}^4k_1}{(2\pi)^4}\,
\frac{{\rm d}^4k_2}{(2\pi)^4}\,\frac{{\rm d}^4k_3}{(2\pi)^4}\,
(2\pi)^4\delta^{(4)}(k_1+k_2+k_3-p)\,.\eeq
In the quasiparticle approximation (\ref{MFAGG}), the associated
collision term reads:
\beq\label{coll}
(G^<\, \Sigma^> - G^> \,\Sigma^<)(p,X)&=&
-\frac{g^4}{6}\int {\rm d}[k_1,k_2,k_3]\,\rho(p,X)
\rho(k_1,X)\rho(k_2,X)\rho(k_3,X)\nonumber\\
&{}&\quad
\Bigl\{N(p,X)[1+N(k_1,X)][1+N(k_2,X)][1+N(k_3,X)]\nonumber\\
&{}&\quad\,\, -\,\,
[1+N(p,X)]N(k_1,X)N(k_2,X)N(k_3,X)\Bigr\}.\eeq
This collision term has the standard Boltzmann structure,
with a gain term and a loss term: it involves
the matrix element squared for binary collisions (which here is simply
  $|{\cal M}_{pk_1\to k_2k_3}|^2 = g^4/6$), together with
statistical factors for the on-shell external particles.

We consider now again the particular case of a single particle excitation
with momentum ${\bf p}\,$. Then, as already discussed,
$N({\bf p},t)\equiv  N(E_p)+\delta N({\bf p},t)$, while
all the other particles are in equilibrium: $N(k_i,X)= N(k_i^0)$.
The collision term (\ref{coll}) takes then the form:
\beq\label{coll1}
\delta N({\bf p},t) \,\frac{-g^4}{6}\int {\rm d}[k_1,k_2,k_3]\,\rho_0(k_1)
\rho_0(k_2)\rho_0(k_3)\left\{[1+N_1][1+N_2][1+N_3]-
N_1 N_2 N_3\right\}\nonumber\\
\,\equiv\,-\delta N({\bf p},t) \Gamma({\bf p}),\qquad\qquad\eeq
where $p_0=E_p$, $\rho_0(k)=2\pi\epsilon(k_0)\delta(k_0^2-E_{k}^2)$
and $N_i\equiv N(k_i^0)$. It can be easily verified that
$\Gamma({\bf p})$ defined as above coincides indeed with the on-shell
discontinuity of the two-loop self-energy in fig.~\ref{sun-set}.

\begin{figure}
\protect \epsfxsize=12.cm{\centerline{\epsfbox{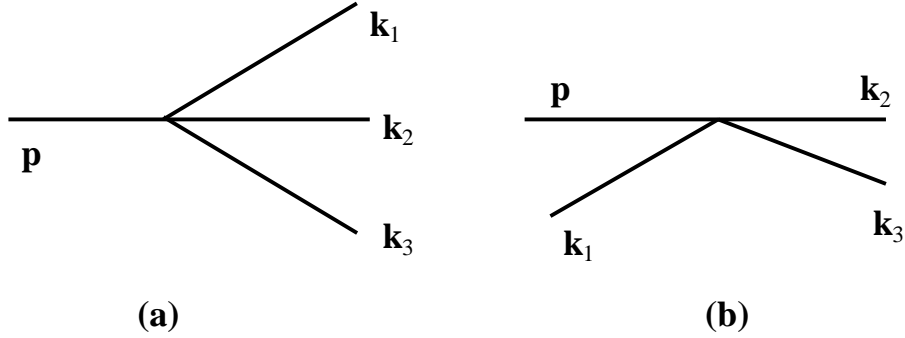}}}
           \caption{Elementary processes leading to the damping
of the single-particle excitation of momentum ${\bf p}$
(cf. eq.~(\ref{coll1})).}
\label{kinfig1}
\end{figure}

By inspection of eq.~(\ref{coll1}), one can identify
the physical processes responsible for the damping.
There are two elementary processes:
the three-particle decay of the incoming field
(see fig.~\ref{kinfig1}.a), and the binary collision
with a particle from the thermal bath
(see fig.~\ref{kinfig1}.b). The statistical factors
corresponding to the 3-body decay are:
\beq\label{proc1}
[1+N(E_1)][1+N(E_2)][1+N(E_3)] \,-\,N(E_1)N(E_2)N(E_3), \eeq
with the first term describing the direct (decay) process,
and the second one representing the inverse (recombination) process.
Because of the 3-particle threshold at $p_0^*=3M$, this decay process
is not effective on the mass-shell $p_0=E_p$, so it does not
contribute to the damping of the single-particle excitation in
eq.~(\ref{expdamp}).
On the other hand, the binary collisions, which are accompanied by
statistical factors of the type:
\beq\label{proc2}
N(E_1)[1+N(E_2)][1+N(E_3)] \,-\,[1+N(E_1)]N(E_2)N(E_3), \eeq
have no kinematical threshold, and contribute indeed to
the on-shell damping rate. The final result for $\gamma$ follows
after performing the phase-space integral in eq.~(\ref{coll1}).
For an arbitrary external momentum ${\bf p}$,
this is quite complicated, and $\gamma(p)$ can be
obtained only numerically \cite{Jeon95,Heinz95}.
However, in the zero momentum limit, an analytical calculation
has been given, with the result \cite{RP92}:
\beq\label{gammap0}
\gamma(p=0)\,=\,\frac{g^2\hat M}{64\pi}\,=\,
\frac{g^3T}{64\pi\sqrt{24}}\,.\eeq

More generally, the Boltzmann equation (\ref{BMFA}) with the collision
term (\ref{coll}) describes a variety of non-equilibrium phenomena
in the weakly coupled scalar field theory and allows one to compute
transport coefficients \cite{Groot80,Daniel85,Mrowcz90,BP91,JY96}.
Its solution accomplishes a non-trivial
resummation of the ordinary perturbation theory,
including in particular that of
ladder diagrams \cite{Jeon93,Jeon95,JY96,BE,USA,Bodeker99}.
The kinetic approach is not only technically
simpler and physically more transparent than the diagrammatic
approach, but also allows for
relatively straightforward extensions
to  gauge theories, as it will be discussed in Sect. 7.

\subsubsection{Time representation and Fermi's golden rule}

When one applies the techniques discussed in the previous
subsection to the calculation of the damping of charged excitations
in hot gauge theories, one is confronted with the difficulty
that the on-shell imaginary part of the self-energy is infrared
divergent, as mentioned in Sect. 1.5. The physical origin of this
problem is the fact that the mean free
path is about the same as the
range of the relevant interactions, so that the particles cannot be
considered  as freely moving (i.e., as on-shell excitations) 
between successive collisions. This invalidates a simple description
in terms of the Boltzmann equation \cite{Heisel93}, or of the
standard perturbation theory in the energy
representation \cite{Pisarski89,Smilga90}.

In preparation for the detailed discussion of this problem in
Sect. 6, here we reformulate the calculation of 
the decay of a single particle excitation using kinetic theory
by using the time representation. 
(See also Refs. \cite{BDV98,BDV00} for a similar construction.)
As in the previous subsection, we consider an excitation
which is obtained by adding, at $t=t_0$, a particle with momentum
${\bf p}$ and energy $E_p=({\bf p}^2 + \hat M^2)^{1/2}$
to a system initially in equilibrium. 
For $t>t_0$, the 2-point functions have the generic structure
in eq.~(\ref{MFAGG}), where we assume that 
$\delta N({\bf k},t)=0$ for any ${\bf k}\ne {\bf p}$,
while for ${\bf p}$ (the momentum of the added particle),
\beq\label{G<+0}
G^<(p,X)=G^<_0(p) + \delta G(p,X),\qquad 
G^>(p,X)=G^>_0(p) + \delta G(p,X),\eeq
where $G^<_0(p)$ and $G^>_0(p)$ are the free equilibrium
two-point functions and
\beq\label{G<+}
\delta G(p,X)\,\equiv\,\delta G(
{\bf p},p_0,t)\,=\,2\pi\delta(p_0-E_p)\,\frac{1}{2E_p}\,
\delta N({\bf p},t),\eeq
with $\delta N({\bf p},t_0)$ describing the initial perturbation.
The only difference with respect to the previous discussion in 
Sect. 2.3.4 is that here we shall not take the limit $t_0\to
-\infty$, but rather keep a finite 
(although relatively large: $t-t_0\gg 1/T$)
time interval $t-t_0$.
This will prevent us from taking the
on-shell limit when evaluating the collision terms.

In order to derive the kinetic equation satisfied by
$N({\bf p},t)$, it is useful to observe, from
eqs.~(\ref{G<+0})--(\ref{G<+}), that
\beq\label{NG<+}
\delta N({\bf p},t)\,=\,2E_p\int \frac{{\rm d}p_0}{2\pi}\,
\delta G^<({\bf p},p_0,t)\,=\,2E_p \,
\delta G^<({\bf p},x_0=y_0=t).\eeq
We thus need the equation satisfied by $\delta G^<({\bf p},x_0,y_0)$
in the equal time limit. The starting point is eq.~(\ref{eqG<})
for $G^<({\bf p},x_0,y_0)$, which for $x_0=y_0=t$ simplifies to
\beq\label{eq11}
\Bigl(\del^2_{x_0}+{\bf p}^2+M^2\Bigr)\Bigl|_{x_0=y_0=t}
G^<({\bf p},x_0,y_0)&=&-i\int_{t_0}^t {\rm d} z_0\,\Bigl\{
\Sigma^>({\bf p},t,z_0)\,G^<({\bf p},z_0,t)\,-
\nonumber\\&{}&\qquad\,\,
-\,\Sigma^<({\bf p},t,z_0)\,G^>({\bf p},z_0,t)\Bigr\}.\eeq
We have neglected here the vertical piece of the contour
(i.e., the third integral in the r.h.s. of eq.~(\ref{eqG<}))
since this becomes irrelevant for sufficiently large times $t-t_0\gg 1/T$.
To perform the gradient expansion in time, we replace
$x_0$, $y_0$ by $s\equiv x_0-y_0$ and $t\equiv (x_0+y_0)/2$,
and proceed as in Sect. 2.3.2, but without introducing the
Wigner transform in time. For instance (with the
momentum variable ${\bf p}$ left implicit)
\beq\label{gradtime}
\int {\rm d} z_0\,\Sigma^>(x_0,z_0)\,G^<(z_0,y_0)
&\equiv& \int {\rm d} z_0\,\Sigma^>(x_0-z_0,(x_0+z_0)/2)
\,\,G^<(z_0-y_0, (z_0+y_0)/2)
\nonumber\\&\simeq&\int {\rm d} z_0\,
\Sigma^>(x_0-z_0,t)\,\,G^<(z_0-y_0,t),\eeq
where in the first line we have rewritten the two-time
functions as functions of the relative and central time variables,
and in the second line we have used the fact that the
off-equilibrium propagators are peaked at small values
of the relative time
(i.e., $|x_0-z_0|\simle 1/T$) to write $(x_0+z_0)/2
\simeq (z_0+y_0)/2 \simeq t$ to leading order in the gradient
expansion. In the equal-time limit $x_0=y_0=t$, the last expression
becomes (with $z_0$ changed into $t'$)
\beq
\int_{t_0}^t {\rm d} t'\,
\Sigma^>(t-t',t)\,G^<(t'-t,t)\equiv
\int_0^{t-t_0} {\rm d} s\,\Sigma^>(s,t)\,G^<(-s,t).\eeq
After considering similarly the equation where the temporal
derivative acts on $y_0$, and taking the difference
of the two equations, we get:
\beq\label{diffst}
2\del_s \del_t G^<({\bf p},s,t)\Bigl|_{s=0}&=&
- i\int_0^{t-t_0} {\rm d} s\,\Bigl\{\Sigma^>(s,t)\,G^<(-s,t)
\,-\,\Sigma^<(s,t)\,G^>(-s,t)\,+\,
\nonumber\\&{}&\,\,\,\,\,\,\,\,\,\,\,+\,\,G^<(s,t)\,\Sigma^>(-s,t)
\,-\,G^>(s,t)\,\Sigma^<(-s,t)\Bigr\},\eeq
where the ${\bf p}$-dependence of the functions in the r.h.s.
is implicit.

Eq.~(\ref{diffst})
is the finite-time generalization of the Boltzmann equation
(\ref{Boltz}). By using the symmetry properties
\beq G^<({\bf p},s,t)=G^>(-{\bf p},-s,t),\qquad 
\Sigma^<({\bf p},s,t)=\Sigma^>(-{\bf p},-s,t),\eeq
together with the isotropy of the equilibrium state
(e.g., $ \Sigma^<_{eq}({\bf p},s,t)=\Sigma^<(p,s)$, with
$p\equiv |{\bf p}|$), one
can easily check that the r.h.s. of this equation vanishes
in thermal equilibrium, as it should. 
Moreover, for the single-particle excitation of interest,
the self-energies can be taken as in thermal equilibrium
(cf. the discussion after eq.~(\ref{b1})), so that
eq.~(\ref{diffst}) reduces to:
\beq\label{diffst1}
2\del_s \del_t \delta G({\bf p},s,t)\Bigl|_{s=0}&=&
- i\int_{-(t-t_0)}^{t-t_0} {\rm d} s\,\Bigl[\Sigma^>(p,s)-
\Sigma^<(p,s)\Bigr]\, \delta G({\bf p},-s,t),\eeq
or, equivalently (cf. eq.~(\ref{G<+})),
\beq\label{b11}
2E_p\frac{\del}{\del t}\,\delta N({\bf p},t)&=&-
\int_{-(t-t_0)}^{t-t_0} {\rm d} s\,\,{\rm e}^{iE_p s}\,\Gamma({\bf p},s)\,
\delta N({\bf p},t),\eeq
with the definition (cf. eq.~(\ref{SIG><k})) :
\beq\label{Gammas}
\Gamma({\bf p},s)\,\equiv \,-[\Sigma^>({\bf p},s)-\Sigma^<({\bf p},s)]
\,=\,\int \frac{{\rm d}p_0}{2\pi}\,{\rm e}^{-ip_0 s}
\,\Gamma(p_0,{\bf p}).
\eeq

We are interested in relatively large time intervals $t-t_0$,
of the order of the mean free time between 
successive collisions in the plasma.
For systems with short range interactions, like the scalar theory
discussed throughout this section, the leading behaviour at large
time is obtained by letting $t\to \infty$ in the integration 
limits in eq.~(\ref{b11}). Then, the unrestricted integral
over $s$ simply reconstructs the on-shell Fourier component
of $\Gamma$,
\beq\label{osGamma}
 \int_{-\infty}^{\infty} {\rm d} s\,\,
{\rm e}^{iE_p s}\,\Gamma({\bf p},s)\,=\,
\Gamma(p_0=E_p, {\bf p})\,,\eeq
so that eq.~(\ref{b11})
reduces to the usual Boltzmann equation (\ref{b2}) describing the
relaxation of single-particle excitations.

Since the integration limits in eq.~(\ref{b11})
involve only the {\it relative} time $t-t_0$, it is clear that
this on-shell limiting behaviour is also obtained by letting
$t_0\to -\infty$, which is indeed how eq.~(\ref{b2}) has been
derived in Sect. 2.3.4. This explains our emphasis on keeping
$t_0$ finite in this subsection. This allows us to treat also
systems with long-range interactions, like gauge theories,
for which the on-shell limit of the self-energy is ill-defined.
Specifically, we shall see in Sect. 6.5 that for gauge theories,
$\Gamma({\bf p},s)$ is only slowly decreasing with $s$
(like $1/s$), so that the unrestricted integral in
eq.~(\ref{osGamma}) is logarithmically divergent.
But even in that case, the finite-time equation (\ref{b11}) 
is still well defined,
and correctly describes the behaviour at large times.
By also using eq.~(\ref{Gammas}), this equation is finally rewritten
as
\beq\label{b13}
E_p\frac{\del}{\del t}\,
\delta N({\bf p},t)\,=\,-\int \frac{{\rm d}p_0}{2\pi}\,
\Gamma(p_0,{\bf p})\,\,\frac{\sin(p_0-E_p)t}{p_0-E_p}\,\,\delta
N({\bf p},t).\eeq
For a fixed large time, the function
\beq
R(t,p_0-E)\,\equiv\,\frac{\sin(p_0-E)t}{p_0-E}\,,\eeq
is strongly peaked
around $p_0=E$, with a width $\sim 1/t$. In the limit
$t\to \infty$, $R(t,p_0-E) \,\to\, \pi \delta(p_0-E)$,
which enforces energy conservation: This limit is known
as Fermi's ``golden rule''.
In the absence of infrared complications, one can use this limit to obtain
the large time behaviour of eq.~(\ref{b13}), and thus obtain eq.~(\ref{b2}).

For gauge theories, however, this na\"{\i}ve large-time
limit leads to singularities, so that
the time dependence of $R(t,p_0-E)$ 
must be kept. The correct kinetic equation rather reads then
\beq\label{b14}
\frac{\del}{\del t}\,\delta N({\bf p},t)\,=\,-2\gamma({\bf p},t)
\,\delta N({\bf p},t)\,,\eeq
with a {\it time-dependent} damping rate:
\beq\label{tgamma}
\gamma({\bf p},t)\,\equiv\,\frac{1}{2E_p}
\int \frac{{\rm d}p_0}{2\pi}\,
\Gamma(p_0,{\bf p})\,\,\frac{\sin(p_0-E_p)t}{p_0-E_p}\,.\eeq
As we shall see in Sect. 6, this quantity is well defined
even in gauge theories, because $1/t$ acts effectively
as an infrared cut-off. 
But this also entails that, in such cases,
$\gamma({\bf p},t)$ remains explicitly time-dependent
even for asymptotically large times, so that the
decay is {\it non-exponential} in time.

\setcounter{equation}{0}
\setcounter{equation}{0}
\section{Kinetic theory for hot QCD plasmas}

With this section, we begin the study of collective excitations of the
quark-gluon plasma. We assume that the  temperature $T$ is high enough
for the condition $g\equiv g(T)\ll 1$ to be satisfied, and proceed with
a weak coupling expansion. As discussed in the introduction, a
convenient way to study the collective phenomena
is to investigate the response of the plasma to ``soft'' external
perturbations with typical Fourier components $P\sim gT$. We shall
consider here external sources which produce excitations with the
same quantum numbers as the plasma constituents.
As we shall see, fermions and bosons play symmetrical
roles in the  ultrarelativistic plasmas, and 
the soft fermionic excitations have a collective nature, 
similar to that of the more familiar plasma waves.
The plasma particles act collectively as {\it induced sources} for
longwavelength {\it average fields}, either gluonic or fermionic,
which will be denoted as $A_\mu^a$, $\Psi$ and
$\bar\Psi$, respectively.
The induced sources can be expressed in terms of 2-point 
Green's functions, and their determination is the main  purpose of this 
section. 

\subsection{Non-Abelian versus non-linear effects}

As emphasized in the Introduction, if we were to  study Abelian plasmas,
the formalism of the linear response theory would be sufficient for our
purpose. In a non Abelian theory, Ward identities, to be discussed 
in Sect. 5.3.3,
force us to go beyond this simple approximation. To see how it comes about,
we examine more closely here the distinction between Abelian and
non Abelian plasmas.

Consider first the response of the QED plasma to a soft
electromagnetic background field, with gauge potentials $A^\mu$.
The response function is the {\it induced current}
$j^\mu_{ind}(x)\equiv \langle j^\mu(x) \rangle$,
where $j_\mu(x)=e\bar\psi(x)\gamma_\mu\psi(x)$ is the 
current operator. In the linear response approximation,
\beq\label{jcorrelator}
 j^\mu_{ind}(x)=\int {\rm d}^4 y \,\Pi^{\mu\nu}(x-y)\,A_\nu(y),
\eeq
with the polarization tensor
\beq\label{KUBOPI}
\Pi^{\mu\nu}(x-y)\,\equiv
-i\theta(x_0-y_0)\langle[j^\mu(x),j^\nu(y)]\rangle\,.
\eeq
Eq.~(\ref{jcorrelator}) is consistent
with the Abelian gauge symmetry because the polarization tensor
is transverse. Indeed, the condition  $\del_\mu \Pi^{\mu\nu}=0$
guarantees that the induced current is conserved, $\del_\mu 
j_{ind}^\mu(x)=0$, and that the expression (\ref{jcorrelator}) 
is gauge invariant (the contribution of a pure gauge potential
$A_\mu=\del_\mu\theta$ cancels out). 

In fact, one can make the gauge invariance
explicit by using the transversality of $\Pi^{\mu\nu}$
to reexpress $j^\mu_{ind}$ in terms of the physical electric
field; going over to momentum space, with
$P^\mu=(\omega,{\bf p})$ and $E^j(P)= i(\omega A^j(P)-p^jA^0(P))$,
we can write (with $j^\mu\equiv j_{ind}^\mu$):
\beq\label{SIGAB}
j^\mu(P)\,=\,\sigma^{\mu j}(P)E^j(P),\qquad\sigma^{\mu j}(P)
\equiv (i/\omega)\Pi^{\mu j}(P),\eeq
where the conductivity tensor $\sigma^{\mu j}(P)$
satisfies $P_\mu\sigma^{\mu j}=0$ and $\Pi^{\mu 0}=-ip^j\sigma^{\mu j}$.
The kinetic theory in Sect. 1.3 provides 
us with an explicit expression for $\sigma^{\mu i}$ 
(see eqs.~(\ref{polarisation2})--(\ref{plasmafrequency})):
\beq\label{DEFSIG}
\sigma^{\mu i}(\omega,{\bf p})\,\equiv\,i\,m_D^2
\int\frac{{\rm d}\Omega}{4\pi}\,\frac{v^\mu v^i}{\omega
- {\bf v\cdot p}+i\eta}\,,\eeq
which turns out to be the correct result
to leading order in $e$ when $P\sim eT$.

The linear relation between the induced current and
the applied gauge potential exhibited in eq.~(\ref{jcorrelator}) stops to be
valid when the first non linear corrections become comparable to the linear
term.  Since $\Pi\propto m_D^2\sim e^2T^2$,
the linear term in $j^\mu_{ind}$ is of order $e^2T^2 A$.
The first non linear correction involves the photon four-point
vertex function and is of order $e^4 A^3$.
Thus, the linear approximation in QED holds as long as $A\simle T/e$,
or $E \sim eTA \simle T^2$. 

Consider now QCD. For sufficiently weak gauge fields $A^\mu_a$,
the linear approximation is valid here as well.
Then, the induced colour current takes the form:
\beq\label{jcorrelator2}
j^{\mu\,a}_{ind}(x)=\int {\rm d}^4 y \, \Pi_{ab}^{\mu\nu}(x,y)
A_\nu^b(y),
\eeq 
where the polarization tensor receives contributions from all
the coloured particles (quarks, gluons, and also ghosts in gauges 
with unphysical degrees of freedom), 
and is diagonal in colour, $\Pi^{\mu\nu}_{ab}
(x,y)=\delta_{ab}\Pi^{\mu\nu}(x,y)$. For inhomogeneities
at the scale $gT$ ($\del_x A\sim gTA$),
and to leading order in $g$, $\Pi^{\mu\nu}$ has the same expression 
as in QED, eq.~(\ref{polarisation2}), but with $m_D^2\sim g^2T^2$.
Thus, the 8 components of the colour current are decoupled
and are individually conserved: $\del_\mu \, j^{\mu}_a =0$.

However, the linear approximation holds in QCD only for fields
much weaker than in QED. This can be seen in various 
ways. For instance, consider the equations of motion for the
soft mean fields, that is, the Yang-Mills equations with the 
induced current $j^{\mu}_a$ as a source in the r.h.s. :
\beq\label{YM}
[D_\nu, F^{\nu\mu}]^a(x)\,=\,j_{ind}^{\mu\,a}(x).\eeq
Since $[D_\mu, [D_\nu, F^{\nu\mu}]]=0$, this equation requires
$j_{ind}^{\mu}$ to be {\it covariantly} conserved, i.e., to
satisfy $[D_\mu,j^\mu]=0$, a condition which is generally not
consistent with the linear approximation (\ref{jcorrelator2}).
In fact, the linearized conservation law $\del_\mu \, j^{\mu}=0$
becomes a good approximation to the correct law $[D_\mu,j^\mu]=0$
only for fields which are so weak that the mean field
term $gA_\mu$ can be neglected within the soft covariant derivative
$D_\mu\/$: $gA_\mu\ll \del_x$.
For $\del_x\sim gT$, this requires $A\ll T$.
But in this limit, all the other non-linear effects in eq.~(\ref{YM})
can be neglected as well, so this equation reduces to
a set of uncoupled Maxwell equations, one for each colour.
In other terms, the linear response approximation for the
induced current is valid only for fields which are so weak
that they are effectively Abelian.

This conclusion is corroborated by an analysis of the current
$j_{ind}^{\mu\,a}\/$. Under gauge transformations of the background 
fields, this must transform
as a colour vector in the adjoint representation (so as to
insure the covariance of eq.~(\ref{YM})). That is,
under the infinitesimal gauge transformation $h(x)= 1+i\theta_a(x)T^a$,
\beq
A_\nu(x)\rightarrow A_\nu(x)-\,\frac{1}{g}\,\del_\nu\theta(x)
-i[A_\nu(x),\theta(x)], \eeq
the current $j^\mu\equiv j^\mu_aT^a$ should transform as
$j^\mu \to j^\mu+\delta j^\mu$, with:
\beq\label{trJ}
\delta j^\mu(x)\,=\,-i [j^\mu(x),\theta(x)].
\eeq
Now, under the same transformation, the variation of the linearized
current (\ref{jcorrelator2}) is instead
\beq\label{deltajmu}
\delta j^\mu(x) \,=\,-i\int {\rm d}^4y\,
\Pi^{\mu\nu}(x,y)\,[A_\nu(y),\theta(y)].\eeq
For a non-local response function $\Pi^{\mu\nu}(x,y)$ this
is different from the correct transformation law (\ref{trJ}).
This suggests that in the presence of a non-Abelian gauge symmetry 
there is an interplay between non-linear and non-local effects.
This can be made more visible by rewriting the
induced colour current in terms of
a ``conductivity'' $\sigma^{\mu i}\,$, as in eq.~(\ref{SIGAB}) :
\beq\label{SIGNAB}
j^{\mu\,a}_{ind}(x)=\int {\rm d}^4 y \,\sigma_{ab}^{\mu i}(x,y)\,
 E_b^i(y).
\eeq
Under a gauge transformation $h(x)=\exp\left(i\theta^a(x) T^a\right)$,
the electric field transforms as a colour vector:
$E^i_a(x)\rightarrow h_{ab}(x)E^i_b(x)$.
In order for the induced current to transform similarly,
the conductivity must transform as:
\beq\label{SIGTRANS}
\sigma^{\mu i}_{ab}(x,y)\,
\rightarrow \,h_{a\bar a}(x)\, \sigma^{\mu i}_{\bar a\bar b}(x,y)\,
h^\dagger_{\bar b b}(y).
\eeq
Since $\sigma$ is generally non-local (see, e.g., eq.~(\ref{DEFSIG})),
this is satisfied only if the conductivity is itself
a functional of the gauge fields, i.e.,  the relation 
(\ref{SIGNAB}) is non-linear.

The particular solution that we shall obtain as the outcome 
of our approximation
scheme satisfies the above requirement in a simple way: 
the dependence of
the conductivity on the gauge fields is simply given by a 
parallel transporter. We have:
\beq\label{SIGU}
\sigma^{\mu i}_{ab}(x,y|A)\,=\, \sigma^{\mu i}(x-y) \,U_{ab}(x,y|A),
\eeq
where $\sigma^{\mu i}(x-y)$ is independent of colour
(actually, it coincides with the Abelian conductivity (\ref{DEFSIG})), and 
\beq\label{UAB}
U_{ab}(x,y|A)\,=\,{\rm
P}\exp\left(-ig\int_y^x{\rm d}z^\mu\, A_\mu(z)\right)\eeq
is the parallel transporter along the straight line
joining $y$ and $x$. Under a gauge transformation,
the parallel transporter becomes
\beq\label{UTRANS}
U_{ab}(x,y|A)\,\rightarrow \,U_{ab}(x,y|A^h)\,=\,h_{a\bar a}(x)
\,U_{\bar a \bar b}(x,y|A)\,h^{-1}_{\bar b b}(y),
\eeq
which insures the correct transformation law
(\ref{SIGTRANS}) for the conductivity tensor.

By expanding the exponential in the Wilson line (\ref{UAB}),
it is possible to express the current (\ref{SIGNAB}) as a formal
series in powers of the gauge potentials:
\beq\label{exp1}
j^{a}_{ind\,\mu} \,=\,\Pi_{\mu\nu}^{ab}A_b^\nu
+\frac{1}{2}\, \Gamma_{\mu\nu\rho}^{abc} A_b^\nu A_c^\rho+\,...
\eeq
The coefficients in this series are the one-particle
irreducible amplitudes of the soft fields in thermal equilibrium
(cf. Sect. 5.1) : $\Pi\sim g^2 T^2$
is the polarization tensor, while the other terms 
(of the generic form $\Gamma^{(n+1)}A^{n}$) are corrections to the 
$(n+1)$-gluon vertices.  The magnitude of the latter
can be estimated as follows: since these terms arise solely 
from expanding the Wilson line (\ref{UAB}), they scale like
$\Gamma^{(n+1)}A^{n}\,\sim\,\Pi\,A\,(glA)^{n-1}$, where $l\sim |x-y|$
is the typical range of the non-locality
of the response, as controlled by $\sigma^{\mu i}(x-y)$
(cf. eq.~(\ref{SIGU})). 
Now, $|x-y| \sim 1/\del_x \sim 1/gT$ (cf. eq.~(\ref{DEFSIG})),
so that the non-linear effects become important when $A\sim T$,
or $E\sim\del A \sim gT^2$. For such fields, $glA \sim 1$,
and all the terms in the expansion (\ref{exp1}) are of the same
order, namely of order $g^2 T^3$.

Repeating the argument in momentum space,
one finds that the $n$-gluon vertex correction scales like
$\Gamma^{(n)}(P)\,\sim\,g^2 T^2 (g/P)^{n-2}$, where $P$
is a typical external momentum. For $P \sim gT$,
such vertices are as large as the corresponding tree 
level vertices, whenever the latter exist. For instance, 
$\Pi\sim g^2 T^2 \sim D_0^{-1}$ (with $D_0^{-1}
\sim P^2$ the tree-level inverse propagator of the gluon),
and similarly $\Gamma^{(3)}\sim g^3(T^2/P)\sim
g^2 T \sim\Gamma_0^{(3)}$ 
(with $\Gamma_0^{(3)}\sim gP$ the tree-level
three-gluon vertex).
Within the approximations that we assume here implicitly,
and which will be detailed in Sect. 3.3, the soft
amplitudes in the r.h.s. of eq.~(\ref{exp1}) are the gluon
``hard thermal loops'' (HTL).

We conclude these general remarks with a few words on
the strategy that we shall follow below. 
Consider the induced colour current $j^{\mu\,a}_{ind}$ 
as an example: This is a non-equilibrium response function, but 
all the coefficients in the expansion (\ref{exp1}) are 
equilibrium amplitudes. This suggests two possible strategies
for computing this current: ({\it i}\/) Within
the {\it equilibrium} formalism in Sect. 2.1, one could evaluate
the coefficients in eq.~(\ref{exp1}) one by one,
by computing loop diagrams at finite temperature. 
({\it ii}\/) Alternatively, one could use the {\it non-equilibrium} 
techniques developed in Sects. 2.2 and 2.3 to compute directly the
induced current in terms of the soft mean fields, by deriving,
and then solving, appropriate equations of motion.

The first strategy has been adopted in the original derivation
of the hard thermal loops from  finite-temperature
Feynman graphs \cite{Klimov81,Weldon82a,Weldon82b,BP90,FT90}.
In this framework, the HTL's emerge as the dominant 
contributions to one-loop amplitudes
with soft external lines ($p\sim gT$) and hard loop momenta
($k\sim T$), and are obtained as the
leading order in an expansion in powers
of $p/k\sim g$. Many of the remarkable properties of 
the HTL's have been identified, 
and studied, within the equilibrium formalism 
\cite{BP90,FT90,TW90,BP90c,EN92,BP92,FT92,JN93,Weldon93,BFT93,Liu93}.

Here, however, we shall rather follow the second strategy, which 
exploits the non-equilibrium formalism to construct directly
the induced current (or other response functions)
\cite{qed,qcd,emt,gauge,BE,USA}. Aside from the fact that is
generates all the soft 
amplitudes at once, this approach has also the advantage 
that the kinematical approximations leading to HTL's, and
which exploit the separation of scales in the problem ($gT\ll T$), 
are more naturally and more economically implemented at the 
level of the equations of motion, rather than on Feynman diagrams.
As in the scalar theory discussed in Sect. 2, these 
approximations will lead from the general Kadanoff-Baym
equations for QCD to relatively simple kinetic equations. 

With respect to the scalar case of Sect. 2.3,
the main new ingredient here,
which is also the main source of technical complications,
is, of course, gauge symmetry. We shall see below that 
it is possible to derive kinetic theory for QCD in a 
 gauge invariant way.  
To this aim, it will be convenient to work with gauge mean
fields as strong as $A^\mu_a \sim T$, for which all the 
non-linear effects associated with gauge symmetry are manifest.
In this case, $gA^\mu \sim gT \sim \del_x$, so that not only
the interactions, but also the soft inhomogeneities, and the
non-linear mean field effects are controlled by powers of the
coupling constant. It is then possible to maintain gauge 
symmetry explicitly via a systematic expansion in powers of $g$.
In particular, by computing the colour current induced by
such fields to leading order in $g$, we 
include all the non-linear effects displayed in eq.~(\ref{exp1}),
and therefore all the gluon HTL's. 

\subsection{Mean fields and induced sources}

At this point, it is convenient to introduce some more formalism:
the so-called ``background field gauge'' \cite{Witt67,Abbott81,HZ87},
which will allow us to preserve explicit gauge covariance with
respect to the background fields $A_\mu^a\/$, $\Psi$ and
$\bar\Psi$, at all intermediate steps. We  stress however
that the choice of this particular gauge is only a convenience:
the final kinetic equations to be 
obtained are independent of the gauge choice.
In fact, these equations have been originally
constructed  in covariant gauges, 
and shown to be independent of the parameter $\lambda$ in the 
gauge fixing term $(\del^\mu A^a_\mu)^2/\lambda$ \cite{qed,qcd}.

\subsubsection{The background field gauge}

The generating functional $Z[j]$ of
a non-Abelian gauge theory may be expressed as
the following functional integral, which
we write in imaginary time:
\beq
\label{ZFP}
Z[j]\,=\,\int{\cal D}A\, \,{\rm det}\left(
\frac{\delta G^a}{\delta \theta^b}\right)
\exp\biggl\{-\int {\rm d}^4 x \biggl(\frac{1}{4}(F_{\mu\nu}^a)^2 +
 \frac{1}{2\lambda}(G^a[A])^2+j_\mu^a A_a^\mu \biggr)
\biggr\},\eeq
where $G^a[A]$ is the gauge fixing term
(for example,  $G^a=\del^\mu A^a_\mu$ for the so-called covariant
gauges, and $G^a=\del^i A^a_i$ for Coulomb gauges),
$\lambda$ is a free parameter (to be referred to as the gauge fixing
parameter) and $\delta G^a/\delta \theta^b$ is the functional
 derivative of $G^a[A]$ with respect to the parameter
$\theta^a(x)$ of the infinitesimal gauge transformations:
\beq\label{gtrinf}
\delta A_\mu^a \,=\,
-\,\frac{1}{g}\,\del_\mu\theta^a + f^{abc}A_\mu^b\theta^c\,=\,
-\,\frac{1}{g}\,[D_\mu, \theta]^a.\eeq
Since the gauge-fixed Lagrangian in eq.~(\ref{ZFP}) 
(including the Faddeev-Popov determinant) is not gauge-invariant, the 
equations of motion derived from it have no simple transformation
properties under the gauge transformations of the external sources
or of the average fields. 
It is however possible to develop a formalism
which guarantees these simple properties. This is  the method of  the
background field gauge  \cite{Witt67,Abbott81}. In this method, one splits
the gauge field into a classical background field $A^a_\mu$, to be later
identified with the
 average field, and a fluctuating quantum field $a_\mu^a$, and one
defines a new generating
functional:
\beq\label{Zbk0}
\tilde Z[j,A]=\int {\cal D}a \,{\rm det}\left(
\frac{\delta \tilde G^a}{\delta \theta^b}\right)
\exp \Biggl\{ -\int {\rm d}^4 x \biggl(\frac{1}{4}\Bigl(F_{\mu\nu}^a
[A+a]\Bigr)^2 +\,\frac{1}{2\lambda}(\tilde G^a[a])^2 +\,j_\mu^b
a_b^\mu \biggr)\Biggr\},\nonumber\\ \eeq
where the new gauge-fixing term $\tilde G^a$ is chosen so as
to be covariant under the gauge transformations of the background 
fields. Specifically, consider the following gauge transformations 
of the various fields and sources:
\beq\label{GT1}
A_\mu\,\to\,h A_\mu h^\dagger -(i/g)h \del_\mu h^\dagger,
\qquad \,\,\,\,
j_\mu\,\to\, h j_\mu h^\dagger,\nonumber\\
a_\mu \,\to\, h a_\mu h^\dagger,\qquad\,\,
\zeta \,\to \,h \zeta h^\dagger,\qquad\,\,
\bar\zeta \,\to \,h^\dagger \bar\zeta h.\eeq
(Note the {\it homogeneous} transformations of the
quantum gauge fields $a_\mu$ and of the ghost fields $\zeta,\,
\bar\zeta$ to be introduced shortly.) Then, the following,
Coulomb-type, gauge-fixing term
\beq\label{FPdet}
\tilde G^a\,\equiv\,[D_i[A],\, a^i]^a\,=\,\del^i a^a_i -
 gf^{abc}A_i^b a^{i\,c},\eeq
is manifestly covariant under the transformations (\ref{GT1}):
$\tilde G^a\to h^{ab} \tilde G^b$. (A gauge-fixing term of the
covariant type can be similarly defined with
$\tilde G^a\equiv[D_\mu[A],\, a^\mu]^a$.)

The Faddeev-Popov determinant in eq.(\ref{Zbk0}) involves the variation of
$\tilde G^a$ in the following gauge transformation:
\beq\label{newgtr}
a_\mu^a \,\to\,a_\mu^a\,-\,\frac{1}{g}\,\del_\mu\theta^a \,+\,
f^{abc}(A_\mu^b + a_\mu^b)\theta^c\,=\, -
\frac{1}{g}\,[D_\mu[A+a],\, \theta]^a.\eeq 
This determinant is written as
a functional integral over a set of anticommuting  ``ghost'' fields
in the adjoint representation, $\zeta^a$ and $\bar\zeta^a\,$:
\beq\label{BKDET}
{\rm det}\left(
\frac{\delta \tilde G^a}{\delta \theta^b}\right)\,=\,
\int{\cal D}\bar\zeta \,{\cal D}\zeta
\,\exp\biggl\{-\int {\rm d}^4 x \,
\bar\zeta^a\Bigl(D_i[A] D^i[A+a]\Bigr)_{ab}\zeta^b\biggr\}.\eeq
We thus obtain:
\beq\label{Zbk1}
\tilde Z[j;A]=\int{\cal D}a\,{\cal D}\bar\zeta\, {\cal D}\zeta\,
\,\exp\Bigl\{-S_{FP}[a,\zeta,\bar\zeta;A] - \int {\rm d}^4x\, 
j_\mu^b a_b^\mu\Bigr\},\eeq
with the  Faddeev-Popov action:
\beq \label{SFP}
S_{FP}[a,\zeta,\bar\zeta;A]=\int{\rm d}^4 x \biggl\{\frac{1}{4}
\Bigl(F_{\mu\nu}^a
[A+a]\Bigr)^2+ \frac{1}{2\lambda}\Bigl(D_i[A] a^i\Bigr)^2 
+\bar\zeta^a\Bigl(D_i[A] D^i[A+a]\Bigr)_{ab}\zeta^b
\biggr\},\nonumber\\\eeq
where $D_\mu[A+a]=\del_\mu + ig(A_\mu + a_\mu)$ is the covariant derivative
for the total field $A_\mu + a_\mu$, and $F_{\mu\nu}^a[A+a]$ is the 
corresponding field strength tensor. 

The essential property of the complete action in eq.~(\ref{Zbk1}),
including the sources, is to be invariant with respect
to the gauge transformations (\ref{GT1}).
Because of this symmetry, the generating functional $\tilde Z[j;A]$ is
invariant under the normal gauge transformations of its arguments, 
given by the first line of eq.~(\ref{GT1}).
To see this, it is sufficient to accompany the gauge transformations
of $j_\mu$ and $A_\mu$  by a change in the 
integration variables $a_\mu$, $\zeta$ and $\bar\zeta$
of the form indicated in the second line of eq.~(\ref{GT1}):
The combined transformations do not modify the full
 action, neither the functional measure (since they  are unitary).
This symmetry of $\tilde Z[j,A]$ guarantees the covariance of the
 Green's functions under the gauge transformations
(\ref{GT1}) of the external field and current, which is the property we
were after. 

Specifically, by functionally differentiating $\tilde W[j,A]\equiv
 -\ln\tilde Z[j,A]$ with respect to the external current $j^\mu_a$,
 one generates the connected Green's functions
of the fields $a_\mu^a$. They depend on the background field
$A_\mu^a$ through the gauge fixing procedure.
Consider, for instance, the average field:
\beq\label{AVA}
\langle a_\mu^b(x)\rangle\,=\,\frac{\delta\tilde W[j,A]}
{\delta j^\mu_b(x)}\,,\eeq 
 We wish to show that, under the gauge transformation (\ref{GT1}),
$\langle a_\mu\rangle \,=\,
\langle a^b_\mu\rangle T^b$ transforms as follows:
\beq\label{amutransfo}
\langle a_\mu\rangle\,\to\,\langle a_\mu\rangle^\prime\,=\,
h\,\langle a_\mu\rangle\,h^{-1}.\eeq
We have:
\beq
\langle a_\mu\rangle^\prime&=&\frac{\delta\tilde W[j',A']}
{\delta j'^\mu_b(x)}\,
\nonumber\\
&=&\tilde Z^{-1}[j^\prime,A^\prime]
\int {\cal D}a^\prime{\cal D}\bar\zeta^\prime {\cal D}\zeta^\prime
\,\,a_\mu^\prime\,
\,\exp\Bigl\{-S_{FP}[a^\prime,\zeta^\prime,\bar\zeta^\prime
;j^\prime,A^\prime]\Bigr\}
\nonumber\\
&=&\tilde Z^{-1}[j,A]
\int {\cal D}a{\cal D}\bar\zeta {\cal D}\zeta\, (h\,a_\mu\,h^{-1})
\,\exp\Bigl\{-S_{FP}[a,\zeta,\bar\zeta;j,A]\Bigr\}
\nonumber\\&=&
h\,\langle a_\mu\rangle\,h^{-1}.\eeq
(Note that the action $S_{FP}$ has been temporarily redefined
so as to include the coupling to the external current.)
In going from the second to the third line,
we have changed the integration variables according to 
 eq.~(\ref{GT1}), and used the invariance
of the functional measure and of the full action under the
transformations (\ref{GT1}).
 Similarly, it is easy to verify that the 2-point function:
\beq\label{DBK}
G_{\mu\nu}^{ab}(x,y)\,\equiv\,\langle{\rm T}
a_\mu^a(x)a_\nu^b(y)\rangle
\,=\,\frac{\delta^2 \tilde W[j,A]}
{\delta j^\mu_a(x)\delta j^\nu_b(y)}\,.\eeq
transforms covariantly:
\beq\label{DTr}
G_{ab}^{\mu\nu}(x,y)&\to&  h_{a\bar a}(x) \,G_{\bar a\bar b}
^{\mu\nu}(x,y)\, h^\dagger_{\bar b b}(y).\eeq
The ghost propagator,
\beq\label{Delta}
\Delta^{ab}(x,y)\,\equiv\,
\langle{\rm T}\,\zeta^a(x)\bar\zeta^b(y)\rangle,\eeq
has the same transformation property.
Similar covariance properties hold for the higher point Green's
functions, and for the various self-energies.

At this point, we require the background field $A_\mu^a$ to
be precisely the average field in the system. 
First, note that, for an arbitrary external current
$j_\mu^a$, the total average field 
is $A_\mu^b + \langle a_\mu^b\rangle$.
To see this, perform a shift of the integration variable
$a_\mu$ in the functional integral (\ref{Zbk0}) of the form
$a_\mu^b\,\to\,a_\mu^b - A_\mu^b$, to get:
\beq\label{Zbk2}
\tilde Z[j,A]\,=\, Z[j;A]\, {\rm exp}\Bigl\{\int j^b_\mu 
A_b^\mu \Bigr\},\eeq where $ Z[j;A]$
is the usual generating functional, eq.~(\ref{ZFP}),
but evaluated in an  unconventional gauge which 
depends on the {\bk} field $A_\mu^a$.
Eq.~(\ref{Zbk2}) implies $\tilde W[j,A]\,=\, W[j;A]\,-\,(j,A)$,
so that the total average field in the system is:
\beq
\frac{\delta W[j;A]}{\delta j^\mu_b(x)}\,=\,
\frac{\delta\tilde W[j,A]}{\delta j^\mu_b(x)}\,+\,A_\mu^b(x)
\,=\, \langle a_\mu^b(x)\rangle \,+\,A_\mu^b(x),\eeq
as anticipated. This becomes equal to $A_\mu^b$ if 
\beq \label{consist0}
\langle a_\mu^b(x)\rangle  \,=\,0.\eeq
This condition,
which has a gauge-invariant meaning since
 the average values of the quantum fields transform homogeneously (see
eq.(\ref{amutransfo})), implies a functional relation between
the external current  and the background field, which we write 
as $j[A]$. The functional:
\beq\label{defGbk}
\Gamma\bigl[A] \,\equiv\,\tilde W[j[A],\, A],\eeq
is the {\it effective action}, whose functional derivatives are 
the one-particle-irreducible amplitudes  in the background field gauge.
By construction, $\Gamma[A]$ is invariant with respect to
the gauge transformations of its argument, but, in general, it  
depends on the gauge-fixing parameter $\lambda$.

 In what follows, we shall not construct
the effective action (\ref{defGbk}) (see however Sect. 5.2), 
but we shall instead  write  directly
the equations of motion for the average fields and the 2-point
functions, and we shall impose on these equations
the consistency condition (\ref{consist0}).
The resulting equations will be  then covariant with respect
to the gauge transformations of the classical mean fields.

Let us now add fermionic fields and sources. The
full generating functional, that we shall use in the rest
of this section, reads then:
\beq
\label{Zbk}
\tilde Z[j,\eta,\bar\eta,A,\Psi,\bar\Psi]\,=\,
\int {\cal D}a {\cal D}\bar\zeta {\cal D}\zeta
 {\cal D}\bar\psi {\cal D}\psi
\,\exp\biggl\{-\,S_{FP}
\,-\,\int\Bigl(j_\mu^b a_b^\mu
+\bar\eta\psi+\bar\psi\eta\Bigr)\biggr\},
\eeq
where
 the Faddeev-Popov action $S_{FP}$ depends on both
the quantum and the background fields:
\beq \label{SFP0}
S_{FP}\equiv S_{cl}(A+a,\Psi+\psi,\bar\Psi+\bar\psi)+
\int\left\{ \,
 \frac{1}{2\lambda}\Bigl(D_i[A] a^i\Bigr)^2 +
\bar\zeta^a\Bigl(D_i[A] D^i[A+a]\Bigr)_{ab}\zeta^b\right\}.
 \nonumber\\\eeq
In the above equation,  $S_{cl}$ is the usual QCD action in imaginary time, 
eq.~(\ref{QCD}), but evaluated for the shifted fields
$A+a$, $\Psi+\psi$, and $\bar\Psi+\bar\psi$.
As in eq.~(\ref{Zbk0}),  the external sources $j^a_\mu$, $\eta$ and
$\bar\eta$ are coupled only to the quantum fields.

Some attention should be paid to the boundary conditions
in the functional integral (\ref{Zbk}). The gluonic fields
to be integrated over are periodic in imaginary time,  with period $\beta$:
$a_\mu(\tau=0)  = a_\mu(\tau=\beta)$. 
The fermionic fields $\psi$, $\bar\psi$
satisfy {\it antiperiodic} boundary conditions (e.g.,
$\psi(\tau=0)  =-\psi(\tau=\beta)$) (see, e.g., 
\cite{BR86,MLB96}). Finally, the ghost fields
$\zeta$ and $\bar\zeta$ are {\it periodic} in spite of
their Grassmannian nature: this is because the Faddeev-Popov determinant
is defined on the space of periodic gauge fields \cite{Bernard74,Hata80}.

The partition function (\ref{Zbk}) is invariant under the
gauge transformations of the background fields and of the
external sources, that is, the transformations (\ref{GT1})
together with: 
\beq\label{GT3} 
 \Psi\,\to\, h\Psi, \qquad &{}&
\bar\Psi \,\to\, \bar\Psi h^{-1},\nonumber\\
\eta\,\to\, h\eta,\qquad &{}&
\bar\eta \,\to\, \bar\eta h^{-1}.\eeq 
Accordingly, the associated Green's functions
are covariant under the same transformations.
Finally, the classical fields $A$, $\Psi$ and $\bar\Psi$ are identified
with the respective average fields by requiring that
(cf. eq.~(\ref{consist0})):
\beq\label{consist}
\langle a_\mu\rangle  \,=\,
\langle\psi\rangle \,= \,  \langle\bar\psi\rangle \,=\,0\,. \eeq

In constructing the kinetic theory below, it will be convenient to 
use the (strict) Coulomb gauge, which offers the most direct 
description of the physical degrees of freedom.
This gauge is defined either 
by eq.~(\ref{SFP0}) with $\lambda\to 0$,
or, which is operationally simpler, by imposing the transversality
constraint
\beq\label{COUL}
D_i[A]\,a^i\,=\,0,\eeq
within the functional integral (\ref{Zbk}).
 In this gauge, the gluons Green's functions are
(covariantly) transverse, that is:
\beq\label{COVTR0}
D^i_x[A] \,G_{i\nu}(x,y)\,=\,0\,,\eeq
and similarly for the higher point functions. 
At tree-level and with $A=0$, the only non-trivial
components of the retarded propagator are:
\beq\label{G0COUL}
G_{00}^{(0)}(k)\,=\,-\,\frac{1}{{\bf k}^2},\qquad
G_{ij}^{(0)}(k)\,=\,-\,\frac{\delta_{ij}-\hat k_i\hat k_j}{
(k_0+i\eta)^2-{\bf k}^2}.\eeq
That is, the electric gluon is static, and the same
is also true for the Coulomb ghost: $\Delta^{(0)}(k)=1/{\bf k}^2$.
Accordingly,
\beq\label{W0}
G^{< \,(0)}_{ij}(k)=(\delta_{ij}-\hat k_i\hat k_j)\,G^<_0(k),\qquad
G^{>\,(0)}_{ij}(k)=(\delta_{ij}-\hat k_i\hat k_j)\,G^>_0(k),\eeq
[where, e.g., $G^<_0(k)=\rho_0(k)N(k_0)$, and
$\rho_0(k)=2\pi \epsilon(k_0)\delta(k^2)$; cf. eq.~(\ref{G><k})],
while all the other components (like $G^{< \,(0)}_{00}$) just vanish.
That is, in this gauge, only the physical transverse gluons
are part of the thermal bath at tree-level. 
This is convenient since ghosts or
electric gluons do not contribute to the polarization effects
to be considered below.
(In gauges with propagating unphysical degrees of freedom,
the contributions from ghost and longitudinal gluons 
cancel each other in the final results, thus leaving only the 
contributions of the transverse gluons \cite{BP90,qcd};
alternatively, the former can be kept unthermalized using the
formalism of Refs. \cite{Landshoff93}.)
Keeping this in mind, we shall completely ignore the ghosts
in what follows.

\subsubsection{Equations of motion for the mean fields}

The mean field equations in the {\bk} field method
are easily derived from the generating functional (\ref{Zbk}),
and read: 
\beq
\label{Acl1}
\Bigl\langle D_{ab}^\nu[A+a] F_{\nu\mu}^b[A+a]\Bigr \rangle
- g\Bigl\langle(\bar\Psi+ \bar\psi) \gamma_\mu
t^a(\Psi+\psi)\Bigr\rangle &=&j_\mu^a(x),\qquad\\
\label{psicl1}
i\,\Bigl\langle\slashchar{D}[A+a](\Psi+\psi)\Bigr\rangle&=&\eta(x),
\eeq
together with the Hermitian conjugate equation for $\bar\Psi$.
Here, $D^\dagger[A]=\buildchar{\del}{\leftarrow}{}
-igA^aT^a$,
and the derivative $\buildchar{\del}{\leftarrow}{}$ acts on the function on
its left. The physically interesting equations are obtained after
imposing the conditions (\ref{consist}), and can be
written compactly as:
\beq
\label{avA1}
\left [\, D^\nu,\, F_{\nu\mu}(x)\,\right ]^a
\,-\,g  \bar\Psi (x)\gamma_\mu t^a \Psi(x)
&=&j_\mu^a(x)+j_\mu^{ind\, a}(x),\\
\label{avpsi1}
i\slashchar{D} \,\Psi(x)&=&\eta(x)+\eta^{ind}(x).
\eeq
Here and in what follows, $D_\mu$ or $F_{\mu\nu}$  denote
the covariant derivative or the field strength tensor
 associated to the  {\bk} field $A_\mu^a(x)$.

The left hand sides
 of the above equations are the same as at tree-level.
All the quantum and medium effects are included in 
{\it the induced sources}
$j_\mu^{ind\,a}$ and $\eta^{ind}$ in the right hand sides.
The induced colour current $j_\mu^{ind\,a}$ may be written as:
\beq\label{jindfb}
j^{ind\,\mu}_a(x)\,=\,j^{\mu}_{{\rm f}\,a}(x)\,+\,j^\mu_{{\rm g}\,a}(x),\eeq
with the two terms representing, respectively, the quark and gluon
contributions:
\beq\label{jindf}
j^{\mu\,a}_{{\rm f}}(x)&=& g \left\langle \bar\psi (x)
\gamma^\mu t^a \psi (x) \right\rangle,\\
j^{\mu\,a}_{\rm  g}(x)&=&
 gf^{abc}\,\Gamma^{\mu\rho\lambda\nu}\left
\langle a^b_\nu\left (D_\lambda a_\rho\right
)^c\right\rangle\,
+\,g^2f^{abc}f^{cde}\left\langle a_\nu^b a^\mu_d
a^{\nu}_e\right\rangle,\label{jindgG}
\eeq
where $\Gamma_{\mu\rho\lambda\nu}\equiv\,2g_{\mu\rho}g_{\lambda\nu}
-g_{\mu\lambda}g_{\rho\nu}-g_{\mu\nu}g_{\rho\lambda}.$
Finally, the induced fermionic source reads:
 \beq
\label{etaind}
\eta^{ind}(x)\,=\,g\gamma^\nu t^a\left\langle a_\nu^a(x)\psi (x)
\right\rangle.
\eeq
In equilibrium,  both the
mean fields and  the induced sources vanish. 
This follows from symmetry: in equilibrium,
the expectation values involve thermal averages over colour singlet states,
and elementary group theory can then be used to prove that, in this case,
all terms on the r.h.s. of eqs.~(\ref{jindfb})---(\ref{jindgG}) indeed vanish.
Similarly, $\eta^{ind}$
is nonvanishing only in the presence of  fermionic mean fields.


We shall compute later in this section the induced sources as functionals
of the average fields. Since we consider both fermionic ($\Psi$ and
$\bar\Psi$) and gauge ($A_\mu^a$) mean fields, it is convenient
to separate the corresponding induced effects by writing:
\beq
j^{ind\,a}_\mu\equiv j^{A\,a}_\mu+j^{\psi\,a}_\mu.\eeq
 The first piece, $j^A_\mu\equiv
j^{ind}_\mu[A_\nu, \Psi=\bar\Psi=0]$, is the colour current
which is induced by  gauge fields alone. The second piece, $j^\psi_\mu$,
denotes the contribution of the fermionic mean fields; in 
general, this is also dependent  on the gauge  fields  $A_\mu^a$.
Similarly, we identify quark and gluon contributions by writing
\beq\label{jAPSI}
j^\mu_{\rm f}=j^{A\,\mu}_{\rm f}+
j^{\psi\,\mu}_{\rm f},\qquad\qquad j^\mu_{\rm g}=j^{A\,\mu}_{\rm g}+
j^{\psi\,\mu}_{\rm g},\eeq
 for the two pieces of the induced current in eq.~(\ref{jindfb}).

\subsubsection{Induced sources and two point functions}

 By inspecting  eqs.~(\ref{jindfb})--(\ref{etaind}), one sees that
 the induced sources are entirely expressed in terms of 2-point functions.
(The only exception is the induced current $j^{\mu\,a}_b$ which also contains
 the 3-point function $\left\langle a_\nu^b a_\mu^d a^{\nu}_e\right\rangle$.
However, the leading contribution to this  3-point function contains at
least two powers of
$g$ more than the other terms, so that it can be  ignored
at leading order.) 

We now introduce specific notations for
the various 2-point functions  which will appear in the forthcoming
developments.
Aside from the usual quark and gluon propagators,
\beq\label{nor}
S_{ij}(x,y)&\equiv&\langle{\rm T}\psi_i(x)\bar\psi_j(y)\rangle
=-\frac {\delta \langle \psi_i(x)\rangle }{\delta\eta_j(y)}, \nonumber\\
G_{\mu\nu}^{ab}(x,y)&\equiv&\langle{\rm T}a_\mu^a(x)a_\nu^b(y)\rangle
=-\frac{\delta\langle a_\mu^a(x)\rangle}{\delta j^\nu_b(y)},\eeq
we shall also need the following ``abnormal'' propagators:
\beq\label{abn}
K_{i\nu}^b(x,y)&\equiv&\langle{\rm T}\psi_i(x)a^b_\nu(y)\rangle
=-\frac{\delta\langle \psi_i(x)\rangle }{\delta j_b^\nu(y)}
=-\frac{\delta \langle a_\nu^b(y)\rangle}{\delta\bar\eta_i(x)},
\nonumber\\
H_{\nu i}^b(x,y)&\equiv&\langle{\rm T}
a^b_\nu(x)\bar\psi_i(y)\rangle=-\frac{\delta
\langle \bar\psi(y)\rangle}{\delta
j_b^\nu(x)}=\frac{\delta \langle a_\nu(x)\rangle}{\delta\eta(y)},
\eeq 
which vanish in equilibrium, and which mix fermionic and bosonic
degrees of freedom.

The time ordered propagators are further separated
into components which are  analytic functions of their time arguments
(see sections 2.1.2 and 2.2.3). For example,
the fermion propagator is written as (with colour indices omitted):
      \beq\label{analytic}
S(x,y) &=&\theta(\tau_x-\tau_y)\langle\psi(x)\bar\psi(y)\rangle -
\theta(\tau_y-\tau_x)\langle\bar\psi(y)\psi(x)\rangle\nonumber\\
&\equiv&\theta(\tau_x-\tau_y)
S^>(x,y)-\theta(\tau_y-\tau_x)S^<(x,y),   \eeq 
where the minus sign appears because of the anticommutation
property of the fermionic fields. In particular, for a free
massless fermion, we have (cf. eq.~(\ref{FMix}))
\beq\label{S0}
 S_0^>(k)\,=\,\slashchar{k}\,\rho_0(k) [1-n(k_0)]\,,\qquad
 S_0^<(k)\,=\,\slashchar{k}\,\rho_0(k) n(k_0)\,.
\eeq
Similar definitions hold
for the other 2-point  functions $G,\,\Delta,\, K$ and $H$, but
without the minus sign.  For instance,
 \beq\label{analyticG}
G_{\mu\nu}^{ab}(x,y)\,=\,\theta(\tau_x-\tau_y)\,G^{>\,ab}_{\,\,\mu\nu}(x,y)
\,+\,\theta(\tau_y-\tau_x)\,G^{<\,ab}_{\,\,\mu\nu}(x,y).\eeq
After continuation to real time, the
functions above have hermiticity properties which generalize
eq.~(\ref{HERM}): e.g., $(S^>)^\dagger(x,y)\\ = \gamma^0S^>(x,y)\gamma^0$
and $(G^>)^\dagger(x,y) = G^> (x,y)$, or, more explicitly,
$(G_{\mu\nu}^{>\,ab}(x,y))^*=G_{\nu\mu}^{>\,ba}(y,x)$.
Note also the following symmetry property, which will be useful later:
\beq\label{symm}
G^{>\,ab}_{\,\,\mu\nu}(x,y)=G^{<\,ba}_{\,\,\nu\mu}(y,x).\eeq

The induced sources in eqs.~(\ref{jindf})--(\ref{etaind}) involve  
products of fields with equal time arguments $\tau_x=\tau_y$. They may be
 expressed in terms of the analytic components of the above propagators by
taking the limit   $\tau_y - \tau_x=\eta\to 0^+$:
\beq\label{jf}
j_{{\rm f}}^{\mu\,a}(x)&=&g\,{\rm Tr}\Bigl(\gamma^\mu
 t^aS^<(x,x)\Bigr),\nonumber\\
j^{\mu\,a}_{\rm  g}(x)&=&i\,g\,\Gamma^{\mu\rho\lambda\nu}\,
{\rm Tr}\,\,T^a\,
 D^x_\lambda\,G_{\rho\nu}^<(x,y) |_{y\to x^+},\nonumber\\
\eta^{ind}(x)&=&g\gamma^\nu t^aK^<_{a\,\nu}(x,x). \eeq
Here, the traces involve both spin and colour indices,
and $y\to x^+$ stands for $\tau_y-\tau_x\to 0^+$
(or for $y_0-x_0 \to i0^+$ after continuation to real time).

Because the induced sources involve products of fields at the same point,
one could expect to encounter ultraviolet divergences when calculating them.
However, this will not be the case in our leading order calculation.
Indeed, as we shall verify later, the dominant contribution
to the induced sources arises entirely from
 the thermal particles; this  contribution is ultraviolet finite,
owing to the presence of the thermal occupation factors.

At this point, it is easy to verify the gauge transformation
properties of the induced sources. 
We have already emphasized that the Green's functions 
are covariant
under the transformations: 
\beq\label{BGT}
A_\mu \to   h A_\mu h^{-1}- ({i}/{g})\,h\del_\mu h^{-1},
\qquad \Psi\,\to\, h\Psi, \qquad\bar\Psi \,\to\, \bar\Psi h^{-1}.\eeq
For instance, the gluon 2-point function transforms according to
eq.~(\ref{DTr}), and, similarly:
\beq\label{GTR1}
S_{ij}(x,y)&\to& h_{ik}(x) \,S_{kl}(x,y)\, h_{lj}^{-1}(y),\nonumber\\
K^{\nu}_{ia}(x,y)&\to&
h_{ij}(x)\, K^{\nu}_{jb}(x,y)\, \tilde h_{ab}(y)\,.\eeq
(We have denoted by $h(x)$ and 
$\tilde h(x)$ 
the elements of the gauge group in the fundamental and the adjoint
representations, respectively.)
It is then easy to see that the induced currents
in eqs.~(\ref{jf}) transform as colour vectors in the
adjoint representation, while $\eta^{ind}_i(x)$ transforms
like $\Psi_i(x)$, i.e., as a colour vector in the fundamental representation.
Accordingly, the mean field equations  (\ref{avA1}) and (\ref{avpsi1}) are
gauge covariant.

\subsection{Approximation scheme}

In this section we develop the approximations that allow
us to construct kinetic equations for the off-equilibrium 
2-point functions in eq.~(\ref{jf}).
These approximations are intended to retain the terms of leading order
in $g$ in the induced sources, given that the coupling constant 
enters not only the interaction vertices, 
but also the space-time inhomogeneities
of the plasma (since $\del_x\sim gT$), and, for the reasons
explained in Sect. 3.1, also the amplitudes of the mean fields:
$gA\sim gT$ (or $F_{\nu\mu}\sim gT^2$),
and $\bar\Psi \Psi\sim gT^3$. The constraint
on $\bar\Psi \Psi$ is introduced for 
consistency with the Yang-Mills equation (\ref{avA1}), together with 
the previous constraint on $A^\mu\/$: this insures, e.g., that
$g  \bar\Psi \gamma_\mu t^a\Psi\sim g^2 T^3$, 
which is of the same order as the terms involving $A^\mu_a$ 
(like $j^{A}_\mu\sim \Pi_{\mu\nu}A^\nu \sim g^2T^2A$) 
within the same equation.

The starting point is provided by the
imaginary-time Dyson-Schwinger equations for the
2-point functions, as obtained by differentiating 
the mean field equations with respect to the external sources.
To this aim, we consider the mean field equations
(\ref{Acl1})--(\ref{psicl1}) for arbitrary values of
the average quantum fields $\langle a_\mu^b\rangle$ etc.,
and use identities like the one listed in the r.h.s.'s
of eqs.~(\ref{nor})--(\ref{abn}). After differentiation,
we set the average values of the quantum fields to zero
(recall eq.~(\ref{consist})). Then, the equations thus obtained
are continued towards real time, by exploiting the analytic properties
of the various Green's functions and self-energy.
The final outcome of this procedure are generalizations
of Kadanoff-Baym equations for QCD (cf. sections 2.2.3 and 7.1).

\subsubsection{Mean field approximation}

The first approximation to be performed
is a {\it mean field approximation} (cf. Sect. 2.3.3), 
which is equivalent to the one-loop approximation of the 
diagrammatic approach.

\begin{figure}
\protect \epsfysize=2.cm{\centerline{\epsfbox{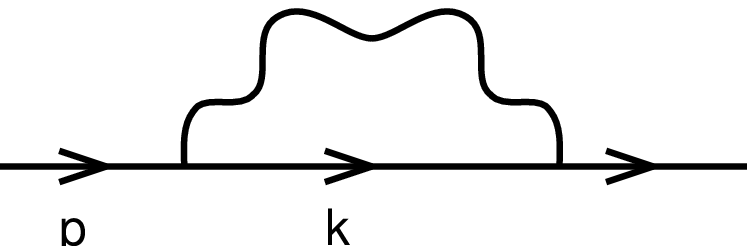}}}
         \caption{The one-loop quark self-energy}
\label{OLSIG}
\end{figure}

To justify this approximation, consider
the equation for the quark propagator $S(x,y)$, obtained
by differentiating eq.~(\ref{psicl1}) with respect to $\eta(y)\/$:
\beq\label{SMA}
-i{\slashchar D}_x\,S(x,y)-
g\gamma^\nu t^a\Psi(x)H^{a}_\nu(x,y)
+\int{\rm d}^4z\,\Sigma(x,z)S(z,y)\,=\,\delta(x-y).\eeq
Here, $\Sigma(x,y)$ is the quark self-energy,
defined as (compare to eq.~(\ref{defSig})) :
\beq\label{oneloopS}
\int{\rm d}^4z\,\Sigma(x,z) S(z,y)\,\equiv\,-\,\frac
{\delta \eta^{ind}(x)}{\delta \eta(y)}\,=\,
g\,\langle{\rm T}\slashchar{a}(x)\psi(x)\bar\psi(y)\rangle_c\,.\eeq
In thermal equilibrium, the first contribution to $\Sigma$ arises
at one-loop order (see fig.~\ref{OLSIG}), and is $\Sigma_{eq}\sim g^2 T$.

\begin{figure}
\protect \epsfxsize=10.cm{\centerline{\epsfbox{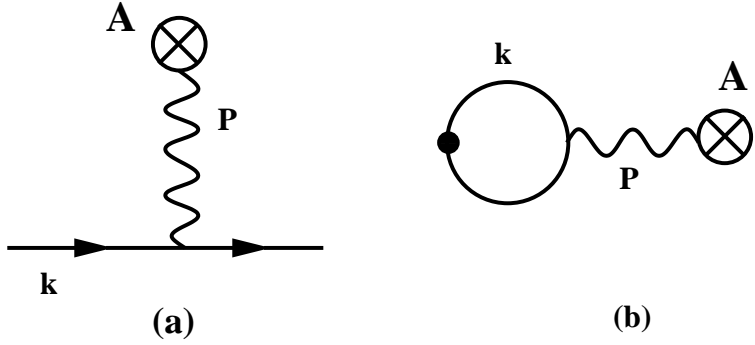}}}
         \caption{Off-equilibrium effects to lowest order: (a)
a mean field insertion in the fermion propagator;
 (b) the corresponding contribution to the induced current.}
\label{MFA1}
\end{figure}

\begin{figure}
\protect \epsfxsize=10.cm{\centerline{\epsfbox{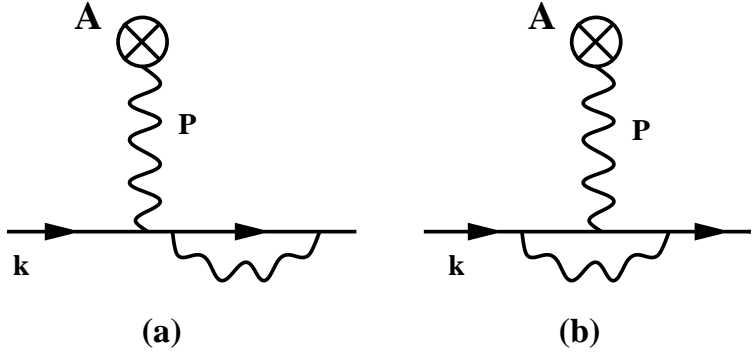}}}
         \caption{One-loop corrections to the single field insertion
 in fig.~\ref{MFA1}.a.}
\label{1loopMF}
\end{figure}

The induced current $j_{\rm f}^\mu$, eq.~(\ref{jf}), involves
 the off-equilibrium deviation of the propagator,
$\delta S\equiv S-S_{eq}$, which can be obtained from
perturbation theory. The diagrams contributing to $\delta S$ 
contain at least one mean field insertion. 
Let us consider insertions of the gauge field
$A^\mu_a$, for definiteness: the lowest order contribution
$\delta S^{(0)}$ is shown in fig.~\ref{MFA1}.a; when
the hard line with momentum $k\sim T$ is closed on itself,
this generates the one-loop contribution to the induced current
displayed in fig.~\ref{MFA1}.b.
The first ``radiative'' corrections to fig.~\ref{MFA1}.a come
from the self-energy term in eq.~(\ref{SMA}) and are displayed in
fig.~\ref{1loopMF}. The diagram ~\ref{1loopMF}.a is obtained 
by inserting the {\it equilibrium} self-energy, fig.~\ref{OLSIG}, in any of 
the external lines in fig.~\ref{MFA1}.a. The diagram ~\ref{1loopMF}.b 
involves the (lowest-order) {\it off-equilibrium} self-energy
$\delta\Sigma^{(0)}$, and is obtained by replacing the internal
fermion propagator  in fig.~\ref{OLSIG} by $\delta S^{(0)}$.
\begin{figure}
\protect \epsfxsize=10.cm{\centerline{\epsfbox{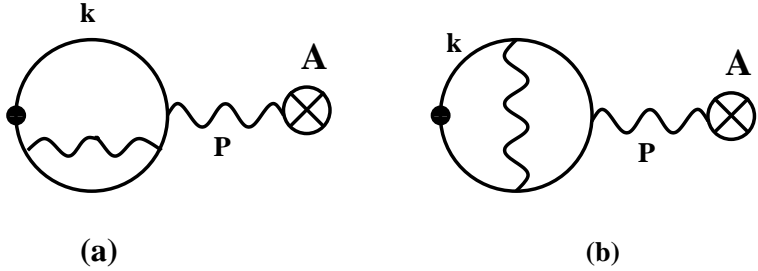}}}
         \caption{One-loop corrections to the induced current
in fig.~\ref{MFA1}.b: (a) self-energy correction; (b) vertex correction.}
\label{2loopjf}
\end{figure}
By closing the external lines in fig.~\ref{1loopMF} on themselves,
one obtains the two-loop corrections to the induced current
shown in figs.~\ref{2loopjf}.a and b. Power counting suggests that
these two-loop corrections in fig.~\ref{2loopjf} can be neglected in leading
order, since they are suppressed by a factor of $g^2$ with respect
to the one-loop contribution in fig.~\ref{MFA1}.b. But 
when both the external and the internal gluon lines in
fig.~\ref{2loopjf} are soft, na\"{\i}ve power 
counting can be altered by infrared effects. For, in that case,
the internal fermion propagators are nearly on-shell and read
$1/(k\cdot P)$, with $k\sim T$ (the hard momentum running along the quark
loop) and $P\simle gT$ (any of the soft momenta carried by the gluons,
or a linear combination of them); the smallness of $P$ gives rise
to an enhancement over the na\"{\i}ve estimates. A careful
analysis shows that the relative magnitude
of the two-loop corrections depends upon the external momentum $P$:
({\it a\/}) if $P\sim gT$, then the two-loop diagrams are indeed suppressed,
but only by one power of $g$ \cite{BP90}; ({\it b\/}) if 
$P\simle g^2T$, then the one- and two-loop contributions
become equally important \cite{Bodeker,BE,Bodeker99}, as are 
also higher loop diagrams to be presented in Sect. 7.

The same conclusion can be reached by analyzing directly
the equations of motion. Consider for instance
the gluon propagator $G(x,y)$ in a soft colour background
field $A^\mu_a(x)$. Its Wigner transform $G(k,X)$ obeys a 
kinetic equation similar to eq.~(\ref{Boltz}) for the scalar 
field, which involves three types of terms: 
a drift term $(k\cdot \del_X)G^<$,
a mean field term, and a collision term $C(k,X)=-(G^>\Sigma^<- \Sigma^>G^<)$.
The relative magnitude of the collision
term is determined by comparing it with the drift term. 
In order to compare  $(k\cdot \del_X)G^<$ with $C(k,X)$, one should
first recall that both vanish in equilibrium (cf. Sect. 2.3.2).
Let us then set $G^<(k,X)=G^<_{eq}(k)+\delta G^<(k,X)$ and 
$\Sigma^<(k,X)\equiv \Sigma^<_{eq}(k)+\delta\Sigma^<(k,X)$. The
drift term becomes
$(k\cdot \del_X)\delta G^<$, while:
\beq\label{OFFCOL}
C(k,X)\,=\,-\Bigl(\Sigma^<_{eq}\,\delta G^>-\Sigma^>_{eq}\,\delta G^<\Bigr)
+\Bigl(\delta\Sigma^>\, G_{eq}^<-
\delta\Sigma^<\,G_{eq}^>\Bigr)\,+\,\dots\,,\eeq
where the dots stand for terms which are quadratic in the off-equilibrium
deviations. Since $\Sigma_{eq}(k)\sim g^2T^2$ is fixed by
the physics in equilibrium, the importance of the self-energy corrections
in the kinetic equation depends upon the scale $\del_X$ of the inhomogeneity:
If $\del_X\sim gT$, then the collision terms are suppressed by one 
power of $g$ and can be neglected to leading order.
If $\del_X\sim g^2T$ or less, the collision terms are
as important as the drift term. (See however Sect. 7, where
``accidental'' cancellations will be discussed which alter
slightly this argument.)

To summarize, when studying the collective dynamics
at the scale $gT$ and to leading order in $g$, we can restrict ourselves
to a mean field approximation where the hard particles 
interact only with the soft mean fields. 
The relevant equations for the 2-point
functions read then (in Coulomb's gauge, cf. eq.~(\ref{COUL})):
 \beq\label{S10}
\slashchar{D}_xS^<(x,y)&=&ig\gamma^\nu t^a\Psi(x)H^{<\,a}_\nu(x,y),\\
\label{K10} \slashchar{D}_x
K^{<\,b}_\nu(x,y)&=&-ig t^a\gamma^\mu\Psi(x)G^{<\,ab}_{\mu\nu}(x,y), \\
\label{K11}
\Bigl(g_{\mu\nu}D^2- D_\mu D_\nu +
2igF_{\mu\nu}\Bigr)_y^{ab}
 K^{<\,\nu}_b(x,y)&=&-gS^<(x,y)\gamma_\mu t^a\Psi(y),\\
\label{D10}
\Bigl(g_{\mu}^\rho D^2 - D_\mu D^\rho
+2igF_{\mu}^{\,\,\,\rho}\Bigr)_x^{ac}G^{<\,cb}_{\rho\nu}
(x,y)&=&
g\bar\Psi(x)\gamma_\mu t^a K_\nu^{<\,b}(x,y)
\,+\, g H_\nu^{<\,b}(y,x)\gamma_\mu t^a\Psi(x).\nonumber\\
&{}&\eeq
They must be supplemented with
the gauge-fixing conditions (cf. eq.~(\ref{COVTR0})):
\beq\label{TR}
D^i_xG_{i\nu}^<(x,y)=0,&{}&\qquad\,\,
G_{\mu j}^<(x,y) D^{j \dagger}_y=0,\nonumber\\
D^i_x H^{<\,a}_i(x,y)=0,&{}&\qquad\,\,D^i_yK^{<\,a}_i(x,y)=0,\eeq
and the initial conditions chosen such that, 
in the absence of the external sources, the system is
in equilibrium:
the mean fields vanish, and the 2-point functions reduce to
the corresponding functions in equilibrium. To the order of 
interest, the latter are the corresponding free functions
(cf. eqs.~(\ref{W0}) and (\ref{S0})):
\beq\label{INITGSK}
G^<(x,y)_{eq}\,\simeq\, G^<_0(x-y),\quad
S^<(x,y)_{eq}\,\simeq \,S^<_0(x-y),\quad
K^<(x,y)_{eq}\,=\,0.\eeq 
Since eqs.~(\ref{S10})--(\ref{D10}) involve only the ``smaller''
components, like $G^<$, we shall often omit the upper indices  ``$<$''
in what follows.

\begin{figure}
\protect \epsfxsize=16.cm{\centerline{\epsfbox{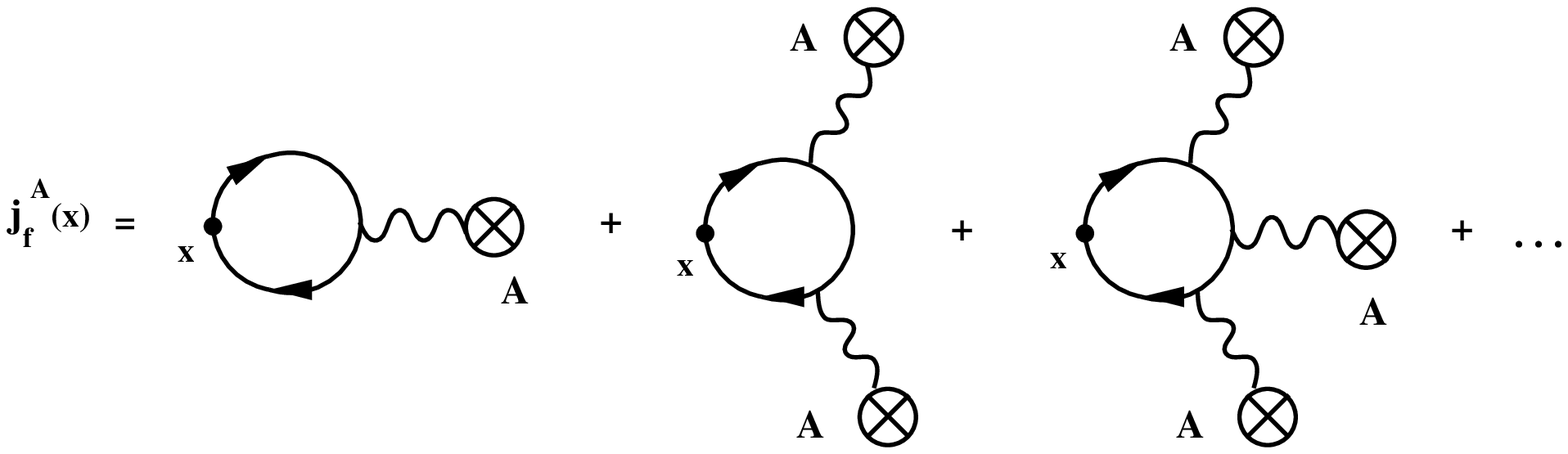}}}
         \caption{The quark current $j_{\rm f}^A$ induced by a colour field $A_\mu$,
 in the one-loop approximation. The blobs represent gauge field insertions.}
\label{jquark}
\end{figure}

\begin{figure}
\protect \epsfxsize=16.cm{\centerline{\epsfbox{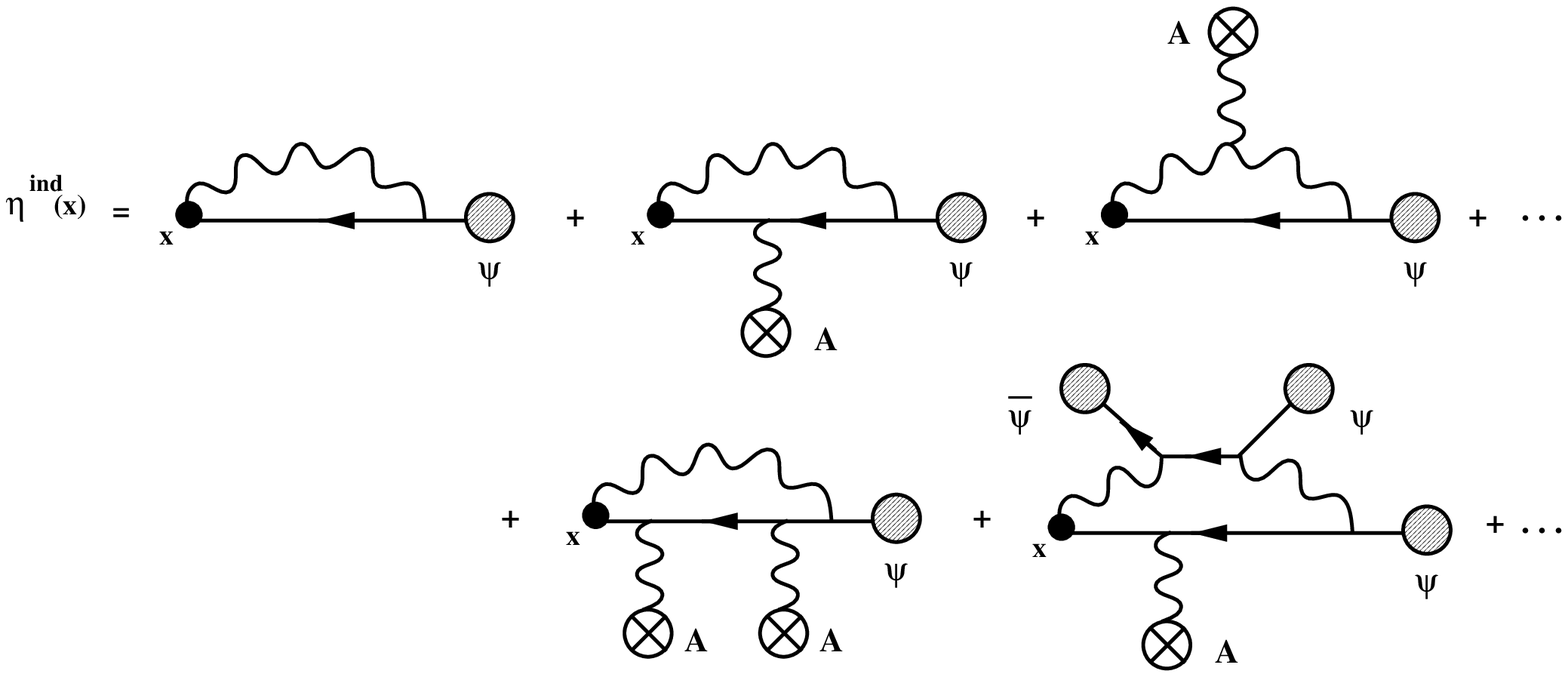}}}
         \caption{Some typical contributions to the
 induced fermionic source $\eta^{ind}$
in the one-loop approximation. The blobs represent either gauge field,
or fermionic field insertions.}
\label{etafig}
\end{figure}

Given the transformation laws in eqs.~(\ref{DTr}) and (\ref{GTR1}),
it is easily seen that eqs.~(\ref{S10})--(\ref{D10})
 are covariant under the gauge transformations
(\ref{BGT}) of the average fields. 
By solving these equations
without further approximations, one would obtain the induced sources
to one-loop order (see figs.~\ref{jquark} and \ref{etafig}
for some corresponding diagrams).
However, by itself, the mean field approximation is not 
a consistent approximation: additional powers of $g$
are hidden in the soft off-equilibrium inhomogeneities, and 
these will be isolated with the help of the gradient expansion.

\subsubsection{Gauge-covariant Wigner functions}

As in the scalar theory in Sect. 2, the equations of motion for the
2-point functions are first rewritten in terms of Wigner functions,
in order to facilitate the gradient expansion. 
If $G_{ab}(x,y)$ is a generic 2-point function, its Wigner
transform reads (cf. eq.~(\ref{G<WIG})):
\beq\label{GWIG}
 {\cal G}_{ab}(k,X)\,\equiv\,\int {\rm d}^4s
\,{\rm e}^{ik\cdot s} G_{ab}\left(X+{s\over 2},X-{s\over 2}\right).\eeq
(From now on, we use calligraphic letters to
denote the Wigner functions.) Unlike $G_{ab}(x,y)$, which is separately
gauge-covariant at $x$ and $y$ 
(see, e.g., eq.~(\ref{DTr})), its Wigner transform
${\cal G}_{ab}(k,X)$ --- which mixes the two points $x$ and $y$ in its definition
(\ref{GWIG}) --- is not covariant. However, it is possible to construct
a gauge covariant Wigner function. Consider first the following  function:
\beq\label{COVG}
\acute G_{ab}(s,X)\,\equiv\,U_{a\bar a}\Bigl(X,X+{s\over 2}\Bigr)
\,G_{\bar a \bar b}\Bigl(X+{s\over 2},X-{s\over 2}\Bigr)\,
U_{\bar b b}\Bigl(X-{s\over 2},X\Bigr),\eeq
where $U(x,y)$ is the non-Abelian parallel transporter,
also referred to as a Wilson line:
\beq\label{PT1} U
(x,y)={\rm P}\exp\left\{ -ig\int_\gamma {\rm d}z^\mu A_\mu(z)\right\},\eeq
In eq.~(\ref{PT1}), $A_\mu^a=A_\mu^aT^a$, $\gamma$ is an arbitrary path
going from $y$ to $x$, and the symbol P denotes 
the path-ordering of the colour matrices in the exponential.
Under the gauge transformations of $A_\mu$, the Wilson
line (\ref{PT1}) transforms as (in matrix notations): 
\beq
 U(x,y) \longrightarrow h(x)\,U (x,y)\,h^\dagger(y),\eeq
so that the function (\ref{COVG}) is gauge-covariant at $X$
for any given $s$:
\beq\label{GSTR}
\acute G(s,X)  \longrightarrow h(X)\,\acute G (s,X)\,h^\dagger(X)\,.\eeq
Correspondingly, its Wigner transform 
\beq\label{GWIGC}
\acute{\cal G}_{ab}(k,X)\,\equiv\,\int {\rm d}^4s
\,{\rm e}^{ik\cdot s} \,\acute G_{ab}(s,X)\eeq
transforms covariantly as well: For any given $k$, $
\acute{\cal G}(k,X)  \longrightarrow h(X)\,\acute{\cal G}(k,X)\,h^\dagger(X)$. 
In principle, any of the two Wigner functions (\ref{GWIG})
and (\ref{GWIGC}) could be used to compute the induced sources (\ref{jf}).
However, only
the new Wigner function in eq.~(\ref{GWIGC}) will satisfy a
{\it gauge-covariant} equation of motion, which makes
its physical interpretation more transparent.

From now on, we shall use systematically gauge-covariant Wigner
functions, denoted
as $\acute {\cal S}(k,X)$, or $\acute{\cal G}_{\mu\nu}(k,X)$,
$\acute {\cal K}(k,X)$. For instance,
\beq\label{KWIGC}
\acute {\cal K}_{ia}^\nu(k,X) \equiv\int {\rm d}^4s
\,{\rm e}^{ik\cdot s} \,U_{ij}\Bigl(X,X+{s\over 2}\Bigr)
\,K^{\nu}_{jb}\Bigl(X+{s\over 2},X-{s\over 2}\Bigr)\,
\tilde U_{ba}\Bigl(X-{s\over 2},X\Bigr),\qquad
\eeq
where $U$ ($\tilde U$) is the parallel transporter
in the fundamental (adjoint) representation. 
Under the gauge transformation (\ref{BGT}),
\beq\label{TRANSK}
\acute {\cal K}_i^a(k,X) &\longrightarrow&
h_{ij}(X)\,\acute {\cal K}^b_j(k,X)\,\tilde h_{ab}(X).\eeq

By using eq.~(\ref{jf}), one can express the induced sources
in terms of these Wigner functions:
\beq\label{jf1}
j_{{\rm  f}}^{\mu\,a}(X)
&=& g\int\frac{{\rm d}^4k}{(2\pi)^4}
\,{\rm Tr}\,\Bigl(\gamma_\mu t^a\acute{\cal S}(k,X)\Bigr),\\
\label{jg1}
j^{\mu\,a}_{\rm  g}(X)&=& g \,\Gamma^{\mu\rho\lambda\nu}
\int\frac{{\rm d}^4k}{(2\pi)^4}\,{\rm Tr}\,T^a
\biggl\{ k_\lambda \acute {\cal G}_{\rho\nu}(k,X)+\frac{i}{2}\Bigl[
D_\lambda^X,\,\acute{\cal G}_{\rho\nu}(k,X)\Bigr] \biggr\},\\
\label{eind1} 
\eta^{ind}(X)&=&g\int\frac{{\rm d}^4k}{(2\pi)^4}\,
 \gamma^\mu t^a\acute {\cal K}_\mu^a(k,X).\eeq

At this stage, the path $\gamma$ in the Wilson line (\ref{PT1})
is still arbitrary. In particular, if $\gamma$ is chosen as
the {\it straight} line joining $x$ and $y$, the transition from
non-covariant to gauge-covariant Wigner functions, e.g.,  from
eq.~(\ref{GWIG}) to  eq.~(\ref{GWIGC}), can be interpreted as the
replacement of the {\it canonical} momentum $\hat k^\mu=i\del_s$ by the 
{\it kinetic} momentum $\hat p^\mu=\hat k^\mu -gA^\mu(X)$
\cite{Heinz83,Elze86,Elze90}.
In fact, most of our results will be independent of
the exact form of $\gamma$. This is so because we need $U(x,y)$ only in
situations where $x$ is close to $y$, as we argue now.

For soft and relatively weak background fields, 
the function $\acute G(s,X)$ remains close to its value in 
equilibrium, so it is peaked at $s=0$, and vanishes  when 
$s \simge 1/T$. Over such a short scale, the mean field $A_\mu$ 
does not vary significantly,
and we can write, for any path\footnote{Strictly speaking,
eq.~(\ref{Uapprox}) is a good approximation provided $\gamma$ 
never goes too  far away from $x$ and $y$, 
that is, provided $|z-x|={\rm O}(1/T)$
for any point $z$ on $\gamma$.}
$\gamma$ joining $x$ and $y$,
\beq\label{Uapprox}
g\int_\gamma {\rm d}z^\mu A_\mu(z)\,\approx\,g\bigl(s \cdot A(X)\bigr),\eeq
up to terms which involve, at least, one soft derivative
$\del_X A\sim gTA$
(and which do depend upon the path). Furthermore, for $s\sim 1/T$, 
$g s\cdot A\sim g$ (since $gA \sim gT$), so we can expand the exponential
in eq.~(\ref{PT1}) in powers of $g$ and get, to leading non-trivial order,
\beq\label{approxPT}
U_{ab}(x,y)\,\simeq\,
\delta_{ab}\,-\,ig\Bigl(s \cdot A_{ab}(X)\Bigr).\eeq
 The present use of the Wilson line should be contrasted
with that in Sect. 3.1, where the parallel transporter
in eq.~(\ref{SIGU}) covers a relatively large space-time 
separation $|x-y|\sim 1/gT$ determined by the inhomogeneity in the
system. In that case, the parallel transporter cannot be expanded
as in eq.~(\ref{approxPT}), for the reasons explained in Sect. 3.1.
However, the corresponding path $\gamma$ is then
fixed by the dynamics of the hard particles (cf. Sect. 4.1.1).

The constraint on the amplitudes of the mean fields entails
a similar constraint on the off-equilibrium deviation
$\delta G\equiv G - G_{eq}\/$: as we shall see later, $\delta G \sim (gA/T)
G_{eq} \sim g G_{eq}$. Thus, by writing
\beq G\equiv G_{eq}+ \delta G,\qquad\qquad
\acute G\equiv G_{eq} +\delta\acute G,\eeq
in eq.~(\ref{COVG}), and recalling that $G_{eq}^{ab}=\delta^{ab}G_{eq}$, 
we can easily obtain the following relation between
$\delta\acute G$ and $\delta G$, valid to leading order in $g$:
\beq\label{COV}
\delta\acute G (s,X)\,\simeq\, \delta G (x,y) \,+\,
ig\Bigl(s \cdot A(X)\Bigr)G_{eq}(s),\eeq
or, after a Wigner transform,
\beq\label{delG}
\delta\acute{\cal G}(k,X)\simeq\delta{\cal G}(k,X)+g(A(X)\cdot\del_k)
G_{eq}(k).\eeq 
For an abnormal Wigner function,
the equilibrium contribution vanishes, so that ordinary
and gauge-covariant Wigner functions coincide to leading order
in $g$ : e.g., $\acute{\cal K}_\nu \simeq {\cal K}_\nu$.
Similar simplifications can be performed on 
eq.~(\ref{jg1}) for the gluonic current, to get:
\beq\label{jb1}
j^{\mu\,a}_{\rm  g}(X) &=&g\int\frac{{\rm d}^4k}{(2\pi)^4}\,{\rm Tr}
\,T^a \Bigl\{- k^\mu \delta\acute {\cal G}_{\nu}^{\,\,\,\nu}(k,X)
+ \delta\acute{\cal G}^{\mu\nu}(k,X)k_\nu \Bigr\},\,\,\,\,\eeq
where the following property has been used: 
\beq D^\mu_x G(x,y)\Big |_{y=x}\,=\,\del^\mu_s 
\acute G(s,X)\Big |_{s=0}.\eeq
Note finally that, within the same approximations, 
the gauge-fixing conditions (\ref{TR}) imply that the
{\it gauge-covariant} gluon Wigner function is (spatially)
transverse, as at tree-level, 
$k^i \delta\acute{\cal G}_{i\nu}=0$, and similarly
$k^i \acute{\cal K}_i=k^i \acute{\cal H}_i=0$.
We can thus write:
\beq\label{GIJ0}
\delta\acute{\cal G}_{ij}(k,X)\,\equiv\,
(\delta_{ij}-\hat k_i\hat k_j)\delta\acute {\cal G}(k,X)\,.\eeq
As we shall verify in Sect. 3.4.1, the spatial components
above are the only
ones to contribute to the induced current to leading order
in $g$ (this is specific to Coulomb's gauge \cite{qcd,BE}).
Thus, finally,
\beq\label{JFIN}
j^{a}_{{\rm g}\,\mu}(X)\,=\,2g\int\frac{{\rm d}^4k}{(2\pi)^4}\,k_\mu {\rm Tr}
\Bigl\{T^a \delta \acute{\cal  G}(k,X)\Bigr\},\eeq
where the overall factor of 2 comes from the sum over 
transverse polarization states.

\subsubsection{Gauge-covariant gradient expansion}

In this section, we show how to extract
the terms of leading order in $g$ in
eqs.~(\ref{S10})--(\ref{D10}). This involves approximations
similar to those already performed in the previous subsection
(in relation with eqs.~(\ref{approxPT}), (\ref{delG}) and (\ref{jb1})),
and which take into account the dependence on $g$ associated
with the soft inhomogeneities ($\del_X\sim gT$), the amplitudes of the
mean fields ($A\sim T$ or $F_{\mu\nu}\sim gT^2$), and 
the magnitude of the off-equilibrium deviations 
$\delta\acute {\cal G}\sim g G_0$. 
Since these approximations are related
by gauge symmetry, we shall refer to them
globally as the {\it gauge-covariant gradient expansion}.

Consider then the gluon 2-point function
$G^{<\,ab}_{\mu\nu}(x,y)$ in the presence of a soft background field
$A^\mu_a$, but without fermionic fields ($\Psi=\bar\Psi=0$).
Like in the scalar theory in Sect. 2.3.2, we start with the
following two Kadanoff-Baym equations for $G\equiv G^<$
(here, in the mean field approximation; cf. eq.~(\ref{D10})) :
\beq\label{MFEQ}
\left(g_\mu^{\,\,\rho}D^2-D_\mu D^\rho+ 2igF_{\mu}^{\,\,\rho}
\right)_xG_{\rho\nu}(x,y)&=&0,\nonumber\\
G_{\mu}^{\,\,\rho}(x,y)\left(g_{\rho\nu}\bigl(D^\dagger\bigr)^2
- D_\rho^\dagger D_\nu^\dagger + 2ig F_{\rho\nu}\right)_y
&=&0,\eeq
and take their difference. This involves, in particular,
\beq\label{Xi}
\Xi(x,y)\equiv D^2_x G(x,y)\,-\,G(x,y) (D^\dagger_y)^2,\eeq
where
$D_x^2\,=\,\del_x^2+2igA\cdot\del_x +ig(\del\cdot A)-g^2A^2\/$,
and Minkowski indices are omitted to simplify the
writing (they will be reestablished when needed).
In this subsection we shall illustrate our
approximations by focusing on $\Xi$.

After replacing the coordinates
$x^\mu$ and $y^\mu$ by $s^\mu$ and
$X^\mu$ (cf. eqs.~(\ref{Rel}) and (\ref{DERIV})),
we have, typically, $s\sim 1/T$, $\del_s\sim T$ and $\del_X \sim gT$. 
We then perform an expansion 
in powers of $\del_X$ and keep only terms 
which involve, at most, one soft
derivative $\del_X$. For instance,
$$A_\mu(X+s/2)\approx A_\mu(X)+ (1/2)(s\cdot \del_X) A_\mu(X).$$
Proceeding as in  Sect. 2.3.2, and paying attention to the colour
algebra, we obtain:
\beq\label{DIFF1}
\Xi(s,X)&=& 2\del_s\cdot\del_X G +2ig\Bigl[A_\mu(X),\del^\mu_s G\Bigr]
+ig\Bigl\{A_\mu(X),\del^\mu_X G\Bigr\}
+ig\Bigl\{(s\cdot\del_X)A_\mu,\del^\mu_s G\Bigr\}\nonumber\\
&{}&+ig\Bigl\{(\del_X\cdot A), G\Bigr\} - g^2\left[A^2(X),G\right]
-\frac{g^2}{2}\left\{(s\cdot\del_X)A^2,G\right\}\,+\,...\,,\eeq
where the right parentheses 
(the braces) denote commutators (anticommutators) of colour matrices,
and the dots stand for terms which involve at least two soft derivatives
$\del_X$.

At this point, we use the fact that $A\sim gT$
and $\delta G\equiv G-G_{eq}\,\sim\, gG_{eq}$ (as it will be verified
a posteriori), with $G_{eq}\approx G_0$ in the present
approximation (cf. eq.~(\ref{INITGSK})). By keeping only terms
of leading order in $g$, one simplifies eq.~(\ref{DIFF1})  to:
\beq\label{DIFF2}
\Xi(s,X)\approx 2(\del_s\cdot\del_X) \delta G +2ig\left[A_\mu,
\del^\mu_s \delta G\right]
+2ig(s\cdot\del_X)A_\mu\,(\del^\mu_s G_0)
+2ig (\del_X \cdot A) G_0,\,\,\eeq
where all the terms in the r.h.s. are of order $g^2 T^2G_0$.
 After a Fourier transform with respect to $s$,
$\,\Xi(s,X)$ becomes $ -i \,\Xi(k,X)$ with:
\beq \label{DIFF3}
\Xi(k,X)\,\approx\,2\Bigl[k\cdot D_X,\, \delta {\cal G}(k,X)\Bigr] +
2 gk^\mu \Bigl(\del_X^\nu A_\mu(X)\Bigr)\del_\nu G_0(k).\eeq
Here, $\delta{\cal G}(k,X)$ is the ordinary Wigner transform of 
$\delta G(x,y)$, eq.~(\ref{GWIG}), but it can be expressed in terms 
of the gauge-covariant Wigner function $\delta\acute{\cal G}(k,X)$ 
with the help of eq.~(\ref{delG}). This finally yields:
\beq \label{KIN}
\Xi_{\mu\nu}(k,X)\,\approx \,2\Bigl[k\cdot D_X, \delta\acute
{\cal G}_{\mu\nu}(k,X)\Bigr] -
 2 gk^\alpha F_{\alpha\beta}(X) \,\del^\beta G_{\mu\nu}^{(0)}(k),\eeq
where the Minkowski indices have been reintroduced.

We recognize here the familiar structure of
the Vlasov equation, generalized to a non-Abelian plasma: 
Eq.~(\ref{KIN}) involves a (gauge-covariant) drift term
$(k\cdot D_X)\delta\acute{\cal G}$, together with
a ``force term'' proportional to the background
field strength tensor. In fact, this ``force term'' involves the
{equilibrium} distribution function $G_0\equiv G^<_0$, so, in this
respect, it is closer to the { linearized} version of the
Vlasov equation, eq.~(\ref{vlasovlinear}). However, 
unlike eq.~(\ref{vlasovlinear}), its non-Abelian counterpart
in eq.~(\ref{KIN}) is still non-linear, because of the presence of the
 covariant drift operator $(k\cdot D_X)$, and because
the non-Abelian field strength tensor is itself non-linear.

\subsection{The non-Abelian Vlasov equations}

In this section, we construct the kinetic equations which
determine the colour current induced by a soft gauge field $A^\mu_a$. 
(The fermionic mean fields $\Psi$ and $\bar\Psi$
are set to zero in what follows.)
According to eqs.~(\ref{jf1}) and (\ref{JFIN}), we need the
equations satisfied by the quark and gluon Wigner functions,
$\delta\acute{\cal S}$ and $\delta\acute{\cal G}_{\mu\nu}$,
in the presence of the background field $A^\mu_a$. 
From the discussion  in the previous subsection, we anticipate that these
equations are non-Abelian generalizations of the (linearized)   Vlasov equation.

\subsubsection{Vlasov equation for gluons}

Since we expect the transverse components 
$\delta\acute{\cal G}_{ij}$ to be the dominant ones, we focus on the 
spatial components ($\mu=i$ and $\nu=j$) of eqs.~(\ref{MFEQ}):
\beq\label{MFIJ}
D_x^2 G_{ij} - D^x_iD^x_0G_{0j}+ 2igF_{i}^{\,\rho}(x)
G_{\rho j}&=&0,\nonumber\\
G_{ij}\Bigl(D_y^\dagger\Bigr)^2 - G_{i0} 
D_{0\,y}^\dagger D_{j\,y}^\dagger + 2igG_{i\rho}F^{\rho}_{\,\nu}(y)
&=&0.\eeq
(We have also used eq.~(\ref{TR}) to simplify some terms
in these equations.) We now take their difference, 
to be succinctly referred to as the {\it difference equation}
(cf. Sect. 2.3.2).
In doing so, we first meet (cf. eqs.~(\ref{Xi}) and (\ref{KIN})):
\beq\label{GIJ1}
D_x^2 G_{ij} -G_{ij}\Bigl(D_y^\dagger\Bigr)^2\,
\longrightarrow \,
2\Bigl[k\cdot D_X, \delta\acute{\cal G}_{ij}\Bigr]
 - 2 gk^\alpha F_{\alpha\beta}(X) \,\del^\beta G^{(0)}_{ij}(k).\eeq
Note the following identity, which will be useful later:
\beq\label{ID1}\lefteqn{ 
k^\alpha F_{\alpha\beta}\del^\beta G^{(0)}_{ij}(k)\equiv
k^\alpha F_{\alpha\beta}\del^\beta [
(\delta_{ij}-\hat k_i\hat k_j)G_0(k)]}\nonumber\\
& &=(\delta_{ij}-\hat k_i\hat k_j)k^\alpha F_{\alpha\beta}\del^\beta
G_0 -k^\alpha F_{\alpha l}\,\frac{k_i\delta_{jl}
+k_j\delta_{il}-2\hat k_i\hat k_j k_l}
{{\bf k}^2}\,G_0.\,\,\,\eeq
This shows that the r.h.s. of eq.~(\ref{GIJ1}) involves also
non-transverse components. These will be canceled by the other
terms in eqs.~(\ref{MFIJ}), as we explain now.

The components $G_{i0}$ and $G_{0j}$ vanish in equilibrium, and remain
small out of equilibrium (see below), 
but nevertheless their contribution to the
difference equation is non-negligible. This is so since the hard 
derivatives $\del^s_0 \del^s_i \sim T^2$ multiplying
these components do not cancel in the difference equation, 
in contrast to what happens in the terms involving
$\delta G_{ij}$ (cf. eq.~(\ref{GIJ1}), where we remember
that $\del_x^2-\del_y^2=2\del_s\cdot\del_X \sim gT^2$).
One can evaluate these components from eqs.~(\ref{MFEQ})
with $\mu=0$ and $\nu=j\/$. This gives \cite{BE} :
\beq\label{GI0}
{\cal G}_{0j}(k,X)\,\approx\,  2ig F_{0l}\,\frac{\delta_{lj}-\hat k_l\hat k_j}
{{\bf k}^2}\,G_0(k),\qquad
{\cal G}_{i0}^{ab}(k,X)={\cal G}_{0i}^{ba}(-k,X),\eeq
which provides the following contribution to the difference equation:
\beq\label{GIJ2}
-\Bigl(D^x_iD^x_0G_{0j}- G_{i0} D_{0\,y}^\dagger D_{j\,y}^\dagger\Bigr)
\,\longrightarrow \, -2gk^0F_{0l}(X)\,\frac{k_i\delta_{jl}
+k_j\delta_{il}-2\hat k_i\hat k_j k_l}
{{\bf k}^2}\,G_0(k).\eeq
Finally, there is a contribution from
the terms involving the field strength tensor in  eqs.~(\ref{MFIJ}).
To the order of interest, it reads:
\beq\label{GIJ3}
-2ig\Bigl(F_{il}(X) G_{lj}^{(0)}(s) - G_{il}^{(0)}(s)
F_{lj}(X)\Bigr)\,\longrightarrow \, -2g G_0(k)\Bigl(
F_{il}\hat k_l\hat k_j + F_{jl}\hat k_l\hat k_i\Bigr).\eeq 
Together, the contributions in eqs.~(\ref{GIJ2}) and (\ref{GIJ3})
precisely cancel the non-transverse piece in the
r.h.s. of eq.~(\ref{GIJ1}), as anticipated.

To conclude, $\delta\acute{\cal G}_{ij}$ is transverse indeed,
as required by the gauge condition (\ref{GIJ0}), and can be written as
$\delta\acute{\cal G}_{ij}=
(\delta_{ij}-\hat k_i\hat k_j)\delta\acute{\cal G}$, with 
$\delta\acute{\cal G}(k,X)$ satisfying:
 \beq\label{VLAS}
\Bigl[k\cdot D_X,\,\delta\acute{\cal G}(k,X)\Bigr]
\,=\,g\,k^\alpha  F_{\alpha\beta}(X)\del^\beta G_0(k).\eeq
Since $D_X \sim gT$ and $g F_{\alpha\beta}\sim 
(D_X)^2 \sim g^2T^2$, it follows that
$\delta\acute{\cal G}\sim (D_X/T)G_0 \sim gG_0$, as anticipated.
(By comparison,
${\cal G}_{0j} \sim (gF_{0i}/T^2)G_0\sim g^2 G_0$
is of higher order in $g$.)

Since $G_0\equiv G^<_0(k)=2\pi \epsilon(k_0)\delta(k^2)
N(k_0)$, and therefore
\beq
k^\alpha  F_{\alpha\beta}\del_k^\beta G_0(k)\,=\,
2\pi\delta(k^2) k^i F_{i0}({\rm d}N/{\rm d}k_0),\eeq
it follows that $\delta\acute{\cal G}(k,X)$ has support only on
the tree-level mass-shell $k^2=0$. By also using
the symmetry property (\ref{symm}), we can write:
\beq\label{GN0}
\delta\acute{\cal G}_{ab}(k,X)
&\equiv&2\pi\delta(k^2)\Bigl\{\theta(k_0)\delta N_{ab}({\bf k},X)
\,+\,\theta(-k_0)\delta N_{ba}(-{\bf k},X)\Bigr\},\eeq
where $\delta N_{ab}({\bf k},X)$ is a density
 matrix satisfying the following, Vlasov-type, equation
  \cite{qcd} (with $v^\mu =(1,{\bf k}/k)$ and $E^i=F^{i0}=E^i_aT^a$):
\beq\label{N}
\left[ v\cdot D_X,\,\delta N({{\bf k}},X)\right]=-\, g\,
{\bf v}\cdot{\bf E}(X)\frac{{\rm d}N}{{\rm d}k}\,.\eeq
This equation implies that $\delta N$ has the same colour
structure as the electric field ${\bf E}(x)$, that is,
$\delta N\equiv\delta N^aT^a$,
with the components $\delta N^a({\bf k},x)$ transforming as
a colour vector in the adjoint representation.
In terms of this density matrix, the induced current 
(\ref{JFIN}) can be finally written as\footnote{The upper
script $A$ is to recall that this is only 
the contribution of the soft gauge fields $A^\mu$
to the current; cf. Sect. 3.2.2.} :
\beq\label{jb1A}
j^{A\,a}_{{\rm g}\,\mu}(X)\,=\,2g\int\frac{{\rm d}^3k}{(2\pi)^3}\,v_\mu
\,{\rm Tr}\Bigl(T^a \delta N({\bf k},X)\Bigr)\,=\,2gN
\int\frac{{\rm d}^3k}{(2\pi)^3}\,v_\mu\,\delta N^a({\bf k},X)
.\eeq
Aside from its covariance under the gauge transformations of
the background field, eq.~(\ref{N}) is also independent of
the gauge-fixing for the quantum fields, as proven
in Ref. \cite{qcd}.

\subsubsection{Vlasov equation for quarks}

We now briefly consider the corresponding equation for
the quark 2-point function $S\equiv S^<$. Starting with
eq.~(\ref{S10}), namely
\beq\label{S2}
\slashchar{D}_x S(x,y) \,=\,0,\qquad
S(x,y)\slashchar{D}^\dagger_y \,=\,0,\eeq
and using  (with $\sigma^{\mu\nu}\equiv (i/2)[\gamma^\mu,\gamma^\nu]$)
\beq\label{D^2}
\slashchar{D}_x\,\slashchar{D}_x=\,D_x^2+\frac {g}{2}\,\sigma^{\mu\nu}F_{\mu
\nu}(x),
\eeq
one obtains the following difference equation:
\beq\label{SD}
D_x^2 S(x,y)-S(x,y)\bigl(D_y^\dagger)^2+\frac{g}{2}\,\Bigl(\sigma^{\mu\nu}
F_{\mu\nu}(x)S(x,y)-S(x,y)\sigma^{\mu\nu}
F_{\mu\nu}(y)\Bigr)=0.\,\,\,\,\eeq 
Then one proceeds with a gauge-covariant gradient expansion,
as in Sect. 3.3.3, which finally yields the following equation
for the covariant Wigner function $\delta\acute{\cal S}(k,X)$ 
(valid to leading order in $g$):
\beq\label{Vlas0}
 \left [k\cdot D_X,\, \delta\acute{\cal S}(k,X) \right ]=
g k\cdot F(X)\cdot \del_k S_0^<
-i\,\frac{g}{4}\,F^{\mu\nu}(X)\Bigl[\sigma_{\mu\nu},\,S_0\Bigr] 
.\eeq
By also using $S_0\equiv S_0^<(k)=\slashchar{k}\tilde\Delta(k)\equiv
\slashchar{k}\rho_0(k)n(k^0)$ (cf. eq.~(\ref{S0})), and
calculating the Dirac commutator $[\sigma_{\mu\nu}, \slashchar{k}]$,
we finally get the simple equation:
\beq\label{Vlas}
 \left [k\cdot D_X,\, \delta\acute{\cal S}(k,X) \right ]=
g\,\slashchar{k}\,k\cdot F(X)\cdot \del_k\tilde\Delta(k),\eeq
which implies that  $\delta\acute {\cal S}$ is
a colour matrix of the form
 $\delta\acute {\cal S} = \delta\acute {\cal S}_at^a$,
with the components $\delta\acute {\cal S}_a$ transforming as
a colour vector. Eq.~(\ref{Vlas}) also shows that
$\delta\acute {\cal S}$ has
the same spin and mass-shell structure
as the free 2-point function $S_0^<$:
\beq\label{gens1}
 \delta\acute{\cal S}(k,X)\,=\,\slashchar{k}\,2\pi\delta(k^2)
\left\{ \theta(k_0) \delta n_+({\bf k}, X)+
\theta(-k_0) \delta n_-(-{\bf k},X)\right\}.
\eeq 
The density matrices
 $\delta n_\pm ({\bf k},X) \equiv\delta n_\pm^a({\bf k},X)\,t^a$
 satisfy the following kinetic equation:
\beq\label{n}
\left[ v\cdot D_X,\,\delta n_\pm({{\bf k}},X)\right]=\mp\, g\,{\bf v}
\cdot{\bf E}(X)\,\frac{{\rm d}n_k}{{\rm d}k},\eeq
which is the non-Abelian version of the Vlasov equation for
quarks. The induced current $j_{\rm f\,\mu}^{A\,a}$ reads finally
(cf. eq.~(\ref{jf1})):
\beq\label{jf1A}
j_{{\rm  f}\,\mu}^{A\,a}(X)
&=&gN_f\int\frac{{\rm d}^3k}{(2\pi)^3}\,v_\mu
\Bigl( \delta
 n_+^a({\bf k},X)-\delta n_-^a({\bf k},X)\Bigr).\eeq

\subsection{Kinetic equations for  the fermionic excitations}

A noteworthy feature of the ultrarelativistic plasmas
is the existence of collective modes
with fermionic quantum numbers.
The associated collective motion involves both quarks
and gluons, whose mutual transformations, over a long
space-time range, can be described 
as a propagating fermionic field $\Psi(x)$ \cite{qed,qcd}. 
We shall now establish the kinetic equations which
determine the corresponding  induced sources  $\eta^{ind}$ and 
$j_\mu^{\psi}$.

\subsubsection{Equation for $\eta^{ind}$}

The fermionic source $\eta^{ind}(x)$ is given by eq.~(\ref{eind1}),
where, to the order of interest, we can replace
the gauge covariant Wigner function $\acute {\cal K}_\mu ^a(k,X)$  
by the ordinary Wigner transform $ {\cal K}_\mu ^a(k,X)$ (cf.
after eq.~(\ref{COV})).
 The kinetic equation for $ {\cal K}_\mu ^a(k,X)$  
is obtained from the equations of motion (\ref{K10}) and (\ref{K11})
for $K^{a}_{\mu}(x,y)$, that is
\beq\label{K12} \slashchar{D}_x
K_\mu^a(x,y)&=& -ig t^a\gamma^\nu\Psi(x)
G^{(0)}_{\nu\mu}(x,y), \\
\label{K13}
\Bigl(g_{\mu\nu}\tilde D^2 -\tilde D_\mu\tilde D_\nu +
2ig \tilde F_{\mu\nu}\Bigr)_y^{ab}
 K^{\nu}_b(x,y)&=&-gS_0(x,y)\gamma_\mu t^a\Psi(y),
\eeq
where $D$ ($\tilde D$) is the covariant derivative
in the fundamental (adjoint) representation. (To simplify notations,
the colour indices for the fundamental representation are not shown 
explicitly; see, e.g., eq.~(\ref{abn}).)
These equations describe a process where a hard gluon scatters off 
a soft fermionic field $\Psi$ and gets ``turned into'' a hard quark,
or a hard antiquark annihilates against $\Psi$ to generate
a hard gluon. 

In the right hand sides of the above equations,
we have replaced the full gluon and fermion propagators $G_{\nu\mu}$
and $S$ by their free counterparts $G^{(0)}_{\nu\mu}$ and
$S_0$. This is correct in leading order
since the off-equilibrium deviations are suppressed by one power of 
$g\/$: e.g., $\delta S\sim gS_0$. (For the deviations induced by
the fermionic fields, this relies on the constraint
$\bar\Psi \Psi\sim gT^3$; see Sect. 3.5.2 below.)
Because of that, these equations
are linear in $\Psi$, although non-linear in $A^\mu_a$.

As for the gluon Wigner function in
in Sect. 3.4.1, here too there is a hierarchy among the components of
$ {\cal K}_\mu ^a(k,X)$: The spatial components
${\cal K}_i^a$ are the large ones, and are transverse:
$k^i{\cal K}_i=0$. The temporal component
${\cal K}_0^a$ is smaller, ${\cal K}_0 \sim g {\cal K}_i$,
so it does not contribute to the induced source $\eta^{ind}$ 
directly; its only role is to remove
the longitudinal component of the r.h.s. of eq.~(\ref{K12}).
(Thus, $K_0$ plays here the same role as $G_{0j}$ in Sect. 3.4.1.)
Then eq.~(\ref{K12}) with $\mu=i$ reduces to:
\beq\label{K14}
(\tilde D_y^2)^{ab} K_i^b(x,y)\,= \,-gS^{(0)}_{ij}(x-y)\gamma^j t^a\Psi(y),
\eeq
where
\beq
S^{(0)}_{ij}(s)\equiv \int\frac{{\rm d}^4k}{(2\pi)^4}\,
{\rm e}^{-ik\cdot s}(\delta_{ij}-\hat k_i\hat k_j) S_0(k).\eeq
To the same order, eq.~(\ref{K12}) simplifies to
\beq\label{K15}
D^2_x K^a_i(x,y)\,\approx\,-ig t^a \gamma^\lambda\gamma^j\Psi(x)
\del^x_\lambda G^{(0)}_{ji}(x-y),\eeq
where the r.h.s. is transverse indeed (cf. eq.~(\ref{W0})).

\begin{figure}
\protect \epsfxsize=10.cm{\centerline{\epsfbox{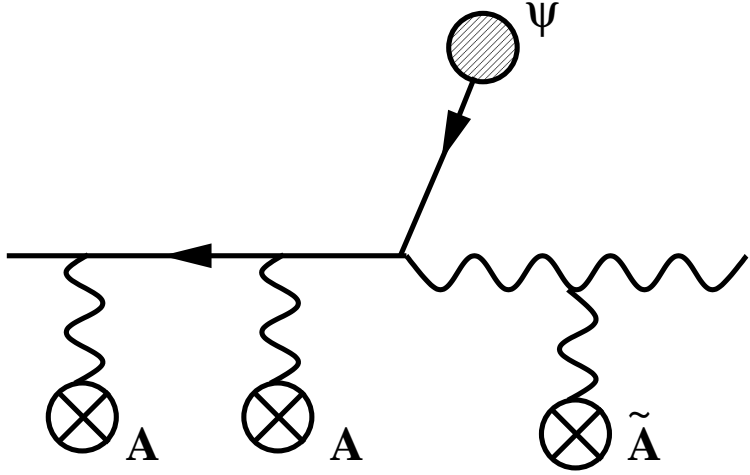}}}
         \caption{Pictorial representation of the solution
${\cal K}^a_i(k,X)$ to eq.~(\ref{KXk}): $A$ ($\tilde A$) denotes
insertions of the gauge mean field in the fundamental (adjoint)
representation.}
\label{Kfig}
\end{figure}

We now subtract eq.~(\ref{K14}) from eq.~(\ref{K15}) and transform
the difference equation as in the previous subsections.
One then finds, for instance,
\beq
\Bigl(D^2_x \delta^{ab}-(\tilde D_y^2)^{ab}\Bigr)K^b_i(x,y)\,\approx \,2
\Bigl((\del_X+igA(X))\delta^{ab} + ig\tilde A^{ab}(X)\Bigr)\cdot \del_s
K^b_i(s,X),\eeq
which, in contrast to eq.~(\ref{DIFF2}), does not involve
the soft derivative $\del_X A$ of the background gauge field
(this is so since $K^a_\mu$ vanishes in equilibrium).
Thus, there will be no ``Lorentz force'' 
in the kinetic equation for ${\cal K}_i^a(k,X)$, which finally reads:
\beq\label{KXk}
 k\cdot\left [(\del_X+igA(X))\delta^{ab} + ig\tilde A^{ab}(X)
\right ]{\cal K}^b_i(k,X)\,=\qquad\qquad\qquad\nonumber\\
\qquad= \,-i\frac {g}{2}\,\rho_0(k)\Bigl(N(k_0)+n(k_0)\Bigr)
\slashchar{k}\gamma^j (\delta_{ij}-\hat k_i\hat k_j)
t^a\Psi(X).\eeq
The differential operator in the l.h.s.
is a covariant derivative acting on vectors
in the direct product of the fundamental and the adjoint 
representation. Thus, eq.~(\ref{KXk}) is consistent with the
transformation law (\ref{TRANSK}) for ${\cal K}^a_i(k,X)$.
In a pictorial representation of the solution ${\cal K}_i^a(k,X)$,
the fundamental gauge field $A^\mu=A^\mu_at^a$ (respectively,
the adjoint field $\tilde A^\mu=A^\mu_aT^a$) in the l.h.s.
of eq.~(\ref{KXk}) is responsible for gauge field insertions
along the external quark leg (respectively, gluon leg) of 
${\cal K}_i^a$ (cf. fig. \ref{Kfig}).

The induced source (\ref{eind1}) involves the quantity 
${\cal {\slashchar K}}(k,X)
\equiv t^a \gamma^\mu {\cal K}_\mu ^a\approx
t^a \gamma^i {\cal K}_i^a$.
To get the equation satisfied by this quantity,
multiply eq.~(\ref{KXk}) by $t^a$ from the left, and use the
identities $t^at^c+if^{abc}t^b=t^ct^a$ and 
$t^at^a=(N^2-1)/2N\equiv C_{\rm f}$ to obtain:
\beq \label{kinK}
\left (k\cdot D_X\right ) {\cal {\slashchar K}}(k,X)
=-ig\frac{d-2}{2}C_{\rm f}\rho_0(k)\Bigl(N(k_0)+n(k_0)\Bigr)\slashchar{k} \Psi(X).
\eeq
The factor $(d-2)=2$ --- which arises via
$\gamma^i\slashchar{k}\gamma^j (\delta_{ij}-\hat k_i\hat k_j)
=(d-2)\slashchar{k}$ --- indicates that only the
transverse gluons are involved in this collective motion.
This is a consequence of the fact that eq.~(\ref{kinK}) is
gauge-fixing independent, as actually proven in Refs. \cite{qed,qcd}.

Note that the adjoint gauge field $\tilde A^{ab}(X)$ has disappeared
in going from eq.~(\ref{KXk}) to eq.~(\ref{kinK}): the latter involves
only the covariant derivative in the fundamental representation,
which is consistent with the fact that
${\cal {\slashchar K}}(k,X)$ must transform in the same way
as $\Psi(X)$ under a gauge rotation of the background fields. 

Since the r.h.s. of eq.~(\ref{kinK}) is proportional to $\rho(k)$,
it follows that:
\beq\label{LWIG}
{\cal {\slashchar K}}(k,X)
=2\pi\delta(k^2)\Bigl\{ \theta(k_0)
\slashchar{\Lambda}
({\bf k}, X)+\theta(-k_0)\slashchar{\Lambda}(-{\bf k},X) \Bigr\},\eeq
with the density matrix
${\slashchar \Lambda}({\bf k},X)$ satisfying
(with $d=4$, and $v^\mu =(1,{\bf k}/k)$):
\beq\label{L}
(v\cdot D_X){\slashchar\Lambda}({\bf k}, X)=
-igC_{\rm f}\,(N_k+n_k)\,{\slashchar v}\,\Psi(X),\eeq
Finally, the fermionic source $\eta^{ind}(X)$ is obtained as:
 \beq\label{etaind1} 
\eta^{ind}(X)=g\,\int\frac{{\rm d}^3k}{(2\pi)^3}\frac{1}{\epsilon_k}
\slashchar{\Lambda}({\bf k},X).\eeq

Eq.~(\ref{L}) is the analog of the Vlasov equation, the fermionic
field playing in the former the same role as the colour electric
field in the r.h.s. of the latter. There is, however,
a noticeable difference: the equilibrium distributions $N_k$ and
$n_k$ enter the r.h.s. of eq.~(\ref{L}), while 
their {variations}, ${\rm d}N/{\rm d}k$ and ${\rm d}n/{\rm d}k$,
enter the r.h.s.'s of the Vlasov equations (\ref{N}) and (\ref{n}),
respectively. This reflects the difference in the mechanism
by which the hard particles react to the propagation
of a colour field, or of a fermionic field.
In the first case, the field slightly changes the momentum
of the hard particles causing a modification of their
distribution functions.
In the second case, the basic mechanism at
work is a conversion of hard gluons into hard fermions,
or vice-versa, the soft fermionic field bringing the necessary
quantum numbers, but no momentum.

The same mechanism acts also in QED, where the
long-range fermionic excitations are described by
eq.~(\ref{L}) with $gC_{\rm f} \to e$, and the non-Abelian
covariant derivative $D_\mu=\del_\mu+ig t^aA^a_\mu$
replaced by the Abelian one, $D_\mu=\del_\mu+ieA_\mu$ \cite{qed}.

\subsubsection{Equations for $j^\psi_\mu$}

Since the background fermionic fields carry also colour, they induce a
colour current $j_\mu^{\psi\,a}$ when acting of the hard particles.
In the general equations (\ref{S10}) and (\ref{D10}) for $S$ and $G$,
the effects of the fermionic fields $\Psi$ and $\bar\Psi$ are mixed
with those of the gauge fields $A^\mu$. It is convenient
to separate these effects, by writing 
\beq
\acute{\cal S}(k,X)= S_0(k)+\delta\acute
{\cal S}^A(k,X)+ \delta {\cal S}^\psi(k,X),\eeq
together with a similar decomposition for $\acute{\cal G}_{\mu\nu}$.
Here, $\delta\acute {\cal S}^A$, and similarly
$\delta\acute {\cal G}_{\mu\nu}^A$,
denote the off-equilibrium deviations induced by the gauge
fields when $\Psi=\bar\Psi=0$, and obey the kinetic equations
established in Sect. 3.4. The other pieces,  $\delta {\cal S}^\psi$ and
 $\delta {\cal G}_{\mu\nu}^\psi$, denote the additional
deviations which emerge in the presence of soft fermionic fields,
and which are responsible for the piece $j_\mu^\psi$ of the
induced current:
\beq\label{psij1}
j_{\mu}^{\psi\,a}(X)&=& 
g\int\frac{{\rm d}^4k}{(2\pi)^4} \,\left\{
{\cal J}^{\psi\,a}_{{\rm f}\,\mu}(k,X)\,+\,{\cal J}^{\psi\,a}_{{\rm g}\,\mu}(k,X)
\right\},\eeq
We have introduced here the following phase-space current densities
(cf. eqs.~(\ref{jf1}) and (\ref{JFIN})):
\beq
{\cal J}^{\psi\,a}_{{\rm f}\,\mu}(k,X)\,\equiv\,
{\rm Tr}\,\gamma_\mu t^a \delta{\cal S}^\psi(k,X),\qquad
{\cal J}^{\psi\,a}_{{\rm g}\,\mu}(k,X)\,\equiv\,
2k_\mu{\rm Tr} \,T^a\delta{\cal G}^{\psi}(k,X),\eeq
in terms of which the kinetic equations pertinent to
$j_{\mu}^{\psi}$ are most conveniently written
  \cite{qed,qcd} (with 
${\cal {\slashchar K}}^a \equiv \gamma^\mu {\cal K}_\mu ^a$, etc.):
\beq\label{kJpsi}
\,\left [k\cdot D_X,\,{\cal J}^\psi_{{\rm f}\, \mu}(k,X) \right ]^a&=&igk_\mu
\,\left \{\bar\Psi(X)\,t^a\,\slashchar{{\cal K}}(k,X)
-\slashchar{{\cal H}}(k,X)\,t^a\,\Psi(X)\right \}\nonumber\\
&-&i\,\frac {g}{2}Nk_\mu\left\{\bar\Psi(X)\,\slashchar{{\cal K}}^a(k,X)
-\slashchar{{\cal H}}^a(k,X)\,\Psi(X)\right \},\\
\label{kJbpsi}
\Bigl[k\cdot D_X,\,{\cal J}^\psi_{{\rm g}\,\mu}(k,X)\Bigr]^a&=&
i\,\frac {g}{2}\,N\,k_\mu\left\{\bar\Psi(X)\,\slashchar{{\cal K}}^a(k,X)
-\slashchar{{\cal H}}^a(k,X)\,\Psi(X)\right \}.
\eeq 
These equations are gauge-fixing independent,
and covariant with respect to the
gauge transformations of the background fields. 
The expressions in their r.h.s.'s are bilinear 
in $\Psi$ and $\bar\Psi$, as necessary for the conservation of the
fermionic quantum number. (The general
equations satisfied by the off-equilibrium propagators
$\delta {\cal S}^\psi$ and $\delta {\cal G}_{\mu\nu}^\psi$
can be found in \cite{qcd}, where it is also verified that,
for $\bar\Psi \Psi\sim gT^3$, these deviations are perturbatively small:
$\delta{\cal S}^\psi \sim gS_0$ and $\delta{\cal G}^{\psi}\sim gG_0$.)

Quite remarkably, the genuinely non-Abelian terms, proportional to $N$,
in the right hand sides
of these two equations cancel in their sum, that is,
in the equation satisfied by the total current
${\cal J}_\mu^\psi\equiv {\cal J}^\psi_{{\rm  f}\,\mu}
+{\cal J}^\psi_{{\rm  g}\,\mu}\/$, which reads simply:
\beq
\label{Jpsi1}
\left [k\cdot D_X,\,{\cal J}^\psi_\mu(k,X) \right ]=igk_\mu\,t^a
\left \{\bar\Psi(X)\,t^a\,\slashchar{{\cal K}}(k,X)
-\slashchar{{\cal H}}(k,X)\,t^a\,\Psi(X)\right \}.
\eeq
Thus, the total current ${\cal J}_\mu^\psi$ is
effectively determined by the {\it Abelian-like} piece of the
quark current ${\cal J}^\psi_{{\rm  f}\,\mu}$ alone (i.e.,
the first term in the r.h.s. of eq.~(\ref{kJpsi})).
Not surprisingly then,  eq.~(\ref{Jpsi1}) has a direct analog
in QED, which reads \cite{qed}
\beq
\label{JpsiQED}
(k\cdot \del_X){\cal J}^\psi_\mu(k,X) \,=\,igk_\mu\,
\left \{\bar\Psi(X)\,\slashchar{{\cal K}}(k,X)
-\slashchar{{\cal H}}(k,X)\,\Psi(X)\right \}.
\eeq
Since the r.h.s. of eq.~(\ref{Jpsi1}) (or (\ref{JpsiQED}))
has support only at $k^2=0$, the current can be expressed
in terms of on-shell density matrices, as follows:
  \beq\label{psifluct}
{\cal J}^{\psi}_{\mu}(k,X)=2k_\mu\,
2\pi\delta(k^2)
\left \{\theta (k^0)\delta n^{\psi}_+({\bf k},X) + \theta (-k^0) \delta
 n^{\psi}_- (-{\bf k},X)\right \}.\eeq
Note that the net contribution to the induced current
${\cal J}_\mu^\psi$ is due to the hard fermions only,
because of the cancellations alluded to before.

The colour density matrices $\delta n_\pm^\psi
=\delta n_{\pm}^{\psi\,a} t^a$ 
satisfy the following kinetic equation (with 
$\bar{\slashchar \Lambda}\equiv {\slashchar \Lambda}^\dagger \gamma^0$) :
\beq\label{vlapsi}
\Bigl[ v\cdot D_X, \,\delta n_\pm^\psi({\bf k},X) \Bigr]^a
=\pm i\,\frac{g}{2\epsilon_k}
\left\{\bar\Psi(X)\,t^a{\slashchar \Lambda}({\bf k},X)-\bar
{\slashchar \Lambda}({\bf k},X)\,t^a\Psi(X)\right \}.
\eeq 
In terms of them, the induced current $j_\mu^\psi$ is finally written as:
\beq\label{Djp1}
j_{\mu}^{\psi\,a}(X)\,=\,2g\int\frac{{\rm d}^3k}{(2\pi)^3}\,v_\mu\,
{\rm  Tr} \,t^a\left(\delta
 n_+^{\psi}({\bf k},X)-\delta n_-^{\psi}({\bf k},X)\right).\eeq

\subsection{Summary of the kinetic equations}

Let us summarize now the kinetic equations which represent
the main result of this section. These are written here for the
various density matrices, and read
(cf. eqs.~(\ref{n}), (\ref{N}), (\ref{L}), and
(\ref{vlapsi})) :
\beq\label{n0}
\left[ v\cdot D_x,\,\delta n_\pm({{\bf k}},x)\right]
&=&\mp\, g\,{\bf v}
\cdot{\bf E}(x)\,\frac{{\rm d}n_k}{{\rm d}k},\\
\label{N0}
\left[ v\cdot D_x,\,\delta N({{\bf k}},x)\right]&=&-\, g\,
{\bf v}\cdot{\bf E}(x)\frac{{\rm d}N_k}{{\rm d}k},\\
\label{L0} 
(v\cdot D_x){\slashchar\Lambda}({\bf k}, x)&=&
-igC_{\rm f}\,(N_k+n_k)\,{\slashchar v}\,\Psi(x),\\
\label{vlapsi0}
\Bigl[ v\cdot D_x, \,\delta n_\pm^\psi({\bf k},x) \Bigr]^a
&=&\pm i\,\frac{g}{2\epsilon_k}
\left\{\bar\Psi(x)\,t^a{\slashchar \Lambda}({\bf k},x)-\bar
{\slashchar \Lambda}({\bf k},x)\,t^a\Psi(x)\right \}.
\eeq 
In these equations, $v^\mu=(1, {\bf v})$, with ${\bf v}={\bf k}/k\/$
a  unit vector which represents the velocity of the hard, and massless,
thermal particles. In writing the equations above,
we have used the lower case letter $x^\mu$ to denote the space-time variable
(rather than the upper case variable $X^\mu$ which was 
introduced earlier for the Wigner transform). This notation, which will
be used  systematically from now on, should not give rise to confusion,
as there is no other space-time variable left. 

As repeatedly emphasized, eqs.~(\ref{n0})--(\ref{vlapsi0}) are
gauge-fixing independent, although they have been derived here by
working in the background-field Coulomb gauge.
This is so since they describe the collective motion
of the hard particles, which,
to the order of interest, are the same as the physical
degrees of freedom of an ideal plasma (quarks, antiquarks,
and transverse gluons).

These equations are also covariant under the gauge
transformations of the soft background fields $A^\mu$, $\Psi$
and $\bar\Psi$, and non-linear in the fields $A^\mu$
which enter the covariant drift term $v\cdot D_x$.
The fermionic density matrix ${\slashchar\Lambda}$ is a colour
vector in the fundamental representation, while all the
other density matrices, which determine the induced colour current,
are adjoint colour vectors. Thus, in the Abelian limit (hot QED),
the equation satisfied by ${\slashchar\Lambda}$ remains non-linear
(with the Abelian covariant derivative $D_\mu=\del_\mu+ieA_\mu$),
while all the other equations involve the ordinary drift operator
$v\cdot \del_x$.

These equations have an eikonal structure: in the presence
of soft background fields, the hard particles follow on the
average straight-line characteristics, although they
may exchange momentum with the soft gauge fields.
The covariant derivative within the drift term induces a colour
precession of the various density matrices, which will become
manifest in the solutions to eqs.~(\ref{n0})--(\ref{vlapsi0}),
to be presented in the next section.

\setcounter{equation}{0}
\section{The dynamics of the soft excitations}
\setcounter{equation}{0}

By solving the kinetic equations, which we shall do in this
section, one can express the induced sources in terms of the soft mean
fields. Then, the Yang-Mills and Dirac equations including these sources
form a closed system of equations which describe the dynamics of the soft
excitations of the plasma. This defines a classical effective theory for
the soft fields, which, as we shall see in the last part 
of this section, can be given a Hamiltonian formulation.
By using this formulation, the calculation of correlation functions at 
large space-time distances can be reduced to averaging
over the initial conditions products of fields obeying the classical 
equations of motion. 
This averaging involves functional integrals which can be calculated
using lattice techniques, which is especially useful in
 applications to non-perturbative problems such as those
mentioned at the end of this section.

\subsection{Solving the kinetic equations}

The kinetic equations (\ref{n0}), (\ref{N0}), (\ref{L0}) and
 (\ref{vlapsi0}) are all first-order differential equations 
which can be solved, at least formally, once the initial
conditions are specified. We shall consider 
retarded boundary conditions and assume that, as $t\to
-\infty$, both the average
fields and the induced sources vanish adiabatically, leaving the system
initially in equilibrium. 

The kinetic equations involve, in
their left hand sides, the covariant line derivative
$v\cdot D_x$ which makes them non-linear with respect to $A^a_\mu$.
If we were to solve these equations  for a fixed
$v^\mu\equiv (1,\,{\bf v})$, we could get rid of the non-linear terms
by choosing the particular (light-cone) axial gauge  $v^\mu A^a_\mu(x)=0$. 
In this gauge,
$(v\cdot D)^{ab}=\delta ^{ab}\,v\cdot\del$ and \\
${\bf v}\cdot{\bf E}^a(x)=
-\,{\bf v}\cdot (\del_0 {\bf A}^a+ \nbfgrad A_0^a)$, as for Abelian fields.
However, in  calculating induced sources like (\ref{jb1A}) or
(\ref{etaind1}), 
 we have to  integrate over all the directions of ${\bf v}$.
It is therefore necessary to solve the kinetic equations in
an arbitrary gauge.

\subsubsection{Green's functions for $v\cdot D_x$}

In order to proceed systematically, we start by defining 
a retarded Green's function by
\beq\label{Gret}
-i\,(v\cdot D_x)\,G_{R}(x,y;{\bf v})&=&\delta^{(4)}(x-y),\eeq
where $D_\mu=\del_\mu+igA_\mu$ is the covariant derivative
in either the fundamental or the adjoint representation, and $G_R$
is a colour matrix in the corresponding representation. A unit matrix in the
appropriate representation is implicit in the right hand side. 
  
In the absence of gauge fields
($A_\mu=0$), eq.~(\ref{Gret}) is
 easily solved by Fourier analysis. We thus get (with $\eta\to 0_+$):
\beq\label{GR0}
G_{R}(x,y;{\bf v})&=&\int\frac{{\rm d}^4 p}{(2\pi)^4}
\,{\rm e}^{-ip\cdot (x-y)} \,\frac {-1}{v\cdot p + i\eta}\nonumber\\
&=&i\,\theta (x^0-y^0)\,\delta^{(3)}
\left({{\bf x}}-{{\bf y}}-{{\bf v}}(x^0-y^0)
\right ).
\eeq
This expression is readily
extended to non vanishing gauge fields. The
corresponding solution to eq.~(\ref{Gret}) can be written as:
\beq\label{GR}
G_{R}(x,y;{\bf v})&=&i\,\theta (x^0-y^0)\,\delta^{(3)}
\left({{\bf x}}-{{\bf y}}-{{\bf v}}(x^0-y^0)
\right )U(x,y)\nonumber\\
&=&i\,\int_0^\infty {\rm d}\tau\,\delta^{(4)}(x-y-v\tau)\,U(x,x-v\tau).\eeq
For later use, we note here also the corresponding advanced
Green's function:
\beq\label{GA}
G_{A}(x,y;{\bf v})&=&-i\,\theta (y^0-x^0)\,\delta^{(3)}
\left({{\bf y}}-{{\bf x}}+{{\bf v}}(x^0-y^0)
\right )U(x,y)\nonumber\\
&=&-i\,\int_0^\infty {\rm d}\tau\,\delta^{(4)}(x-y+v\tau)\,U(x,x+v\tau).\eeq
In these equations,  $U(x,y)$  is the parallel transporter (\ref{PT1})
along the straight line $\gamma$ joining $x$ and $y$. In particular,
\beq\label{U}
U(x,x-v\tau)=P\exp\left\{ -ig\int_0^\tau {\rm d}t\,\, v\cdot A(x-v(\tau -
t))
\right\},\eeq
where the path joining $x-v\tau$ to $x$ is parameterized by $(t, {\bf
x}(t))$ with ${\bf x}(t)={\bf x}-{\bf v}\tau+{\bf v} t$. In order to verify,
for instance, that (\ref{GR})
is the correct  solution to eq.~(\ref{Gret})
 we may use the following formula for the line-derivative of the parallel 
transporter:
\beq\label{derU}
(v\cdot \del_x)\,U(x,y)\Big |_{y=x-v\tau}= -ig\,v\cdot A(x)\,U(x,x-v\tau).
\eeq 

Under a gauge transformation
$A_\mu \longrightarrow   h A_\mu h^{-1}-({i}/{g})\,h\del_\mu h^{-1}$,
the above Green's functions transform as 
$G_{R,\,A}(x,y;v) \longrightarrow h(x)G_{R,\,A}(x,y;v) h^{-1}(y)$, a
property  which may also be verified directly on the equations
(\ref{Gret}). Thus, the solutions 
(\ref{GR})--(\ref{GA}) are related  to 
 the solutions (\ref{GR0})  by  the gauge  transformation
which connects  the axial gauge $v\cdot A=0$ to an arbitrary gauge.

\subsubsection{The induced colour current}

In order to solve eqs.~(\ref{n0}) and (\ref{N0}),
it is convenient to express first the quark and gluon
density matrices  $\delta n_\pm$ and $\delta N$ in
terms of  new functions,
$W_a^\mu(x,{\bf v})$, solutions of:
\beq\label{eqw}
\left[ v\cdot D_x,\, W^\mu(x,{\bf v})\right]\,=\,F^{\mu\nu}(x)\,v_\nu.\eeq
Here we use  matrix notations, with
$W^\mu\equiv W_a^\mu t^a$ for quarks, and $W^\mu\equiv W_a^\mu T^a$ for
gluons. It follows from eq.~(\ref{eqw}) that the quantities $W_a^\mu(x,{\bf
v})$ satisfy
\beq\label{transW} v_\mu\,W^\mu_a(x,{\bf v})\,=\,0, \eeq
so that  $W_a^0 =v^iW_a^i$. From the equations above,
and eqs.~(\ref{n0})--(\ref{N0}) we have 
\beq\label{dnN}
\delta n_\pm^a({\bf k}, x)=\mp\, g W^0_a(x,{\bf v})\,\frac{{\rm d}n}
{{\rm d}k}\,,\qquad
\delta N^a({\bf k}, x)=- gW^0_a(x,{\bf v})\,\frac{{\rm d}N}{{\rm d}k}\,.\eeq
As already mentioned in Sect. 1.2, the
functions $W_a^0$ measure the
local distortions of the momentum distributions
(see also Sect. 4.1.4).

Eq.~(\ref{eqw}) is  easily solved with the  help of the Green's functions
introduced above. Using the retarded Green's function (\ref{GR}) one gets:
\beq\label{W}
W_\mu^a(x,{\bf v})&=&-i\int {\rm d}^4y \,G_{R}^{ab}(x,y,{\bf v})\,
F_{\mu\nu}^b(y)\,v^\nu
\nonumber \\ &=& \int_0^\infty {\rm d}
\tau\, U_{ab}(x,x-v\tau)\, F_{\mu\nu}^b
(x-v\tau)\,v^\nu\,,\eeq
or, in matrix notations,
\beq\label{W1}
W^\mu(x,{\bf v})\,=\, \int_0^\infty {\rm d}\tau\, U(x,x-v\tau)\,
F^{\mu\nu} (x-v\tau)\,v_\nu\,U(x-v\tau,x).\eeq

Once the solution of the kinetic equation is known, one can  calculate
the  induced current $j^A_\mu\equiv j^A_{{\rm f}\,\mu}+
j^A_{{\rm b}\,\mu}$
in closed form. By inserting eqs.~(\ref{dnN})
in the expressions (\ref{jf1A}) and (\ref{jb1A}), and performing the
integration over $k=|{\bf k}|$, one obtains
\beq\label{j1A}
j_\mu^{A\,a}(x)\,=\,m_D^2\int\frac{{\rm d}\Omega}{4\pi}
\,v_\mu\,W^a_0(x,{\bf v}).\eeq
Here, the  integral $\int 
{\rm d}\Omega$ runs over all the directions of the unit vector ${\bf v}$,
and $m_D$ is the { Debye screening mass} (cf. Sect. 4.3.2 below), 
\beq\label{omegap}
m_D^2 &\equiv&-\,\frac{g^2}{2\pi^2}\int_{0}^\infty 
{\rm d}k \,k^2\,\left\{
2N\,\frac{{\rm d} N_k}{{\rm d} k}
\,+\,2N_{\rm f}\frac{{\rm d} n_k}{{\rm d} k}
\right\}\nonumber\\
&=&(2N+N_f)\,\frac{g^2T^2}{6}\,.\eeq
The induced current (\ref{j1A}) is covariantly conserved,
\beq\label{cjA}
\Bigl[ D^\mu, j_\mu^{A}(x)\Bigr]&=&0,\eeq
as is most easily seen using eq.~(\ref{j1A}) and (\ref{eqw}):
\beq
\Bigl[ D^\mu, j_\mu^{A}(x)\Bigr]\,\propto\,
\int\frac{{\rm d}\Omega}{4\pi}
\,\left[ v\cdot D_x,\, W^0(x,{\bf v})\right]
\,=\,\int\frac{{\rm d}\Omega}{4\pi}\,{\bf v\cdot
E}(x)\,=\,0\,.\eeq
According to  eqs.~(\ref{W}) and (\ref{j1A}), the 
induced current can also be written as: 
\beq\label{jAR}
j_\mu^{A\,a}(x)
\,=\,m_D^2\int\frac{ {\rm d}\Omega}{4\pi}
\,v_\mu \int_0^\infty {\rm d}
\tau\, U_{ab}(x,x-v\tau)\, {\bf v}\cdot{\bf E}^b
(x-v\tau)\,.\eeq
It transforms as a colour vector in the
adjoint representation. The parallel
transporter  in eq.~(\ref{jAR}), which ensures this property,
also makes it a {non-linear}  functional
of the gauge fields.

\subsubsection{The fermionic induced sources}

The retarded solution to eq.~(\ref{L0}) for $\slashchar{\Lambda}(k,x)$ 
reads
\beq
\slashchar{\Lambda}(k,x)\,=\,-gC_{\rm f}(N_k+n_k)
\,{\slashchar{v}}\int {\rm d}^4y\,G_{R}(x,y;v)\,\Psi(y),\eeq
where $G_R$ is now in the fundamental representation.
When this is inserted in eq.~(\ref{etaind1}),
we obtain the fermionic source with retarded conditions:
\beq\label{eindR}
\eta^{ind}(x)&=& -i\omega_0^2\int\frac{{\rm d}\Omega}{4\pi}
\,\slashchar{v}
\int_0^\infty {\rm d}\tau\,U(x,x-v\tau)\Psi(x-v\tau).\eeq
Here,
\beq\label{omega0}
\omega_0^2={g^2C_{\rm f}\over 8\pi^2} \int_0^\infty
{\rm d}k\,k\,
\left( n_k+N_k\right)\,=\,{g^2C_{\rm f}
\over 8} T^2,
\eeq
is the plasma frequency for fermions \cite{Klimov81,Weldon82b}.

In the Abelian case, $\eta^{ind}$ looks formally the same as
in eq.~(\ref{eindR}),  but with $\omega^2_0=e^2T^2/8$. 
That is, both in QCD and in QED the
fermionic source $\eta^{ind}$ is linear in the
fermionic field $\Psi$, but non-linear in the gauge 
fields $A^\mu$. As explained in Sect. 3.1, the non-linearity is a consequence
of the gauge symmetry together with the non-locality
of the response functions. The presence of the parallel
transporter in eq.~(\ref{eindR}) ensures that $\eta^{ind}$ 
 transforms in the same way as $\Psi$  under
gauge rotations.

After similarly solving eq.~(\ref{vlapsi0}), one obtains
the colour current $j_\mu^{\psi}=j_\mu^{\psi\,a}t^a$ as
\beq\label{jfR}
j_\mu^{\psi}(x)
&=&g\,\omega_0^2\, t^a\int\frac{{\rm d}\Omega}{4\pi}\,v_\mu\int_0^\infty
{\rm d}t\int_0^\infty {\rm d}s\nonumber\\ &&\qquad
\bar\Psi(x-vt)\,\slashchar{v}\,
U(x-vt,x)\,t^a\,U(x,x-vs)\Psi(x-vs).\eeq
It satisfies the following continuity equation
($\bar\eta^{ind}\equiv \eta^{ind\,\dagger}\gamma_0$)
\beq\label{cjp}
\Bigl[D^\mu,\,j_\mu^\psi\,\Bigr]=ig\,t^a\Bigl(\bar\Psi\, t^a\eta^{ind}-
\bar\eta^{ind}t^a\Psi\Bigr).\eeq
In QED, the corresponding current $j_\mu^{\psi}$ reads
\beq\label{jfqed}
j_\mu^{\psi}(x)
&=&e\,\omega_0^2\int\frac{{\rm d}\Omega}{4\pi}\,v_\mu\int_0^\infty
{\rm d}t\int_0^\infty {\rm d}s\nonumber\\ &&\qquad
\bar\Psi(x-vt)\slashchar{v}
U(x-vt,x-vs)\Psi(x-vs).\eeq

\subsubsection{More on the structure of the kinetic equations}

The
equations of motion for a classical particle of mass $m$ and charge $e$,
 moving in an electromagnetic
 background field, may be given two equivalent forms. In terms
of the {\it kinetic} momentum ${k^\mu}$, related to the
velocity of the particle  ($k^\mu=k^0 v^\mu$), we have
\beq\label{lorentz}
\frac{{\rm d} k^\mu}{{\rm d}t}\,=\,e\,F^{\mu\nu}\,v_\nu.\eeq
In terms of the {\it canonical} momentum $p^\mu=k^\mu+eA^\mu$, 
the equation reads
\beq\label{LEC}
\frac{{\rm d}p^\mu}{{\rm d}t}\,=\,e\,v^\nu\del^\mu A_\nu.\eeq
While the kinetic momentum $k^\mu$ is independent of the choice of the  
gauge, and eq.~(\ref{lorentz}) is  manifestly gauge covariant,
the canonical momentum $p^\mu$ depends on the gauge, as obvious from
eq.~(\ref{LEC}).

We show now that  the kinetic  equations for colour (or charge) oscillations
  can be also written in two different forms, whose interpretation
is similar to that of the above equations (\ref{lorentz})
and (\ref{LEC}).
We consider first  a  QED plasma. Then eq.~(\ref{eqw}) reduces to
\beq\label{abw}
( v\cdot \del_x)\, W^\mu(x,{\bf v})\,=\,F^{\mu\nu}(x)\,v_\nu,\eeq
where, we recall,  the velocity ${\bf v}$ is a constant unit vector.
This may be rewritten as
\beq\label{wlor}
\frac{{\rm d}}{{\rm d}t}\, 
W^\mu(t,{\bf x}(t),{\bf v})\,=\,F^{\mu\nu}(t,{\bf x}(t))\,v_\nu,
\eeq
where ${\rm d}/{\rm d}t$ is the total time derivative along the
 characteristic  ${\bf x}(t)={\bf x}_0+{\bf v}t$.
This is the same as eq.~(\ref{lorentz}) for constant velocity in its r.h.s.
Thus,  $e\,W^\mu(x,{\bf v})$ may be interpreted as
the kinetic 4-momentum acquired  by a charged particle 
following, at constant velocity ${\bf v}$,  a straight line trajectory which
goes through ${\bf x}$ at time $t$. (For $W^0(x,{\bf v})$, this
interpretation has been already given in Sect. 1.2.)
 Then, the condition (\ref{transW}) simply reflects
the fact that the energy transferred by the
field, $eW^0$, coincides with the mechanical work done by the Lorentz force,
$e{\bf v}\cdot\Delta{\bf k}=e v^i W^i$.

In the non-Abelian case,  the fluctuations
$\delta  n_\pm$ and $\delta N$ are matrices in colour space, that is,
 $W^\mu=W^\mu_a T^a$. The colour
vector of components $W^\mu_a$ precesses 
in the background gauge field. This precession
is induced by the covariant derivative in eq.~(\ref{eqw}). Viewing this
precession along the characteristic as an additional source of
time-dependence for the  colour vector $W^\mu_a$, one can write
\beq\label{wnab}
\frac{d}{dt}\, W_a^\mu(t,{\bf x}(t);v)\,\equiv \,\Bigl[\bigl(\del_t+{\bf
v}\cdot {\nbfgrad}\bigr)\delta_{ac}\,-\,gf_{abc}\,(v\cdot
A_b)\Bigr]\,W^\mu_c,\eeq so that eq.~(\ref{eqw}) may be given a form
similar to eq.~(\ref{wlor}).

A different form of the kinetic equation, which corresponds to 
eq.~(\ref{LEC}) for the canonical momentum, is obtained
by defining
\beq\label{defa}
a^\mu(x,{\bf v})\equiv A^\mu(x)\,+\,W^\mu(x,{\bf v}).\eeq
Since
\beq 
F^{\mu\nu}\,v_\nu\,=\,\del^\mu(v\cdot A)\,-\,\Bigl[v\cdot D, A^\mu\Bigr],\eeq
eqs.~(\ref{eqw}) and (\ref{defa}) give immediately
\beq\label{eqa}
\bigl[v\cdot D,\, a^\mu\bigr]\,=\,\del^\mu(v\cdot A).\eeq
In the Abelian case, eq.~(\ref{eqa}) can be rewritten in a form
 analogous to eq.~(\ref{LEC}) for $p^\mu$:
\beq\label{La}
\frac{{\rm d}}{{\rm d}t}\, 
a^\mu(t,{\bf x}(t);v)\,=\,(v\cdot \del_x)\,a^\mu\,=\,
\del^\mu(v\cdot A),\eeq
showing that $e a^\mu$ is the
change in the {\it canonical} momentum $p^\mu\equiv k^\mu+eA^\mu$
of a charged particle following the trajectory ${\bf x}(t)={\bf x}_0
+{\bf v}t$. A similar interpretation holds for QCD, with
$a^\mu_a$ a colour vector subject to the
precession described by eq.~(\ref{wnab}).

The relevance of these new functions follows from the fact that 
$a^\mu(x,{\bf v})$ is a gauge potential of zero field strength
\cite{gauge}, i.e.,
\beq\label{0f}
\del_\mu a_\nu\,-\,\del_\nu a_\mu\,+\,ig\,\bigl[a_\mu, a_\nu\bigr]\,=\,0.\eeq
A particular projection of eq.~(\ref{0f})  has been
obtained by Taylor and Wong  \cite{TW90} by enforcing the
gauge invariance of the effective action $\Gamma_{A}$, eq.~(\ref{Sb}).
The zero field strength condition is 
at the origin of interesting formal developments relating 
the effective action of
 the HTL's to the eikonal of a Chern-Simon theory 
\cite{EN92,JN93,Nair93}.

Finally, by combining eq.~(\ref{j1A}) with $W^0=-A^0+a^0$
together with eq.~(\ref{eqa}) for $a^0$, one obtains
the following expression for the induced current:
\beq\label{jA2}
j_\mu^{A\,a}(x)\,=\,-\delta_{\mu 0}\,m_D^2\,A_0^a\,+\,
m_D^2\int\frac{{\rm d}\Omega}{4\pi}
\,v_\mu \int_0^\infty {\rm d}\tau\, U_{ab}(x,x-v\tau)(v\cdot 
\dot A_b(x-v\tau)),\eeq
where $\dot A_\mu\equiv \del_0 A_\mu$.
This expression  will be useful later.

\subsection{Equations of motion for the soft fields}

 By solving the kinetic equations
we have expressed the induced sources in
 terms of the soft average fields. The equations for  the mean fields
become then a closed system of equations describing the 
dynamics of long wavelength excitations ($\lambda\sim 1/gT$) of the plasma.
These equations, which generalize the usual Dirac and Yang-Mills equations,
read:
\beq\label{4avpsi}
i\slashchar{D} \Psi(x)\,=\,-\,
i\omega_0^2\int\frac{{\rm d}\Omega}{4\pi}\,\slashchar{v}
\int_0^\infty {\rm d}\tau\,U(x,x-v\tau)\Psi(x-v\tau)
,\eeq
and
\beq
\label{4avA}
\left [\, D^\nu,\, F_{\nu\mu}(x)\,\right ]^a 
-g  \bar\Psi (x)\gamma_\mu t^a \Psi(x)\,=\,j_\mu^{\psi\, a}(x)
\qquad\qquad\qquad\qquad
\qquad\qquad\nonumber \\\qquad
 - m_D^2\int\frac{ {\rm d}\Omega}{4\pi}
\,v_\mu \int_0^\infty {\rm d}
\tau\, U_{ab}(x,x-v\tau)\, {\bf v}\cdot{\bf E}^b
(x-v\tau).\eeq 
The colour current $j_\mu^{\psi\, a}$, not written explicitly
here, can be found in eq.~(\ref{jfR}).

These equations have a number of  noteworthy properties:

(a) They are gauge-covariant; this follows
immediately from the covariance of the various induced currents. They are
also gauge-fixing independent, i.e., they are independent
of the choice of the gauge in the quantum generating functional
(\ref{ZFP})). 

(b) The induced sources in the r.h.s. of these equations are non-local, and
non-linear. As already mentioned, the non linearity is a consequence of the
gauge symmetry and the non-locality.  

(c) The induced sources in the r.h.s. are of
the same order in
$g$ as the tree-level terms in the  l.h.s. of the equations. That is, the
propagation of the soft modes is non-perturbatively renormalized by their
interactions with the hard particles. 
For mean fields as strong as allowed,
i.e. $F\sim gT^2$ and $\bar\Psi\Psi\sim gT^3$,
{ all} the non-linear terms in
eqs.~(\ref{4avpsi})--(\ref{4avA}) are of the same order.

(e) The linearized versions of the above equations read, 
in momentum space,
\beq\label{LIN}
\Bigl(p^2 g_{\mu\nu}-p_\mu p_\nu + \Pi_{\mu\nu}(p)\Bigr)\,
A^\nu(p)&=&0,\nonumber\\
\Bigl(-\slashchar{p} + \Sigma(p)\Bigr)\,\Psi(p)&=&0,\eeq
where $\Pi_{\mu\nu}(p)$ and $\Sigma(p)$ are the self-energies
for soft gluons and, respectively, soft fermions in 
the HTL approximation (the retardation prescription on
these self-energies is implicit here). They are given 
by eq.~(\ref{polarisation2}) (with $m_D^2$ from eq.~(\ref{omegap}))
in the case of $\Pi_{\mu\nu}$, and,  respectively, 
by eq.~(\ref{sigmaultra}) below in the case of $\Sigma$. 
Eqs.~(\ref{LIN}) define the excitation energies of the collective 
modes which carry the quantum numbers of the elementary constituents,
gluons or quarks. These modes have been first
studied in Refs. \cite{Klimov81,Weldon82a,Weldon82b}, and will
be discussed in the next subsection (see also Refs.
\cite{Pisarski89a,Pisarski89c,BIO95,MLB96} for more details).

\subsection{Collective modes, screening and Landau damping}

The collective behaviour at the scale $gT$ results in
{plasma waves, as well as}
screening and dissipative phenomena, which are encoded 
in the mean field equations (\ref{4avpsi})---(\ref{4avA}),
or their linear version (\ref{LIN}).
This section is devoted to a brief presentation of these
collective phenomena.

\subsubsection{Collective modes}

The collective plasma waves are propagating solutions to
eqs.~(\ref{4avpsi})---(\ref{4avA}). In the weak field, or
Abelian limit, to which we shall restrict ourselves in this
subsection, these are solutions to the linearized equations (\ref{LIN}).
That is, they are eigenvectors, with zero eigenvalues,
of the matrices in the left hand sides of these equations.
Note that some of the solutions to the first equation correspond
to spurious excitations coming from the lack of gauge fixing.
In order to proceed systematically and identify the physical
degrees of freedom, we recognize that the matrices in the
l.h.s. of eqs.~(\ref{LIN}) are nothing but the inverse
propagators in the HTL approximation.
Such propagators --- to be generally referred to as
the ``HLT-resummed propagators'', and denoted by ${}^*\!G_{\mu\nu}$ 
for gluons, and ${}^*\!S$ for fermions --- 
are constructed in detail in Appendix B.

We shall mostly use the gluon propagator in Coulomb's gauge,
where ${}^*\!G_{\mu\nu}$ has the following non-trivial 
components (compare with eq.~(\ref{G0COUL})), corresponding
to longitudinal (or electric) and transverse (or magnetic)
degrees of freedom:
\beq\label{DSTAR}
{}^*\!G_{00}(\omega, {\bf p})\equiv {}^*\!\Delta_L(\omega,p),\qquad
{}^*\!G_{ij}(\omega, {\bf p})\equiv
(\delta_{ij}-\hat p_i\hat p_j){}^*\!\Delta_T(\omega,p),\eeq
where (with $\omega\to \omega+i\eta$ for 
retarded boundary conditions) :
 \beq\label{effd0}
{}^*\!\Delta_L(\omega,p)\,=\,\frac{- 1}{p^2 + \Pi_L(\omega,p)},\qquad
{}^*\!\Delta_T(\omega,p)\,=\,\frac{-1}{\omega^2-p^2 -\Pi_T(\omega,p)}\,,\eeq
and the electric ($\Pi_L$) and magnetic ($\Pi_T$)
polarization functions are defined as:
\beq\label{PISC}
\Pi_L(\omega, p)\,\equiv\, -\Pi_{00}(\omega, p)\,,\qquad
\Pi_T(\omega,p)\,\equiv\,
\frac{1}{2}\,(\delta^{ij}-\hat p^i \hat p^j)
\Pi_{ij}(\omega,{\bf p})\,.\eeq
Explicit expressions for these functions can be found in eqs.~(\ref{plpt}).

\begin{figure}
\protect \epsfxsize=13.cm{\centerline{\epsfbox{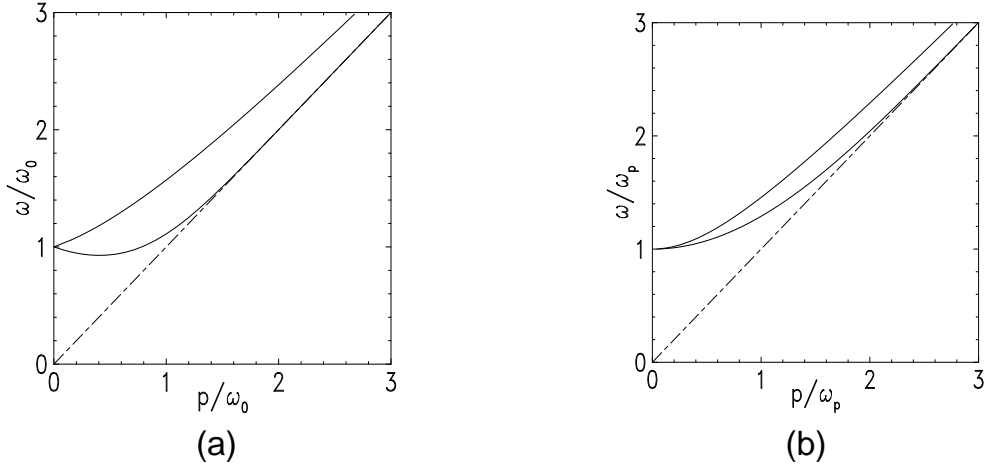}}}
	 \caption{Dispersion relation for soft excitations
in the linear regime: (a) soft fermions; the upper branch is the 
``normal'' fermion, with dispersion relation $\omega_+(p)$, while
the lower branch, with the characteristic plasmino minimum, 
is the ``abnormal'' mode, with energy $\omega_-(p)$;
(b) soft gluons (or linear plasma waves), with
the upper (lower) branch corresponding to transverse (longitudinal)
polarization.}
\label{disp}
\end{figure}


The dispersion relations for the modes are obtained from
the poles of the propagators, that is,
\beq
{p}^2+\Pi_L(\omega_L,p)=0,\qquad \omega_T^2=p^2+\Pi_T(\omega_T,p),
\eeq
for longitudinal and transverse excitations, respectively.
The solutions to these equations,
$\omega_L(p)$ and $\omega_T(p)$,  are displayed in fig.~\ref{disp}.b.
The longitudinal mode is the analog of the familiar plasma oscillation. It
corresponds to a collective oscillation of the hard particles,
and disappears when $p\gg gT$.
Both dispersion relations are time-like ($\omega_{L,T}(p)>p$), 
and show a gap at
zero momentum (the same for transverse and longitudinal modes 
since, when $p\to 0$, we recover isotropy).
As we shall see in Sect. 4.3.3, there is no Landau
damping for the soft modes in the HTL approximation.
Rather, these modes get attenuated via collisions in the plasma,
a mechanism which matters at higher order in $g$ and which will
be discussed in Sect. 6.

For small $p\ll m_D$, the dispersion relations read:
\beq
\omega_T^2\,=\,\omega_{pl}^2+
{3\over 5}\,p^2+\cdots\qquad\qquad \omega_L^2\,=\,
\omega_{pl}^2+{6\over 5}\,p^2+\cdots
\eeq 
where $\omega_{pl}\equiv m_D/{\sqrt 3}$ is the plasma frequency
(the gap in fig.~\ref{disp}.b).
For large momenta, $p\gg m_D$, one has 
\beq\label{masymp}
\omega^2_T\, \simeq\,p^2 +m_D^2/2,\qquad\qquad
\omega^2_L\,\simeq\, p^2(1+4x_L),\eeq
where
$ x_L\equiv \exp\left\{ -(2p^2/m_D^2)-2\right\}.$
Thus, with increasing momentum, the transverse
branch becomes that of a relativistic particle
with an effective mass $m_{\infty}\equiv m_D/\sqrt{2}$
(commonly referred to as the ``asymptotic mass'').
Although, strictly speaking, the HTL approximation
does not apply at hard momenta, the
above dispersion relation $\omega_T(p)$ remains nevertheless correct 
for  $p\sim T$ where it coincides with the light-cone limit 
of the full one-loop result \cite{Kraemmer94} :
\beq\label{minfty}
m_\infty^2\,\equiv\,\Pi_T^{1-loop}(\omega^2=p^2)\, =\,
\frac{m_D^2}{2}.\eeq
The longitudinal branch, on the other hand,
approaches the light cone exponentially, but, as already
mentioned, it disappears from the spectrum, as its residue
is exponentially suppressed \cite{Pisarski89a}.

We turn now to soft fermionic excitations. The corresponding 
HTL is easily obtained from eq.~(\ref{eindR}) with $A^\mu\to 0$,
and reads (cf. Sect. 5.3.1) :
\beq \label{sigmaultra}
\Sigma(\omega ,{\bf p})\,=\, \omega_0^2 
\int \frac{{\rm d} \Omega}{4\pi}\frac{\slashchar{v}}{\omega 
-{\bf v}\cdot{\bf p}+i\eta}\,.
\eeq
Let us first consider in more detail what happens in the long
wavelength limit, $p\to 0$. From
eq.~(\ref{sigmaultra}) one gets then:
\beq
\Sigma(\omega,p=0)\,=\,\frac{\omega_0^2}{\omega}\,\gamma_0.
\eeq
The spectrum at $p=0$ is thus obtained from the poles of
\beq
{}^*\!S(\omega)\,=\,\frac{\gamma_0}{-\omega+\omega_0^2/\omega}\,.
\eeq
For each  eigenstate of $\gamma_0$, corresponding to
the eigenvalues $\pm 1$, there are two poles, at $\omega=\pm \omega_0$.
Thus, the pole at $\omega= \omega_0$ is double degenerate,
and similarly for the pole at $\omega= -\omega_0$.
This degeneracy is removed by a small (zero-temperature) mass,
or, as we shall see, by a finite momentum
$p$, which leads to the split spectrum shown in fig.~\ref{disp}.a. 

Consider then the HTL-resummed propagator ${}^*\!S$ at finite momentum.
This has the following structure (cf. Sect. B.2.3):
\beq\label{SFL}
{}^*\!S(\omega, {\bf  p})\gamma_0\,=\,{}^*\!\Delta_+(\omega,p)
\Lambda_+ (\hat {\bf p})\,+\,{}^*\!\Delta_-(\omega,p)
\Lambda_-(\hat {\bf p})\,,\eeq
with the functions ${}^*\!\Delta_\pm$ given by eq.~(\ref{DPM}),
and the matrices $\Lambda_{\pm}\equiv
(1 \pm \gamma^0 \bfgamma\cdot\hat{\bf p})/{2}$ are projectors.
Charge conjugation exchanges the poles of 
${}^*\!\Delta_+$ and ${}^*\!\Delta_-$. ${}^*\!\Delta_+$ has 
{two} poles, one at positive $\omega$, with energy
$\omega_+(p)$,  and another one at negative $\omega$, with energy
$-\omega_-(p)$; these go over to $\pm \omega_0$ as $p\to 0$.
Correspondingly, ${}^*\!\Delta_-$ has poles at
$\omega_-$ and $-\omega_+$.

In the limit $p\gg \omega_0$, the branches $\pm\omega_+(p)$
describe the ``normal'' (anti)fermion, with a dispersion relation
\beq\label{FERMasymp}
\omega_+^2(p)\,\simeq\,p^2 + M_{\infty}^2\,,\qquad M_{\infty}^2
\,\equiv\,2\omega_0^2\,,\eeq
describing the propagation of a massive particle with the
``asymptotic'' mass $M_{\infty}=\sqrt{2}\omega_0$.
In the same limit, the ``abnormal'' branches $\pm\omega_-(p)$
disappear from the
spectrum since their residues are exponentially suppressed
\cite{Pisarski89c}. Incidentally, eq.~(\ref{FERMasymp}) is also
correct for $p\sim T$, where it coincides with the full one-loop
result \cite{Kraemmer94}.

For generic soft momenta, on the other hand, all the four branches
are present in the spectrum, and describe collective excitations
in which the hard quarks get converted into hard gluons, or vice-versa,
giving rise to longwavelength
oscillations in the  number density of the hard fermions.
In particular, for $p\ll \omega_0$,
\beq
\omega_+(p)\,\simeq\,\omega_0\,+\,{p\over 3}\,+\cdots,\qquad
\omega_-(p)\,\simeq\,\omega_0\,-\,{p\over 3}\,+\cdots,\eeq
so that the ``abnormal'' (or ``plasmino'') branch is actually
decreasing at small $p$, down to a minimum at $p=p_{min}
\approx .408 \,\omega_0\,$, and then it increases and approaches
$\omega=p$, as shown in fig.~\ref{disp}.a.
The origin of this peculiar behaviour is a collective phenomenon whereby  the
single particle strength at small momentum $p$ is split  by the coupling of the
soft modes to the hard particles  which form a continuum of states with energy
$|\omega|<p$. Due to this coupling, a fraction of the {\it anti}--fermion
strength, initially at energy
$\omega=-p$, is pushed up to the positive energy
$\omega=\omega_-(p)$, producing the abnormal branch. For small $p$ the
behaviour of $\omega_-(p)$ is therefore that of a negative energy state: it
decreases as $p$ increases. This physical interpretation is made  explicit by the
construction of the plasmino state
 at zero temperature and  large 
chemical potential in Ref.~\cite{BOllie}.

We note finally that particular solutions of the {\it non-linear}
equations (\ref{4avA}) have also been found, in 
Refs. \cite{prl2,planew,BIO95}. These solutions describe
non-linear plane-waves propagating through the plasma,
and represent truly non-Abelian collective excitations.

\subsubsection{Debye screening}

The screening of a static chromoelectric field by the plasma
constituents is the natural non-Abelian
generalization of the Debye screening, a familiar phenomenon 
in classical plasma physics \cite{PhysKin}.
In coordinate space, screening means that the range of the gauge
interactions is reduced as compared to the vacuum.
In momentum space, this corresponds to a softening of
the infrared behaviour of the various $n$-point functions.

Consider a static colour field,
that is, a field configuration which, at least in some particular
gauge, can be represented by time-independent
gauge potentials $A^a_\mu({\bf x})$.
For such fields, the expression (\ref{jA2}) of the colour
current reduces to its first, local, term 
(a static colour charge density):
\beq\label{jstat}
j_\mu^{A\,a}({\bf x})
\,=\,-\,\delta_{\mu 0}\,m^2_D A_0^a({\bf x})\,.\eeq
The equations of motion (\ref{4avA}) simplify accordingly:
\beq\label{snon}
[D_i, E^i({\bf x})]\,+\,m_D^2A^0({\bf x})\,=\,0,
\nonumber\\ \epsilon^{ijk}[D_j, B^k]\,=\,i\,g\,[A^0, E^i].\eeq
They differ from the corresponding equations
in the vacuum only by the presence of a
``mass'' term $m_D^2\,A^2_0$ for the electrostatic fields.

To appreciate the role of this mass, consider first the
Abelian case, where the equations above are linear:
\beq\label{Abst}
(-\Delta\,+\,m^2_D)\,A_0({\bf x})&=&\rho({\bf x}),\nonumber\\
\Delta A^i\,-\,\nabla^i(\bfgrad\cdot {\bf  A})&=&0.\eeq
The first equation, in which we have added an external source 
with charge density $\rho({\bf x})=Q\,\delta^{(3)}({\bf x})$,
is easily solved by Fourier transform and yields
the familiar screened Coulomb potential:
\beq\label{VR}
A_0(r) = Q \int\frac{{\rm d}^3p}{(2\pi)^3}
\,{{\rm e}^{i {\bf p} \cdot {\bf r}}
\over p^2 + m^2_D } \,=\,Q\,\frac{{\rm e}^{-m_D r}}{4\pi r}\,.
\eeq
The second equation (\ref{Abst}) shows that the static magnetic
fields are {\it not} screened, which can be related to the fact
that plasma particles carry no magnetic charges.

The same conclusion can be reached by an analysis of
the effective photon (or gluon) propagators (\ref{effd0})
in the static limit $\omega\to 0$. Eqs.~(\ref{plpt}) imply
\beq\label{static1} \Pi_L(0,p)\,=\,m_D^2\,,\qquad\,\,\,\,
\Pi_T(0,p)=0,\eeq and therefore:
 \beq\label{DSTAT}
{}^*\!\Delta_L(0,p)\,=\,\frac{- 1}{p^2 + m_D^2}\,,\qquad
{}^*\!\Delta_T(0,p)\,=\,\frac{1}{p^2}\,,\eeq
which clearly shows that the Debye mass acts as an infrared cut-off $\sim gT$
in the electric sector, 
while there is no such cut-off in the magnetic sector.

In non-Abelian plasmas, 
Debye screening may be accompanied by interesting
non-linear effects.  A particular solution to eqs.~(\ref{snon}) is
\cite{Liu93} 
\beq\label{WY}
A_0^a\,=\,{\cal A}\,\hat r^a\,\frac{{\rm e}^{-m_Dr}}{r}\,,\qquad
\,\, A_i^a\,=\,\epsilon^{aij}\hat r^j\,\frac{1}{r}\,,\eeq
where ${\cal A}$ is a constant and $\hat r^i=x^i/r$.
This solution is a superposition of the Wu-Yang  magnetic monopole
\cite{WY76} and a screened electrostatic field. More generally, it has been
shown in Ref. \cite{Liu93} that all the ($SU(2)$-radially symmetric)
solutions which are regular at infinity
approach at the origin the monopole solution (\ref{WY}).
That is, all such solutions show electric screening, but there is no
sign of magnetic screening, in spite of the non-Abelian coupling
between electric and magnetic fields in eqs.~(\ref{snon}).
Moreover, there are no finite energy solutions (no 
static solitons), in complete analogy to what happens
in  the vacuum (i.e., for $m_D=0$) \cite{Deser76}.

\subsubsection{Landau damping}

For time-dependent fields, there exists a different screening mechanism
associated to the energy transfer to the plasma constituents. 
In Abelian plasmas, this mechanism is known as
{\it Landau damping} \cite{PhysKin}.
For simplicity, let us start with this Abelian case, and compute the
mechanical work done by a longwavelength electromagnetic field acting
on the charged particles. The rate of energy transfer
has the familiar expression \cite{PhysKin}:
\beq\label{ratew}
\frac{{\rm d}\,{\rm E}_W(t)}{{\rm d}\,t}\,=\,\int {\rm d}^3{\bf x}
\,{\bf E}(t,\,{\bf x})\cdot {\bf j}(t,\,{\bf x}),\eeq
where  $j^i(p)=\Pi_R^{i\nu}(p)A_\nu(p)$ is the induced current.
Consider, for instance, a periodic electric field of 
the form $ {\bf E} (t,\,{{\bf p}})\,=\,{\bf E}({\bf p})\,
\cos\omega_0t$. From eq.~(\ref{ratew}), one can compute the average 
energy loss over one period $T_0=2\pi/\omega_0$, with the following result:
\beq
\label{erate}
\left\langle \frac{{\rm d}\,{\rm E}_W}{{\rm d}
\,t}\right\rangle &=&\frac{1}{2\omega_0}
\int \frac{{\rm d}^3 p}{(2\pi)^3}\,E^i(-{\bf p})\left (
- {\rm Im}\,\Pi_R^{ij}(\omega_0,\,{\bf p})\right )\,E^j({\bf p})
\nonumber\\&=&\frac{\pi m_D^2}{2} 
\int\frac{{\rm d}\Omega}{4\pi}\,
\delta (\omega - {\bf  v}\cdot{\bf  p})
\,\Big |{\bf v}\cdot{\bf E}({\bf p})\Big |^2,\eeq 
where we have used the following expression for the imaginary part
of the retarded polarization tensor (cf.  eq.~(\ref{polarisation2}))
\beq\label{ImPi}
{\rm Im}\, \Pi_R^{\mu\nu}(\omega, {\bf  p})\,=\,-\,
\pi m_D^2\,\omega\int
\frac{{\rm d}\Omega}{4\pi}\,\,v^{\mu}v^{\nu}
\,\delta (\omega - {\bf  v}\cdot{\bf  p})\,.\eeq
The expression in eq.~(\ref{erate})
is non-negative: on the average, the energy is
transferred from the electromagnetic field to the particles.
The $\delta$-function in Eq.~(\ref{erate}) shows that the particles which
absorb energy are those moving in phase with the field
(i.e., the particles whose velocity component along  ${\bf p}$ is equal
to the field phase velocity: ${\bf v}\cdot \hat{\bf p} =\omega/p$).
Since in ultrarelativistic plasmas ${\bf v}$ is a unit vector,
 only {\it space-like} ($|\omega| < p$) fields
are damped in this way.

To see how this mechanism leads to screening, 
consider the effective photon (or gluon) propagator in the hard thermal
loop approximation (cf. eq.~(\ref{effd0})), and focus on the magnetic
propagator. For small but non-vanishing frequencies the corresponding
polarization function  $\Pi_T(\omega,p)$ is dominated by its imaginary
part, which vanishes linearly as $\omega\to 0$ (see eq.~(\ref{ImPi})),
in contrast to the real part which vanishes quadratically.
Specifically,  the second equation (\ref{plpt}) yields:
\beq\label{IMPIT}
\Pi_T(\omega \ll p)\,=\,-i\,\frac{\pi}{4}\,m_D^2\,\frac{\omega}{p}
+\,{\rm O}(\omega^2/p^2)\,,
\eeq
and therefore
\beq\label{Vt}
{}^*\!\Delta_T(\omega\ll p)\simeq\,\frac{1}
{p^2-i\,(\pi \omega/4p)\,m_D^2}.\eeq
In the computation of scattering cross sections, 
the relevant matrix element squared is proportional to
(see, e.g., eq.~(\ref{G2L}))
\beq\label{deltaT0} |{}^*\!\Delta_T(\omega,p)|^2\,\simeq\,
\frac{1} {p^4 + (\pi m_D^2 \omega/4p)^2}\,,\eeq
which shows that Im$\,\Pi_T(p)$ acts as
a frequency-dependent IR cutoff at momenta
$p\sim (\omega m_D^2)^{1/3}$. That is, as long as the frequency
$\omega$ is different from zero, the soft momenta are 
dynamically screened by Landau damping \cite{Baym90}.

Dynamical screening occurs also for the longitudinal interactions,
but in this case it is less important, since Debye
screening dominates at small frequency.

Furthermore, in the case of QCD, the study of Landau damping
is complicated by non-linear effects. The non-Abelian expression
for the rate of mechanical work (see eq.~(\ref{rateNA}) below)
involves the non-linear colour current (\ref{jAR}); accordingly, 
all the $n$-point HTL amplitudes (self-energy and vertices) 
develop imaginary parts (see Sect. 5).
Moreover, the Landau damping is also operative for soft
{\it fermions},  both in QCD and in QED; this is described,
e.g., by  the imaginary part of the
fermion self-energy, eq.~(\ref{sigmaultra}).

\subsection{Hamiltonian theory for the HTL's}

There exists a concise and elegant formulation of
the effective theory for the soft fields dynamics as a Hamiltonian theory
\cite{Nair93,emt,baryo}. At first sight, this may be surprising
since the  corresponding equation of motion, namely eq.~(\ref{4avA}),
(the fermionic fields are set to zero in this section):
\beq\label{MFAYM}
\left [\, D^\nu,\, F_{\nu\mu}(x)\,\right ]^a 
\,=\,m^2_D\int\frac{ {\rm d}\Omega}{4\pi}
\,v_\mu \int_0^\infty {\rm d}
\tau\, U_{ab}(x,x-v\tau)\, {\bf v}\cdot{\bf E}^b
(x-v\tau)\,,\eeq
is non-local in space and time, and also dissipative:
Because of Landau damping,  energy is transferred between
the particles and the background fields. However,  at the expense
of keeping the field $W^a_0(x,{\bf v})$ as a soft degree of freedom
(summarizing the effects of the plasma particles), one can rewrite
eq.~(\ref{MFAYM}) as
to the following system of {\it local} equations (cf. eqs.~(\ref{eqw}) and
(\ref{j1A})):
\beq\label{ava2}
\left [\, D^\nu,\, F_{\nu\mu}(x)\,\right ]^a &=&
m^2_D\int\frac{{\rm d}\Omega}{4\pi}
\,v_\mu\,W^a_0(x,{\bf v}),\nonumber\\
\left[ v\cdot D_x,\, W_0(x,{\bf v})\right]^a&=&{\bf v\cdot E}^a(x).\eeq
The  fields  $A^\mu_a(x)$ and $W^a_0(x,{\bf v})$
will be regarded as independent degrees of freedom in the following.

\subsubsection{The energy of the colour fields}

In order to obtain the Hamilton function for these degrees of freedom, we start
by computing the energy ${\rm E}$ carried by the longwavelength colour excitations
\cite{Weldon93,BFT93,Nair93,emt}. We can write:
\beq\label{entot}
{\rm E}\,=\,
{\rm E}_{YM}(t)\,+\,{\rm E}_W(t)\,\equiv\,\int {\rm d}^3 {\bf x}\,
{\cal E}(t,\,{\bf x}),\eeq
where ${\rm E}_{YM}(t)$ is the energy stored in the colour fields
at time $t$, 
\beq\label{enloc}
{\rm E}_{YM}(t)\,=\,\int {\rm d}^3 {\bf x}\,
\frac{1}{2}\Bigl ({\bf E}^a(t,\,{\bf x})\cdot{\bf E}^a(t,\,{\bf x})\,+\,
{\bf B}^a(t,\,{\bf x})\cdot{\bf B}^a(t,\,{\bf x})\Bigr),\eeq
while ${\rm E}_W(t)$ is the polarization energy, that is,
the energy transferred by the colour field to the plasma constituents,
as mechanical work. Energy conservation d${\rm E}/{\rm d}t=0$,
together with the first equation (\ref{ava2}), imply
\beq\label{rateNA}
\frac{{\rm d}\,{\rm E}_W(t)}{{\rm d}\,t}\,=\,\int {\rm d}^3 {\bf x}
\,{\bf E}^a(t,\,{\bf x})\cdot {\bf j}_a(t,\,{\bf x}),\eeq
where ${\bf j}_a$ is the induced current (\ref{j1A}):
\beq
{\bf j}_a(t,\,{\bf x})\,=\,m^2_D\int\frac{{\rm d}\Omega}{4\pi}
\,{\bf v}\,W^a_0(x,{\bf v}).\eeq
We recognize in eq.~(\ref{rateNA})  the non-Abelian generalization of
eq.~(\ref{erate}). By using the equation of motion for $W^a_0(x,{\bf v})$
(i.e. the second equation (\ref{ava2})), we can write:
\beq\label{work}
{\rm E}_W(t)&=&\int_{-\infty}^t dt^\prime \,\int {\rm d}^3 {\bf x}'
\,{\bf E}^a(t^\prime,\,{\bf x'})\cdot {\bf j}_a(t^\prime,\,{\bf x'})
\nonumber\\&=&m^2_D\int\frac{{\rm d}\Omega}{4\pi}
\int_{-\infty}^t dt^\prime \,\int {\rm d}^3{\bf x}'\,
\,W^a_0(t^\prime,\,{\bf x'},{\bf v})
\left[ v\cdot D_{x'},\, W_0(t^\prime,\,{\bf x'},{\bf v})\right]^a
\nonumber\\&=&\frac{m^2_D}{2}\int\frac{{\rm d}\Omega}{4\pi}
\int_{-\infty}^t dt^\prime \,\int {\rm d}^3 {\bf x}'
\,(v\cdot \del_{x'}) \left(W^a_0\, W_0^a\right)
.\eeq
The integral over $t^\prime$ can now be done (we assume the fields to
vanish at spatial infinity and at $t\to -\infty$), and yields:
\beq\label{ennon}
{\rm E}_W(t)\,=\,\frac{m^2_D}{2}
\int {\rm d}^3 {\bf x}\int\frac{{\rm d}\Omega}{4\pi}\,
\,W_a^0(x,{\bf v})\,W_a^0(x,{\bf v}).\eeq
Together, eqs.~(\ref{entot}), (\ref{enloc}) and (\ref{ennon})
express the energy associated with the propagation of a longwavelength
colour wave  in a hot QCD plasma \cite{emt}. Clearly, this quantity is
positive definite, which reflects the stability of the plasma 
with respect to longwavelength colour oscillations \cite{Nair93}.
 
The energy flux density, or Poynting vector, of the propagating
colour waves can be computed  similarly, with the result \cite{emt}
\beq\label{Poyn}
{\bf S}(x)\,=\,{\bf E}^a(x)\times{\bf B}^a(x)\,+\,\frac{m_D^2}{2}
\int\frac{d\Omega}{4\pi}\,{\bf v}\,W_a^0(x,{\bf v})\,W_a^0(x,{\bf v}).\eeq 
Then, the energy conservation can be also written in local form, as 
\beq\label{EP}
\del_0\,{\cal E}(x)\,+\,\del_i\,S^i(x)\,=\,0,\eeq
where ${\cal E}(x)$ is the energy density
in eq.~(\ref{entot}).
Note, however, that the above expressions are local only when
expressed in terms of both the gauge fields $A^\mu_a(x)$ and
the auxiliary fields $W^0_a(x,{\bf v})$.

\subsubsection{Hamiltonian analysis}

In the
temporal gauge $A^a_0=0$, eqs.~(\ref{ava2}) become (with
$W_0^a(x,{\bf v})\equiv W^a(x,{\bf v})$ in what follows)
\beq\label{CAN}
\del_0 A^a_i &=&-E^a_i,\nonumber\\
-\del_0 E^a_i +\epsilon_{ijk}(D_j B_k)^a &=&
m_D^2\int\frac{{\rm d}\Omega}{4\pi}\,v_i\,W^a(x,{\bf v}),\nonumber\\
\left(\del_0 + {\bf v\cdot D}\right)^{ab} W_b&=&{\bf v \cdot E}^a,\eeq
together with Gauss' law
(the $\mu=0$ component of the first equation (\ref{ava2})):
\beq\label{GAUSS}
G^a({\bf x})\equiv
({\bf D\cdot E})^a\,+\,m_D^2\int\frac{{\rm d}\Omega}{4\pi}\,W^a(x,{\bf v})
\,=\,0.\eeq
This last equation contains no time derivative and should therefore be
regarded as a constraint. 

We show now that eqs.~(\ref{CAN}) can be given a Hamiltonian structure. 
To this aim, consider the conserved energy in
eq.~(\ref{entot}) which we denote here by $H$:
\beq\label{H}
H\,=\,\frac{1}{2}\int {\rm d}^3{\bf x}\left\{{\bf E}^a\cdot{\bf E}^a\,+\,
{\bf B}^a\cdot{\bf B}^a\,+\,m_D^2
\int\frac{{\rm d}\Omega}{4\pi}\,W^a(x, {\bf v})\,W^a(x, {\bf v})\right\}.\eeq
This expression is independent of the choice of gauge. However, in the
gauge $A_0^a=0$, we
can make it act as a Hamiltonian, that is, as the generator of the time
evolution. 
{As independent degrees of freedom, we choose the
vector potentials $A^i_a({\bf x})$, the electric fields
$E_a^i({\bf x})$, and the density matrices $W^a({\bf x,v})$.
(Note that the latter are fields  on the extended
configuration space $E^3\times S^2$, 
with $E^3$ being the ordinary three-dimensional coordinate 
space and $S^2$ the unit sphere  spanned by ${\bf v}$.)
Then, following Nair \cite{Nair93}, we organize this as
a Hamiltonian system by introducing the
following Poisson brackets:}
\beq\label{PB}
\left\{E^a_i({\bf x}), A^b_j({\bf y})\right\}&=&-\,\delta^{ab}\delta_{ij}
\delta^{(3)}({\bf x-y})\,,\nonumber\\
\left\{E^a_i({\bf x}), W^b({\bf y,v})\right\}&=&v_i\,\delta^{ab}
\delta^{(3)}({\bf x-y})\,,\nonumber\\
m^2_D\left\{W^a({\bf x,v}), W^b({\bf y,v'})\right\}&=&
\left(gf^{abc}W^c+({\bf v\cdot D}_x)^{ab}\right)\delta^{(3)}({\bf x-y})
\delta({\bf v},{\bf v}^\prime)\,.\eeq
Here, $\delta({\bf v},{\bf v}^\prime)$ is the delta function
on the unit sphere, normalized such that
\beq
\int\frac{{\rm d}\Omega}{4\pi}\,\delta({\bf v},{\bf v}^\prime)
\,f({\bf v})\,=\,f({\bf v}^\prime),\eeq
and all  other  brackets are assumed to vanish.
We also assume the standard properties for the Poisson brackets,
namely antisymmetry, bilinearity and Leibniz identity:
$\{AB,C\}=A\{B,C\}+\{A,C\}B$. It is then straightforward to verify that
(a) the Poisson brackets (\ref{PB}) satisfy the Jacobi identity (a 
necessary  consistency condition) and (b) the equations of motion
(\ref{CAN}) follow as canonical equations for the Hamiltonian (\ref{H}).
For instance,
$\del_0 W^a=\{H,W^a\}$, and similarly for $E_i^a$ and $A_i^a$.

The Hamiltonian in eq.~(\ref{H}) is remarkably simple:  
it is quadratic in the auxiliary fields $W^a_0$. Up to the colour indices,
this piece would be the same in QED. Thus, all the non-Abelian
complications are encoded in the Yang-Mills piece of $H$ and in the
non-trivial Poisson brackets (\ref{PB}).

\subsubsection{Effective classical thermal field theory}

We shall now use the above Hamiltonian formulation of
the HTL effective theory to write down a generating functional
for the thermal correlations of the soft fields, in the classical
approximation. As emphasized in Sect. 2.2.5, the classical
approximation is correct only at soft momenta, so we shall introduce an
ultraviolet cutoff  $\Lambda$, with $gT \ll \Lambda \ll T$, to eliminate the
hard ($k\simge T$) fluctuations from the classical theory.
Correspondingly, $\Lambda$ will act as an infrared cutoff
for the hard, quantum, modes (see below).

We denote by  ${\cal E}^a_i({\bf x})$, 
${\cal A}^a_i({\bf x})$ and ${\cal W}^a({\bf x,v})$ the initial
conditions for the HTL equations of motion (\ref{CAN}).
The 
partition function reads as follows (compare with eq.~(\ref{ZCLS})):
\beq\label{ZCL}
Z_{cl}\,=\,
\int {\cal D}{\cal E}^a_i\,{\cal D}{\cal A}^a_i\,{\cal D}{\cal W}^a\,
\delta({\cal G}^a)\,{\rm e}^{-\beta{\cal H}},\eeq
where ${\cal G}^a$ and ${\cal H}$ are expressed in terms of the 
initial fields as in eqs.~(\ref{GAUSS}) and (\ref{H}).
Eq.~(\ref{ZCL}) can be rewritten as
\beq\label{ZRED}
Z_{cl}\,=\,\int {\cal D}{\cal A}^a_0\,{\cal D}{\cal A}^a_i\,
\exp\left\{-\frac{\beta }{2}\int{\rm d}^3x \,\Bigl(
{\cal B}^a_i{\cal B}^a_i + (D_i {\cal A}_0)^a
(D_i {\cal A}_0)^a+ m_D^2{\cal A}_0^a{\cal A}_0^a
 \Bigr)\right\},\eeq
where the temporal components ${\cal A}_0^a$ of the gauge fields 
have been reintroduced
as Lagrange multipliers to enforce Gauss' law, and the Gaussian  
functional integrals over ${\cal E}^a_i$ and ${\cal W}^a$ have
been explicitly performed.
In particular, the integral over the ${\cal W}'$s
has generated the screening mass for the electrostatic fields,
as expected. We recognize in eq.~(\ref{ZRED}) the static limit 
(\ref{SEFFST}) of the HTL action.

More generally, time-dependent correlations 
of the soft fields are obtained by averaging  products of fields $A^i_a(t,{\bf
x})$ obeying eqs.~(\ref{CAN}). These correlations  can be generated from
\beq\label{Z}
Z_{cl}[J^a_i]\,=\,
\int {\cal D}{\cal E}^a_i\,{\cal D}{\cal A}^a_i\,{\cal D}{\cal W}^a\,
\delta({\cal G}^a)\,
\exp\left\{-\beta{\cal H}\,+\,\int{\rm d}^4x J^a_i(x) A^a_i(x)\right\},\eeq
where $A^i_a(t,{\bf x})$ is the solution to eqs.~(\ref{CAN})
(in particular, $A^a_i(t_0,{\bf x})={\cal A}^a_i({\bf x})$), and
the external current $J^a_i$ is introduced as a device to
generate the correlations of interest via functional differentiations,
but does not enter the equations of motion for the fields.
 It can be verified
\cite{baryo} that the phase-space measure  ${\cal D}{\cal E}^a_i{\cal
D}{\cal A}^a_i{\cal D}{\cal W}^a$  is invariant under the time evolution
described by eqs.~(\ref{CAN}), so that $Z_{cl}[J]$ is independent of the 
initial time $t_0$, as it should.
Since the dynamics is also gauge-invariant, 
it is sufficient to enforce Gauss' law at $t=t_0$,
as we did in eq.~(\ref{Z}).

As a simple, but still non-trivial, check of eq.~(\ref{Z}),
let us consider the Abelian limit, where the equations of 
motion can be  solved analytically, and the functional integral
in eq.~(\ref{Z}) can be  computed exactly, since Gaussian \cite{baryo}.
We know already the result that we want to obtain:
This should read (compare with eq.~(\ref{GCL0})) :
\beq\label{ZAB}
Z_{cl}[J_i]&=&Z_{cl}[0]\,\exp\left\{-\frac{1}{2}\int{\rm d}^4x \int{\rm d}^4y
\, J^i(x) \,G_{ij}^{cl}(x-y) J^j(y)\right\},\\
\label{DCL}
G_{ij}^{cl}(x-y)
&\equiv&\int\frac{{\rm d}^4k}{(2\pi)^4}\,{\rm e}^{-i k\cdot(x-y)}\,
\,{}^*\!\rho_{ij}(k) N_{cl}(k_0),\eeq
where $N_{cl}(k_0)= T/k_0$, 
and ${}^*\!\rho_{ij}(p)$ is the photon spectral density
in the HTL approximation and in the temporal gauge
(cf. eqs.~(\ref{rhos})--(\ref{rhos0})) :
\beq\label{RHOAX}
{}^*\!\rho_{ij}(\omega,{\bf k})=
(\delta_{ij}-\hat k_i\hat k_j)\,{}^*\!\rho_T(\omega,k)
\,+\,\frac{k_i k_j}{\omega^2}\,\,{}^*\!\rho_L(\omega,k)\,.\eeq
The 2-point function (\ref{DCL}) is the classical
limit of the corresponding quantum correlator, which reads
\beq\label{specD0}
{}^*\!G_{\mu\nu}(x,y)
=\int\frac{{\rm d}^4k}{(2\pi)^4}\,{\rm e}^{-ik\cdot(x-y)}
\,\,{}^*\!\rho_{\mu\nu}(k)\Bigl[\theta(x_0-y_0)+N(k_0)\Bigr]\,.\eeq
In the classical limit $N(k_0)\approx T/k_0 \gg 1$, so that the 
2-point functions $G^>$ and $G^<$ reduce to
the unique classical correlator $G^{cl}$ (see sections
2.1.4 and 2.2.5). For instance, in the transverse sector,
\beq \label{DDCL}
{}^*\!G^>_T(\omega,k)\,\simeq\,{}^*\!G^<_T(\omega,k)\,\simeq\,
\frac{T}{\omega}\,\,{}^*\!\rho_T(\omega,k)\,=\,
G_T^{cl}(\omega,k).\eeq
It is our purpose here to verify that eqs.~(\ref{ZAB})--(\ref{RHOAX})
emerge indeed from the computation of
the Abelian version of the functional integral (\ref{Z}).

To this aim, we have to solve first
the linearized equations of motion (\ref{CAN}) :
\beq\label{EAB}
\Bigl[(\del_0^2 - \bfgrad^2)\delta^{ij}+\del^i\del^j\Bigr]
A^j(x)&=&m_D^2\int\frac{{\rm d}\Omega}{4\pi}\,v^i\,W(x,{\bf v}),\nonumber\\
\left(\del_0 + {\bf v}\cdot\bfgrad \right)W(x,{\bf v})&=&{\bf v \cdot E}
(x),\eeq
with the initial conditions 
\beq\label{INIT}
A^i(t_0,{\bf x})={\cal A}^i({\bf x}),\qquad
\dot A^i(t_0,{\bf x})= -{\cal E}^i({\bf x}),\qquad
W(t_0,{\bf x,v})={\cal W}({\bf x,v}).\eeq
To simplify the presentation  we shall limit ourselves here to the 
transverse sector,
and consider the transverse projection of the first
equation  (\ref{EAB}),
\beq\label{MAXT0}
(\del_0^2+{\bf k}^2)A^i_T\,=\,
m_D^2\int\frac{{\rm d}\Omega}{4\pi}\,v_T^i\,
W({\bf k, v}),\eeq
with ${\bf k \cdot A}_T = 0$
and $v_T^i=(\delta^{ij}-\hat k^i\hat k^j)v^j$.  { We choose as initial
conditions  ${\cal A}_T^i=0$ and 
${\cal E}^i_T=0$, but let ${\cal W}({\bf x,v})$ be arbitrary.
That is, for the gauge fields, we choose as initial values the
corresponding average values in thermal equilibrium.
All the fluctuations are generated by the randomly chosen
initial conditions for $W({\bf x, v})$, that is, by the
longwavelength initial
fluctuations in the charge density of the hard fermions.
These fluctuations will generate time-dependent gauge fields,
via the equations of motion (\ref{EAB}).

Consider then the solution
$W(x,{\bf v})$ to the Vlasov equation (i.e., the second equation
(\ref{EAB})) which we write as:}
\beq\label{DA}
W\,=\,W_{ind}+W_{fl},\eeq
where $W_{ind}$ is the solution to the Vlasov equation with zero
initial condition: $W_{ind}(t_0,{\bf x,v})=0$),
and $W_{fl}$ is the fluctuating piece, solution
of the homogeneous equation
\beq\label{eqWFL}
\left(\del_0 + {\bf v}\cdot\bfgrad \right)W_{fl}&=&0,\eeq
with the initial condition $W_{fl}(t_0,{\bf x,v})
={\cal W}({\bf x,v})$.
It follows that (for $x_0 > t_0$):
\beq\label{WFLA}
W_{ind}(x,{\bf v})&=&-i\int {\rm d}^4 y\,\theta(y_0-t_0)\,
G_R(x,y|{\bf v})\,\,{\bf v \cdot E}(y),\nonumber\\
W_{fl}(x,{\bf v})&=&{\cal W}({\bf x-v}(x_0-t_0),{\bf v}),\eeq
where $G_{R}(x,y|{\bf v})$ is the retarded Green's function 
in eq.~(\ref{GR0}).
Eq.~(\ref{DA}) implies a similar decomposition for the current:
$j^i=j^i_{ind} + \xi^i$, with
\beq\label{XIA}
j^i_{ind}(x)
&=&-\int {\rm d}^4 y\,\theta(y_0-t_0)\,\Pi_R^{ij}(x-y)A^j(y)\nonumber\\
\xi^i(x)&=&m_D^2\int\frac{{\rm d}\Omega}{4\pi}\,v^i
\,{\cal W}({\bf x-v}(x_0-t_0),{\bf v}),\eeq
where $\Pi_R^{ij}$ is the (retarded)
HTL polarization tensor given in eq.~(\ref{polarisation2}).

The Maxwell equation (\ref{MAXT0}) now becomes
\beq\label{MAXT1}
(\del_0^2+{\bf k}^2)A^i_T\,+\,
\int_{t_0}^\infty {\rm d}y_0\,
\Pi_T(x_0-y_0,{\bf k})A^i_T(y_0)=\xi_T^i,\eeq
which should be compared with the equation that we had before,
namely the first equation (\ref{LIN}): the crucial difference 
is the presence
of the fluctuating current $\xi^i(x)$ in the right hand side,
which is independent of the gauge fields, 
and acts as a ``noise term'' to induce the thermal
correlations of the classical electromagnetic fields.
Specifically, eqs.~(\ref{ZCL}) and (\ref{H}) imply that the 
initial conditions ${\cal W}$ are Gaussian random variables 
with zero expectation value and the following, local, 
2-point correlation:
\beq\label{WN0}
\langle {\cal W}({\bf x, v})\,{\cal W}({\bf y, v}^\prime)\rangle=
(T/m^2_D)\,\delta^{(3)}({\bf x-y})\delta({\bf v},{\bf v}^\prime).
\eeq
This immediately implies:
\beq\label{XIXI}
\langle \xi^i(x) \xi^j(y)\rangle\,=\,m^2_D T
\int\frac{{\rm d}\Omega}{4\pi} v^i v^j
\delta^{(3)}({\bf x-y}-{\bf v}(x_0-y_0)),\eeq
which is non-local; that is, the fluctuating  current 
$\xi^i(x)$ is {\it not} a ``white noise''. For $t_0\to -\infty$,
eq.~(\ref{MAXT1}) can be easily  solved by Fourier transform to yield
(recall that
${\cal A}_T^i={\cal E}_T^i=0$):
\beq\label{ATT}
A_T^i(k)\,=\,\,{}^*\!\Delta_T(k)\,\xi_T^i(k),\eeq
where ${}^*\!\Delta_T(k)$ is the
retarded magnetic propagator in the HTL approximation,
cf. eq.~(\ref{effd0}). The gauge field correlation
induced by the ``noise term''  $\xi^i(x)$ is finally obtained as:
\beq
\langle A_T^i(k) A_T^{j\,*}(p)\rangle 
\,=\, (2\pi)^4 \delta^{(4)}(p+k)\,(\delta^{ij}-\hat k^i\hat k^j)\,
\tilde G_T^{cl}(k), \eeq
with
\beq\label{DT0}
\tilde G_T^{cl}(\omega,k)&\equiv&m^2_D T\,|{}^*\!\Delta_T(k)|^2\,
\int\frac{{\rm d}\Omega}{4\pi} \,\,{\bf v}_T^2\,\,2\pi\,
\delta(\omega-{\bf v \cdot k})\nonumber\\&=&
- 2\,(T/\omega)\,|{}^*\!\Delta_T(k)|^2\,\,
{\rm Im} \,\Pi_T(\omega,k),\eeq
where in writing the second line we have recognized
${\rm Im} \,\Pi_T$ from eq.~(\ref{ImPi}).
This can be rewritten as
\beq
\tilde G_T^{cl}(\omega,k)\,=\, (T/\omega)\, \beta_T(\omega,k),\eeq
where $\beta_T$ is the off-shell (or Landau damping) piece of the
transverse photon spectral density in the HTL approximation
(cf. eqs.~(\ref{rhos0}) and (\ref{RHOLT}) in the Appendix).

Thus, by averaging over the initial conditions ${\cal W}$
alone, one has generated the Landau 
damping piece of the magnetic propagator.  Similarly, by averaging
over the initial fields ${\cal A}_T^i$ and ${\cal E}_T^i$,
one generates also the pole piece\footnote{This identification of the
Landau damping spectral density $\beta_T$
with the average over ${\cal W}$, and of the
pole spectral density ${}^*\!\rho_T^{\,pole}$
with the average over ${\cal A}^i$ and ${\cal E}^i$,
holds only in the limit $t_0\to -\infty$ \cite{baryo}.},
${}^*\!\rho_T^{\,pole}\propto
\delta(\omega^2 -\omega_T^2(k))$
 \cite{baryo}. Altogether, this gives the classical transverse 
2-point function in the expected form
(cf. eq.~(\ref{DDCL})):
\beq\label{DT}
G_T^{cl}(\omega,k)\,=\, (T/\omega)\,\,{}^*\!\rho_T(\omega,k).
\eeq
An entirely similar result holds for the longitudinal
2-point function as well \cite{baryo}, which completes the
 verification of the result announced in
 eqs.~(\ref{ZAB})--(\ref{RHOAX}). 
One thus sees, on the example of the 2-point function in QED,
that the physics of HTL's is correctly reproduced
by the classical theory.

We note that in this calculation the intermediate scale
$\Lambda$ did not play any role. In QCD, however, because the
equations of motion for the soft modes are non-linear, the 
average over the initial conditions generate soft thermal loops
which are linearly ultraviolet divergent (as for the scalar
theory in Sect. 2.1.4; recall, especially, 
eq.~(\ref{MDCL1}) there). In this case, 
the intermediate scale $\Lambda$, with
$gT\ll \Lambda \ll T$, is necessary. 
As compared with the classical theory in Sect. 2.2.5, the new
complication here is that the cutoff procedure must be implemented in
a way consistent with gauge symmetry. Furthermore, in numerical solutions
of the equations of motion using the lattice techniques, an 
additional complication arises from the fact that
 the lattice regularization breaks
 rotational and dilation symmetries.
As a consequence, the ultraviolet divergences of the lattice theory
cannot be all absorbed into just one parameter, a ``lattice
Debye mass''. In spite of many efforts 
\cite{McLerran,Arnold97,baryo,Nauta99}, 
no complete solution to such problems has been found.

Nevertheless, the HTL effective theory has
been already implemented on a lattice \cite{BMR99}
(see also Refs. \cite{Hu,Moore98,Rajantie99}), and applied to
the calculation of the anomalous baryon number violation rate
in a high-temperature Yang-Mills theory (cf. Sect. 1.6). Remarkably,
the results obtained in this way, even without any matching, 
appear to be rather insensitive to lattice artifacts, and are
moreover consistent with some previous numerical calculations
\cite{Hu} (where the HTL's are simulated via classical 
coloured ``test particles'' \cite{Kelly94}), and also with
the theoretical predictions in Refs. \cite{ASY97,Bodeker}.

\setcounter{equation}{0}
\setcounter{equation}{0}
\section{Hard thermal loops}

In the previous section we have obtained explicit expressions
for the induced sources as functionals of the mean fields. 
These functionals may be used
to obtain, by successive differentiations with respect to the fields,
the effective propagators and vertices for the soft fields. 
The resulting expressions are the so-called   ``hard thermal loops'' (HTL)
\cite{Pisarski89,BP90,FT90,TW90}, i.e.,  the leading order
 thermal corrections to the one-particle-irreducible 
(1P-I) amplitudes with soft external lines.

These amplitudes may be used as building blocks to improve 
perturbative calculations through various resummation schemes.
These will be discussed in the last section of this chapter,
which contains also digressions on the limitation of weak
coupling calculations and how these can be overcome using lattice
calculations.

\subsection{Irreducible amplitudes from induced sources}

By taking the derivatives of the effective action
$\Gamma\bigl[A,\Psi,\bar\Psi\bigr]$ with respect to its field arguments,
we obtain the equations of motion for the mean fields
{(all the subsequent formulae hold for time arguments along a complex
contour of the type discussed in Sect. 2) }:
\beq\label{1G}
j^\mu=-\frac{\delta\Gamma}{\delta A_\mu},\qquad
\eta=-\frac{\delta\Gamma}{\delta \bar\Psi},\qquad
\bar\eta=\frac{\delta\Gamma}{\delta \Psi},\eeq
where $j^\mu,\eta,\bar\eta$ are external sources. 
By writing $\Gamma=S_{cl}+\Gamma_{ind}$, 
where $S_{cl}$ is the classical action (\ref{QCD}), we get:
\beq\label{defj}
j_a^\mu(x)&=&\left [\, D_\nu,\, F^{\nu\mu}(x)\,\right ]^a
-g  \bar\Psi (x)\gamma^\mu t^a \Psi(x)\,-\,
\frac{\delta \Gamma_{ind}}{\delta A_\mu^{a}(x)},
\nonumber\\\eta(x)&=&i\slashchar{D} \,\Psi(x)\,-\,
\frac{\delta \Gamma_{ind}}{\delta \bar\Psi(x)}.\eeq
Then, a  comparison with eqs.~(\ref{avA1})--(\ref{avpsi1}) allows us to 
identify the induced sources as the first derivatives of $\Gamma_{ind}\,$:
\beq\label{Sind}
j^{ind}_{\mu\,a}(x)=\frac{\delta \Gamma_{ind}}{\delta A^\mu_{a}(x)},\qquad
\eta^{ind}(x)=\frac{\delta \Gamma_{ind}}{\delta \bar\Psi(x)},\qquad
\bar\eta^{ind}(x)=-\frac{\delta \Gamma_{ind}}{\delta \Psi(x)}.\eeq
Accordingly, all the $n$-point 1P-I Green's functions with $n\ge 2$
can be obtained by  differentiating the 
 induced sources with respect to the mean fields.
 Unless otherwise specified, we set the fields
 to zero after differentiation.
That is, we compute the {equilibrium} 1P-I Green's functions.

For instance, the gluon 1P-I 2-point function
(which coincides with the gluon inverse propagator) is obtained 
as
\beq
(D^{-1})_{\mu\nu}^{ab}(x,y)\,=\,\frac{\delta^2 \Gamma}{\delta A^\mu_a(x)
\delta A^\nu_b(y)}\,=\,(D_0^{-1})_{\mu\nu}^{ab}(x,y)+
\Pi_{\mu\nu}^{ab}(x,y),\eeq
where:
\beq\label{S0D0}
(D_0^{-1})_{\mu\nu}^{ab}(x,y)\,=\,\delta^{ab}\left (-g_{\mu\nu}\del^2+
(1-\lambda^{-1})\del_\mu\del_\nu\right )\delta_C (x-y)\eeq
is the  corresponding free propagator written here in a covariant
gauge with gauge fixing term $(\del\cdot A^a)^2/2\lambda$, and
\beq\label{PIdef}
\Pi_{\mu\nu}^{ab}(x,y)=\frac{\delta j_\mu^{ind\,a}(x)}{\delta 
A_b^\nu(y)}\eeq
is the gluon polarization tensor.  For fermions we
write similarly:
\beq
S^{-1}(x,y)\,=\,\frac{\delta^2\Gamma}{\delta\Psi(y) 
\delta \bar\Psi(x)}\,=\,S_0^{-1}+\Sigma,\eeq
with the free propagator $S_0^{-1}(x,y)=-i{\slashchar {\del}}_x\,\delta(x-y)$
and the self-energy 
\beq\label{SPI} 
\Sigma(x,y)=\frac{\delta\eta^{ind}(x)}{\delta\Psi(y)}\,.\eeq
More differentiations yield the irreducible (or proper) vertices.
For instance, the quark-gluon vertex is
\beq
\frac{\delta^3 \Gamma}{\delta\Psi(z)\delta\bar\Psi(y)\delta A_\mu^a(x)}=
g\gamma^\mu t^a\delta_C(x-y)\delta_C(y-z)+g\Gamma_a^\mu(x,y,z),\eeq
whose induced part is obtained either from the induced color
current $j^{ind}_\mu$, or from the fermionic source $\eta^{ind}$, according to
\beq\label{Vqg}
g\Gamma^a_\mu(x,y,z)=\frac{\delta^2 j^{ind}_{\mu\,a}
(x)}{\delta\Psi(z)\delta\bar\Psi(y)}
=\frac{\delta^2\eta^{ind}(y)}{\delta
A_\mu^a (x)\delta\Psi(z)}\,.\eeq
Similarly, the proper three-gluon vertex is obtained as
\beq\label{V3g}
g\Gamma_{\mu\nu\rho}^{abc}(x,y,z)=\frac{\delta^2 j_\mu^{ind\,a}(x)}
{\delta A_b^\nu(y)\delta A_c^\rho (z)}\,.\eeq

The induced piece $\Gamma_{ind}$ of the effective action
depends in general on the specific form of the gauge fixing term $G^a[A]$
in the generating functional (\ref{ZFP}), and, within a given
class of gauges (i.e. for a given $G^a[A]$), on the 
gauge parameter $\lambda$. Moreover,
as a functional of the classical fields $A^\mu_a$, $\Psi$
and $\bar\Psi$,  $\Gamma_{ind}$ is generally not invariant under 
the gauge transformations of its arguments.

However, in the HTL approximation,
the induced sources are both
independent of the {quantum} gauge fixing,
and also covariant under the gauge transformations of the {classical}
fields. Besides, the induced current is covariantly
conserved, $[D^\mu, j^{ind}_\mu]=0$, cf.
eqs.~(\ref{cjA}) and (\ref{cjp}). 
Together, these conditions guarantee
that the HTL effective action (to be denoted as $\Gamma_{HTL}$)
is both gauge-fixing independent, and invariant
under the gauge transformations of its field arguments.
The latter property can be easily verified as follows: Under
the infinitesimal gauge transformation (we omit the fermionic fields,
for simplicity)
\beq
A_\nu\rightarrow A_\nu + \delta A_\nu,\qquad 
\delta A_\nu\,=\,-\,\frac{1}{g}\,[D_\nu,\theta], \eeq
the induced action changes as (cf. eq.~(\ref{Sind}))
\beq
\delta \Gamma_{ind} = \int_C {\rm d}^4x \,j^{ind}_{\nu\,a} \delta A_a^{\nu}\,
=\,-\int_C {\rm d}^4x \,[D^\nu,\,j^{ind}_{\nu}]_a,\eeq
where the second equality follows after an integration by parts
(the surface term has been assumed to vanish). Clearly, the
gauge invariance of $\Gamma_{ind}$ requires that $j^{ind}_{\nu}$
is covariantly conserved, a condition satisfied indeed
at the HTL level, i.e. for $\Gamma_{ind}=\Gamma_{HTL}$.

\subsection{The HTL effective action}

Since the induced sources are known explicitly in terms of the
classical fields $A^\mu_{a}$, $\Psi$ and  $\bar\Psi$,
it is tempting to try and use eq.~(\ref{Sind}) to 
construct also the effective action in explicit form. 
Note, however, that the induced sources are known
only for {\it real-time} arguments,
and for retarded (or advanced) boundary conditions
(cf. Sect. 4.1).
Thus, the contour action cannot be derived by simply
integrating eqs.~(\ref{Sind}) with the induced sources
in Sect. 4.1. 
Still, if we temporarily ignore the boundary
conditions, it is possible to write down a functional which
generates these induced sources,
and summarizes in a compact form many of the 
remarkable features of the HTL's.

Consider QED first, and let us construct the effective action which
generates the electromagnetic induced current $j^\mu_{ind}=
\Pi^{\mu\nu} A_\nu$. Since this is linear in $A^\mu$,
the first eq.~(\ref{Sind}) can be trivially integrated
to give (with $\Gamma_{A}$ denoting the purely photonic piece
of $\Gamma_{HTL}$) :
 \beq\label{SAP}
\Gamma_{A}&=&\frac{1}{2}
\int\frac{{\rm d}^4p}{(2\pi)^4}\,A^\mu(-p)
\Pi_{\mu\nu}(p) A^\nu(p)
\nonumber\\ &=&\frac{1}{4}\,m_D^2
\int\frac {{\rm d}\Omega}{4\pi}
\int\frac{{\rm d}^4p}{(2\pi)^4}\,
F_{\lambda\mu}(-p) \,\frac{v^\mu v_\nu}{(v\cdot p)^2}\,
F^{\nu\lambda}(p)\,,\eeq 
where the second line follows by using 
eq.~(\ref{polarisation2}) for $\Pi^{\mu\nu}$ and some simple
algebraic manipulations.
The expression (\ref{SAP}) is well defined only for fields $F^{\nu\lambda}$
which carry time-like momenta, $|\omega| > p$, for which the
denominator $(v\cdot p)^2$ is non-vanishing. 
As discussed in Sect. 4.3.3, these
are the fields which propagate without dissipation.
The polarization tensor obtained by differentiating
$\Gamma_{ind}$ twice (cf. eq.~(\ref{PIdef}))
comes out necessarily symmetric:
\beq
\Pi^{\mu\nu}(x-y)\,=\,\frac{\delta^2 \Gamma_{ind}}{\delta A_\mu(x) 
\delta A_\nu(y)}\,=\,\Pi^{\nu\mu}(y-x)
\,,\eeq
or equivalently: $\Pi^{\mu\nu}(p)=\Pi^{\nu\mu}(-p)$.
This symmetry property is satisfied by the contour self-energy
$\Pi_{\mu\nu}^{\,C}(x,y)$, but it is inconsistent with the retarded
prescription (one rather has $\Pi_R^{\nu\mu}(y-x)=\Pi_A^{\mu\nu}(x-y)$).
However, under the conditions for which $\Gamma_A$ is well defined,
the boundary conditions play no role.

Eq.~(\ref{SAP}) admits a
straightforward generalization to QCD.
Specifically, the induced color current
in eq.~(\ref{j1A}) can be formally rewritten as:
\beq\label{NONDIS}
j_\mu^{A\,a}(x)\,=\,m_D^2\int\frac{{\rm d}\Omega}{4\pi}
\int {\rm d}^4 y \,
\langle x, a|\frac {v_\mu}{v\cdot D}|y, b\rangle
\,{\bf v \cdot E}^b(y)\,.\eeq
This can be generated, via eq.~(\ref{Sind}),
by the following ``action'' (see Ref. \cite{qcd} for an
explicit proof) :
\beq\label{Sb}
\qquad \Gamma_A&\equiv&\frac{1}{2}\,m_D^2\int \frac{{\rm d}
\Omega}{4\pi}\int {\rm d}^4 x
\int {\rm d}^4 y \,\,{\rm Tr} \left [ F_{\lambda\mu}(x)
\langle x|\frac {v^\mu v_\nu}{-(v\cdot D)^2}|y\rangle
 F^{\nu\lambda}(y)\right ].\eeq
Formally, this functional is obtained
from the Abelian action (\ref{SAP}) by simply replacing
the ordinary derivative $(v\cdot \del)^2$ with the
covariant one $(v\cdot D)^2$ \cite{BP92}.

For time-independent fields $A^\mu_a({\bf x})$,
eq.~(\ref{Sb}) reduces to a (screening) mass term for
the electrostatic potentials (cf. eq.~(\ref{jstat})):
\beq\label{Gstatic}
\Gamma_A^{static}\,=\,\frac{1}{2}\,m^2_{D}\, A_0^a({\bf x})
A_0^a({\bf x})\,.\eeq
Within the imaginary time formalism, this provides an
effective three-dimensional action for soft ($k\sim gT$) 
and static ($\omega_n=0$) Matsubara modes:
\beq\label{SEFFST}
\Gamma_{static}\,=\,\beta\int {\rm d}^3x\,\left\{
\frac{1}{4}\, F_a^{ij}F_{ij}^a\,+\,\frac{1}{2}\,(D^iA_0^a)^2
\,+\,\frac{1}{2}\,m^2_{el}\,A_0^aA_0^a\right\},\eeq
This coincides, as expected, with the leading-order result
of the dimensional reduction (cf. Sect. 2.1.4) in QCD
(see Refs. \cite{Kajantie94,Braaten94}, and references therein).

Let us finally add the fermionic fields. 
The HTL effective action is written
as $\Gamma_{HTL}=\Gamma_A +  \Gamma_{\psi}$, with 
$\Gamma_{\psi}$ satisfying:
\beq\label {delSf}
\delta \Gamma_{\psi}/\delta \bar\Psi(x)=\eta^{ind}(x),\qquad\qquad
\delta \Gamma_{\psi}/{\delta A^\mu_{a}(x)}=j_\mu^{\psi\, a}(x).\eeq
After rewriting $\eta^{ind}$ in eq.~(\ref{eindR}) as follows:
\beq\label{NONETA}
\eta^{ind}(x)&=& \omega_0^2\int\frac{{\rm d}\Omega}{4\pi}
\int {\rm d}^4 y\,\langle x|\frac
{\slashchar{v}} {i(v\cdot D)}|y\rangle \Psi(y)\,,\eeq
 it becomes clear that
the first equation (\ref{delSf}) is satisfied by
\beq\label{Sf}
\Gamma_\psi&=& \omega_0^2\int\frac{{\rm d}\Omega}{4\pi}\int
{\rm d}^4 x\int {\rm d}^4 y\,
\bar\Psi(x)\langle x|\frac
{\slashchar{v}} {i(v\cdot D)}|y\rangle \,\Psi(y)\,.
\eeq 
It is then simply to verify that the above $\Gamma_{\psi}$
provides also the correct current $j_\mu^{\psi}(x)$ of eq.~(\ref{jfR}).

The above construction of the HTL effective action from kinetic
theory follows closely Refs.  \cite{qcd}. Originally,
this action has been derived by Taylor and Wong \cite{TW90} 
(although in a form different from eq.~(\ref{Sb})),
by exploiting the properties of the hard thermal loops,
in particular, their gauge symmetry (cf. Sect. 5.3. below).
The manifestly gauge invariant action in eq.~(\ref{Sb})
has been first presented in Refs. \cite{BP92,FT92}. (See also
\cite{Zahed}.)

\subsection{Hard thermal loops}

By differentiating the expressions for the
induced sources obtained in Sect. 4.1, it is straightforward
to construct the 1P-I amplitudes of the soft fields. 
Given the boundary conditions that we have
chosen in solving the kinetic equations, this procedure
naturally generates the corresponding retarded amplitudes.

\subsubsection{Amplitudes with one pair of external fermion lines}

From eq.~(\ref{eindR}), the soft quark self-energy  in 
a background gauge field $A_\mu$ is obtained as
\beq\label{SigmaA}
\Sigma(x,y)=\frac{\delta\eta^{ind}(x)}{\delta\Psi(y)}
=-\omega_0^2\int\frac{{\rm d}\Omega}{4\pi}
\,{\slashchar{v}}\,G_{R}(x,y;v).\eeq
For $ A_\mu=0$, this reduces to:
\beq\label{Sigma}
\Sigma(p) =\omega_0^2\int\frac{{\rm d}\Omega}{4\pi}\, \frac{\slashchar{v}}
{v\cdot p+i\eta},\eeq
where the small imaginary part $i\eta$ implements the 
retarded conditions. 
The angular integral in eq.~(\ref{Sigma}) is performed
in   Appendix B. 

Since $\eta^{ind}$ is  linear in $\Psi$, there is no polarization
 amplitude with more than one pair of soft  external fermions.
On the other hand, eq.~(\ref{eindR}) is non-linear 
in the gauge mean fields (through the parallel transporter),
and it generates an infinite series of
vertex functions between a quark pair and 
any number of soft gluons (or photons).  
To be specific, we define the correction to the amplitude 
between a quark pair and $n$ soft gluons by
\beq\label{2fNg}
g^n \Gamma^{a_1...a_n}_{\mu_1...\mu_n}(x_1,...,x_n;y_1,y_2)&=&
\frac {\,\,\,\,\delta^n}{\delta A_{a_n}^{\mu_n}(x_n)...\delta A_{a_1}
^{\mu_1}(x_1)}\,\Sigma(y_1,y_2),\eeq
with $\Sigma(y_1,y_2)$  given by eq.~(\ref{SigmaA}).
In doing these differentiations, we use the formula
\beq\label{derG}
\frac{\delta\,G_{R}(x_1,x_2;v)}{\delta A_{a}^\mu(y)}\,=\,-g\,v_\mu\,
G_{R}(x_1,y;v)\,t^a\,G_{R}(y,x_2;v),\eeq
 which follows from eqs.~(\ref{GR}) and (\ref{U}).
The normalization we choose for the amplitudes (\ref{2fNg}) is such that 
$\Gamma^{(n)}$  depends on $g$ only through $\omega_0^2$.
In all the amplitudes  (\ref{2fNg}),
 $y_1^0$ is the largest time, while $y_2^0$ is
the smallest one. The relative chronological ordering of the $n$ gluon
lines is arbitrary, and, in fact, the amplitudes are totally symmetric
under their permutations. Up to minor changes due to the color algebra, all
the amplitudes obtained in this way coincide with the corresponding abelian
amplitudes \cite{qed}.

We give now the explicit expressions for the
amplitudes involving  a quark pair and one or two gluons. 
After one differentiation with respect to $A_\mu$, eq.~(\ref{SigmaA})
yields the quark-gluon vertex correction:
\beq
\label{delG1}
\Gamma_{\,\mu}^{a}(x;y_1,y_2)&=&\omega_0^2\gamma^\nu
\int\frac{{\rm d}\Omega}{4\pi}\,v_\mu v_\nu\,
G_{R}(y_1,x;v)\,t^a\,G_{R}(x,y_2;v).\eeq
The time  arguments above satisfy $y_1^0\ge x^0\ge y_2^0$.
For $A=0$, we define the Fourier transform of  $\Gamma^a_\mu$ by
\beq\label{GMF}
\lefteqn{(2\pi)^4\,\delta^{(4)}(p+k_1+k_2)\,
\Gamma^a_{\mu}(p;k_1,k_2)\equiv
\qquad\qquad}\nonumber\\
& &\int {\rm d}^4x \,{\rm d}^4y_1\, {\rm d}^4y_2
 \,\exp\Bigl\{i(p\cdot x+ k_1\cdot y_1+ k_2\cdot y_2)\Bigr\}
\Gamma^{a}_{\mu}(x;y_1,y_2),\eeq
and we get
\beq\label{4gamma}
\Gamma_{\mu}^a(p;k_1,k_2)\,=\,-t^a
\omega_0^2\gamma^\nu \int\frac{{\rm d}\Omega}{4\pi}\frac{v_\mu\,
v_\nu}{(v\cdot k_1+i\eta)(v\cdot k_2-i\eta)}\equiv t^a\Gamma_\mu
(p;k_1,k_2).\eeq
Since all the external  momenta are  of the order $gT$,
 $g\Gamma_\mu\,\sim\,g\omega_0^2/k^2\,\sim\,g$ is
 of the same order as the bare vertex $g\gamma_\mu$.
Thus, the complete quark-gluon vertex at leading order in $g$ is
$g\,t^a\,{}^*\! \Gamma_\mu$, where ${}^*\! \Gamma_\mu\equiv \gamma_\mu + 
 \Gamma_\mu$.

Consider now the vertex between a quark pair and two gluons.
This vertex does not exist at tree level, and in leading order
it arises entirely from the hard thermal loop. We have: 
\beq\label{2f2gG}
\Gamma_{\mu\nu}^{ab}(p_1,p_2;k_1,k_2)&=&
-\omega_0^2\gamma^\rho \int\frac{{\rm d}\Omega}{4\pi}\frac{v_\mu
\,v_\nu\,v_\rho }{(v\cdot k_1+i\eta)(v\cdot k_2-i\eta)}\nonumber\\
&&\left\{ \frac {t^at^b}{v\cdot (k_1+p_1)+i\eta}+
 \frac {t^bt^a}{v\cdot (k_1+p_2)+i\eta}\right\}.\eeq

Alternatively, we can derive the amplitudes (\ref{2fNg}) 
from the expression (\ref{jfR}) for the induced current  
$j_\mu^\psi$. The resulting amplitudes will obey different
boundary conditions since the time
argument of $j^\psi_\mu(x)$ is now the largest one.
It is convenient to rewrite eq.~(\ref{jfR}) as
\beq\label{jfRA}
j_\mu^{\psi}(x)&=&g t^a\,\omega_0^2 \int\frac{{\rm d}\Omega}{4\pi}\,v_\mu
\int {\rm d}^4y_1 {\rm d}^4y_2\nonumber\\&&
 \bar\Psi(y_1) \,{\slashchar{v}}\,G_{A}(y_1,x;v)\,t^a\,
G_{R}(x,y_2;v)\,\Psi(y_2),\eeq
where  the definitions (\ref{GR}) and (\ref{GA}) have been used.
Then, the correction to the quark-gluon vertex is (recall the
first equality in eq.~(\ref{Vqg}))
\beq\label{GRA}
\Gamma_{\,\mu}^{a}(x;y_1,y_2)&=&\omega_0^2\gamma^\nu
\int\frac{{\rm d}\Omega}{4\pi}\,
v_\mu v_\nu\,G_{A}(y_1,x;v)\,t^a\,G_{R}(x,y_2;v),\eeq
where now the time arguments satisfy $x^0\ge max(y_1^0,y_2^0)$, the
chronological order of $y_1$ and $y_2$ being arbitrary (compare, in this 
respect, with eq.~(\ref{delG1}) above).
 For $A=0$, we have
\beq\label{4gamma2}
\Gamma_{\mu}^a(p;k_1,k_2)&=&-t^a
\omega_0^2\gamma^\nu \int\frac{{\rm d}\Omega}{4\pi}\frac{v_\mu
\,v_\nu}{(v\cdot k_1-i\eta)(v\cdot k_2-i\eta)},\eeq
which differs from (\ref{4gamma}) solely by the $i\eta$'s in the denominators
reflecting the respective boundary conditions. If we further differentiate 
 eq.~(\ref{GRA}) with respect to $A_\mu$,
we generate amplitudes of the type (\ref{2fNg}), in which $x_1^0$ is  the
largest time.

\subsubsection{Amplitudes with only gluonic external lines}

The amplitudes involving only soft gluons may be derived from the
induced current $j_\mu^A$ given in
eqs.~(\ref{jAR}) or (\ref{jA2}). A first
differentiation in eq.~(\ref{jA2}) yields (cf. (\ref{PIdef}))
\beq\label{PIHTL}
\label{Pi}\Pi^{ab}_{\mu\nu}(p)=
m_D^2\,\delta^{ab}
\left \{-\delta^0_\mu\delta^0_\nu \,+\,p^0 \int\frac{{\rm d}\Omega}{4\pi}
\frac{v_\mu\, v_\nu} {v\cdot p+i\eta}\right\}.
\eeq
This coincides with 
the electromagnetic polarization
tensor (\ref{polarisation2}) derived in Sect. 1.3.

A second differentiation 
of  eq.~(\ref{jA2}) with respect to $A$ yields
the three-gluon vertex (recall eq.~(\ref{V3g})). 
With the Fourier transform defined as in
eq.~(\ref{GMF}),  we obtain
\beq \label{htl3}
\Gamma^{abc}_{\mu\nu\rho}(p_1,p_2,p_3)
=i f^{abc}\,m_D^2\int
\frac{{\rm d}\Omega}{4\pi}\frac {v_\mu v_\nu v_\rho}{v\cdot 
p_1 +i\eta}
\left \{\frac{p_3^0}{v\cdot p_3-i\eta}\,-\,\frac{p_2^0}{v\cdot  p_2-i\eta}
\right \},\eeq
where  the imaginary parts in the denominators correspond to the  time
orderings $x_1^0\ge x_2^0\ge x_3^0$ for the first term inside the parentheses,
and $x_1^0\ge x_3^0\ge x_2^0$ for the second term.
We can rewrite this more symmetrically as $\Gamma^{abc}_{\mu\nu\rho}
\equiv if^{abc}\Gamma_{\mu\nu\rho}$ with
\beq \label{htl3g}
\lefteqn{
\Gamma_{\mu\nu\rho}(p_1,p_2,p_3)
\,=\,\frac{m_D^2}{3}\int
\frac{{\rm d}\Omega}{4\pi}\,v_\mu v_\nu v_\rho
\,\Biggl\{\frac{p_1^0-p_2^0}{(v\cdot p_1+i\eta)(v\cdot
 p_2-i\eta)} }\nonumber\\
& &\mbox{}+\frac{p_2^0-p_3^0}{(v\cdot p_2-i\eta)(v\cdot
 p_3-i\eta)}+\frac{p_3^0-p_1^0}{(v\cdot p_3-i\eta) (v\cdot
 p_1+i\eta)}\Biggr\}.\eeq
This vanishes for zero external frequencies (static external gluons),
in agreement with eq.~(\ref{Gstatic}).
For $p_i\sim gT$, $g\Gamma_{\mu\nu\rho}\,\sim\,g^2T\,\sim\, gp_i$
is of the same order as the corresponding tree-level vertex.

\subsubsection{Properties of the HTL's}

Originally, the ``hard thermal loops'' have been identified in one-loop
diagrams in thermal equilibrium. The self-energy corrections
(\ref{Sigma}) and  (\ref{Pi}) have been obtained first
by Klimov \cite{Klimov81} and by Weldon \cite{Weldon82a,Weldon82b}. 
These early
works have been put in a new perspective by Braaten and Pisarski
\cite{Pisarski89,BP90,BP90c}, and by Frenkel, Taylor and Wong
\cite{FT90,TW90}, who recognized that, in hot gauge theories,
both the propagators and the vertex functions
receive thermal contributions of order $T^2$
in the limit of high temperature and soft external momenta.

It is worth emphasizing that a HTL is just {\it a part}
 of the corresponding one-loop correction,
namely that part which arises by integration over
{\it hard} loop momenta (this is the origin of the
name ``hard thermal loop''), 
and  after performing  kinematical approximations allowed by the
smallness of the external momenta  
$p_i\simle gT$ with respect to the  hard loop
momentum $k\sim T$. In fact, if we consider 
that all the $p_i$'s are precisely of the
order $gT$, then the HTL is the leading order piece in the expansion
in powers of $g$ which takes into account the assumed
$g$-dependence of the external four-momenta \cite{BP90}.
Alternatively, since all the {\HTL} are proportional to $T^2$,
they can be also obtained as the leading order terms in the high-temperature
expansion  of the one-loop amplitudes
 \cite{Klimov81,Weldon82a,Weldon82b,FT90}.

When compared to the general one-loop corrections
in vacuum, the {\HTL} have remarkably
simple features which we examine now.

\noindent (i) {\it Independence with respect to gauge fixing.}
The {\HTL} are gauge-fixing
independent for arbitrary values of their external momenta. 
For the self-energies, this has been noticed already in
Refs. \cite{Klimov81,Weldon82a,Weldon82b}.
For higher-point vertex functions, it has been verified
either by explicit calculations in various gauges \cite{FT90,BP90}, 
or by induction \cite{BP90}. An explicit proof
for all the HTL's has been given within kinetic theory, in Ref.
\cite{qcd}. As emphasized in Sect. 3, the
gauge-fixing independence reflects the fact that
only the physical, on-shell, excitations of the ideal
quark-gluon plasma contribute to the collective motions
to the order of interest.

\noindent  (ii){\it Ward identities.} 
The {\HTL} are connected by simple Ward identities, similar to those
satisfied by the {\it tree-level} propagators and vertices, 
or by the QED amplitudes.
The first such identities can be easily read off 
the equations in the previous subsection. For instance,
eqs.~(\ref{Sigma}), (\ref{4gamma}) and (\ref{2f2gG}) imply
\beq\label{IW1}
p^\mu\Gamma_\mu(p;k_1,k_2)&=&\Sigma(k_1)- \Sigma(k_1+p)\nonumber\\
p_1^\mu\,\Gamma_{\mu\nu}^{ab}(p_1,p_2;k_1,k_2)&=& if^{abc}
\Gamma_\nu^c(p_1+p_2;k_1,k_2) \nonumber\\
&+&\Gamma^b_\nu(p_2;k_1,k_2+p_1)t^a-
t^a\Gamma^b_\nu(p_2;k_1+p_1,k_2),\eeq
while eqs.~(\ref{Pi}) and (\ref{htl3g}) yield:
\beq\label{Wtr}
p^\mu\,\Pi_{\mu\nu}(p)&=&0,\nonumber\\
p_1^\mu\Gamma_{\mu\nu\rho}(p_1,p_2,p_3)&=&\Pi_{\nu\rho}
(p_3)-\Pi_{\nu\rho}(p_2)\,.\eeq
These identities follow directly from the conservation laws 
for the induced colour current, and ultimately express the 
fact that the HTL effective action is invariant under the
gauge transformations of its field arguments.
For instance, by successively differentiating  eq.~(\ref{cjA})
for $j^A_\mu$ 
with respect to $A_\mu$ one obtains Ward identities relating
HTL's with gluonic external lines, like those in
eq.~(\ref{Wtr}). Similarly, identities like those in eq.~(\ref{IW1})
can be obtained by  differentiating  eq.~(\ref{cjp}) for $j^\psi_\mu$.

\noindent  (iii) {\it Non local structure.}
The specific non-locality of the HTL's, in $1/(v\cdot p)$
(where $v^\mu$ is the velocity of the hard particle
around the loop, and $p^\mu$ a linear combination of the
external momenta) finds its origin in the
(covariant) drift term in the kinetic equations, and
reflects the eikonal propagation
of the hard particles in the soft background fields.
In particular, the limit
$v\cdot p \to 0$ may lead to singularities in the HTL's.
We distinguish two types of such singularities: {\it a\/})
infrared divergences when the external momenta tend to zero
(for instance, note the singular behaviour of the
transverse gluon self-energy $\Pi_T(\omega,p)$
in eq.~(\ref{IMPIT}) as $\omega,p\to 0$ with $\omega \ll p$);
{\it b\/}) collinear divergences for light-like ($P^2\equiv \omega^2-
p^2=0$) external momenta, in which case the angular integration
over ${\bf v}$ (like, e.g., in eqs.~(\ref{Sigma}), (\ref{PIHTL}) or
(\ref{htl3g})) leads to
logarithmic singularities (note the logarithmic branching
points at $\omega=\pm p$ in the gluon self-energies 
in eqs.~(\ref{plpt})--(\ref{Q}),
or in the fermion self-energies in eqs.~(\ref{Sigma1})--(\ref{S-ab})).

\begin{figure}
\protect \epsfxsize=13.cm{\centerline{\epsfbox{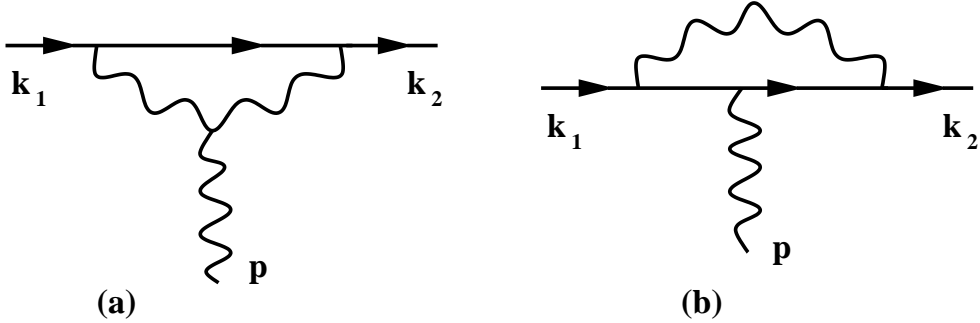}}}
         \caption{One-loop corrections to the quark-gluon vertex,
with the external gluon attached to the internal gluon line
(a), respectively to the internal fermion line (b).}
\label{1loop3G}
\end{figure}
\noindent (iv) {\it Cancellations. } In the diagrammatic
calculation  of the HTL's, one has observed some
``accidental'' compensations with interesting consequences:

\noindent
--- For instance, in QED, the only photon HTL is the polarization
tensor: while HTL-like contributions show up also in individual 
diagrams with more external photons ($n\ge 3$), these contributions 
appear to cancel each other when all the diagrams contributing to a
given vertex function in the HTL approximation
are added together \cite{BP92}. 

\noindent
--- Also, in QCD, interesting cancellations
occur when computing HTL's with quark and gluon external
lines \cite{TW90}. The simplest example is provided by the
quark-gluon vertex $\Gamma^\mu_a$. To one loop order,
the vertex correction is obtained from the two diagrams
in fig.~\ref{1loop3G}. Both these diagrams contain {\HTL}, namely,
\beq\label{G3a}
\Gamma_{(a)\,\mu}^a(p;k_1,k_2)\,=\,t^a\,
\frac{g^2 T^2}{8}\,\frac{N}{2}\,
 \int\frac{d\Omega}{4\pi}\frac{v_\mu\,{\slashchar v}}
{(v\cdot k_1)(v\cdot k_2)}\,, \nonumber\\
\Gamma_{(b)\,\mu}^a(p;k_1,k_2)\,=\,-t^a\,
\frac{g^2 T^2}{8}\,\left(C_{\rm f}+\frac{N}{2}\right)\,
 \int\frac{d\Omega}{4\pi}\frac{v_\mu\,{\slashchar v}}
{(v\cdot k_1)(v\cdot k_2)}\,.\eeq
 When combining the two contributions, the terms
proportional to $N/2$ cancel, so that we are left only with the term
proportional to $C_{\rm f}$, which originates from diagram ~\ref{1loop3G}.b.
Similar cancellations occur also  for the diagrams with more 
external gluons. As a consequence, so that
the corresponding {\HTL} are all
 proportional to $C_{\rm f}$ \cite{TW90}
(as one can verify on eqs.~(\ref{4gamma}) or (\ref{2f2gG})).
If, at a first glance, these cancellations may seem accidental,
note however that they are essential to fulfil the Ward 
identities (\ref{IW1}). 

\noindent All these compensations find a clear interpretation
at the level of the kinetic equations. They reflect the fact
that the only non-linear effects which persist
in the present approximations are those 
required by gauge symmetry:

\noindent --- In QED,
the electromagnetic current $j^A_\mu$ is
linear in the gauge fields, and also gauge-invariant (cf. Sect. 3.1); 
thus, there is no room for purely multi-photon HTL vertices.

\noindent --- In QCD, the cancellation of HTL-like contributions 
associated with soft gluon insertions in diagrams with fermionic 
legs corresponds to the disappearance of the adjoint background
field $k\cdot \tilde A(X)$ in going from eq.~(\ref{KXk}) for 
${\cal K}^a_i(k,X)$ to eq.~(\ref{kinK}) for ${\cal {\slashchar K}}(k,X)$,
and also to the compensation of the terms proportional to
$N$ between the two components (fermionic and gluonic) of the
induced current $j_\mu^\psi$ (cf. eqs.~(\ref{kJpsi})--(\ref{Jpsi1})).
Such compensations are imposed by gauge symmetry;
they ensure, e.g., that
${\cal K}^a_i(k,X)$ transforms as a fundamental colour vector
(i.e., like $\Psi(X)$) under a gauge rotation of the background fields. 

\noindent (v) {\it Non-perturbative character.} 
For external momenta of order $gT$, the {\HTL}
are of the same order in $g$ as the corresponding
tree-level amplitudes, whenever the latter exist.
That is, the effects induced by the collective motion 
at the scale $gT$ are { leading order} effects,
and not just perturbative corrections.
This observation is the basis of the resummation
programme proposed by Braaten and Pisarski \cite{Pisarski89,BP90},
to be discussed in the next subsection.

\subsection{HTL and beyond}

In high-temperature gauge theories, the na\"{\i}ve
perturbation theory breaks 
down at the soft scale $gT$, because of the large collective effects.
This  crucial observation, 
due to Pisarski \cite{Pisarski89b,Pisarski89} 
led subsequently Braaten and Pisarski \cite{Pisarski89,BP90,BP90c} to propose
a reorganization of the perturbative expansion where the hard thermal loops
are included at the tree level.

In some respects, the resummation of hard thermal loops can be
seen as a generalization of the resummation of ring 
diagrams in the computation of the correlation energy for a high-density
electron gas, by Gell-Mann and Brueckner \cite{GMB}. 
Many other examples can be found in the literature,
both in non-relativistic many-body physics 
 \cite{Abrikosov63,FW71,BR86,NO}, and in relativistic plasmas
\cite{Akhiezer60}--\cite{Carrington92}.

\subsubsection{The Braaten-Pisarski resummation scheme}

At a formal level, the resummed theory is defined by the effective action
$\Gamma_{eff}=S_{cl}+\Gamma_{HTL}$ where
$S_{cl}$ is the classical action for QCD, eq.~(\ref{QCD}), and
$\Gamma_{HTL}$ is the generating functional of HTL's
described in Sects. 5.1 and 5.2 (cf. eqs.~(\ref{Sb}) and (\ref{Sf})).
In practice, this means that the Feynman rules
to be used for the soft fields are defined so as
to include the HTL self-energies and vertices.
On the other hand, the bare Feynman rules are to be applied
for the hard fields \cite{Pisarski89,BP90}. Indeed,
the leading corrections to the self-energy of a hard field are 
${\cal O}(g^2)$,
while the corrections to a vertex in which any leg is hard are, at most,
${\cal O}(g)$ \cite{BP90}; these are truly perturbative corrections,
and do not call for  resummation.
Thus, when computing a Feynman graph, one is instructed to use
the bare (thermal) propagators for all the internal lines which carry hard 
momenta, and the bare vertices for all the interaction vertices which involve,
at least, one pair of hard fields. But for the soft internal lines, 
and the vertices with only soft external legs, one must use
{\it effective} propagators and vertices.

The effective quark and gluon propagators ${}^*\!S$ and ${}^*\!G_{\mu\nu}$
are obtained by inverting 
\beq\label{DS*}
{}^*\!G_{\mu\nu}^{-1}(p)\,=\, G_{0\,\mu\nu}^{-1}(p)\,+\,\Pi_{\mu\nu}(p),
\qquad {}^*\!S^{-1}(p) \,=\, S_0^{-1}(p)\,+\, \Sigma(p)\,,
\eeq
and are given explicitly in Appendix B (cf. Sects. B.1.3 and B.2.3). 

Consider now the effective vertices connecting (soft) quarks and gluons.
The three-particle vertex reads:
\beq\label{Gqg*}
{}^*\!\Gamma_{\mu}^a(p;k_1,k_2)\,=\,t^a \gamma_\mu \,+\,
\Gamma_{\mu}^a(p;k_1,k_2)\,\equiv\, t^a\Gamma_\mu
(p;k_1,k_2)\,,\eeq
with $\Gamma_{\mu}^a$ given by eq.~(\ref{4gamma}). It
satisfies the following Ward identity:
\beq\label{ABWI}
p^\mu\,{}^*\!\Gamma_\mu(p;k_1,k_2)\,=\,{}^*\!S^{-1}(k_1)- {}^*\!S^{-1}(k_1+p),
\eeq
which follows from eqs.~(\ref{IW1}) and (\ref{DS*}).
At tree-level, $\Gamma_{\mu}^{0}\equiv \gamma_\mu$ is the only
quark-gluon vertex. In the effective theory, on the other hand,
we have an infinite series of new vertices, which are related through Ward
identities, and which connect a quark-antiquark pair to any number of gluons.
For instance, the corresponding four-particle (2 quarks-2 gluons) vertex reads
 ${}^*\!\Gamma_{\mu\nu}^{ab}(p_1,p_2;k_1,k_2) =
\Gamma_{\mu\nu}^{ab}(p_1,p_2;k_1,k_2)$ (cf. eq.~(\ref{2f2gG})),
and is related to the three-particle vertex
in  eq.~(\ref{Gqg*}) by the second Ward identity (\ref{IW1}).
A similar discussion applies to the effective vertices with only gluon
legs. 

When performing perturbative calculations, one has to fix the
gauge. This only affects the form of the gluon
propagator (recall that the HTL's are gauge-fixing independent).
Also, in gauges with ghosts, one must use bare Feynman rules for
the ghost propagator and vertices. Indeed, it can be verified
that there are no HTL corrections for the amplitudes with 
ghost external lines \cite{BP90,FT90,qcd}, a property which
reflects the gauge-invariance of the HTL effective action.

Consider now the systematics of the resummed perturbation theory.
Since the HTL's are now included at the tree-level 
of the effective theory, one must be careful to avoid overcounting.
A standard procedure consists in
adding and subtracting the action
$\Gamma_{HTL}$ to the bare action $S_{cl}$, by writing
\beq\label{rearr}
S_{cl}\,\equiv \,(S_{cl}+\Gamma_{HTL})-\Gamma_{HTL}
\,= \,\Gamma_{eff}+\delta S.\eeq
In the effective expansion, the tree-level amplitudes are
generated by  $\Gamma_{eff}\equiv S_{cl}+\Gamma_{HTL}$, while
the reminder $\delta S\equiv 
-\Gamma_{HTL}$ is treated perturbatively as a counterterm
(i.e., a quantity which is formally of one-loop order)
to ensure that the HTL's are not double counted.

\begin{figure}
\protect \epsfxsize=12.5cm{\centerline{\epsfbox{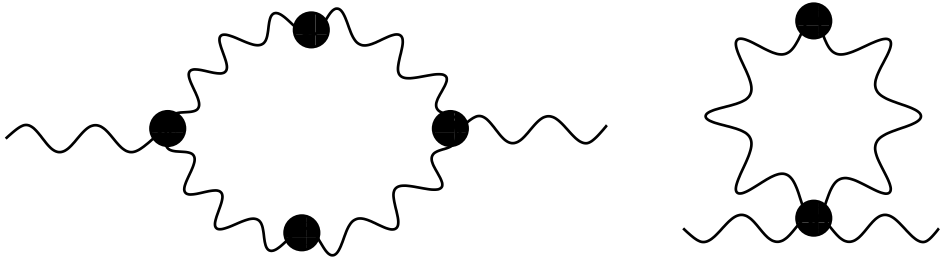}}}
	 \caption{Effective one-loop diagrams contributing to the
soft gluon self-energy to next-to-leading order. All the lines in
these diagrams are soft, so all the propagators and vertices are
defined to include the corresponding HTL's.}
\label{sgluon}
\end{figure}
In practice, this requires a systematic separation of
soft and hard momenta. As an example, consider
the first correction to the soft gluon self-energy
beyond the HTL of eq.~(\ref{Pi}).
By power counting, this is of order $g^3 T^2$, 
and involves three types of contributions \cite{BP90}: 
({a}) one-loop diagrams with soft loop momentum (fig. \ref{sgluon});
({b}) one-loop diagrams with hard internal momentum and
 with the HTL subtracted;  ({c}) two-loop diagrams with
only hard internal momenta. In case  ({a}), all the momenta
(internal and external) are soft, so one has to use effective
propagators and vertices (cf. fig. \ref{sgluon}).
In case  ({b}), the subtraction of the HTL
is ensured by the corresponding counterterm in $-\Gamma_{HTL}$.
This calculation has been done
in Refs. \cite{Schulz94,Schulz95} where
in particular the next-to-leading order correction to the
plasma frequency in QCD has been obtained:
  $\delta\omega_{pl}^2=\eta g{\sqrt N}\omega_{pl}^2$,
where $\omega_{pl}^2=g^2NT^2/9$ is the leading order result
(we consider here a purely Yang-Mills plasma), and the
coefficient $\eta\approx -0.18$ is found to be
gauge-fixing independent, as expected from general arguments \cite{KKR90}.
(See also Ref. \cite{RP92} for a similar calculation in
the scalar theory with quartic self-interactions.)

\subsubsection{Some applications of HTL-resummed perturbation theory}

We shall briefly mention here some other applications
of the HTL-resummed perturbation theory. More  can be found in the textbook by
LeBellac \cite{MLB96}, and also in the review paper by Thoma \cite{ThomaQGP}.

The simplest version of the HTL resummation applies
to the calculation of { static} quantities
(like the thermodynamical functions, or the time-independent
correlations) in the imaginary-time formalism.
Then, only the internal lines with zero Matsubara
frequency ($\omega_n=0$) can be soft, and require resummation
\cite{AE,Arnold94}.
Since in the static limit the HTL's collapse to the Debye mass term
(cf. eq.~(\ref{Gstatic})), the resummation then
reduces to including $m_D^2$ in the electric propagator.
By using this technique, the free energy has been computed up to order 
$g^5$ for massless scalar $\phi^4$ theory
\cite{Saa92,PS}, Abelian gauge theories \cite{CoPa,Pa95},
and QCD \cite{Arnold94,KZ95}. These results have been reobtained
by using the dimensionally-reduced effective theory
in Refs. \cite{BN96,Anders96}.

The whole machinery of the HTL resummation comes into play when
considering {\it dynamical} quantities, like time-dependent correlators.
The most celebrated example in this respect is the calculation of the 
quasiparticle { damping rate} $\gamma$ (see Sect. 6
below). The original attempts to compute $\gamma$ 
have met with various conceptual problems which have been
a major stimulus for several interesting progress in hot gauge theories.
It was this problem, dubbed for some time the ``plasmon puzzle'',
which triggered  the discovery of the hard thermal loops, 
and the study of their remarkable properties.
An historical account of this subject, together with references to
previous work, can be found in Ref. \cite{Pisarski91}.
The HTL-resummed calculation of $\gamma$ for excitations with
zero momentum (gluons or fermions) is presented in 
\cite{BP90,BP90a,KKM,BP92b}.

Quite generally, one expects
the predictions of the effective theory
to be different from those of the bare theory for all the quantities
which are sensitive to soft momenta. In particular, most of
the { logarithmic} infrared divergences of the
bare expansion are eliminated by the resummation of the HTL's.
An important example in this sense, which is also the earliest
one to have shown the role of dynamical screening in removing 
IR divergences, is the calculation of the viscosity in hot QCD,
by Baym {\it et al} \cite{Baym90}. This example is quite
generic for the transport phenomena based on momentum-relaxation processes
(other examples are the charge and quark diffusivities)
\cite{Thoma91a,Heisel92,Heisel94a}. By contrast, the 
transport coefficients for { colour} remain IR sensitive
even after the inclusion of the HTL's \cite{Gyulassy93,Heisel94,Bodeker},
as it will be explained in Sect. 7 below.

Here are some more examples of applications of the HTL 
perturbation theory to the calculation of dynamical quantities:
the calculation of the collisional energy-loss of charged or colored 
partons \cite{Thoma91,BThoma91,Mrowcz91,Koike92}, 
the Primakoff production of axions from a QED 
plasma \cite{Altherr91,BY91,Altherr93a}, and
the photon production by a quark-gluon plasma,
for both hard \cite{Kapusta91,Baier92,Thoma94},
or soft \cite{Baier94,Niegawa97} photons.
In the particular case of the photon production rate,
it is the Landau damping of a soft fermion
which provides infrared finiteness \cite{Kapusta91,Baier92}
(in the bare perturbation theory, there is a logarithmic divergence
associated with the exchange of a massless quark).
Another example where the resummation enters in
a decisive way is the calculation of the production
rate of soft dileptons in a hot quark-gluon plasma \cite{Yuan90}.
Note, however, that collinear divergences
identified in higher orders \cite{Baier94,Gelis98}
raise doubts about the consistency of the original calculations
in  \cite{Yuan90,Kapusta91,Baier92}. In spite of significant recent
work and progress \cite{Baier94,RFlechsig95,Gelis98,AGKZ00,AGZ00},
the complete calculation of the production rates for photons and
dileptons in the quark-gluon plasma remains an open problem.

\subsubsection{Other resummations and lattice calculations}

If the resolution of the ``plasmon puzzle'' was one of the first,
and most remarkable, successes of the HTL perturbation theory,
it is still in relation with the damping rate calculation
that the limits of the HTL resummation first emerged.
Simultaneously with the first successful calculation of the
damping rates for excitations with zero momentum, it was found that
for finite-momentum excitations, infrared divergences
remain even after including the HTL's \cite{Pisarski89,Smilga90},
 \cite{Burgess92}--\cite{Rebhan95}. 
As we shall see later, in Sect. 6, this difficulty is related to the 
fact that the HTL's do not
play any role in the { static} magnetic sector.

The non-perturbative contributions of the magnetic fluctuations,
which occur at ${\cal O}(g^6)$ in the free energy
(cf. Sect. 1.1), occur already at ${\cal O}(g^4)$ in
the static magnetic self-energy $\Pi_T(0,p)$. Various theoretical arguments
 \cite{GPY81,BMuller93,Nair95,BP95,JPi96,KN96} predict that
(static) magnetic screening should be nonperturbatively generated at the
scale $g^2T$, and this is indeed confirmed by lattice calculations
\cite{Karsch97,Karsch98,CKP00}. 
For practical purposes, this may be represented as a simple magnetic mass
$\Pi_T(0,p)\approx m^2_{mag} \sim (g^2T)^2$, although the precise nature
of the screening mechanism is not yet fully understood.
By contrast, in Abelian gauge theories it can be proven that,
to all orders in the coupling constant, 
there is no static magnetic screening \cite{sqed}.

\begin{figure}
\protect \epsfxsize=10.cm{\centerline{\epsfbox{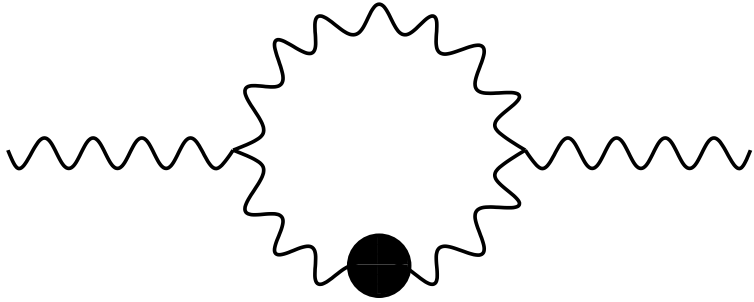}}}
	 \caption{Effective one-loop contribution, of ${\cal O}(g^3)$,
to the electric polarization
function $\Pi_L(0,p)$ in QCD. All the lines in this diagram are static. The
external lines, as well as the internal line marked by a blob,
are electric and dressed by the Debye mass.
The other internal one is magnetic and therefore massless.}
\label{QCDelec}
\end{figure}

Similarly, the next-to-leading order contribution to
the Debye mass in QCD, of ${\cal O}(g^3)$, is logarithmically IR divergent,
but the coefficient in front of the logarithm can be 
computed perturbatively \cite{Rebhan93}, from the one-loop effective diagram
in fig.~\ref{QCDelec}. This yields the positive correction
$\delta m_D^2 \simeq  2\alpha NT m_{D}\ln (1/g)$, where $\alpha=g^2/4\pi$,
$m_D$ is the LO Debye mass, eq.~(\ref{omegap}), and the logarithm has been
generated as $\ln(m_{D}/m_{mag}) \simeq \ln (1/g)$ to logarithmic accuracy
(see also Ref. \cite{BN94}). In fact, as shown in
Ref. \cite{debye}, a similar problem occurs in the Abelian context of
scalar QED, where no magnetic mass is expected.
 There, the IR divergence of perturbation theory
has been cured via an all-order resummation of soft photon effects in
the vicinity of the mass-shell \cite{debye}. 

A systematic framework for the non-perturbative calculation
of the thermodynamical quantities and of the static correlations
is provided by finite-temperature lattice QCD \cite{DeTar}.
By using four-dimensional lattice simulations, the QCD pressure 
has been computed for a pure SU$(3)$ Yang-Mills theory
in Refs. \cite{Boyd,Okamoto}, and, more recently, also for
QCD with two and three light quarks \cite{KLP00}.
The electric and magnetic screening masses have been
similarly computed in Refs. \cite{Karsch95,CKP00}.

Since lattice calculations 
are easier to perform in lower dimensions, their efficiency
can be increased by using dimensional reduction.
The corresponding effective theory for QCD (or
the electroweak theory) is a three-dimensional SU$(N)$ 
gauge theory coupled to a massive adjoint ``Higgs'' (the electrostatic
field $A^0_a$ of the original theory in $D=4$) with all the interactions
permitted by the symmetries in the problem \cite{Kajantie94,Braaten94,BN96}. 
The parameters in this effective theory 
(the mass of the scalar field and the strengths of the various
effective interactions) can be obtained by
matching the soft correlation
functions calculated in the original theory and
the effective theory, to the order of interest
 \cite{Braaten94,BN95b,BN96,Kajantie96}.
To lowest order, this yields the effective theory in eq.~(\ref{SEFFST}).

The combination of dimensional reduction and three-dimensional
lattice calculations has allowed for systematic studies of the
phase transition in the electroweak theory
\cite{Kajantie94,Kajantie96,Kajantie98,Rumm98,Laine98}
and its minimal supersymmetric extension \cite{LR98}, and 
of the static long-range correlations in high-temperature QCD
\cite{Karsch97,Karsch98,Kajantie97,LP98,CKP00}.
In QCD, estimates have been obtained in this way 
for the non-perturbative ${\cal O}(g^6)$-contribution to the free energy
\cite{Karsch97}, for the magnetic screening mass \cite{Karsch98,CKP00}, 
and for the Debye mass \cite{Kajantie97,LP98}.
(See also \cite{AY95} for a non-perturbative definition of
the Debye mass, which has been used for the numerical calculations
in \cite{LP98}.)
Whenever lattice calculations in both $D=3$ (with 
dimensional reduction) and $D=4$ are available, 
the results agree reasonably well (see Refs. \cite{Karsch98,HP99,CKP00}
for explicit comparisons). The Debye mass found on the lattice
is very well fitted by the following formula \cite{LP98} :
\beq
m_D^{lattice}\,=\,m_D\,+\,\alpha NT\,\biggl(\ln\frac{m_D}{g^2T}+7.0\biggr)
\,+\,{\rm O}(g^3T).\eeq
It differs significantly from the lowest order perturbative
prediction (the HTL value $m_D$) up to
temperatures as high as $T\sim 10^7T_c$.
This shows that the Debye mass, as well as other long-range
correlators, receive at most temperatures of interest important
non-perturbative contributions
from the longwavelength fluctuations in the plasma.

Returning to the thermodynamical functions which are dominated
by the hard degrees of freedom, one may expect such
non-perturbative contributions to be quantitatively small. And indeed
the lattice data \cite{Boyd,Okamoto,KLP00}
show a (slow) approach of the ideal-gas result from below with
deviations of not more than some 10-15\% for temperatures a few times 
the deconfinement temperature.
Recent lattice calculations using dimensional reduction 
provide further evidence that the total, non-perturbative,
contribution of the soft modes to the free energy is rather small
 \cite{Laine00}.

This being said, it is worth reminding that na\"{\i}ve
perturbation theory is inadequate to describe the thermodynamics
of the quark-gluon plasma.
Already the next-to-leading order perturbative correction,
the so-called plasmon effect which is of order $g^3$
\cite{Kapusta79},  signals the inadequacy of the conventional
thermal perturbation theory because,
in contrast to the leading-order correction of ${\cal O}(g^2)$, 
it leads to
a free energy in excess of the ideal-gas value. In fact,
for the ${\cal O}(g^3)$ effect to be less important that the
${\cal O}(g^2)$ negative correction, the QCD coupling constant must be 
as low as $\alpha \simle 0.05$, which would correspond 
to temperatures as high as $\simge 10^5 T_c$.

This suggests a further reorganization of perturbation theory
where more information on the plasma quasiparticles is included
already at tree-level, and this
is the place where the HTL come back into the game.
 A possible strategy 
is the so-called ``screened perturbation theory''\cite{KPP97}
where the HTL-resummed Lagrangian $\Gamma_{eff}$ in eq.~(\ref{rearr})
is now used at {all} momenta, soft and hard.
The efficiency of this method in improving
the convergence of perturbation theory has been demonstrated 
in the context of scalar field theories, via calculations up
to two-loop \cite{KPP97,CH98} and three-loop \cite{ABS00} order.
Recently, this scheme has been extended to QCD \cite{ABS99},
where, however, only one-loop calculations have been presented
so far. A problem with this approach is that, at any {finite} 
loop order, the UV structure of the theory is modified: new 
(eventually temperature-dependent) divergences
occur and must be subtracted, thus introducing a new source of
renormalization scheme dependence \cite{ABS00,Rebhan00}.
In gauge theories, this is further complicated by
the non-locality of the HTL's \cite{ABS99}.

An alternative approach has been worked out in Refs. \cite{BIR99,BIR00}
and uses the HTL's only in the kinematical regimes where they are
accurate. This approach is based on a self-consistent 
(``$\Phi$-derivable'' \cite{Baym62})
two-loop approximation to the thermodynamic potential, but 
focuses on the {\it entropy} which has the simple form
(given here for a scalar field) :
\beq\label{entropy}
S\,=\,-\int\frac{{\rm d}^4k}{(2\pi)^4}\,\frac{\partial 
n}{\partial T}\, \left\{{\rm
Im}\ln G^{-1}\,-\,{\rm Im}\Pi\,{\rm Re}G\right\}\,
\eeq
This effectively one-loop expression is correct
up to terms of loop-order 3 (i.e., of ${\cal O}(g^4)$ or higher)
provided $G$ and $\Pi$ are the {\it self-consistent}
one-loop propagator and self-energy \cite{Baym62}. 
Thus, any explicit two-loop interaction
contribution to the entropy has been absorbed into the spectral
properties of quasiparticles.
Remarkably, this holds equally true for fermionic
\cite{VB98} and gluonic \cite{BIR99,BIR00} interactions.
The expression (\ref{entropy}) is manifestly UV finite,  
the statistical factors providing an ultraviolet  cut-off.

Based on the formula (\ref{entropy}), 
approximately self-consistent calculations 
have been proposed \cite{BIR99,BIR00}
where the self-energy
$\Pi$ is determined in HTL-resummed perturbation theory
(with manifestly gauge-invariant results), but the
entropy is evaluated exactly, by numerically integrating
eq.~(\ref{entropy}) with this approximate self-energy.
The results compare very well with the lattice data for all temperatures
above $T\simeq 2.5T_c$ (with $T_c$ the critical temperature for
the deconfinement phase transition). This method has been applied 
successfully also to plasmas with non vanishing baryonic density
(i.e., with a non-zero chemical potential) \cite{BIR99,BIR00},
for which lattice calculations are not yet available.

\setcounter{equation}{0}
\section{The lifetime of the quasiparticles}

Because of their interactions with the particles in the  thermal bath,
all the excitations of a plasma have finite lifetimes.
In the weak coupling regime, one expects these lifetimes to be long.
This was indeed verified in Sect. 2.3.4 for 
a scalar theory with a quartic interaction. However, in gauge theories, the
perturbative calculation of the lifetimes is plagued with infrared
divergences,  both in Abelian and
non-Abelian  plasmas \cite{Pisarski89,Smilga90}
\cite{Burgess92}--\cite{Rebhan95}. Although screening
corrections contribute to cure much of the problem, these are not enough.
In this section we shall identify the physical origin of the problem and
show how it can be solved, at least in the case of QED, by an all order
resummation of soft photon effects.  This resummation  is
reminiscent of the Bloch-Nordsieck calculation at zero temperature
\cite{BN37,Bogoliubov}, and calls upon kinematical
approximations which have been met several times along this
review. The calculation that we shall present is also interesting from the
point of view of kinetic theory, as it provides an example where coherence
effects between successive scatterings  need to be taken into
account, thus preventing a simple description via a Boltzmann equation.

\subsection{The fermion damping rate in the Born approximation}

In this subsection,
we  compute the damping rate of a hard electron
($p\sim T$) in a hot QED plasma, to leading order in $e$.  
 The damping is caused by collisions involving a photon exchange with
the electrons of the heat bath. 
\begin{figure}
\protect\epsfxsize=6.cm{\centerline{\epsfbox{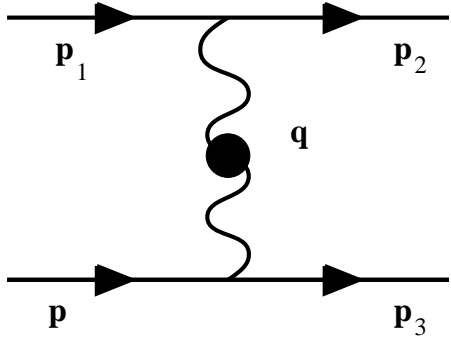}}}
         \caption{Fermion-fermion elastic scattering in the resummed
Born approximation. As usual the blob on the photon propagator represents
the screening correction, in the hard thermal loop approximation.}
\label{Born}
\end{figure}
The relevant Feynman graph is depicted in fig.~\ref{Born}. The 
collision  rate is obtained by integrating the  corresponding matrix
element squared  ${|{\cal M}|^2}$ over 
the thermal phase space for the scattering partners.
At the order of interest, we can treat the (hard) external fermion lines 
as  free  massless  Dirac particles.
On the other hand, since, as we shall verify later, the scattering rate is
dominated by {\it soft} momentum transfers, $q\simle eT$, it is
 essential to include the screening corrections on the 
photon line. We are thus led to the following expression for
the damping rate:
\beq\label{gammaB}
\gamma_p&=&\frac{1}{4\varepsilon}
\int {\rm d}\tilde p_1 \,{\rm d}\tilde p_2\,{\rm d}\tilde p_3\,
(2\pi)^4\delta^{(4)}(p+p_1-p_2-p_3)\nonumber\\
&{}&\,\,\,\,\Bigl\{n_1(1-n_2)(1-n_3)+(1-n_1)n_2 n_3\Bigr\}\,
{|{\cal M}|^2},\eeq
where  the matrix element squared ${|{\cal M}|^2}$ is computed
with the effective photon propagator ${}^*\!G_{\mu\nu}(q)$ given by 
eq.~(\ref{DSTAR}).

The other notations in eq.~(\ref{gammaB}) are as follows:
 all the external particles are on their mass-shell
(i.e., $\varepsilon=p$ and $\varepsilon_i=p_i$ for $i=1,\,2,\,3$),
and $\int {\rm d}\tilde p_i \equiv
 \int {\rm d}^3p_i/((2\pi)^3\,2\varepsilon_i)$.
The statistical factors $n_i\equiv n(\varepsilon_i)$
take care of the Pauli principle for the two 
processes (direct and inverse) associated with the diagram of
fig.~\ref{Born}. Note that, for fermions, the rates of these two processes
 have to be {\it added} together to give the 
depopulation of the  state with momentum $p^\mu$
\cite{KB62}.  Except for this change of sign,  the expression
(\ref{gammaB}) of the damping rate has the same structure as that 
 for a scalar particle obtained from eqs.~(\ref{coll1}) and
(\ref{proc2}).

In the  regime where  $q\simle eT \ll p_i$, 
 the matrix element simplifies to \cite{damping}:
\beq \label{MBORN}
|{\cal M}|^2\simeq 64\, e^4 p^2 p_1^2\,\Big|\,{}^*\!\Delta_L(q_0,q)
+ ({\bf v \times \hat q})\cdot ({\bf v_1 \times \hat q})\,
 {}^*\!\Delta_T(q_0,q)\Big|^{\,2},\eeq
 where ${\bf v \equiv \hat p}$, ${\bf v_1 \equiv \hat p_1}$,
and ${}^*\!\Delta_{L,\,T}(q_0,q)$ are respectively 
the electric ($l$) and the magnetic ($t$) photon propagators in the
HTL approximation, as defined in eq.~(\ref{effd0}).
By using eq.~(\ref{MBORN}), and  performing some of the momentum
integrals in  eq.~(\ref{gammaB}), we can rewrite $\gamma_p$ 
as a double integral over the energy $q_0$ and the magnitude $q=|{\bf q}|$
of the momentum of the virtual photon:
\beq\label{G2L}
\gamma \simeq\, \frac{e^4 T^3}{12}\,
\int_{\mu}^{\infty}{\rm d}q  \int_{-q}^q\frac{{\rm d}q_0}{2\pi}
\left\{ |{}^*\!\Delta_L(q_0,q)|^2\,+\,\frac{1}{2}\left(1-\frac{q_0^2}{q^2}\right)^2
|{}^*\!\Delta_T(q_0,q)|^2
\right\}\,.\eeq
Note that, as a result of our kinematical 
approximations, the damping rate has become 
independent of $p$.
The integration limits on $q_0$, namely $|q_0|\le q$, arise from
the kinematics: the exchanged photon is necessarily space-like.
Finally, the momentum integral is infrared divergent, which
is why we have introduced the lower cutoff
$\mu$. 

If we were to use  a  bare photon propagator in (\ref{G2L}),
i.e. 
$|\Delta_L(q_0,q)|^2= 1/q^4$ and
$|\Delta_T(q_0,q)|^2= 1/(q_0^2-q^2)^2$, one would find that the resulting 
  $q$-integral is quadratically divergent:
\beq\label{G2L0}
\gamma\simeq  \frac{e^4T^3}{8\pi} \,
\int_{\mu}^{\infty}\frac{{\rm d}q}{q^3}\,
\propto \frac{e^4 T^3}{\mu^2}\,.\eeq
This divergence  is softened by screening effects which are different in
the longitudinal (electric) and in the transverse (magnetic) channel. In
the  electric sector, the Debye screening provides a natural IR cutoff, 
namely the electric mass
$ m_D\sim eT$. Accordingly,
the electric contribution to $\gamma$ is finite,
and of the order  $\gamma_L \sim e^4 T^3/m_D^2
\sim e^2 T$. In the magnetic sector,
the  dynamical screening  due to Landau damping 
  is not sufficient to completely
remove the IR singularity in $\gamma_T$. 
A logarithmic divergence remains, which we now analyze.

The leading IR contribution to $\gamma_T$ 
comes from small photon momenta, $ q \simle m_D$, where we can
 use the approximate expression (\ref{Vt}) to write:
\beq\label{G2LR}
\gamma_T \simeq \frac{e^4 T^3}{24}
\int_{\mu}^{\infty}{\rm d}q  \int_{-q}^q\frac{{\rm d}q_0}{2\pi}
\,\frac{1}{q^4 + (\pi m_D^2 q_0/4q)^2}\,.
\eeq
In general, eq.~(\ref{Vt}) holds only for sufficiently low frequencies
$q_0\ll q\,$; but it can  nevertheless be used to study the IR
divergence of $\gamma_T$ since, in the limit of small momenta
$q\ll m_D$, 
$|{}^*\!\Delta_T(q_0,q)|^2$ is strongly peaked at $q_0=0$,
with a width $\Delta q_0\sim q^3/m_D^2 \ll q$.
In fact, when $ q\to 0$,
\beq\label{deltaT} |{}^*\!\Delta_T(q_0,q)|^2\,\simeq\,
\frac{1} {q^4 + (\pi m_D^2 q_0/4q)^2}\longrightarrow \frac{4}{q
m_D^2}\,\delta(q_0)\,,
\eeq
and this limiting behaviour is sufficient to extract the IR-divergent
piece of  eq.~(\ref{G2LR}) which reads:
\beq\label{gammaTF}
\gamma_T\simeq\,
\frac{e^2T}{4\pi} \int_{\mu}^{m_D}\frac{{\rm d}q}{q}\,=\,
\frac{e^2T}{4\pi}\,\ln\frac{m_D}{\mu}.\eeq
 We have introduced the upper
 cutoff $m_D\sim eT$  to
 approximately account for the correct UV behaviour
of the integrand in eq.~(\ref{G2LR}): namely, as $q\gg m_D$,
the integrand is decreasing like $m_D^2/q^3$, so that the 
$q$-integral is indeed cutoff at $q\sim m_D$.
(Incidentally, the final result in eq.~(\ref{gammaTF}) is 
the same as the {\it exact} result for $\gamma=\gamma_T+\gamma_L$
obtained by evaluating
the integrals in eq.~(\ref{G2L}) with a sharp IR momentum
cutoff equal to $\mu$ \cite{TBN}.)

Thus the logarithmic divergence is due to collisions involving the
exchange of very soft, {quasistatic} ($q_0\to 0$), 
magnetic photons,
which are not screened by plasma effects.
This situation is quite generic: in both QED and QCD, the
IR complications which remain after the resummation of the HTL's
are generated by very soft magnetic photons, or gluons, with momenta
$q\ll gT$ and frequencies $q_0\simle q^3/m_D^2 \ll q$ (see also the
discussion at the end of Sect. B.1.4 in the appendix, and
fig.~\ref{rho2} there).

Note also that, if we ignore temporarily this IR problem, both
the electric and the magnetic damping rates are of order $e^2T$,
rather than $e^4T$ as one would naively expect
by looking at the diagram in fig. \ref{Born}. This situation,
 sometimes referred to as {anomalous damping} \cite{Smilga90},
 is a consequence of the strong sensitivity of the cross section
to the IR behavior of the photon propagator.
By comparison,  the other processes contributing to
the damping of the fermion, namely the Compton
scattering and the annihilation process, are less IR singular
because they involve the exchange of
a virtual {fermion}; as a result, these contributions are indeed
of order $e^4 T$. 

Note finally that there is no IR problem in the calculation
of the damping rate at zero temperature and large chemical
potential \cite{LBM96,BO96}. In that case too, the dominant
contribution to $\gamma$ comes from
the exchange of soft magnetic photons (or gluons in QCD).
In the vicinity of the Fermi surface, $\gamma$
is proportional to $|E-\mu|$, where $E$ is the fermion energy and
$\mu$ the chemical potential. 
(The electric photons alone would give a contribution
proportional to $(E-\mu)^2$, a behaviour familiar in
nonrelativistic Fermi liquids \cite{Brown}.)

\subsection{Higher-order corrections}

While the above calculation of the interaction rate in the (resummed) Born
approximation is physically transparent, for the analysis of the
higher order corrections it is more convenient
to obtain $\gamma$ from the imaginary part of the retarded self-energy.
To lowest order, one can write 
\beq\label{damping}
\gamma_p\,=\, -\,\frac{1}{4p}\,{\rm tr}\left(
{\slashchar p}\, {\rm Im}\,{}^*\!\Sigma_R(p_0+i\eta,{\bf p})\right)\Big |_
{p_0=p}\,,\eeq 
with ${}^*\!\Sigma(p)$ given by the (resummed) 
one-loop diagram in fig.~\ref{RESF}.
This diagram is evaluated in Appendix B, where we also
verify that the resulting expression for $\gamma$ 
(eq.~(\ref{damping})) coincides  with the interaction 
rate obtained above, in eq.~(\ref{G2L}). 
 
\begin{figure}
\protect \epsfxsize=9.cm{\centerline{\epsfbox{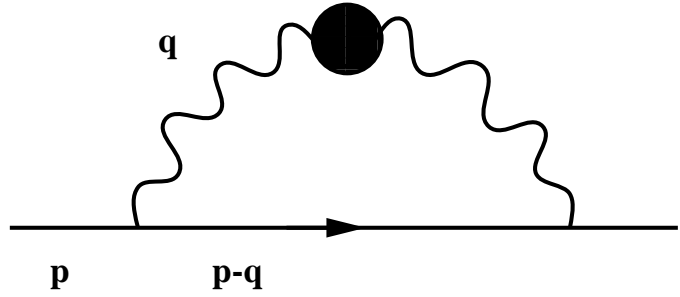}}}
         \caption{The resummed one-loop self-energy of a hard fermion}
\label{RESF} \end{figure}

\begin{figure}
\protect \epsfxsize=16.cm{\centerline{\epsfbox{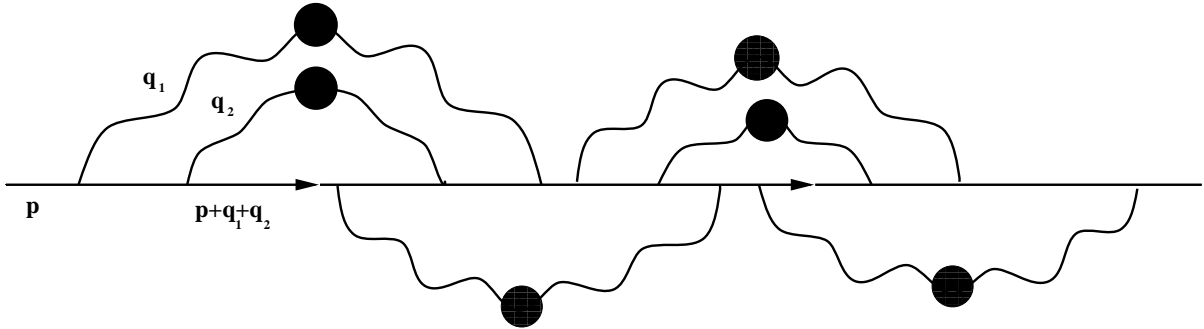}}}
         \caption{A generic $n$-loop diagram (here, $n=6$)
which is responsible for infrared divergences in perturbation theory.
All the photon lines are soft and dressed by the hard thermal loop.
The fermion line is hard and nearly on-shell.}
\label{NDAMP}
\end{figure}

Let us turn now to higher order contributions to $\Sigma$, and focus on
those diagrams which can be obtained by dressing the fermion
propagator by an arbitrary number of soft photon lines. An example of
such a diagram is given in fig.~\ref{NDAMP}. We refer to this class of
diagrams as to the ``quenched approximation'' (no fermion loops are
included except for the hard thermal loops dressing the soft photons
lines). One can  verify
\cite{lifetime,damping} that the leading
infrared contribution to $\gamma_p$ comes from these diagrams where the
internal fermion lines are hard and nearly on-shell. 
The individual contributions of these diagrams to the damping rate
contain power-like infrared divergences. An explicit calculation to
two-loop order can be found in Appendix C of Ref. \cite{damping}.

As we shall see,
it is possible to  resum all these leading IR contributions and obtain a
finite result. This is most conveniently done by formulating
the perturbation theory in the {\it time} (rather than the {\it energy}) 
representation. As we shall see, the inverse of the time acts
effectively  as an IR cutoff, needed to account for coherence
effects between successive scatterings. In the energy representation,
one essentially assumes that the particles return on their mass shell 
after each scattering, and this assumption is not satisfied in the
present case where the typical mean free path is comparable to the range
of the relevant interactions. We come back to this in the discussion later.

Because of the aforementioned
coherence effects, it is furthermore convenient to
consider approximations for the propagator rather than 
for the corresponding self energy. 

Consider then the 
contour  propagator
\beq\label{SC1}
-iS(x-y)\,\equiv\,\langle {\rm T}_C \psi(x)
\bar\psi(y)\rangle\,=\,\theta_C(x_0,y_0)\,S^>(x-y)\,-\,
\theta_C(y_0,x_0)\,S^<(x-y),\eeq
where the time variables $x_0$ and $y_0$ lie on a
contour $C$ in the complex time plane, as explained in section 2.
In the ``quenched'' approximation (in the sense of fig.~\ref{NDAMP}),
$S(x-y)$ is given by the following functional integral:
\beq\label{FunctionalS}
S(x-y)= Z^{-1}\int  {\cal D} A \,G(x,y|A)\,{\rm e}^{iS_C[A]},\eeq
where $G(x,y|A)$ is the tree-level propagator
in the presence of a background electromagnetic field, that is,
the solution of the  Dirac equation:
\beq\label{SA0}
-i{\slashchar D}_x  G(x,y|{A})\,=\,\delta_C(x,y),\eeq
with antiperiodic boundary conditions: 
\beq\label{BC1} G(t_0,y_0|A)=- \,G(t_0-i\beta,y_0|A),\eeq
and similarly for $y_0$. 
Furthermore,
$S_C[A]$ is the effective action for soft photons in the
HTL approximation (which we write here 
in a covariant gauge) :
\beq\label{SEFF} 
S_C[A]&=&\int_C {\rm d}^4x\left\{-\frac{1}{4}\,F_{\mu\nu}F^{\mu\nu}
-\frac{1}{2\zeta}(\del\cdot A)^2\right\}
\,+\,\int_C {\rm d}^4x\int_C {\rm d}^4y\,\,\frac{1}{2}\,A^\mu(x)
\Pi_{\mu\nu}(x,y)A^\nu(y)\nonumber\\
&\equiv&\int_C {\rm d}^4x\int_C {\rm d}^4y\,\,\frac{1}{2}\,A^\mu(x)
\,{}^*\!G_{\mu\nu}^{-1} (x-y) A^\nu(y).\eeq
The gauge fields to be integrated over
in eq.~(\ref{FunctionalS}) satisfy the periodicity condition
$A_\mu(t_0,{\bf x})=A_\mu(t_0-i\beta,{\bf x})$. 
Correspondingly, the photon  propagator satisfies
the KMS condition (cf. eq.~(\ref{KMSG})) :
\beq {}^*\!G_{\mu\nu}(t_0-y_0)\,=\,
{}^*\!G_{\mu\nu}(t_0-y_0-i\beta),\eeq
and can be given the following spectral representation
(cf. eq.~(\ref{GC0})):
\beq\label{specD1}
{}^*\!G_{\mu\nu}(x-y)=-i\int\frac{{\rm d}^4q}{(2\pi)^4}\,{\rm e}^{-iq\cdot(x-y)}
\,\,{}^*\!\rho_{\mu\nu}(q)\Bigl[\theta_C(x_0,y_0)+N(q_0)\Bigr]\,,\eeq
where ${}^*\!\rho_{\mu\nu}(q)$ is the photon spectral density in the HTL 
approximation, eqs.~(\ref{rhos})--(\ref{rhos0}),
and $N(q_0) = 1/\bigl({\rm e}^{\beta q_0}-1\bigr)$.

The resulting fermion propagator, given by
 eq.~(\ref{FunctionalS}), satisfies the
 KMS condition:
\beq\label{KMS11}
S(t_0-y_0)\,=\,-\,S(t_0-y_0-i\beta),\eeq
and can be given the following spectral representation:
\beq\label{specS1}
S(x-y)&=&i\int\frac{{\rm d}^4p}{(2\pi)^4}\,{\rm e}^{-ip\cdot(x-y)}\,
\acute\rho(p)\Bigl[\theta_C(x_0,y_0)-n(p_0)\Bigr]
\,,\eeq
where $\acute\rho(p)$ is the fermion spectral density in the present
approximation, and $n(p_0) = 1/\bigl({\rm e}^{\beta p_0}+1\bigr)$.

\begin{figure}
\protect \epsfxsize=10.cm{\centerline{\epsfbox{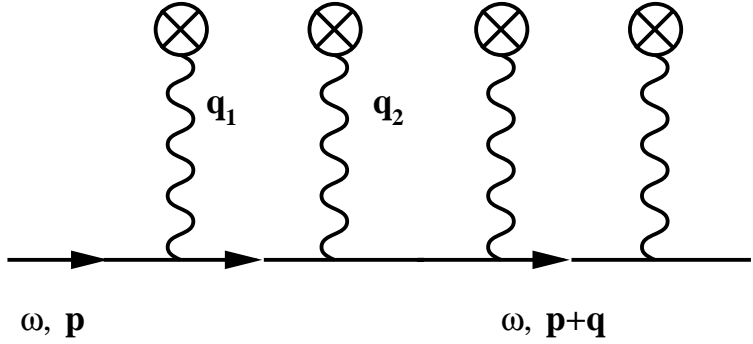}}}
         \caption{ A typical diagram contributing to $G(x,y|A)$
to order $e^n$ in perturbation theory (here, $n=4$).
This diagram involve $n=4$ photon field insertions,
and $n+1=5$ bare fermion propagators $S_0$ (including
the external lines).}
\label{BNfig}
\end{figure}

To illustrate the previous equations, we display in fig.~\ref{BNfig}
a typical diagram contributing to $ G(x,y| A)$ in perturbation
theory. This diagram involves $n$ photon insertions and contributes
to order $e^n$. By integrating over the fields $A_\mu(x)$ in
eq.~(\ref{FunctionalS}) one effectively closes the external photon lines
in  fig.~\ref{BNfig} into effective 
photon propagators. In this way, one generates all the Feynman graphs such
as those illustrated in fig.~\ref{NDAMP}, that is, all the diagrams 
of the quenched approximation.

\subsection{The Bloch-Nordsieck approximation}

As suggested by the previous analysis, a quasiparticle  decays mainly
through collisions involving the exchange of  soft ($q\simle eT$)
 virtual photons  with the hard fermions of the heat bath.
When the quasiparticle  is  hard
($p\sim T$), we can perform  kinematical approximations similar to those 
 widely used in relation with soft photon effects
(see, e.g., Refs. \cite{Bogoliubov,Weinberg}).
The main outcome of these  approximations is the  
replacement of the Dirac equation (\ref{SA0}) by the following equation,
known as the Bloch-Nordsieck (BN) equation \cite{BN37,Bogoliubov} :
\beq\label{Gret1}
-i\,(v\cdot D_x)\,G(x,y|A)&=&\delta_C(x,y),\eeq
where $v^\mu  =(1,{\bf v})$ and 
 ${\bf v}$ is a fixed parameter, to be identified
with the velocity of the hard quasiparticle (here ${\bf v}$ is a unit
vector).

To get some justification for this approximation one may 
analyze  the perturbative solution  of (\ref{SA0}) in the relevant
kinematical domain. Consider then,  in the energy-momentum representation, 
 a hard ($p\sim T$) electron,   nearly
on-shell ($p_0\simeq p$), and  propagating through a soft
($q \simle eT$) electromagnetic background field.
A typical Feynman diagram  contributing to the  the Dirac propagator,
solution of equation (\ref{SA0}), is displayed   in fig.~\ref{BNfig}. In
such a diagram, the free propagator, when expanded near the mass shell,
takes the form:
\beq\label{S00p}
S_0(p_0+q_0, {\bf p+q})&=&-\,\frac {(p_0+q_0)\gamma_0 -{\bf (p+q)}
\cdot {\bfgamma}}
{(p_0+q_0)^2-({\bf p+q})^2}\nonumber\\
&\simeq &\frac {\gamma_0 -{\bf v}\cdot {\bfgamma}}{2}\,
\frac {-1}{p_0+q_0 - {\bf v\cdot (p+ q)}}
\equiv h_+({\bf v})\,G_0(p_0+q_0, {\bf p+q})
\,,\nonumber\\\eeq
where $q^\mu=(q_0,{\bf q})$ is a linear combination of the 
external photon momenta, ${\bf v = p}/p$, $v^\mu=(1,{\bf v})$.
The matrices  $h_+({\bf v})$ coming from the 
fermion propagators combine with the various
photon-fermion vertices to   give a global contribution:
\beq \label{IMU}
 h_+( {\bf v})\gamma^{\mu_1}
h_+({\bf v})\gamma^{\mu_2}\,...\,
h_+({\bf v})\gamma^{\mu_{n}}h_+( {\bf v})
\,=\,v^{\mu_1}\,v^{\mu_2}\,...\,
v^{\mu_{n}} h_+( {\bf v})\,,\eeq
for $n$ external photon lines.
Note  the  factorization of  the matrix $h_+({\bf v})$
which is independent of the photon momenta and plays no dynamical
role. Thus, within the present kinematical approximations, the  diagram
in fig.~\ref{BNfig} could as well have been evaluated with
\beq\label{G02}
G_0(p+q)\,=\,\frac{-1}{(p_0+q_0)-{\bf v}\cdot ({\bf p+q})}\,,\eeq
as the fermion propagator 
and $\Gamma^\mu=v^\mu$ as the photon-fermion vertex .
We recognize here the Feynman rules generated by the 
BN equation (\ref{Gret1}), provided we identify in the latter
the vector ${\bf v}$ with the velocity of the hard particle.

The Bloch-Nordsieck equation (\ref{Gret1}) defines a Green's function
of the covariant derivative $v\cdot D$, and in this sense
it is formally identical to eq.~(\ref{Gret}) in Sect. 4.1.
However, the solutions of  eqs.~(\ref{Gret1}) and (\ref{Gret})
differ because of the respective boundary conditions. In Sect. 4.1,
  eq.~(\ref{Gret}) is solved for retarded (or advanced) boundary conditions.
In principle, the thermal BN equation (\ref{Gret1}) is
to be solved with antiperiodic boundary conditions (cf. eq.~(\ref{BC1})).

However, as discussed in Refs.~\cite{damping,TBN} these antiperiodic
boundary conditions, which greatly complicate the solution of the BN
equation, are not needed to obtain  the dominant
behaviour at large time of the fermion propagator: this   is indeed
identical to that of a  {\it test particle}, by which 
we mean a particle which is distinguishable from the plasma particles,
 and is therefore not part of the thermal bath. 
The propagator of a test particle has only one analytic
component, namely $S^>$ ($S^<$ vanishes since the
thermal bath acts like the vacuum for the field operators
of the test particle).
Therefore, for real times
$x_0,\, y_0\in C_+$, the contour propagator of a test particle coincides
with the retarded propagator (cf. eqs.~(\ref{A}) and (\ref{SC1})):
\beq
S(x-y)\,=\,i\theta(x_0-y_0)S^>(x-y)\,=\,S_R(x-y).\eeq
 
The resulting propagator 
$S\equiv S_R$ is still given by eq.~(\ref{FunctionalS}), but now
$G\equiv G_R$ obeys retarded conditions  and is therefore given
explicitly  by eq.~(\ref{GR}): it depends on the background field only
through the parallel transporter (\ref{U}). Accordingly,  the functional
integration in (\ref{FunctionalS}) is straightforward and yields:
\beq\label{SRT}
S_R(t,{\bf p})&=&i\theta(t)\, {\rm e}^{-it({\vp})}\,\Delta(t),\eeq
where the quantity 
\beq\label{Delta0}
\Delta(t)&\equiv& Z^{-1}\int {\cal D}{A}\,
\,U(x,x-vt) \,\,{\rm e}^{iS_C[A]}\nonumber\\
&=&{\rm exp}\left\{-\,\frac{e^2}{2}\int_0^t {\rm d}s_1 
\int_0^t {\rm d}s_2\, v^\mu\, \,{}^*\!G_{\mu\nu}
({\bf v}(s_1-s_2))\, v^\nu
\right \}\eeq
contains all the non-trivial time dependence. The $s_1$ 
and $s_2$ integrations in eq.~(\ref{Delta0}) can
be performed by using the spectral representation (\ref{specD1}).
We then obtain (omitting an irrelevant phase factor):
\beq\label{Delta2}
\Delta(t)\,=\,{\rm exp}\left \{-e^2
\int \frac{{\rm d}^4q}{(2\pi)^4} \,\,\tilde\rho(q)\,N(q_0)\,\,
\frac{1-  {\rm cos}\,t({v\cdot q})}{(v\cdot q)^2}\,
\right\},
\eeq
where 
\beq
\tilde \rho(q)\equiv
v^\mu\,{}^*\!\rho_{\mu\nu}(q)v^\nu.
\eeq

It is interesting to observe that this result can be also obtained in the
framework of the classical field theory of Sect. 4.4.3. 
Indeed, the integral over $q$ in eqs.~(\ref{Delta0}) and
(\ref{Delta2}) being dominated by soft momenta, one can replace
there $N(q_0)$ by $T/q_0$, so that, when written in the temporal 
gauge $A^0=0$, the Feynman propagator ${}^*\!G_{ij}$ in these equations
reduces to the classical correlator $G_{ij}^{cl}$ 
in eq.~(\ref{DCL}). Then, eq.~(\ref{Delta0}) is effectively
the same as eq.~(\ref{ZAB}) with the eikonal current density
\beq
J^i(z)\,=\,ev^i\int_0^t {\rm d}s\, \delta^{(4)} (z-x+v(t-s)).\eeq
Thus, the result (\ref{Delta0}) can be seen as the result of the
classical averaging over the initial 
conditions for the HTL effective theory, that is,
\beq\label{Delta0cl}
\Delta(t) &=& Z_{cl}^{-1}
\int  {\cal D}{\cal E}_i\,{\cal D}{\cal A}_i\,{\cal D}{\cal W}\,
\delta({\cal G}^a)\,
\,U(x,x-vt|A_{cl}) \,{\rm e}^{-\beta{\cal H}},\eeq
where 
\beq
U(x,x-vt|A_{cl})\,=\,{\rm exp}\Bigl\{i\int {\rm d}^4z \,J^i(z)
A_{cl}^i(z)\Bigr\},\eeq
and $A_{cl}^i(x)$ is the solution to the classical 
 equations of motion (\ref{EAB})
with the initial conditions $\{{\cal E}_i,\,{\cal A}_i,\,{\cal W}\}$.

\subsection{Large-time behaviour}

We are now in a position to study the large-time behaviour of the
fermion propagator, as described
by  the function $\Delta(t)$. Let us first observe
that for a fixed time
$t$, the function:
\beq\label{Asymp}
f(t,{v\cdot q})&\equiv &\frac{1-{\rm cos}\,t(v\cdot q)}{(v\cdot q)^2}\,,
\eeq
in eq.~(\ref{Delta2}) is strongly peaked around ${v\cdot q}\equiv
q_0-{\bf v\cdot q}=0$, 
with a width $\sim 1/t$. In the limit $t\to\infty$,
$f(t,{v\cdot q})\,\to\,\pi t \delta (v\cdot q)$.
In the absence of infrared complications, we could use this
limit in eq.~(\ref{Delta2}) to obtain 
 $\Delta(t\to \infty)\,\sim\,{\rm e}^{-\gamma t}$, with:
\beq\label{naive}
\gamma &\equiv&\pi e^2\int \frac{{\rm d}^4q}{(2\pi)^4} \,\,
\tilde\rho(q)\,N(q_0)\,
\delta (v\cdot q)\,.\eeq
We recognize in eq.~(\ref{naive}) the one-loop damping rate 
(see eq.~(\ref{g4})). We know, however, that $\gamma$ is 
infrared divergent (cf. eq.~(\ref{G2LR})), so a different strategy must be
used to extract the large time behaviour of $\Delta(t)$.

In the Coulomb gauge, the photon spectral density reads
(cf. eqs.~(\ref{rhos})--(\ref{RHOLT})) :
\beq\label{Coul}
\tilde \rho(q_0,{\bf q})\,=\,{}^*\!\rho_L(q_0,q)\,+
\,\left(1- ({\bf v}\cdot \hat{\bf q})^2\right)\,
{}^*\!\rho_T(q_0,q).\eeq
The infrared problems come from the magnetic sector and, more
precisely, from the IR limit, $q\to 0$, where we can use the approximation:
\beq\label{rhot00}
{}^*\!\rho_T(q_0,q) N(q_0)\,\simeq\, 
\frac{\pi}{2}\, \frac{m_D^2 \, qT}
{q^6\,+\, (\pi m_D^2 q_0/4)^2}\,\longrightarrow\,
T\,\frac{2\pi}{q^2}\,\delta(q_0)\,\,\,\,\,\, {\rm as} 
\,\,\,\, q\to 0.\,\,\eeq
(Since ${}^*\!\rho_T(q_0,q)=2{\rm Im}\,{}^*\!\Delta_T(q_0+i\eta,q)$,
the above equation is, of course, equivalent to eq.~(\ref{deltaT}) for the
magnetic propagator.) 
In the computation of $\Delta(t)$, it is convenient
to isolate the singular behaviour in eq.~(\ref{rhot00}) by writing
(with $N(q_0)\simeq T/q_0$):
\beq\label{sept}
{}^*\!\rho_T(q_0,q) N(q_0)\,\equiv\,
2\pi T\delta(q_0)\left(\frac{1}{q^2}\,-\,\frac{1}{q^2+m_D^2}\right)\,+\,
\frac{T}{q_0}\,\nu_T(q_0,q).\eeq
A contribution $\propto 1/(q^2+m_D^2)$ has been subtracted from 
the singular piece
--- and implicitly included in $\nu_T(q_0,q)$ --- 
to avoid spurious ultraviolet 
divergences: written as they stand, both terms in the
 r.h.s. of eq.~(\ref{sept}) give UV-finite contributions. In fact, it
turns out \cite{TBN} that with this particular choice of a regulator,
the non-singular contribution to $\Delta(t)$, i.e., that coming from
\beq
\tilde\rho_{non-sing}\,\equiv\,{}^*\!\rho_L\,-\,
({\bf v}\cdot \hat{\bf q})^2\,{}^*\!\rho_T\,+\,\nu_T,
\eeq
is precisely zero, so that the net contribution arises entirely
from the piece proportional to $\delta(q_0)$
in eq.~(\ref{sept}). The latter is easily evaluated as
\beq\label{FIR}
\ln\Delta(t)&=&- g^2 T\int \frac{{\rm d}^3q}{(2\pi)^3}
\left(\frac{1}{q^2}\,-\,\frac{1}{q^2+m_D^2}\right)\,
\frac{1-  {\rm cos}\,t({\vq})}{({\vq})^2}\nonumber\\
&=& -\alpha Tt\Bigl(
 \ln(m_D t)\,+\,(\gamma_E-1)
\,+\,{\cal O}(g,\,1/m_D t)\Bigr),\eeq
where $\alpha=e^2/4\pi$ and $\gamma_E=0.5772157$ is Euler's constant. 

\begin{figure}
\protect \epsfxsize=10.cm{\centerline{\epsfbox{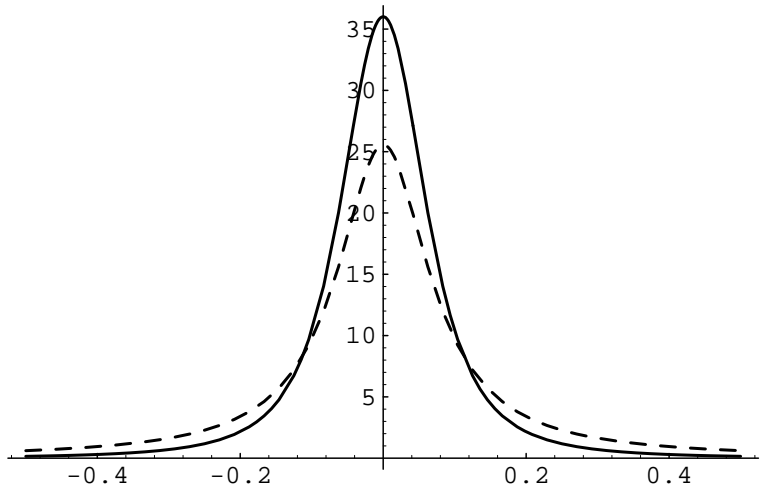}}}
         \caption{The spectral density $\acute\rho(\varepsilon)$ (full line)
and the lorentzian $\rho_{L}(\varepsilon)$ (dashed line) for $e=0.08$.
 All the quantities are made adimensional
 by multiplication with appropriate powers of $m_D$ (e.g.,
$\varepsilon/m_D$ is represented on the abscissa axis, and $m_D\rho$ 
on the vertical axis).}
\label{ac2}
\end{figure}

Thus, at very large times, the decay of the retarded propagator is not
exponential. In fact,
$\Delta(t)$ is decreasing faster than an  exponential. It follows that the 
Fourier transform 
\beq\label{SRE}
S_R(\omega, {\bf p})\,=\,
\int_{-\infty}^{\infty} {\rm d}t \,{\rm e}^{-i\omega t}
S_R(t,{\bf p})\,=\,
i\int_0^{\infty}{\rm d}t
\,{\rm e}^{it(\omega- {\bf v\cdot p}+i\eta)}\,\Delta(t),\eeq
exists for {any} complex (and finite) $\omega$.  
In contrast to what one would expect from perturbation theory,
the quasiparticle propagator has no pole, nor any other kind of
singularity, at the mass-shell. However, the associated spectral density
\beq\label{rhoD}
\acute\rho(\omega, {\bf p})= 2\,{\rm Im}\,S_R(\omega, {\bf p})
= 2 \int_0^{\infty}{\rm d}t\,\cos\,t(v\cdot p)\,\Delta(t),\eeq
(with $v\cdot p= \omega -{\vp}$) retains the shape
of a {resonance}  strongly peaked around the perturbative mass-shell 
$\omega \sim {\vp}$, 
with a typical width of order $\sim e^2T \ln(1/e)$.
(See fig.~\ref{ac2}, where we also represent, for comparison,
the Lorentzian spectral function
$\rho_{L}(\varepsilon)={2\gamma}/(\varepsilon^2 + \gamma^2)$,
with $\varepsilon= v\cdot p$ and  $\gamma= \alpha T \ln(1/e)$.)

As the previous analysis shows, the leading logarithmic behaviour at 
large times --- i.e., the term $\ln(m_D t)$ in eq.~(\ref{FIR}),
which is the counterpart of the IR divergence $\ln(m_D/\mu)$ in
the energy representation --- is due to scattering involving the exchange
of quasistatic magnetic photons. Thus, in the path integrals
in eqs.~(\ref{Delta0}) and (\ref{Delta0cl}), the dominant contribution
comes from integration over static magnetic field configurations.
To isolate this contribution in eq.~(\ref{Delta0cl}),
it is enough to replace there the classical solution $A_{cl}^i(t, {\bf x})$
by its initial condition ${\cal A}_i({\bf x})\,$; this gives
(the integrals over ${\cal E}_i$ and ${\cal W}$ contribute only
to the normalization factor):
\beq\label{DeltaDR}
\Delta(t)&\approx&
Z^{-1}\int  {\cal D}{\cal A}_i\,\,U(x,x-vt|{\cal A}_i) \,
\,{\rm e}^{-\beta\int {\rm d}^3 z\,\frac {1}{4}\,
{\cal F}^2_{ij}}\,.\eeq
The same integral would have been obtained by restricting
the integration in
the quantum path-integral in eq.~(\ref{Delta0})
to the static Matsubara modes of the transverse
fields. It can be easily verified that the integral (\ref{DeltaDR})
yields indeed the asymptotic behaviour 
exhibited in eq.~(\ref{FIR}):
$\Delta(m_Dt\gg 1) \simeq {\rm exp}\{-\alpha Tt \ln(m_D t)\}$.
(The scale $m_D\sim gT$ enters this calculation as an ad-hoc upper
momentum cutoff, which is necessary since the integral
in eq.~(\ref{DeltaDR}) has a spurious, logarithmic, ultraviolet 
divergence, due to the reduction to the static modes \cite{lifetime}.
In the full calculation including also the non-static modes
and leading to eq.~(\ref{FIR}), this cutoff has been
provided automatically by the screening effects at the scale $gT$.)

We conclude this subsection with two comments:
First, we notice that the previous analysis has 
been extended to massive test particles \cite{TBN},
and also to soft quasiparticles, that is,
to collective fermionic excitations with typical momenta $p\sim eT$
\cite{damping}. It has also been shown that the result presented here is
gauge invariant \cite{TBN}. Second, we note that a
similar problem  has been investigated in a
completely different context, that of the propagation of
electromagnetic waves in random media \cite{ADZ80} (see also \cite{JPB96}).

\subsection{Discussion}

At this point, a few comments on the results that we have obtained are
called for. 
In particular we have earlier alluded to the fact that
the inverse of the time acts effectively as an infrared cutoff. We
wish to  see now more explicitly how this occurs, both in the calculation
of the propagator, and in that of the damping rate from kinetic
theory. 

Consider the one-loop correction to the
retarded propagator $S_R(t,{\bf p})$; for $t>0$, this is given by:
\beq\label{LT}
\delta S_R^{(2)}(t,{\bf p})
\,=\,-\int_0^t {\rm d}t_1\,\int_0^{t_1} {\rm d}t_2\,
S_0(t-t_1, {\bf p})\Sigma_R^{(2)}(t_1-t_2, {\bf p}) S_0(t_2, {\bf p}),\eeq
where $S_0(t, {\bf p})$ is the free retarded propagator and
$\Sigma_R^{(2)}(t, {\bf p})$ is the retarded one-loop self-energy.
Since, in the BN approximation, 
$S_0(t,{\bf p})=i\theta(t) {\rm e}^{-it({\bf v\cdot p})}$, 
we immediately obtain 
$\delta S_R^{(2)}(t)= - S_0(t)\,\delta\Delta(t)$,
with
\beq\label{DSR1}
\delta\Delta(t,{\bf p})&\equiv& i\int_{0}^t {\rm d}t'\, (t-t')\,
 {\rm e}^{it'({\bf v\cdot p})}\,\Sigma_R^{(2)}(t',{\bf p})\nonumber\\
&\simeq& it\int_{0}^t {\rm d}t' 
\, {\rm e}^{ipt'}\,\Sigma_R^{(2)}(t',{\bf p}),\eeq
where the last, approximate, equality holds in the large time limit.
The above expression
is well defined although the limit $t\to \infty$ of the integral over  $t'$
(which is precisely the on-shell self-energy 
$\Sigma_R^{(2)}(\omega = p)$) does not exist.
In fact \cite{TBN}
\beq\label{SRT1}
\Sigma_R^{(2)}(t, {\bf p})\,\simeq\,
-i\alpha T \,
\frac{{\rm e}^{-ipt}}{t}\,\,\,\qquad {\rm for}\,\,\,\,\,\,\,
t\gg \frac {1}{m_D}\, \eeq
does not decrease fast enough with time to have a Fourier transform.
However, 
\beq\label{LT2}
\delta\Delta (t,{\bf p})\,\simeq\,
\alpha T t\int_{1/m_D}^t
 \frac{{\rm d}t'}{t'}\,=\,-\alpha Tt \ln (m_D t) \eeq
is finite and, as shown in Refs. \cite{damping,TBN}, 
this second order  correction to the
retarded propagator exponentiates in an all-order calculation:
\beq\label{DLT0}
S_R(t,{\bf p})\,\propto\, {\rm exp}\Bigl( -\alpha Tt \ln m_D
t\Bigr)\,\,\,\qquad {\rm for}\,\,\,\,\,\,\,
t\gg \frac {1}{m_D}\,.\eeq
In other terms, the full BN result in eq.~(\ref{Delta2}) is nothing
but the exponential of the one-loop correction to the propagator
in the time representation: $\Delta(t)=\exp\{-\delta \Delta(t)\}$.

We shall recover this mechanism of exponentiation from a different point of
view, that of kinetic theory. As shown at  the end of Sect. 2, the
single-particle excitation with momentum
${\bf p}$ can be described as an off-equilibrium deviation 
$\delta N({\bf p},t)\equiv N({\bf p},t) - N(p)$ in the 
distribution function, which  obeys 
 eq.~(\ref{b14}).  
Here, we compute the time-dependent
damping rate $\gamma({\bf p},t)$  for an electron, to leading order
in perturbation theory. According to eq.~(\ref{tgamma}) we need the
discontinuity
\beq
\Gamma(p_0,{\bf p})\,=\, -\,{\rm tr}\left(
{\slashchar p}\, {\rm Im}\,{}^*\!\Sigma_R(p_0+i\eta,{\bf p})\right)\,,\eeq 
which can be extracted from  eqs.~(\ref{SL}) and (\ref{ST}) in 
Appendix B. For large enough times $t\gg 1/m_D$, we need this
quantity only in the vicinity of the mass-shell ($|p_0- p| \ll m_D$),
where it reads (compare to eq.~(\ref{naive})):
\beq\label{GammaQED}
\frac{1}{2p}\,\Gamma(p_0,{\bf p})&=&2\pi e^2\int \frac{{\rm d}^4q}{(2\pi)^4} \,
\tilde\rho(q)\,N(q_0)\,\delta (p_0-p-q_0+{\bf v\cdot q})
\nonumber\\
&=&\frac{e^2 T}{2\pi}\,\ln\frac{m_D}{|p_0-p|}\,,
\eeq
up to terms which vanish as $p_0\to p$.
When inserted into eq.~(\ref{tgamma}), this yields:
\beq\label{tgammaQED}
\gamma(t)\,=\,
e^2\int \frac{{\rm d}^4q}{(2\pi)^4}\,\tilde\rho(q)\,N(q_0)\,
\,\frac{\sin(v\cdot q)t}{v\cdot q}\,,\eeq
which is independent of $p$. The naive infinite-time limit of
this expression, using  \\ $\sin(v\cdot q)t/
(v\cdot q)\,\to\, \pi\delta(v\cdot q)$, coincides with the usual
one-loop damping rate in eq.~(\ref{naive}), and is IR divergent.
But for any finite $t$, the expression of $\gamma(t)$ given by
eq.~(\ref{tgammaQED}) is well defined and can be used in eq.~(\ref{b14})
to get:
\beq\label{deltaNQED}
\delta N({\bf p},t)\,=\,{\rm e}^{-2\int_0^t {\rm d}t'\gamma(t')}
\,\delta N({\bf p},0)\,,\eeq
with (cf. eq.~(\ref{tgammaQED})) 
\beq
\int_0^t {\rm d}t'\gamma(t')\,=\,e^2
\int \frac{{\rm d}^4q}{(2\pi)^4} \,\,\tilde\rho(q)\,N(q_0)\,\,
\frac{1-  {\rm cos}\,t({v\cdot q})}{(v\cdot q)^2}\,.\eeq
A short comparison with eq.~(\ref{Delta2}) reveals that:
\beq
\Delta (t)\,=\,{\rm exp}\left \{-
\int_0^t {\rm d}t'\gamma(t')\right\}.\eeq

Note that, in both approaches, the final  result 
emerges as the exponential of  a one-loop correction
in the time representation. In this one-loop correction the inverse of
the time plays the role of an infrared cut-off, 
and perturbation theory in the {time} representation
can be applied for sufficiently small times (here,
$t\simle 1/gT$). Another approach leading to a similar exponentiation
of the  one-loop result is the so-called ``dynamical 
renormalization group'' developed in Refs. \cite{BVHS98,BDV98,BDV00}.

\subsection{Damping rates in QCD}

Let us now turn to  QCD.
As already mentioned, the one-loop calculations of damping rates
 are afflicted by the same IR problem as in QED. However,  in
QCD we may expect these  divergences to be screened at
the scale $g^2 T$  by the self-interactions of the magnetic gluons.
That is, we expect the quasiparticle (quark or gluon) propagator
to decay exponentially  according to 
\beq\label{qLT2}
\Delta(t\to \infty)\simeq {\rm exp}\left\{
- \,C_{\rm r}\,\frac{g^2T}{4\pi}\,t\,\Bigl(\ln \frac{m_D}{m_{mag}}
\,+\,{\cal O}(1)\Bigr)\right\},\eeq
where $C_{\rm r}$ is the Casimir
factor of the appropriate color representation
(i.e.,   $C_{\rm f} =(N^2-1)/2N$ for a quark, and $C_{\rm g}=N$ for 
a gluon), and  $m_{mag}\sim g^2T$ is the ``magnetic mass''.
(We consider here a hard quasiparticle,  with momentum $p\simge T$.
Except for the magnetic mass, the leading term displayed
in eq.~(\ref{qLT2}) is determined by the one-loop calculation.)

Note however that the magnetic mass matters
only at  times $t\simge 1/g^2T$.
For  intermediate times, $1/gT \ll t \ll 1/g^2T$, relying on the analogy
with the Abelian problem, we may expect a non-exponential decay law:
\beq\label{qIT2}
\Delta(1/gT \ll t\ll 1/g^2T)\simeq {\rm exp}\left\{
- \,C_r\,\frac{g^2T}{4\pi}\,t\,\Bigl(\ln(m_D t)
\,+\,{\cal O}(1)\Bigr)\right\}.\eeq
To verify the behaviour in eqs.~(\ref{qLT2})--(\ref{qIT2}),
a non-perturbative analysis is necessary. 
As shown in Ref.~\cite{TBN}, the BN approximation leads to the following
functional representation of the  quasiparticle propagator (compare with
eq.~(\ref{Delta0})) :
\beq\label{Delta0N}
\Delta(t)&\equiv& Z^{-1}\int {\cal D}{A}\,{\rm Tr} \left(\,
U(x,x-vt) \right)\,\,{\rm e}^{iS_C[A]},\eeq
where $U(x,x-vt)$ is the non-Abelian parallel transporter, eq.~(\ref{U}),
and $S_C[A]$ is the contour effective action for soft gluons in
the HTL approximation. Now, 
in QCD the action is not quadratic and 
the functional integral (\ref{Delta0N}) cannot be computed analytically.

A possible continuation would be to treat the soft gluons in the classical
approximation, and thus replace eq.~(\ref{Delta0N}) by the
non-Abelian version of eq.~(\ref{Delta0cl}),
to be eventually computed on a classical lattice (cf. Sect. 4.4.3).
A more economical proposal \cite{TBN} is to use ``dimensional
reduction'', as in eq.~(\ref{DeltaDR}). This should be enough to
generate the non-perturbative magnetic screening, and therefore
the leading logarithm $\ln(m_D/m_{mag})\simeq \ln(1/g)$
of the asymptotic behaviour in eq.~(\ref{qLT2}) (but not also
the constant term of ${\cal O}(1)$ under the logarithm
which is sensitive to the non-static modes).

\begin{figure}
\protect \epsfxsize=11.cm{\centerline{\epsfbox{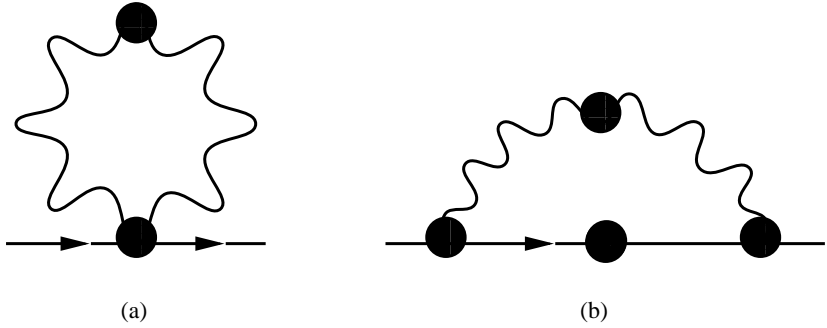}}}
         \caption{One-loop diagrams for the soft fermion self-energy in the
effective expansion.}
\label{figeff}
\end{figure}

To conclude, let us recall that for quasiparticles with
zero momentum the damping
rates are finite and of order $g^2T$.
The IR problems are absent in this case since
the magnetic interactions do not contribute.
The corresponding damping rates have been computed
for  both gluons \cite{BP90a} and fermions \cite{KKM,BP92b},
to one-loop order in the effective theory. Since all the external and
internal lines are soft, the corresponding diagrams involve
resummed propagators and vertices (see, e.g.,  fig.~\ref{figeff}
for the case of a soft external fermion). The resulting damping rates were
shown to be gauge-fixing  independent, a property which
relies strongly on the Ward identities satisfied by the HTL's
\cite{BP90,BP90a}.

\setcounter{equation}{0}
\section{The Boltzmann equation for colour excitations}

As we have already argued, as long as we are interested in the collective
excitations with wavelength $\sim 1/gT$ we can ignore, in leading order in $g$,
collisions among the plasma particles.  However 
collisions  become a dominant effect
for colour excitations with wavelength $\sim 1/g^2T$, and colorless excitations
with wavelength $\sim 1/g^4T$. In this section we focus on excitations
with wavelength $\sim 1/g^2T$ to which we shall refer as  {\it ultrasoft}
excitations. 

We shall then reconsider briefly  the approximations which led us in Sect. 3 to
kinetic equations,  and study the dynamics of hard particles ($k\sim T$) in
the background of ultrasoft fields $A^\mu_a(X)$ such that $\del_X\sim gA\sim
g^2T$. As we shall see, in leading order, the role of the soft degrees of freedom
($k\sim gT$)  is merely to mediate collisions between the plasma particles. The
resulting kinetic equation  is a {\it Boltzmann equation},
whose solution implicitly resums an infinite number of diagrams of perturbation
theory. These diagrams generalize the HTL's to the case where the external
lines are ultrasoft and are called ``ultrasoft amplitudes''. 

In order to specify
the separation between soft and ultrasoft fields, we shall introduce an
intermediate scale
$\mu$ such that  $g^2T\ll\mu\ll gT$. The ultrasoft
amplitudes depend logarithmically on this scale which
plays the role of an infrared cutoff in their calculation.

An alternative description of the ultrasoft dynamics relies on the fact that it
is essentially that of classical fields. Already the soft modes are classical,
and to leading order their  dynamics is entirely contained in the classical
equations of motion given in Sect. 4. Furthermore, in order to calculate
correlation functions in real time, one can use the hamiltonian formulation of
Sect. 4.4 in order to perform the necessary averages over the initial
conditions.  Then, the non-perturbative dynamics can be
studied for instance
via classical lattice simulations.
 The  effective theory presented in Sec. 4.4.3
 turns out to have  a relatively strong (linear) dependence
upon its ultraviolet cutoff, which may lead to  lattice artifacts.
However, as
suggested by  B\"odeker \cite{Bodeker},  by integrating out the soft
modes  in classical perturbation theory, one obtains an effective theory
for the ultrasoft fields which is only logarithmically sensitive
to the scale $\mu$ introduced above, and could therefore be better suited for
numerical calculations.
This effective theory involves the Boltzmann equation
alluded to before supplemented by a noise term which, as we shall see,
is related to the collision term by the fluctuation-dissipation theorem.

\subsection{The collision term}

For simplicity, throughout this section we shall restrict ourselves
to a Yang-Mills plasma without quarks, and use the background field
Coulomb gauge, as defined in eq.~(\ref{COUL}). 
The Kadanoff-Baym equations for the gluon 2-point
function $G^<_{\mu\nu}(x,y)$ read (compare to eqs.~(\ref{eqsG})):
\beq\label{KBYM1}
\left(g_\mu^{\,\rho}D^2-D_\mu D^\rho+ 2igF_{\mu}^{\,\rho}
\right)_xG^<_{\rho\nu}(x,y)&=&\nonumber\\
\int {\rm d}^4z\,\Bigl\{
g_{\mu\lambda}\Sigma_{R}^{\lambda\rho}(x,z)\,G^<_{\rho\nu}(z,y) 
&+&\Sigma_{\mu\rho}^{<}(x,z)G_{A}^{\rho\lambda}
(z,y)g_{\lambda\nu}\Bigr\},\eeq
and 
\beq\label{KBYM2}
G_{\mu}^{<\rho}(x,y)\left(g_{\rho\nu}\Bigl(D^\dagger\Bigr)^2
- D_\rho^\dagger D_\nu^\dagger + 2ig F_{\rho\nu}\right)_y
&=&\nonumber\\
\int {\rm d}^4z\,\Bigl\{
g_{\mu\lambda}G_{R}^{\lambda\rho}(x,z)\,\Sigma^<_{\rho\nu}(z,y) 
&+&G_{\mu\rho}^{<}(x,z)\Sigma_{A}^{\rho\lambda}
(z,y)g_{\lambda\nu}\Bigr\}.\eeq
The subsequent analysis of eqs.~(\ref{KBYM1})--(\ref{KBYM2})  
proceeds as in Secs. 3.3--3.4: we construct the difference of the
two equations, introduce gauge-covariant Wigner functions,
and then perform a gauge-covariant gradient expansion which
is controlled by powers of $g^2$ (since $D_X\sim g^2T$). 
The new feature is
the emergence of the collision term, coming from the
terms involving self-energies in 
eqs.~(\ref{KBYM1})--(\ref{KBYM2}). 

To leading order accuracy, we can restrict
ourselves to a { quasiparticle approximation}, in the sense
of Sect. 2.3.4; that is, we can ignore the off-shell effects
for the hard particles (here, the transverse gluons), together with the
Poisson brackets generated by the gradient expansion of the self-energy terms
(cf. Sect. 2.3.2). This means that, to leading non-trivial order,
the (gauge-covariant) Wigner functions conserve the same
structure as in the mean field approximation (cf. Sect. 3.4.1),
namely:
\beq\label{GMFAQCD}
\acute{\cal G}_{ij}^<(k,X)&=&(\delta_{ij}-\hat k_i\hat k_j)\,
[G^<_0(k)+\delta\acute{\cal G}(k,X)]\,,\eeq
with (compare with eq.~(\ref{GN0}))
\beq\label{GN01}
\delta\acute{\cal G}_{ab}(k,X)
=-\rho_0(k) W_{ab}(k,X) \frac{{\rm d}N}{{\rm d}k_0}.\eeq
A similar equation holds for $\acute{\cal G}_{ij}^>$ with 
$G^<_0$ replaced by $G^>_0$.

The final kinetic equation is conveniently written as an equation
for $\delta\acute{\cal G}(k,X)$, and reads (in matrix notations):
 \beq\label{bolt}
2\Bigl[k\cdot D_X,\,\delta\acute{\cal G}(k,X)\Bigr]
\,-\,2gk^\mu  F_{\mu\nu}(X)\del^\nu_k G_0^<(k)\,=\,
C(k,X),\eeq
with the collision term $C(k,X)$ (a colour matrix
with elements $C_{ab}(k,X)$) to be constructed now.
To this aim, consider a typical convolution term in
the r.h.s. of eq.~(\ref{KBYM1}) or (\ref{KBYM2}); to leading
order in the gradient expansion, this yields
(with Minkowski indices omitted, for simplicity):
\beq\label{PBQCD}
\int {\rm d}^4z\,G(x,z)\,\Sigma(z,y) 
\,\longrightarrow {\cal G}(k,X) \Sigma(k,X)\,+\,\,...\,,\eeq
where as compared to eq.~(\ref{AOB}) we have neglected the
Poisson bracket term. 
By collecting all the terms coming from the r.h.s. of 
eqs.~(\ref{KBYM1}) and (\ref{KBYM2}), and paying attention to the 
ordering of the
colour matrices,  we obtain:
\beq\label{COL0}
C(k,X)&=&i\Bigl(\Sigma_R {\cal G}^< - {\cal G}^< \Sigma_A
+\Sigma^< {\cal G}_A - {\cal G}_R \Sigma^<\Bigr)\nonumber\\
&=&-\,\frac{1}{2}\Bigl(\{{\cal G}^>,\,\Sigma^<\} - \{\Sigma^>,\,
{\cal G}^<\}\Bigr)
-i[{\rm Re} \Sigma_R,\,{\cal G}^<] +i[{\rm Re}\,{\cal G}_R,\,\Sigma^<],\eeq
where $[\,,\,]$ and $\{\,,\,\}$ stand here for colour commutators
and anticommutators, respectively.
In writing the second line above, we have also used the relations
(\ref{NEQAG}) and (\ref{RETWIG}).

We now proceed with further approximations. Since
$A^\mu\sim gT$, $gF^{\mu\nu}\sim g\del_X A^\mu \sim g^4T^2$, and
eq.~(\ref{bolt}) implies that
$\delta {\cal G}^< \sim g^2 G_{eq}^<$. Similarly,  writing
$\Sigma^<=\Sigma^<_{equ}+\delta \Sigma^<$, one finds 
$\delta \Sigma^< \sim g^2\Sigma_{eq}^<$ . Thus, we can linearize
$C(k,X)$ in eq.~(\ref{COL0}) with respect to the off-equilibrium fluctuations.
Since the equilibrium two-point functions are diagonal in colour
(e.g.,  $G_{eq}^{ab}=\delta^{ab}G_{eq}$),
the two commutators in eq.~(\ref{COL0}) vanish after
linearization, while the anticommutators yield:
\beq\label{COL}
C(k,X)\,\simeq\,-\,\Bigl(G_{eq}^>\delta\Sigma^< \,+\,
\delta{\cal G}^>\Sigma^<_{eq}\Bigr)\,+\,\Bigl(\delta\Sigma^>G_{eq}^<
\,+\,\Sigma^>_{eq}\delta{\cal G}^<\Bigr).\eeq
It is straightforward to rewrite this
in a manifestly gauge-covariant way. To this aim, 
it is enough to replace the non-covariant
fluctuations  $\delta {\cal G}$ and $\delta \Sigma$ by the
corresponding gauge-covariant expressions $\delta\acute{\cal G}$ and
$\delta\acute \Sigma$ (cf. eq.~(\ref{delG})):
\beq\label{SIGG}
\delta{\cal G}(k,X)=\delta\acute{\cal G}(k,X)-g(A(X)\cdot\del_k)G_{eq}(k),
\eeq
and similarly for $\delta \Sigma(k,X)$. One then gets:
\beq\label{COLC}
C(k,X)&=&-\,\Bigl(G_{eq}^>\,\delta\acute\Sigma^< \,+\,
\delta\acute {\cal G}^>\Sigma^<_{eq}\Bigr)\,+\,\Bigl(\delta\acute
\Sigma^>G_{eq}^< \,+\,\Sigma^>_{eq}\,\delta\acute {\cal G}^<\Bigr).\eeq
This turns out to be the same expression as above, eq.~(\ref{COL}), except
for the replacement of ordinary Wigner functions by gauge-covariant ones: 
The corrective terms in eq.~(\ref{SIGG}) cancel out in  
$C(k,X)$ since they contribute a term  proportional
to the collision term in equilibrium, which is zero:
\beq g(A(X)\cdot\del_k)\Bigl(G_{eq}^>\,\Sigma^<_{eq}
\,-\,G_{eq}^<\,\Sigma^>_{eq}\Bigr)\,=\,0.\eeq
Actually, eq.~(\ref{COLC}) can be simplified even further:
 to the order of interest,
$G_{eq}^>\simeq G_0^>$, $G_{eq}^<\simeq G_0^<$ and eq.~(\ref{GMFAQCD}) implies
that,
$\delta\acute{\cal G}^< \simeq \delta\acute {\cal G}^>=
 \delta\acute{\cal G}$. It follows that 
\beq\label{COLLIN}
C(k,x) \simeq
-\Gamma(k)\,\delta\acute{\cal G}(k,x)\,+\,
\Bigl(\delta\acute\Sigma^>(k,x) \,G_0^<(k)\,-\,
\delta\acute\Sigma^<(k,x)\,G_0^>(k)\Bigr),\eeq
with $\Gamma(k)=\Sigma^<_{eq} - \Sigma^>_{eq}$.

\begin{figure}
\protect \epsfxsize=11.cm{\centerline{\epsfbox{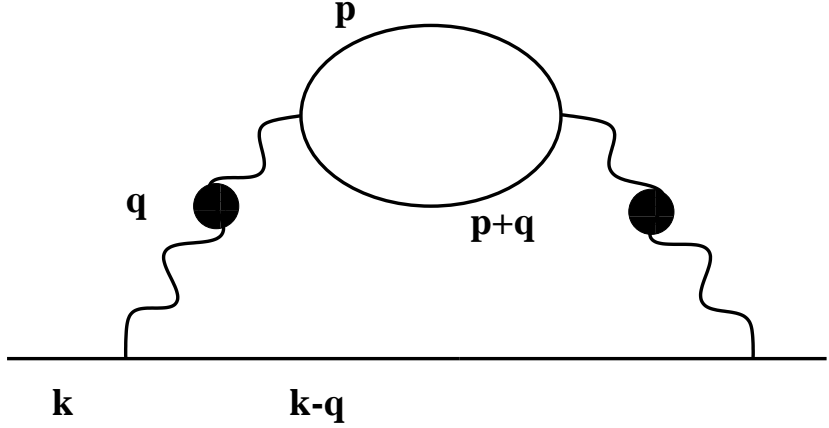}}}
         \caption{Self-energy describing collisions
in the (resummed) Born approximation. All the lines represent
off-equilibrium propagators. The continuous lines refer to the
hard colliding particles in fig.~\ref{Born}. The wavy lines
with a blob denote soft
gluon propagators dressed by the hard thermal loops.}
\label{S11}
\end{figure}

\begin{figure}
\protect \epsfxsize=14.5cm{\centerline{\epsfbox{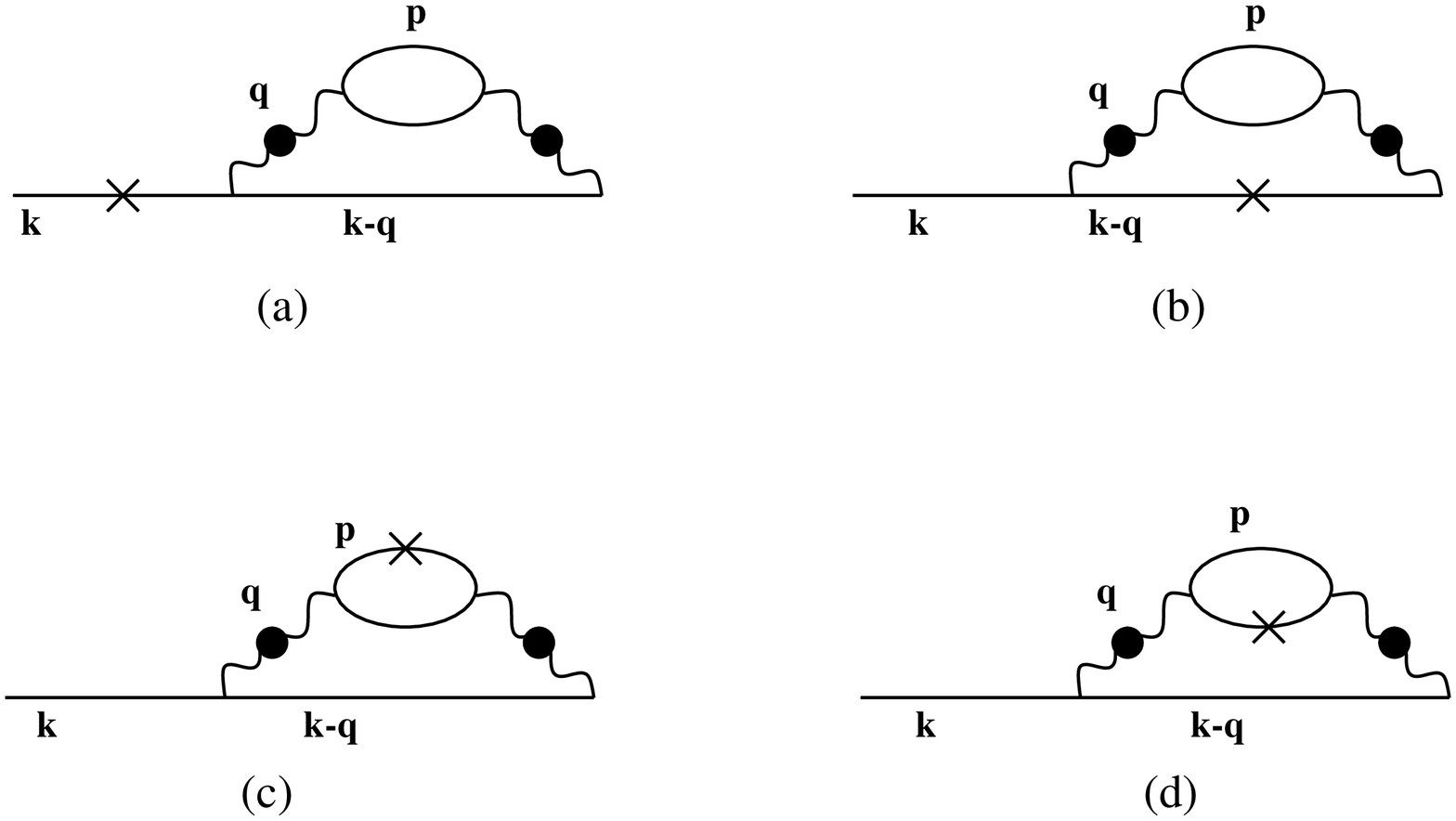}}}
         \caption{Pictorial representation of the linearized collision term.
Each one of the four diagrams correspond to off-equilibrium fluctuations in one
of the colliding fields (the one which is marked with a cross).
All the unmarked propagators are in equilibrium.}
\label{FLS}
\end{figure}

The
structure of the collision term is independent of the specific form of
the collisional self-energies $\delta\acute\Sigma^>$
and $\delta\acute\Sigma^<$, and is solely a consequence
of the (gauge-covariant) gradient expansion. However, for consistency
with the gradient expansion, 
these self-energies have to be computed to leading order in $g^2$.
As we shall verify, these are obtained from the two-loop diagram in 
fig.~\ref{S11}.
After  linearization the collision term may be represented by the four processes
displayed in fig.~\ref{FLS}, where each diagram involves a fluctuation denoted
by a cross while all other propagators are equilibrium propagators.

The collision term associated to the diagrams in fig.~\ref{FLS}
is constructed in detail in Ref. \cite{BE} and can be written as follows:
\beq\label{LINCOL}
C_{ab}(k,X)&=&
-\int {\rm d}{\cal T}\left| {\cal M}_{pk\to p'k'}\right|^2
\,N(k_0)\,N(p_0)\,
[1+N(k^\prime_0)]\,[1+N(p^\prime_0)]\nonumber\\&{}&\qquad\times
\Bigl\{N\Bigl(NW_{ab}(k,X) - (T^aT^b)_{cd}
W_{cd}(k^\prime,X)\Bigr)\,+\nonumber\\&{}&\qquad\,\,\,\,\,\,\,\,+\,
(T^aT^b)_{c\bar c}(T^cT^{\bar c})_{d\bar d}\Bigl(W_{\bar d d}(p,X)
-W_{d\bar d}(p',X)\Bigr)\Bigr\}.\eeq
In this equation, $\left| {\cal M}_{pk\to p'k'}\right|^2\propto g^4$
is the matrix element squared corresponding to the 
one-gluon exchange depicted in fig.~\ref{Born}, and
${\rm d}{\cal T}$ is a compact notation for the 
measure of the phase-space integral:
\beq\label{PHI}
\int {\rm d}{\cal T}\,\equiv\,\beta
\int\frac{{\rm d}^4p}{(2\pi)^4} \int\frac{ {\rm d}^4 q}{(2\pi)^4}\,
 \rho_0(k)\rho_0(p)\rho_0(p+q)\rho_0(k-q).\eeq
The four terms within the braces in eq.~(\ref{LINCOL})
are in one to one correspondence with the diagrams
\ref{FLS}.a, b, c and d.

The collision term (\ref{LINCOL}) is formally of order $g^4$,
since proportional to $\left| {\cal M}\right|^2$. 
However, because of the sensitivity of the phase space integrals to soft
momenta, it may
be enhanced  and become of order $g^2$. The mechanism here is similar to that
leading to the anomalous damping rate discussed in Sect. 6. And in fact the
first term in eq.~(\ref{LINCOL}) (the one involving $W_{ab}(k,x)$) 
is the same as
$-\Gamma(k)\,\delta\acute{\cal G}(k,x)$ (cf. eq.~(\ref{COLLIN})), that is,
it is proportional to the damping rate of a hard excitation.
Whether the collision integral is of order $g^2$ or $g^4$ depends however on
subtle cancellations which are studied in the next subsection.

\subsection{Coloured and colourless excitations}

The construction of the collision term  $C_{ab}(k,x)$ in eq.~(\ref{LINCOL}) 
involves two kinds of gradient expansions  \cite{BE}:
one in powers of 
$D_x/k \sim g^2$, and another 
in powers of $\del_x/q$, where $q$ is the momentum exchanged in the collision.
The latter assumes that the range of the interactions (as measured
by $1/q$) is much shorter than the range of the inhomogeneities
$\sim 1/\del_x$. It is this approximation that makes the collision term
 { local} in $x$.
As we shall argue now, this is a good approximation for colourless
fluctuations, but is only marginally correct for coloured ones.

For colourless fluctuations,
 $\delta\acute G_{ab} =\delta_{ab}\delta \acute G$, and
$W_{ab}=\delta_{ab} W$. The various colour traces
in eq.~(\ref{LINCOL}) are then elementary (e.g., $(T^aT^b)_{cc}=
N\delta_{ab}$), and yield $C_{ab}=\delta_{ab}C$, with
\beq\label{CLESS}
C(k,x)&=&
-N^2\int {\rm d}{\cal T}\left| {\cal M}_{pk\to p'k'}\right|^2
\,N_0(k_0)\,N_0(p_0)\,[1+N_0(k^\prime_0)]\,[1+N_0(p^\prime_0)]
\nonumber\\&{}&\qquad\times
\Bigl\{W(k,x)-W(k^\prime,x)+W(p,x)-W(p^\prime,x)\Bigr\}.\eeq
What is remarkable about eq.~(\ref{CLESS}) is that the
 phase-space integral is dominated by relatively
{\it hard} momentum transfers $gT\simle q\simle T$, 
even though each of the
four individual terms in the r.h.s. is actually dominated by
soft momenta. This is a consequence of
the cancellation of the leading infrared contributions among the
various terms \cite{BE}. For instance,  
for soft $q$, $W(k',x)\equiv W(k-q,x)
\approx W(k,x)$, so that the IR contributions to the first two terms
in eq.~(\ref{CLESS}) cancel each other. 
A similar cancellation occurs between the last two
terms in eq.~(\ref{CLESS}), namely $W(p,x)$ and $W(p',x)$.
Thus, in order to get the leading IR ($q\ll T$) behaviour
of the full integrand in eq.~(\ref{LINCOL}) one needs to  expand
$W(k',x)$ and $W(p',x)$ in powers of $q$. By doing so, one 
generates, in leading
order, an extra factor of  $q^2$ in the integrand  which removes the most severe
IR divergences in the collision integral. (This is the familiar factor
$1-\cos\theta\approx q^2/2$,  with $\theta$ 
the scattering  angle, which characterizes 
the transport cross sections.) 
As a result, the integral
in eq.~(\ref{CLESS}) leads to  relaxation rates typically of order 
$ g^4T\ln(1/g)$, where the logarithm
originates from screening effects at the scale $gT$.
Such rates control the transport 
coefficients like the shear viscosity \cite{Baym90,Heisel94a}
or the electric conductivity \cite{Baym97}. 
(See also Refs. \cite{Smilga90,Kraemmer94,Petit98}
where similar cancellations are identified via diagrammatic calculations
in Abelian gauge theories.) Under such conditions, 
the effects of the collisions become important for
inhomogeneities at the scale $g^4 T$; the inequality $\del_x \ll q$ is
then very well satisfied because $q$ is here relatively hard, $gT\simle q\simle
T$. However, the fact that the relaxation 
rates are not saturated by small angle
scattering  implies that to calculate them, even to leading order in $g$, 
one has to consider { all} the collisions with one particle exchange
(including, e.g.,  Compton scattering).
In terms of self-energy diagrams for the collision term, this 
means that one has to include { all} the two-loop diagrams
contributing to
$\Sigma^>$ and $\Sigma^<$, and not only the diagram in fig.~\ref{S11}.

The situation is different for { colour} relaxation. The longwavelength
colour excitations are described by a density matrix $W(k,x)$  in the adjoint
representation: $W(k,x)\equiv W_a(k,x)T^a$. The colour algebra
in eq.~(\ref{LINCOL}) can then be performed by using the following identities:
\beq\label{TRACE}
{\rm Tr} (T^aT^bT^c) \,=\,if^{abc}\,\frac{N}{2}\,,\qquad
(T^aT^b)_{c\bar c}(T^cT^{\bar c})_{d\bar d}(T^e)_{\bar d d}
\,=\,if^{abe}\,\frac{N^2}{4}\,.\eeq
The resulting collision term is $C\equiv C_a T^a$, with
\beq\label{CCOL}
C_a(k,x)&=& -N^2\int {\rm d}{\cal T}\left| {\cal M}_{pk\to p'k'}\right|^2
\,N_0(k_0)\,N_0(p_0)\,[1+N_0(k^\prime_0)]\,[1+N_0(p^\prime_0)]
\nonumber\\&{}&\qquad\times
\left\{W_a(k,x)-\,\frac{1}{2}\,
W_a(k^\prime,x)-\,\frac{1}{4}\Bigl(W_a(p,x)+W_a(p^\prime,x)
\Bigr)\right\}.\eeq
The cancellation of the leading
infrared contributions no longer takes place and we can simply set $k'=k$ and
$p'=p$ in  eq.~(\ref{CCOL}) which then simplifies to \cite{ASY98,BE}:
\beq\label{CSIMP}
C_a(k,x)\simeq -\,\frac{N^2}{2}
\int {\rm d}{\cal T}\left| {\cal M}_{pk\to p'k'}\right|^2
\,\frac{{\rm d}N}{{\rm d}k_0}\,\frac{{\rm d}N}{{\rm d}p_0}\,
\left\{W_a(k,x)\,-\,W_a(p,x)\right\}.\eeq
For soft $q$, the matrix element $|{\cal M}|^2$ 
has been already evaluated in eq.~(\ref{MBORN}):
\beq\label{COLM1}
|{\cal M}|^2\,=\,16g^4\varepsilon_k^2\varepsilon_p^2\,
 \Big|{}^*\!{\Delta}_l(q)
+ ({\bf \hat q \times v})\cdot ({\bf \hat q \times v}^\prime)\,\,
{}^*\!{\Delta}_t(q)\Big|^2,\eeq
where ${\bf v}\equiv \hat{\bf k}$ and ${\bf v}^\prime \equiv \hat{\bf p}$.
The phase-space measure (\ref{PHI}) can be similarly simplified.
This eventually yields a simpler equation for 
$W_a(k,x)$ which, remarkably, is consistent with $W_a(k,x)$ being
independent of the magnitude $|{\bf k}|$ of the hard momentum,
as in the HTL approximation (cf. eq.~(\ref{dnN})). We thus write:
\beq\label{WONSHELL}
W_{a}(k,x)= g\left\{\theta(k_0) W_{a}(x,{\bf v})-\theta(-k_0)W_{a}
(x,-{\bf v})\right\},\eeq 
where a factor of $g$ is introduced to keep in line
with the normalization in eq.~(\ref{dnN}). 
In particular, the induced colour current preserves the
structure in eq.~(\ref{j1A}).

Finally.
the Boltzmann equation reads \cite{BE}:
\beq\label{W12}
(v\cdot D_x)^{ab}W_b(x,{\bf v})&=&{\bf v}\cdot{\bf E}^a(x)-m_D^2
\frac{g^2 N T}{2}\int\frac{{\rm d}\Omega'}{4\pi}
\,\Phi({\bf v\cdot v}^\prime)\Bigl\{W^a(x,{\bf v})-
W^a(x,{\bf v}^\prime)\Bigr\}.\nonumber\\&{}&\eeq
The angular integral above runs over all the directions of the unit
vector ${\bf v}^\prime$, and $m_D$ is the Debye mass, $m_{D}^2 =
 g^2 T^2 N/3$ for the pure Yang-Mills theory. Furthermore:
\beq\label{PHII}\Phi({\bf v\cdot v}^\prime)\equiv (2\pi)^2
\int\frac{{\rm d}^4 q}{(2\pi)^4}\,
\delta(q_0- {\bf q\cdot v})
\delta(q_0- {\bf q\cdot v}^\prime)
\Big|{}^*\!{\Delta}_l(q)+ ({\bf \hat q \times v})\cdot ({\bf \hat q \times
v}^\prime)\, {}^*\!{\Delta}_t(q)\Big|^2,\,\nonumber \\
  \eeq
with the two delta functions expressing the energy conservation
at the two vertices of the scattering process in fig.~\ref{Born}.

 The  collision term in eq.~(\ref{W12}) is local in
$x$, but non-local in ${\bf v}$. As it stands, it is infrared divergent. 
At this point, one should recall that we are eventually interested in the
effective theory for the ultrasoft fields which can be separated from the soft
degrees of freedom that we are ``eliminating'' by   an
intermediate scale
$\mu$ such that $g^2T\ll\mu\ll gT$.
This scale acts as an IR cutoff for the collision term, which is
therefore finite, but logarithmically dependent on $\mu\,$. For instance,
the damping rate of a hard gluon, given by the first term 
(local in ${\bf v}$) of the collision integral is :
\beq\label{gamma1}
m_D^2\,\frac{g^2 N T}{2}\int\frac{{\rm d}\Omega'}{4\pi}
\,\Phi({\bf v\cdot v}^\prime)\,=\,\frac{\Gamma(k_0=k)}{4k}\,=
\,\gamma\,,\eeq
(Up to a colour factor, 
this is the same equation as eq.~(\ref{G2L}). Note also that
a cancellation has
taken place making the contribution of the damping rate to
the collision integral only half of what it
would normally give; compare, in this respect, eqs.~(\ref{W12}) and
(\ref{gamma1}) above to eq.~(\ref{b2}) in Sect. 2.3.4.)
The integral over ${\bf v}'$ can be analytically computed,
with the simple result \cite{TBN,USA} (with $\alpha=g^2/4\pi$)
\beq\label{G1loop}
\gamma\,=\, \alpha NT\ln\frac{m_D}{\mu}\,.\eeq
Note that for 
colour excitations at the scale $g^2 T$, the inequality $\del_x \ll q$ is only
marginally satisfied since the  
collision term  is logarithmically 
sensitive to all momenta $\mu\simle q\simle g T$.

We have no such a simple exact result for the full quantity
$\Phi({\bf v\cdot v}^\prime)$, but it is nevertheless
straightforward to extract its $\mu$
dependence from eq.~(\ref{PHII}): this is obtained by retaining
only the singular piece of the matrix element for magnetic
scattering, namely using  eq.~(\ref{deltaT}) to get:
\beq\label{LLPHI}
\Phi({\bf v\cdot v}^\prime)\,\simeq\,
\frac{2}{\pi^2 m_D^2}\,
\frac{({\bf v\cdot v}^\prime)^2}{\sqrt{1-({\bf v\cdot v}^\prime)^2}}\,
\ln\,\frac{m_D}{\mu}\,,
\eeq

By using eq.~(\ref{gamma1}), the Boltzmann equation for 
colour relaxation is finally written as:
\beq\label{W10}
(v\cdot D_x)^{ab}W_b(x,{\bf v})&=&{\bf v}\cdot{\bf E}^a(x)-
\gamma\Bigl\{W^a(x,{\bf v})\,-\,\langle W^a(x,{\bf v})\rangle
\Bigr\},\eeq
where we have introduced the following compact notation: for an 
arbitrary function of ${\bf v}$, say $F({\bf v})$, we denote by
$\langle F({\bf v})\rangle$ its angular average with weight
function $\Phi({\bf v\cdot v}^\prime)$ (that is, its average
with respect to the scattering cross section):
\beq\label{AVF}
\langle F({\bf v})\rangle\,\equiv\,\frac{
\int\frac{{\rm d}\Omega'}{4\pi}\,\Phi({\bf v\cdot v}^\prime)
F({\bf v}')}{\int\frac{{\rm d}\Omega'}{4\pi}\,
\Phi({\bf v\cdot v}^\prime)}\,,\eeq
which is still a function of ${\bf v}$. 
{}From eq.~(\ref{W10}), it is clear that the quasiparticle damping rate
$\gamma$ sets also the time scale for colour relaxation:
 $\tau_{col}\sim 1/\gamma \sim 1/(g^2T\ln(1/g))$ 
\cite{Gyulassy93}.

We conclude this subsection by noticing that eq.~(\ref{W10})
is invariant under the gauge transformations of the
{ background field}, and also with respect to
the choice of a gauge for the  fluctuations  with momenta $k\simge \mu$.
 In Ref. \cite{BE},
eq.~(\ref{W12}) was derived in Coulomb gauge, but we expect it
to be gauge-fixing independent 
since it involves only the off-equilibrium fluctuations of the
(hard) transverse gluons, together with the (gauge-independent)
matrix element squared (\ref{COLM1}). Finally we note that the
Boltzmann equation (\ref{W12}) (with the collision term 
approximated as in eq.~(\ref{LLPHI})) has been also obtained
in Ref. \cite{Manuel99} by using a classical transport theory of
colour particles \cite{Wong,Heinz83,Kelly94}.

\subsection{Ultrasoft amplitudes}

By solving the Boltzmann equation,
one can obtain $W_a(x,{\bf v})$, and thus
the induced current $j^\mu_a(x)$ as a functional of the fields $A^\mu_a(x)$,
which can be expanded in the form:
\beq\label{exp}
j^{a}_\mu \,=\,\Pi_{\mu\nu}^{ab}A_b^\nu
+\frac{1}{2}\, \Gamma_{\mu\nu\rho}^{abc} A_b^\nu A_c^\rho+\,...
\eeq
The coefficients $\Pi_{\mu\nu}^{ab}$,
$\Gamma_{\mu\nu\rho}^{abc}$, etc., in this expansion are
one-particle-irreducible amplitudes for the ultrasoft fields
in thermal equilibrium (cf. Sect. 5.1), and will be referred to as the
{\it ultrasoft amplitudes} (USA) in what follows. 
These are the generalizations of the HTL's to the case 
where the external legs carry momenta of order $g^2T$ or less.
Specifically, these are the leading contributions of the hard and 
soft degrees of freedom to the amplitudes of the ultrasoft fields.

\subsubsection{General properties}

The ultrasoft amplitudes share many of the remarkable 
properties of the HTL's: {\it i}) They are gauge-fixing independent
(like the Boltzmann equation itself),
{\it ii}) satisfy the simple Ward identities shown
in eq.~(\ref{Wtr}) (these follow from the conservation
law $D_\mu \, j^\mu=0$ for the current), and {\it iii}) 
reduce to the usual  Debye mass $m_D^2$ 
(cf. eq.~(\ref{jstat})) in the static limit $\omega \to 0$.
To verify this last point, it is convenient to use 
the decomposition (\ref{defa}) for $W^a(x,{\bf v})$:
\beq\label{CALA}
W^a(x,{\bf v})\,\equiv\,-A_0^a(x)\,+\,{\cal A}^a(x,{\bf v}).\eeq
The first term does not contribute to the collision term since
$A_a^0(x)\,-\,\langle A^0_a(x)\rangle\,=\,0$. One is then left with  the
following equation for ${\cal A}^a(x,{\bf v})$:
\beq\label{W11}
(v\cdot D_x)^{ab}{\cal A}_b(x,{\bf v})&=&\del^0(v\cdot A^a)
\,-\,\gamma\Bigl\{{\cal A}^a(x,{\bf v})\,-\,\langle 
{\cal A}^a(x,{\bf v})\rangle \Bigr\}.\eeq
Thus, for time-independent fields
$A_\mu^a({\bf x})$, the homogeneous eq.~(\ref{W11}) with
retarded boundary conditions admits the trivial solution
${\cal A}^a(x,{\bf v})=0$, and therefore
\beq\label{jstat1}
j_\mu^{a}({\bf x})
\,=\,-\,\delta_{\mu 0}\,m^2_D A_0^a({\bf x})\,,\eeq
as in the HTL approximation (cf. eq.~(\ref{jstat})):
 all the ultrasoft vertices
with $n\ge 3$ external lines vanish, while $\Pi_{\mu\nu}
(\omega=0,{\bf p})=-\delta_{\mu 0}\delta_{\nu 0}\,m_D^2$.
Thus, to this order, the physics of the static Debye screening
is not affected by the collisions among the hard particles.
This is consistent with the results in \cite{Rebhan93,BN94}
according to which the first correction to $m_D^2$,
of order $g^3T^2\ln(1/g)$, is due to soft and ultrasoft loops
(cf. fig.~\ref{QCDelec}).

Consider now time-dependent  fields. In order to analyze the solutions of the 
Boltzmann equation (\ref{W10}), it is convenient
to write the collision term as an operator acting on
$W^a(x,{\bf v})\,$:
\beq\label{CLAN}
C_a(x,{\bf v})&=&
-\gamma\Bigl\{W_a(x,{\bf v})\,-\,\langle W_a(x,{\bf v})\rangle\Bigr\}
\nonumber\\&\equiv&-\int{\rm d}^4 x'\int\frac{{\rm d}\Omega'}{4\pi}\,
{\cal C}_{ab}(x,x';{\bf v,v'})\,W^b(x',{\bf v}')\,\equiv\,-\,({\cal C}
W)_a(x,{\bf v}),\eeq
with the following kernel, which is non-local but symmetric
in ${\bf v}$ and ${\bf v}'$:
\beq\label{COLKER}
{\cal C}_{ab}(x,x';{\bf v,v'})=-\,
\,\frac{\delta \,C^a(x,{\bf v})}
{\delta\, W^b(x',{\bf v'})}=\delta^{ab}\delta^{(4)}(x-x')\,
\gamma\,\Bigl\{\delta^{(2)}({\bf v,v'})-\Bigl\langle
\delta^{(2)}({\bf v,v'})\Bigr\rangle\Bigr\}.\,\,\eeq
Then, the solution to the Boltzmann equation (\ref{W10}) can
be formally written as:
\beq\label{Wsol}
W(x,{\bf v})\,=\,\int{\rm d}^4 x'\int\frac{{\rm d}\Omega'}{4\pi}\,
\Bigl\langle x,{\bf v}\Big |\frac{1}{v\cdot D+{\cal C}}\Big |
x',{\bf v'}\Bigr\rangle\,{\bf v}'\cdot{\bf E}(x')\,\,.\eeq
This expression exhibits in particular the role of the collisions
in smearing out the divergences of the HTL's at $v\cdot D \to 0$
(cf.  Sect. 5.3.3).

\subsubsection{The ultrasoft polarization tensor and its
diagrammatic interpretation}

The solution (\ref{Wsol}) can be used to
derive an expression for the ultrasoft polarization tensor 
$\Pi_{\mu\nu}$. To this aim, one needs the induced current
only to linear order in $A^\mu$, so one can replace
$v\cdot D$ by $v\cdot\del$ in eq.~(\ref{Wsol}), and use 
the momentum representation. It is also convenient to
use the decomposition (\ref{CALA}), so as to obtain
the tensor $\Pi_{\mu\nu}$ in a manifestly symmetric form (compare with
eq.~(\ref{Pi}):
\beq\label{PIUSA}
\Pi^{ab}_{\mu\nu}(P)\,=\,m_D^2\,\delta^{ab}
\left \{-\delta_{\mu 0}\delta_{\nu 0} \,+\,\omega
 \int\frac{{\rm d}\Omega}{4\pi}\int\frac{{\rm d}\Omega'}{4\pi}\,
\Bigl\langle v\Big|v_\mu\,\frac{1}{v\cdot P+i{\cal C}}\,v_\nu\Big|
v'\Bigr\rangle\right\}.
\eeq

A diagrammatic interpretation of this formula is obtained by formally 
expanding out the collision term. One thus obtains 
$\Pi_{\mu\nu}\,=\,\Pi_{\mu\nu}^{(0)}+\Pi_{\mu\nu}^{(1)}+\ldots\,$,
where $\Pi_{\mu\nu}^{(0)}$ is the HTL given by eq.~(\ref{Pi}), and
\beq\label{PI1} \Pi_{\mu\nu}^{(1)}(P)&=&-i\gamma \omega m_D^2
\int\frac{{\rm d}\Omega}{4\pi}\,
\frac{v_\mu} {v\cdot P}\left\{
\frac{v_\nu}{v\cdot P}\,-\,\left\langle
\frac{v_\nu}{v\cdot P}\right\rangle\right\}\nonumber\\&=&
-i \omega m_D^4\,\frac{g^2 N T}{2}\int\frac{{\rm d}\Omega}{4\pi}
\int\frac{{\rm d}\Omega'}{4\pi}
\,\Phi({\bf v\cdot v}^\prime)\,
\frac{v_\mu} {v\cdot P}\left\{
\frac{v_\nu}{v\cdot P}\,-\,
\frac{v^{\prime}_\nu}{v'\cdot P}\right\},\eeq
where in the second line we have used the definition
(\ref{AVF}) of the angular averaging together with eq.~(\ref{gamma1})
for $\gamma$. For time-like momenta $\Pi_{\mu\nu}^{(0)}$
is a real quantity while the first-order iteration
in eq.~(\ref{PI1}) is purely imaginary, reflecting the dissipative role of the
collisions.

\begin{figure}
\protect \epsfxsize=8.5cm{\centerline{\epsfbox{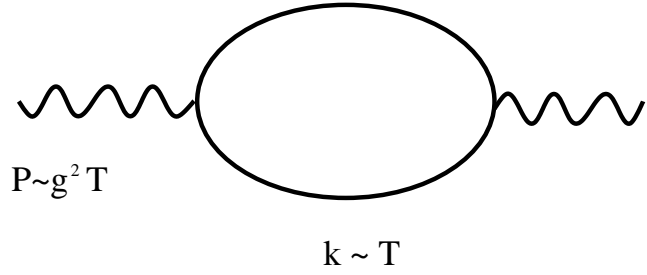}}}
         \caption{A generic one-loop diagram contributing
to the HTL $\Pi_{\mu\nu}^{(0)}$.
The internal continuous lines denote hard transverse gluons;
the external wavy line is an ultrasoft gluon.}
\label{FIGPI}
\end{figure}

Higher order iterations $\Pi_{\mu\nu}^{(N)}$,
 proportional to $\gamma^N$,
can be written down similarly.
Note however that, for $P\sim g^2 T$, the contribution in
eq.~(\ref{PI1})  is of the same order in $g$ as the HTL (\ref{Pi}). 
Thus, the iterative expansion is only formal.
It is only used here for the comparison with 
perturbative calculations of $\Pi_{\mu\nu}$ in terms of Feynman diagrams,
and to identify the nature of the resummations  achieved by the Boltzmann
equation \cite{BE,USA} (see also Refs. \cite{Jeon93,Jeon95,JY96}).
\begin{figure}
\protect \epsfxsize=13.5cm{\centerline{\epsfbox{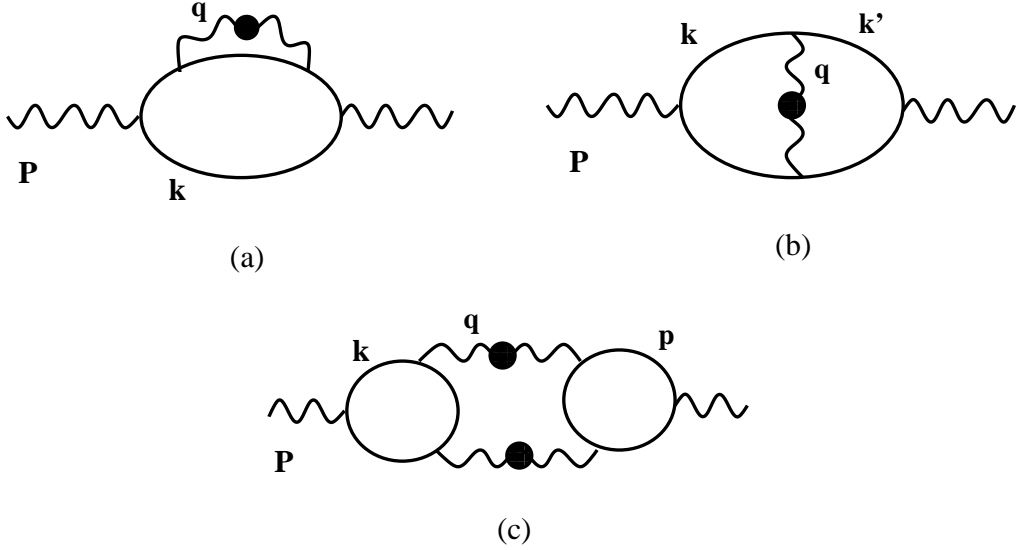}}}
         \caption{The diagrams contributing to the first iteration
$\Pi_{\mu\nu}^{(1)}(P)$ of the polarization tensor, eq.~(\ref{PI1});
the continuous lines are hard transverse gluons; the internal wavy
lines are soft gluons with momenta $g^2 T \simle q \simle gT$ 
(with the blob denoting HTL resummation).}
\label{PI2}
\end{figure}

Thus the zeroth order iteration
$\Pi_{\mu\nu}^{(0)}$ is the  HTL,
which is the  leading order contribution in an expansion in powers of $P/k$ of
the one loop diagram of fig.~\ref{FIGPI}.
The first order iteration $\Pi_{\mu\nu}^{(1)}$ is obtained
via a similar expansion from the three diagrams displayed in
figs.~\ref{PI2} \cite{Bodeker99}.
The internal wavy lines in these diagrams are soft gluons
dressed with the HTL. In the language of the Boltzmann
equation, these are the soft quanta
exchanged in the collisions between the hard particles 
(the latter being represented by the
continuous lines in figs.~\ref{PI2}). 
As shown in \cite{BE,USA},  these are precisely the
diagrams generated by the first iteration of the collision term in
eq.~(\ref{LINCOL}).
\begin{figure}
\protect \epsfxsize=14.cm{\centerline{\epsfbox{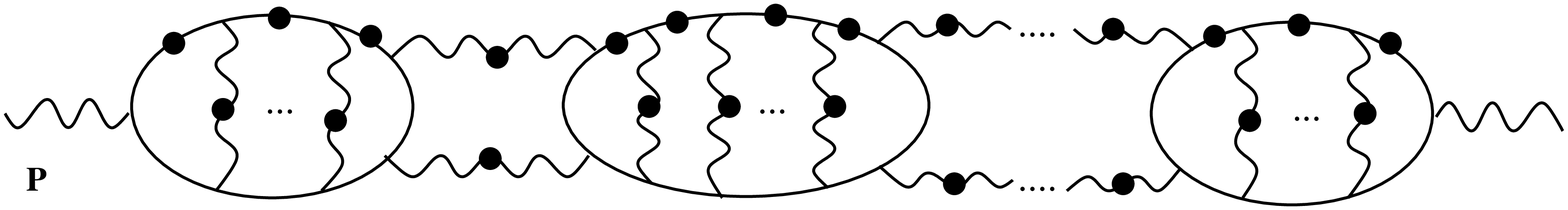}}}
         \caption{A generic ladder diagram contributing to the
ultrasoft polarization tensor, as obtained from the Boltzmann equation.}
\label{LAD}
\end{figure}

\begin{figure}
\protect \epsfxsize=14.cm{\centerline{\epsfbox{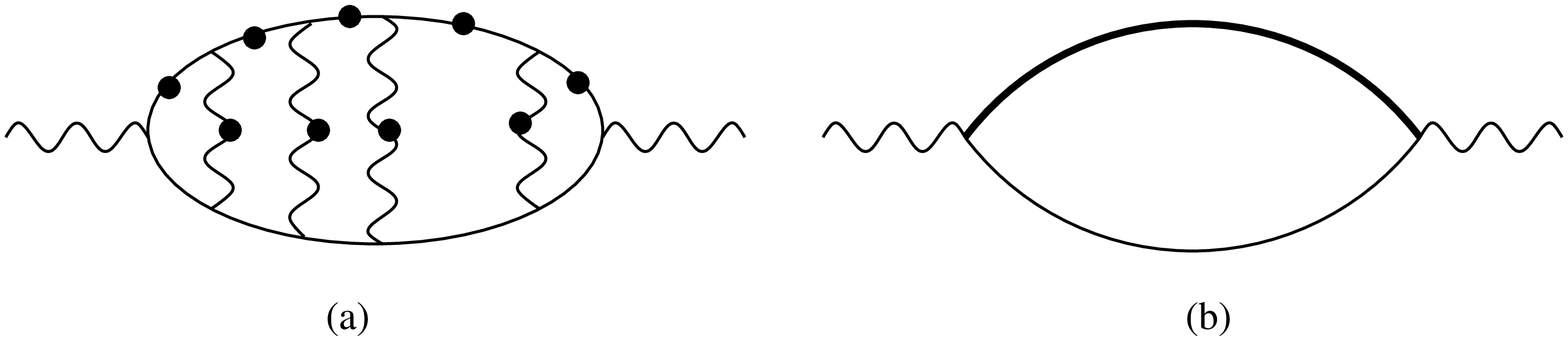}}}
         \caption{(a) A ladder diagram
generated by iterations of the first
two pieces, $W_a(k,x)$ and $(-1/2)W_a(k^\prime,x)$,
 of the collision term (\ref{CCOL}); the smooth lines
are eikonal propagators with a damping rate $2\gamma$.
(b) The sum of all the ladders in (a);
the thick line is an eikonal propagator 
with a damping rate $\gamma$.}
\label{F}
\end{figure}

The higher order iterations $\Pi_{\mu\nu}^{(N)}$ with $N\ge 2$ 
can be similarly given a diagrammatic interpretation, by
iterating the diagrams for the collision
term in fig.~\ref{FLS} \cite{BE}. A typical diagram contributing
to $\Pi_{\mu\nu}$ which is obtained in this way is shown in fig.~\ref{LAD}.
The continuous lines with a blob represent the following { dressed}
eikonal propagator (compare with eq.~(\ref{GR0})):
\beq\label{DEIK0}
\frac{-1}{v\cdot P + 2i\gamma}
\,,\eeq
as obtained after resumming the self-energy corrections
to the hard propagators, i.e., by iterating the self-energy
insertion in fig.~\ref{PI2}.a, or, equivalently,
the first piece $W_a(k,x)$ of the collision term (\ref{CCOL}).
The continuous lines without a blob in fig.~\ref{LAD}
are thermal correlation functions like $G^>_0$ and 
$G^<_0$, or derivatives of them. The vertex corrections
(the wavy lines, or ladders) { inside}
any of the hard loops in fig.~\ref{LAD} are generated by iterating 
the second piece, $(-1/2)W_a(k^\prime,x)$, of the collision term
(\ref{CCOL}). The net effects of these vertex corrections
is to replace $2\gamma$ by $\gamma$ in the eikonal propagator
(\ref{DEIK0}). This relies on the approximation
$W_a(k,x)- (1/2)W_a(k^\prime,x)\approx (1/2)W_a(k,x)$ (which has
been used in going from eq.~(\ref{CCOL}) to (\ref{CSIMP})), 
and is illustrated in fig.~\ref{F}, where the thick internal 
line denotes the following eikonal propagator:
\beq\label{AEIK0}
\frac{-1}{v\cdot P + i\gamma}
\,.\eeq
Finally, the wavy lines relating {\it different}
 hard loops in fig.~\ref{LAD} are generated by iterating 
the diagrams in figs.~\ref{FLS}.c and ~\ref{FLS}.d, or, equivalently,
the last two pieces $(-1/4)(W_a(p,x)+W_a(p^\prime,x))\approx
(-1/2)W_a(p,x)$ of the collision term (\ref{CCOL}).

A similar diagrammatic interpretation holds for the
$n$-point ultrasoft vertices 
(see Ref. \cite{USA} for more details).

\subsubsection{Leading log approximation and colour conductivities}

A simple approximation where 
the polarization tensor can be calculated in closed
form is the  ``leading-logarithmic approximation'' 
\cite{Bodeker,ASY98,Bodeker99}, which relies on the
following observation: for colour inhomogeneities
at the scale $g^2T$, the collision term, which is 
of order $\gamma\sim g^2T\ln(1/g)$, wins over the drift term
 $v\cdot D_x\sim g^2T$ by a ``large'' logarithm $\ln(1/g)$.
Thus, to leading logarithmic accuracy (LLA), one can neglect
the drift term in the l.h.s. of eq.~(\ref{W10}) and, for consistency,
use  the approximation (\ref{LLPHI}) 
in the collision term. Then the Boltzmann equation (\ref{W10})
reduces to:
\beq\label{W2}
{\bf v}\cdot{\bf E}^a(x)\,=\,\gamma 
\Bigl\{W^a(x,{\bf v})\,-\,\langle W^a(x,{\bf v})\rangle
\Bigr\}.\eeq
The electric field in this equation is assumed to be transverse. Neglecting the
drift term is in general not allowed for longitudinal fields; in particular we
have seen that the collision term vanishes for static longitudinal fields. 

 The equation (\ref{W2}) has
the following solution:
\beq\label{LLW}
W^a({\bf v})\,=\,\frac{{\bf v}\cdot {\bf E}^a}{\gamma}\,.
\eeq 
This is easily verified: the above
 $W^a({\bf v})$ is an odd function of ${\bf v}$, 
while the approximate $\Phi ({\bf v\cdot v}^\prime)$ in
eq.~(\ref{LLPHI}) is even, so that 
$\langle W^a({\bf v}) \rangle\approx 0$ to LLA. 
(Note that this  would not hold 
with the whole collisional cross-section in eq.~(\ref{PHII}).)
After insertion in eq.~(\ref{j1A}), the approximation (\ref{LLW})
for $W^a({\bf v})$ generates the following, local, colour current:
\beq\label{JLL}
{\bf j}^a =\sigma{\bf E}^a\,,\qquad {\rm with}\quad
\sigma\,\equiv\,\frac{m_D^2}{3\gamma}\,=\,
\frac{4\pi T}{9}\,\frac{1}{\ln(m_D/\mu)}\,.\eeq
Although $\sigma$ 
is not really a physically measurable quantity, at the level
of approximation at which we are working it behaves as such.
One could therefore expect it to be independent of the 
arbitrary scale $\mu$ separating
soft and ultrasoft degrees of freedom.  
For that to happen, however, one needs to
perform a complete calculation at the scale $g^2T$, that is,
one needs to compute the  contributions of the ultrasoft fields
themselves. These are given by loop diagrams of the ultrasoft
effective theory, with 
$\mu$ acting then as an ultraviolet (UV) cutoff. Because the UV divergences in
the effective theory are only logarithmic (this can be verified by power
counting), the loop corrections in the effective theory will
lead to terms proportional to 
$\ln \mu$. Without doing any calculation, one expects these terms 
to cancel the $\mu$-dependence in the ultrasoft amplitudes, leaving
in place of $\mu$ the natural 
scale of the effective theory, that is, $g^2T$.
To LLA, the constant term under $\ln(1/g)$ can be neglected.
Thus, in LLA, the color conductivity is obtained by simply replacing
$\ln(m_D/\mu)\approx \ln(1/g)$ in eq.~(\ref{JLL})
 \cite{Gyulassy93,Heisel94,Bodeker,ASY98}.

We should notice here an important difference 
with  the HTL's.  Recall that the HTL's, obtained after integrating
out the hard ($p\sim T$) modes, are { leading order} 
amplitudes at the scale $gT$ in a strict expansion in powers of
$g$. Moreover, to this order, they are independent of
the scale $\Lambda$ separating $T$ from $gT$.
This is so since, if computed with an infrared cutoff $\Lambda$,
the HTL's would depend { linearly} of this scale
(see, e.g., eq.~(\ref{MDCL1})), and the corresponding dependence
would be suppressed by a factor $\Lambda/T$ as compared to
the leading order contribution, of order $g^2 T^2$.
Because of that, in writing the HTL's we have generally omitted 
their explicit dependence on the separation scale $\Lambda$.

The ultrasoft amplitudes, on the other hand, depend
logarithmically on the separation scale $\mu$ ($g^2T\ll \mu\ll gT$),
so $\mu$ has to be kept explicitly as an IR cutoff when computing
the USA's. This logarithmic dependence also
implies that the contributions of the ultrasoft fields to the
respective amplitudes are of
the { same} order in $g$ as the USA's themselves;
thus, the latter are { not} dominant quantities, but only
part of the full amplitudes at the scale $g^2T$.
Now, the remaining contributions, due to the
interactions of the ultrasoft fields, are fully non-perturbative,
and can in general be obtained only from a numerical calculation
using for instance the lattice techniques.
Thus, in order to compute ultrasoft physical correlations
{ already to leading order}, one
has to perform a numerical calculation within the
effective theory. A possible way to do that will be discussed
in the next subsection.

\subsection{The Boltzmann-Langevin equation: noise and correlations}

In order
to compute thermal correlations of the ultrasoft fields in real time,
like, e.g., $\langle A^i(x) A^j(y)\rangle$, one could 
rely on the fluctuation-dissipation theorem (see Sect. 2.1.3)
in order to construct
the thermal correlators from the retarded response functions;
these can in turn be obtained from
the solution of the Boltzmann equation with retarded 
boundary conditions (cf. eq.~(\ref{exp})).
For instance, the two-point function is obtained as
(in the classical approximation where $G^>\approx G^< \approx G_{cl}$):
\beq \label{FDTD}
G_{cl}^{\mu\nu}(x,y)\equiv
\langle A^\mu(x) A^\nu(y)\rangle\,=\,
\int\frac{{\rm d}^4k}{(2\pi)^4}\,\,{\rm e}^{-i k\cdot(x-y)}\,\,
{T\over k_0}\,\,\rho^{\,\mu\nu}(k),\eeq
where $\rho^{\mu\nu}=2{\rm Im} \,G_R^{\mu\nu}$,
and $G_R^{-1}=G_0^{-1}+\Pi_R$, with $\Pi_R^{\mu\nu}$ 
(formally) constructed in Sec. 7.3.2.
Similar representations can be written for 
the thermal self-energy:
\beq\label{FDTPI}
\Pi^{\mu\nu}(x,y)\,=\,\langle j^\mu(x)j^\nu(y)\rangle
\,=\,
\int\frac{{\rm d}^4k}{(2\pi)^4}\,\,{\rm e}^{-i k\cdot(x-y)}\,\,
{T\over k_0}\,\,\Bigl(-2{\rm Im} \,\Pi_R^{\,\mu\nu}(k)\Bigr),\eeq
and also for the higher order n-point correlators.
In practice, however, this strategy is not useful because 
the  Boltzmann equation in general can only
be solved numerically, and it is not easy to extract the spectral
function; besides, the correlators one needs to evaluate do not necessarily
have the simple form of a product of gauge fields (see, e.g.,
eq.~(\ref{Baryorate})). 

An alternative procedure relies on the fact
 that the ultrasoft dynamics is that of
classical fields which obey the equations of motion discussed in Sect. 4.
Correlation functions can then be obtained by averaging over the
initial conditions appropriate products of the fields which solve
the classical equations of motion (cf. Sect. 4.4).
Following B\"odeker~\cite{Bodeker}, let us split the classical fields  into soft
and ultrasoft components ($A\to A+a$, $W\to W+w$, etc), where capitals 
(lower cases) denotes ultrasoft (soft) components.  The equations for the
ultrasoft fields take the form:
\beq\label{avaUS}
D_\nu F^{\mu\nu}&=&m^2_D\int\frac{{\rm d}\Omega}{4\pi}
\,v^\mu\,W(x,{\bf v}),\nonumber\\
(v\cdot D_x)^{ab} W_b&=&{\bf v}\cdot {\bf E}^a
+g f^{abc} \left(v\cdot a^b w^c\right)_{\!\mu},
\eeq
where 
$\left(a^b w^c\right)_{\!\mu}$ means that only the ultrasoft components (with
$k\simle \mu$) have to be kept in the product of fields. 
The soft fields $a$ and
$w$ obey themselves equations of motion which relates them to the ultrasoft
background fields. In leading order the self-interactions of the soft
fields can be neglected and their equations of motion read simply:
\beq\label{avasoft}
\del^\nu f_{\nu\mu}^a(x)&=&
m^2_D\int\frac{{\rm d}\Omega}{4\pi}
\,v_\mu\,w^a(x,{\bf v}),\nonumber\\
v\cdot\del_x w^a(x,{\bf v})&=&{\bf v}\cdot{\bf e}^a(x)+h^a(x,{\bf v}),\eeq
where $f^{\nu\mu}=\del^\nu a^\mu -\del^\mu a^\nu$ and 
$h^a$ describes the coupling between soft and 
ultrasoft fields:
\beq
h^a(x,{\bf v})\,\equiv\,gf^{abc}\left[ v\cdot A^b(x)w^c(x,{\bf v})+
v\cdot a^b(x)W^c(x,{\bf v})\right]\,.\eeq
Both the soft and the ultrasoft fields
are thermal fluctuations whose typical amplitudes have been estimated
in Sect. 1.2 as $a\sim g^{1/2}T$ and $A \sim gT$.
The corresponding estimates for the fields $w$ and $W$ follow from
eq.~(\ref{WN0}), which gives $W\sim (Tk^3/m_D^2)^{1/2}$ where $k$ is a
typical spatial gradient: thus for $k\sim gT$, $w\sim g^{1/2} T$ and for
$k\sim g^2T$, $W\sim g^2T$. Since $e\sim g^{3/2} T^2$, while $h\sim gAw \sim
g^{5/2} T^2$ at most, 
eqs.~(\ref{avasoft}) can be solved perturbatively in
$h$ for arbitrary initial conditions for the soft fields $a$ and $w$.
By inserting the corresponding solutions into eqs.~(\ref{avaUS}) and
performing the average over the initial conditions for the
soft fields, one recovers the Boltzmann equation from the second
equation (\ref{avaUS}) \cite{Bodeker}. Thus, it does not make any
difference whether one eliminates the soft fields in the quantum theory or in
the classical one. However, as noticed by B\"odeker, the equation of motion for
$W$ contains a source term independent of the ultrasoft fields: the
term $(aw)$ where $a$ and $w$ are the solution of their equations of motion in
the absence of ultrasoft background, i.e. the solution of the second equation
(\ref{avasoft}) with $h=0$. The average over the soft initial conditions of this
terms vanishes, but it is present for arbitrary conditions, and it has a
non-vanishing correlator. Such a term plays the role of a noise term and can be
used in a Boltzmann-Langevin equation to effectively perform the
averaging over the ultrasoft initial conditions. Note that this
averaging is not a trivial task here: 
While the effective theory at the scale $gT$ could be put in
a Hamiltonian form, thus providing the Boltzmann weight for the
initial conditions, no such a simple structure exists in the
effective theory at the scale $g^2T$.

Having identified the strategy that we wish to follow,
we can go back to general principles and use
the fluctuation-dissipation theorem together with the known
structure  of the collision term in the Boltzmann equation (\ref{W10})
in order to 
deduce the statistics of the noise term in the Boltzmann-Langevin equation.
Consider then such an equation:
\beq\label{WL0}
(v\cdot D_x)^{ab}W_b(x,{\bf v})\,+\,
\gamma\Bigl\{W^a(x,{\bf v})\,-\,\langle W^a(x,{\bf v})\rangle
\Bigr\}\,=\,{\bf v}\cdot{\bf E}^a(x)\,+\,\nu^a(x,{\bf v}).\eeq
where $\nu_a(x,{\bf v})$ is the noise term, to be constrained
by the collision term in eq.~(\ref{CLAN}).
The latter has the following properties:
It is ({\it i}) linear in the colour distribution  $W^a(x,{\bf v})$, 
({\it ii})  local in space and time, ({\it iii})  diagonal in colour,
({\it iv}) non-local in the velocity ${\bf v}$, 
and ({\it v}) independent of the gauge mean fields $A^\mu_a$.
Correspondingly, the noise can be chosen as Gaussian, ``white'' 
(i.e., local in $x^\mu=(t,{\bf x})$),
and colourless (i.e., diagonal in colour), but non-local in ${\bf v}$.
That is, its only non-trivial correlator is the two-point
function $\langle \nu_a(x,{\bf v})\nu_b(x',{\bf v}')\rangle$,
which is of the form:
\beq\label{XIXI0}
\langle \nu_a(x,{\bf v})\nu_b(x',{\bf v}')\rangle
\,=\, {\cal J}({\bf v,v'})\,\delta^{ab}\,\delta^{(4)}(x-x'),\eeq
with ${\cal J}({\bf v,v'})$ independent
of the gauge fields, and therefore also independent of $x$
(since the background field is the only source of inhomogeneity
in the problem). 

The following steps are quite
similar to the discussion in Sec. 4.6.3.
According to eq.~(\ref{WL0}), there are two sources
for colour excitations $W_a(x,{\bf v})$ at the scale $g^2T$: 
the mean field ${\bf E}^a(x)$
and the noise term $\nu_a(x,{\bf v})$.
The full solution can thus be written as:
\beq
W^a(x,{\bf v})\,=\,W_{ind}^a(x,{\bf v})\,+\,
{\cal W}^a(x,{\bf v}),
\eeq
where $W^{ind}_a(x,{\bf v})$ is the solution in the absence
of noise (i.e., the solution to the Boltzmann equation (\ref{W10})),
and ${\cal W}^a(x,{\bf v})$ is a fluctuating piece satisfying
\beq\label{CWL0}
(v\cdot D_x)^{ab}{\cal W}_b(x,{\bf v})\,+\,
\gamma\Bigl\{{\cal W}^a(x,{\bf v})\,-\,\langle 
{\cal W}^a(x,{\bf v})\rangle
\Bigr\}\,=\,\nu^a(x,{\bf v}).\eeq
Thus, ${\cal W}^a(x,{\bf v})$ is proportional to $\nu$, and
generally also dependent upon the mean field $A^\mu$.
The colour current is correspondingly decomposed as:
\beq\label{jind}
j_\mu^a(x)\,=\,m_D^2\int\frac{{\rm d}\Omega}{4\pi}
\,v_\mu\,\Bigl(
W^{ind}_a(x,{\bf v})\,+\,{\cal W}_a(x,{\bf v})\Bigr)
\,\equiv\,j_{\mu}^{ind\,a}(x)\,+\,\xi_\mu^a(x),
\eeq
with $\xi_\mu^a(x)$ denoting the fluctuating current,
which acts as a noise term in the Yang-Mills 
equation\footnote{Note that current conservation becomes more
subtle in the presence of the noise \cite{Bodeker}.} :
\beq\label{YMnoise}
 (D^\nu F_{\nu\mu})^a(x)&=&j_\mu^{ind\,a}(x)\,+\,\xi_\mu^a(x),
\eeq 
and generates the thermal correlators
 $\langle A^\mu(x) A^\nu(y)\,\cdots\,A^\rho(z)\rangle$ of the
ultrasoft fields. The correlators of $\xi_\mu$ can be
obtained from the original two-point function
$\langle \nu\nu\rangle$ in eq.~(\ref{XIXI0}) by solving 
the equations of motion. A priori, this is complicated by
the non-linear structure of the equations.

However, since the unknown function  ${\cal J}({\bf v,v'})$  
is independent of the background field $A^\mu_a$, it can be obtained
already in the weak field limit, that is, by an analysis of the
linearized equations of motion. 
In this limit, $j^\mu_{ind}=\Pi^{\mu\nu}_R A_\nu$, with
$\Pi^{\mu\nu}_R$ as constructed in Sec. 7.3.2, while the 
fluctuation-dissipation theorem (\ref{FDTPI}) implies:
\beq\label{zz}
\langle \xi^\mu_a(P)\xi^{\nu\,*}_b(P')\rangle\,=\,
(2\pi)^4\delta^{(4)}(P+P')\,\delta_{ab}\,
\left(-2\,\frac{T}{\omega}\,\right)
{\rm Im}\,\Pi^{\mu\nu}_R(\omega,p).\eeq
Furthermore,  ${\cal W}$ satisfies the
linearized version of  eq.~(\ref{CWL0})
(in momentum space, and with colour indices omitted): 
\beq\label{WM}
(v\cdot P){\cal W}(P,{\bf v})\,+\,i
\gamma\Bigl\{{\cal W}(P,{\bf v})\,-\,\langle 
{\cal W}(P,{\bf v})\rangle
\Bigr\}\,=\,i\nu(P,{\bf v}),\eeq
which is formally solved by
\beq
{\cal W}(P,{\bf v})\,=\,\int\frac{{\rm d}\Omega_1}{4\pi}\,
\langle v|\frac{i}{v\cdot P+i{\cal C}}|
v_1\rangle\,\nu(P,{\bf v}_1)\,,\eeq 
where (cf. eq.~(\ref{COLKER}))
\beq{\cal C}({\bf v,v'})\,=\,
\gamma\,\Bigl\{\delta^{(2)}({\bf v,v'})-\Bigl\langle
\delta^{(2)}({\bf v,v'})\Bigr\rangle\Bigr\}.\,\,\eeq
This, together with eqs.~(\ref{XIXI0}) and
(\ref{jind}), implies
\beq\label{zz1}
\langle \xi_\mu(P)\xi_\nu^*(P')\rangle&=&
(2\pi)^4\delta^{(4)}(P+P')\,
m_D^4
\int\frac{{\rm d}\Omega}{4\pi}\int\frac{{\rm d}\Omega_1}{4\pi}
\int\frac{{\rm d}\Omega'}{4\pi}\int\frac{{\rm d}\Omega_2}{4\pi}\,
\nonumber\\&{}&
v_\mu\,v^{\prime}_\nu\,\langle v|\frac{i}{v\cdot P+i{\cal C}}|
v_1\rangle\,{\cal J}({\bf v}_1,{\bf v}_2) \,
\langle v'|\frac{-i}{v\cdot P-i{\cal C}}|
v_2\rangle\,\eeq 
which is to be compared with eq.~(\ref{zz}) and the
known expression for the imaginary part of the retarded
polarization tensor (cf. eq.~(\ref{PIUSA}) :
\beq
{\rm Im}\,\Pi^{\mu\nu}_R(\omega,p)
&=&-\omega m_D^2
\int\frac{{\rm d}\Omega}{4\pi}\int\frac{{\rm d}\Omega_1}{4\pi}
\int\frac{{\rm d}\Omega'}{4\pi}\int\frac{{\rm d}\Omega_2}{4\pi}\,
\,v_\mu\,v^{\prime}_\nu\,\nonumber\\&{}&
\langle v|\frac{1}{v\cdot P+i{\cal C}}|
v_1\rangle\,\,{\cal C}({\bf v}_1,{\bf v}_2) \,
\langle v'|\frac{1}{v\cdot P-i{\cal C}}|
v_2\rangle\,.\eeq 
Clearly, the above equations are consistent with each other
provided
\beq {\cal J}({\bf v,v'})&=&\frac{2T}{m_D^2}\,\,{\cal C}({\bf v,v'})\,,\eeq
which is the result obtained by Bo\"deker.
Incidentally, the above derivation together with
eq.~(\ref{COLKER}) show that the noise correlator
(\ref{XIXI0}) admits the following representation
\beq\label{XIXIFIN}
\langle \nu_a(x,{\bf v})\nu_b(x',{\bf v}')\rangle
\,=\,-\,\frac{2T}{m_D^2}\,\frac{\delta \,C^a(x,{\bf v})}
{\delta\, W^b(x',{\bf v'})},\eeq
which is consistent with some general properties discussed
in Ref. \cite{CHu99}.

The inclusion of thermal fluctuations via a local noise term,
like in eq.~(\ref{WL0}), is very convenient for numerical
simulations. In order to compute the thermal correlators
of the ultrasoft fields, it is now sufficient to solve the coupled
system of Boltzmann-Langevin and Yang-Mills equations only once,
i.e., for a single set of (arbitrary) initial
conditions, but for large enough times.
Thus, in order to compute, e.g., the two-point function
$\langle A^i(t,{\bf x}) A^j(t',{\bf x}')\rangle$,
it is enough the take the product of the solution
$A^i(t,{\bf x})$ with itself at two space-time points
$(t_1,{\bf x}_1)$ and $(t_2,{\bf x}_2)$ such that $t_1-t_2=t-t'$,
${\bf x}_1-{\bf x}_2={\bf x}-{\bf x}'$, and
$t_1$ and $t_2$ are large enough.

In practice, the only numerical calculations within
this effective theory \cite{Moore98}
have been performed until now in the leading logarithmic approximation, 
where the theory drastically simplifies \cite{Bodeker}: 
it then reduces to a local
stochastic equation for the magnetic fields, of the form
(below, the cross product stands for both the vector product,
and the colour commutator):
\beq\label{localUSA}
{\bf D}\times {\bf B}&=&\sigma {\bf E}\,+\,\bfxi,\nonumber\\
\langle \xi_a^i(x)\xi_b^j(y)\rangle
&=& 2T\sigma\,\delta_{ab}\,\delta^{ij}\,\delta^{(4)}(x-y),\eeq
with the colour conductivity in the LLA (cf. eq.~(\ref{JLL})):
$\sigma=\omega_{pl}^2/\gamma_0$ and $\gamma_0=\alpha NT\ln
(1/g)$. Note that, in this approximation, the noise term $\xi_a^i(x)$
in the Yang-Mills equations is both white and Gaussian. This
is to be contrasted with the general noise 
$\xi_\mu^a(x)$ in eqs.~(\ref{CWL0})--(\ref{jind}) which
in general is a non-linear functional of the gauge fields,
and has an infinite series of non-local $n$-point correlators
(see, e.g., the two-point function in eq.~(\ref{zz1})).

Besides being local, the effective theory in eq.~(\ref{localUSA})
is also ultraviolet finite \cite{Bodeker,Zinn}, thus allowing for
numerical simulations which are
insensitive to lattice artifacts \cite{Moore98}.
It turns out, however, that this LLA is not very accurate
when applied to the computation of the hot baryon number violation rate:
the numerical result in \cite{Moore98} is only about $20\%$
of the corresponding result obtained in lattice
simulations of the full HTL effective theory \cite{MHM98,BMR99}.

Recently, it has been argued \cite{Arnold00,AY99} that the local
form (\ref{localUSA}) of the ultrasoft theory remains valid also 
to ``next-to-leading logarithmic accuracy'' (NLLA), 
i.e., at the next order in an expansion in powers of the
inverse logarithm $\ln^{-1}\equiv 1/\ln(1/g)$.
The only modification refers to the value of the
parameter $\sigma$, which now must be computed to NLLA.
This in turns involves the matching between two calculations:
the expansion of the solution to the Boltzmann equation
to NLLA \cite{USA,AY99}, and a perturbative calculation
within the ``high energy'' sector of the ultrasoft theory
\cite{AY99} (by which we mean loop diagrams with internal momenta
of order $g^2T\ln(1/g)$ which must be computed with 
USA-resummed propagators and vertices).
The complete result for $\sigma$ to NLLA has been 
obtained in Refs. \cite{AY99} (see also \cite{Guerin00}). Quite remarkably,
by using this result within the simplified effective theory 
(\ref{localUSA}), one obtains an estimate
for the hot sphaleron rate which is rather close
(within $20\%$) to the HTL result in Refs. \cite{MHM98,BMR99}.

\setcounter{equation}{0}
\setcounter{equation}{0}
\section{Conclusions}

This report has been mostly concerned with the longwavelength
collective excitations of ultrarelativistic plasmas, with emphasis
on the high temperature, deconfined phase of QCD, where the coupling
``constant'' is small, $g(T)\ll 1$, and the basic degrees of freedom
are those which are manifest in the Lagrangian,
i.e., the quarks and the gluons. In this regime, the dominant
degrees of freedom, the plasma particles, have typical momenta of order $T$.
Other important degrees of freedom are soft,
collective excitations, with typical momenta $\sim gT$. In this work we have
constructed a theory for those collective excitations which carry the
quantum numbers of the elementary constituents.

The effective theory for the collective excitations takes the familiar form
of coupled Yang Mills and Vlasov equations. The separation of
scales between hard and soft degrees of freedom  leads to kinematical
simplifications which allowed us  to reduce the Dyson Schwinger equations for
the Green's functions to kinetic equations for the plasma particles, by
performing a gradient expansion compatible with gauge symmetry. The
resulting theory is gauge invariant. In this construction, the gauge
coupling plays an essential role  which was not recognized in previous works
aiming at developing a kinetic theory for the quark-gluon plasma. Aside from
the fact that it measures the strength of the coupling, $g$ also
characterizes the typical momentum of the collective modes
($\sim gT$), and also the amplitude of the field
oscillations.

The solution of the kinetic equations provides
the source for Yang Mills fields, the so-called induced
current, which  can be regarded as the generating functional
for the hard  thermal loops. These are the dominant
contributions of the one loop amplitudes at high temperature
 and soft external momenta, and they  need to be resummed on soft
internal lines when doing perturbative calculations. Such resummations may
be taken into account by a reorganization of  perturbation theory,
sometimes called ``HTL perturbation theory'', and this has led to  numerous
applications. Note however that since the  HTL effective action  is non
local, high order calculations within that scheme may become rapidly
difficult.

Recent progress indicates that HTL may
also be useful in thermodynamical calculations. In the most naive approach,
  HTL effective action reduces, for the calculation of static quantities,
to the three dimensional effective action for QCD. This however is of limited
use for analytical calculations because of the infrared divergences of the
three dimensional theory. However other schemes exist  which  allow to
take into account the full spectral information on the quasiparticles which
is correctly coded in the HTL. In particular it was shown
recently that self-consistent calculations of the entropy of a quark-gluon
plasma  are able to reproduce accurately the lattice data for
$T\simge 2.5T_c$. This suggests that the quasiparticle picture of the
quark-gluon plasma remains valid even in regimes where the coupling is not
small.

Further support of this picture is provided by the
calculation of the  quasiparticle damping rate.  This calculation  plaid an
important role in the development of the subject and led in particular to
the identification of the hard thermal loops. However HTL are not sufficient
to obtain the full answer.  We have seen that further
resummations are necessary in order to eliminate the IR divergences left
over by the HTL resummations. Theses divergences signal the necessity to
take into account coherence effects related to the fact that particles
never come quite on shell between collisions. We
have presented various ways to do this. But in spite of the fact that the
damping of single particle excitations is unconventional, it remains small
in weak coupling.

The calculation of the damping rate provides one example of a calculation
where one needs to take into account the effect of soft thermal fluctuations.
Because these  can be treated as classical fields, a possible way to proceed
is to use  the classical theory developed in
this paper. Then the integration of soft
fluctuations amounts to averaging over the initial conditions for the
classical fields, with a Boltzmann weight which was
explicitly given. In some applications, it is necessary to go beyond and to
consider explicitly the effective theory for ultrasoft fluctuations,
obtained after integrating not only the hard degrees of freedom, but also
the soft ones. In this case, no Hamiltonian could be found. Rather the
kinetic theory for the hard particles is a Boltzmann equation, and the
averaging over the ultrasoft initial conditions is done via a noise term in
a Boltzmann Langevin equation.  

\vskip 1.cm
{\large {\bf Acknowledgements}}

During the preparation of this work, we have benefited from useful
discussions and correspondence with many people,
 whom we would like to thank:
P. Arnold, G. Baym, D. B\"odeker, D. Boyanovsky,
E. Braaten, P. Danielewicz,
H. de Vega, F. Guerin, U. Heinz, H. Heiselberg,
K. Kajantie, F. Karsch,
M. Laine, M. LeBellac, L. McLerran,
G. Moore, S. Mr\'owczy\'nski, B. M\"uller, J.Y. Ollitrault, R. Parwani,
R. Pisarski, A. Rebhan, K. Rummukainen, D. Schiff, M. Shaposhnikov,
D. Schiff, A. Smilga, A. Weldon, and L. Yaffe.
We address special thanks to Tony Rebhan for his careful reading
of the manuscript and his numerous remarks.
Part of this work was done while one of us (E.I.) was a fellow
in the Theory Division at CERN, which we thank for hospitality and support.

\setcounter{equation}{0}

\appendix
\section{Notation  and conventions}

\parindent 0pt 

In order to facilitate the reading of this report,
we summarize here our notation and conventions,
and list some of the most important symbols indicating
for each of them where it is first introduced.

We shall always use the Minkowski metric (even within the
imaginary time formalism),
\beq g^{\mu\nu}=g_{\mu\nu}={\rm diag}(1,-1,-1,-1),\eeq
and natural units,  $\hbar=c=1$.

We consider a
SU($N$) gauge theory with $N_f$ flavours of massless
quarks, whose Lagrangian is:
\beq\label{LQCD}
{\cal L}\,=\,-\,\frac{1}{4}F_{\mu\nu}^aF^{\mu\nu\,a}\,+\,
\bar\psi_i(i\slashchar{D})^{ij}\psi_j\,.\eeq
The corresponding action reads $S=\int{\rm d}^4x\,{\cal L}$.
In this equation, the colour indices for the adjoint representation  ($a$, 
$b$, ...) run from 1 to $N^2-1$, while those for the
fundamental representation ($i$, $j$, ...) run from 1 to $N$. 
The sum over the quark flavours is implicit, and so will be also
that over colours whenever this cannot lead to confusion.
The generators of the gauge group in different representations are taken to
 be Hermitian and traceless. They are denoted by
$t^a$ and $T^a$, respectively, for the fundamental and the adjoint 
representations, and are normalized thus:
\beq
\label{gen}
{\rm Tr}\,(t^at^b)\,=\,\frac {1}{2}\,\delta^{ab},\qquad\qquad
{\rm Tr}\,(T^aT^b)\,=\,N\delta^{ab}.
\eeq
It follows that:
\beq 
 (T^a)_{bc}=-if^{abc},\quad {\rm Tr} (T^aT^bT^c) =if^{abc}\,
\frac{N}{2}\,,\quad T^aT^a=N,\quad t^at^a=C_{\rm f}\equiv\,
\frac{N^2-1}{2N},\eeq
where $f^{abc}$ are the structure constants of the group:
\beq
\label{lie}
[t^a,t^b]=if^{abc}t^c.
\eeq
We use, without distinction, upper and lower positions for the color
indices.   

Furthermore, in eq.~(\ref{LQCD}), $\slashchar{D}
\equiv \gamma^\mu D_\mu$, with the
usual Dirac matrices $\gamma^\mu$  satisfying
$\{\gamma^\mu, \,\gamma^\nu\} = 2 g^{\mu\nu}$, and
 $D_\mu\equiv \del_\mu+igA_\mu^a t^a$
the covariant derivative in the fundamental representation.

 More generally, we shall use
the symbol $D_\mu$ to denote the covariant derivative in {\it any} of the
group representations, i.e. $D_\mu\,=\, \del_\mu+igA_\mu$,
where $A_\mu$ is a colour matrix, i.e.,
$A_\mu\equiv A_\mu^a t^a$ in the fundamental representation
and $A_\mu\equiv A_\mu^a T^a$ in the adjoint representation.
For any matrix $O(x)$ acting in a representation of the
colour group, we write
\beq \label{adder} [D_\mu,
O(x)]\equiv \del_\mu O(x)+ig[A_\mu(x),O(x)]. \eeq
The gauge field strength tensor $F_{\mu\nu}^a$ is defined as:
\beq
F_{\mu\nu}\equiv [D_\mu, D_\nu]/(ig) =F_{\mu\nu}^at^a,\quad {\rm with}
\,\,\,\,F_{\mu\nu}^a=\del_\mu A_\nu^a -
\del_\nu A_\mu^a - gf^{abc}A_\mu^b A_\nu^c\,.\eeq
The electric and magnetic fields are:
\beq E^i_a\,=\,F^{i0}_a\,,\qquad
B^i_a\,=\,-(1/2)\epsilon^{ijk}F^{jk}_a\,.\eeq

The gauge transformations are implemented by unitary matrices
$h(x)=\exp(i\theta^a(x) t^a)$, and read:
\beq\label{BGT1}
A_\mu \,\rightarrow\, h A_\mu h^{-1}\,-\,\frac{i}{g}\,h\del_\mu h^{-1},
\qquad \psi\,\rightarrow\, h\psi, \qquad
D_\mu\psi\,\rightarrow\,h D_\mu\psi.\eeq
Whenever we need to distinguish between
various representations of the colour group, we use a tilde
to denote quantities in the adjoint representation.
For instance, we write $\tilde D_\mu=\del_\mu+igA_\mu^a T^a$, 
and $\tilde h(x)=\exp(i\theta^a(x) T^a)$, so that, in
the gauge transformation (\ref{BGT1}),
\beq 
E^i_a(x)\,\rightarrow\, \tilde h_{ab}(x)E^i_b(x), \qquad
B^i_a(x)\,\rightarrow\, \tilde h_{ab}(x)B^i_b(x).\eeq
If, in the same transformation, the matrix $O(x)$ transforms
covariantly, $O(x)\rightarrow h(x)O(x)h^{-1}(x)$, then the same
holds for its covariant derivative  $[D_\mu,
O(x)]\rightarrow h(x)[D_\mu, O(x)]h^{-1}(x)$.

When working in the imaginary-time formalism, we write
$x^\mu=(x_0, {\bf x})=(t_0-i\tau,{\bf x})$, with $t_0$ real and arbitrary,
and $0\le \tau\le \beta;$ therefore, $\del_0=i\del_\tau$ and
${\rm d}x_0=-i{\rm d}\tau$. We define the Euclidean action by writing
${\rm e}^{iS}\,\equiv\,{\rm e}^{i\int{\rm d}^4x\,{\cal L}}\,
\equiv\,{\rm e}^{-S_E}$, with
\beq
\label{QCD}
S_{E}\,=\,\int_0^\beta {\rm d}\tau {\rm d}^3x\left\{ 
\frac{1}{4}F_{\mu\nu}^aF^{\mu\nu\,a}+
\bar\psi(-i\slashchar{D})\psi
 \right\}.
\eeq
Unless it  may induce confusion, we generally omit the subscript
$E$ on the imaginary-time quantities.

{The thermal occupation numbers for gluons, $N(k_0)$, and quarks,
 $n(k_0)$, are written as:
\beq\label{BF}
N(k_0)\,=\,\frac{1}{{\rm e}^{\beta k_0}\,-\,1},\qquad\qquad
n(k_0)\,=\,\frac{1}{{\rm e}^{\beta (k_0-\mu)}\,+\,1},\eeq
where $\beta\equiv 1/T$, and $\mu$ is a chemical potential.
In fact,, we consider here mostly a plasma
with $\mu=0$, but many of the results are easy
to generalize to arbitrary $\mu$.}

We now pursue with an enumeration of the symbols used
in the text, presented here in alphabetical order.
In parentheses we indicate the sections where they are
introduced.

\begin{list}%
{1--4~}{}

\item[{$A^\mu_a(x)$:}] the gauge vector potential, usually
identified with the soft classical background field, i.e.,
the {\it gauge mean field} (1.1, 3.1, 3.2.1).

\item[{$a^\mu_a(x)$:}] the (typically hard)
quantum fluctuations of the gauge field (3.2.1).

\item[{$\beta =1/T$:}] the inverse temperature.

\item[{$C_a(x, {\bf v})$, $C_{ab}(x,x';{\bf v},{\bf v}')$ :}]
the collision term for the colour Boltzmann equation and its kernel
(7.3.1).

\item[{${\cal D}$,  ${\cal D}_j$ :}] 
the statistical density operator, in or out of thermal
equilibrium (2, 2.2.1).

\item[{$D_\mu[A]=\del_\mu+igA_\mu :$}] the covariant derivative
with the gauge field $A^\mu_a$.

\item[{$\eta^{ind}(x)$ :}] the induced fermionic source
in QED or QCD (3.2.2).

\item[{${\cal E}^a_i$, ${\cal A}^a_i$, ${\cal W}^a$;
${\cal G}^a$, ${\cal H}$ :}] initial conditions for the classical
equations of motion in the HTL effective theory; the associated
Gauss' operator and Hamiltonian (4.4.3).

\item[{$\delta f({\bf k},X)$, $\delta n({\bf k},X)$, 
$\delta N({\bf k},X)$ :}] 
off-equilibrium density matrices for hard electrons (1.2),
quarks (3.4.2) and gluons (3.4.1).

\item[{$F_{\mu\nu}^a=\del_\mu A_\nu^a -
\del_\nu A_\mu^a - gf^{abc}A_\mu^b A_\nu^c\,:$}] the gauge field 
strength tensor.

\item[{$G^a$, $\tilde G^a$ :}] gauge-fixing terms in the 
quantum path integral for a gauge theory (3.2.1).

\item[{$G$, $G_0$, $G^{(n)}$, $G_{\mu\nu}$ :}] 
time-ordered Green's functions
(2-point, free, $n$-point), in real- or complex-time, for
scalars (2.1.1, 2.2.2) and  gluons (3.2.1).

\item[{$G_R$, $G_A$ :}]  {\it i\/)} retarded and advanced
2-point functions, in or out of thermal equilibrium,
for scalars (2.1.2, 2.2.2) or gluons (7.1); 
{\it ii\/)} retarded and advanced
Green's functions for the drift operator in the kinetic equations
(4.1.1).

\item[{$G_{cl}$ :}]  the classical thermal 2-point correlation,
for scalars (2.2.4) or gluons (4.4.3, 7.4).

\item[{$G^<$, $G^>$ :}] analytic 2-point functions,
 for scalars (2.1.2, 2.2.2) and gluons (3.2.3).

\item[{$G^<(k,X)$, $G^>(k,X)$, $G_R(k,X)$, $G_A(k,X)$
:}] various Wigner functions for scalars (2.3.1).

\item[{${\cal G}^<(k,X)$, ${\cal G}^>(k,X)$, ${\cal G}_R(k,X)$, 
${\cal G}_A(k,X)$ :}] various Wigner functions for gluons (3.3.2, 7.1).

\item[{${}^*G_{\mu\nu}$, ${}^*\Delta_L$, ${}^*\Delta_T$ :}]
the HTL-resummed propagator for soft gluons, and its longitudinal
 and transverse components (4.3.1, B.1.3).

\item[{$\Gamma$ :}] the discontinuity of the self-energy
 (2.1.3, 2.3.2).

\item[{$\gamma$, $\gamma({\bf p},t)$
 :}] the quasiparticle damping rate (2.1.3, 2.3.4, 2.3.5, 6.1).

\item[{$\Gamma$, $\Gamma_{ind}$, $\Gamma_{HTL}$, $\Gamma_A$,
 $\Gamma_\Psi$ :}] the effective action, its induced piece,
the HTL effective action and its various components (5.5.1, 5.5.2).

\item[{$H$, $H_0$, $H_1$, $H_j(t)$ :}] {\it i\/)} 
the Hamiltonian for a generic field theory (total, free,
interacting, in the presence of an external source) (2, 2.1,
2.2.1, 2.2.3); {\it ii\/)} the Hamiltonian for the HTL
effective theory (4.4.3).

\item[{$h(x)=\exp(i\theta(x))$,}]
with $\theta\equiv \theta^aT^a$ or $\theta\equiv \theta^a t^a$:
a SU($N$) gauge transformation.

\item[{${\cal I}({\bf v},{\bf v}')$ :}] noise-noise correlator
for the colour Boltzmann-Langevin equation (7.4).

\item[{ $j_\mu(x)$:}] the external source driving the system
out of equilibrium (2, 2.2), and also 
the argument of the generating functional $Z[j]$
(2.2.2 and 3.2.1).

\item[{ $j^{ind}(x)$ :}] the induced source in the scalar field
theory (2.2.3).

\item[{ $j^{ind}_\mu(x)$, $j^{ind\,a}_\mu(x)$ :}] the
 induced electromagnetic (1.2) and  colour (3.2.2) currents.

\item[{$j^{A}_\mu$, $j^{\psi}_\mu$
$j^\mu_{\rm f}$,  $j^\mu_{\rm g}$ :}] the colour current induced
by the gauge ($A$) or fermionic ($\psi$) mean fields
acting on hard quarks (f) or gluons (g) (3.2.2). 

\item[{$K_{\nu}^a$,  $H_{\nu}^a$, ${\cal K}^{a}_\nu$,
${\cal H}^a_\nu$:}] the ``abnormal'' propagators and their
Wigner transforms (3.2.3, 3.3.2).

\item[{$\Lambda$ :}] separation scale between hard and soft
 momenta (2.1.4, 4.4.3).

\item[{${\slashchar \Lambda}({\bf k},X)$ :}] ``abnormal''
density matrix (3.5.1).

\item[{ $m_D$, $\omega_{pl}$, $\omega_{0}$, $m_\infty$, $M_\infty$ :}]
Debye mass, plasma frequencies (for gluons and fermions),
asymptotic masses (for gluons and fermions) (4.1.2, 4.1.3, 4.3.1).

\item[{ $m_{mag}$ :}] the magnetic screening mass (5.4.3).

\item[{$\mu$ :}] {\it i\/)} infrared cutoff (1.1, 6.1);
{\it ii\/)} separation scale between soft and ultrasoft momenta
(7).

\item[{$N_k$, $n_k$ :}] thermal occupation factors 
for single-particle bosonic, or fermionic, states (1.1).

\item[{$\nu^a$, $\xi_\mu^a$ :}] colour ``noise terms'', for the 
Boltzmann-Langevin (7.4) and, respectively, Yang-Mills
(4.4.3, 7.4) equations.

\item[{$\phi$, $\phi_0$, $\Phi$,  $\Phi_{cl}$ :}]
 scalar fields: the quantum field (2), 
its static Matsubara mode (2.1.4), the 
average field (2.2.1, 2.2.3), the classical field (2.2.4).

\item[{$\Phi({\bf v\cdot v}^\prime)$ :}] collisional cross-section
for hard particles with velocities ${\bf v}$ and ${\bf v}^\prime$
(7.2).

\item[{$\Pi_{\mu\nu}$, $\Pi_{\mu\nu}^{ab}$ :}]
the photon (1.2) or gluon (4.2, 5.3.2, 7.3.2, B.1)
polarization tensor.

 \item[{$\psi$, $\bar\psi$ :}] the (typically hard)
 quantum fermionic fields (3.2.1).

\item[{$\Psi$, $\bar\Psi$ :}] the soft fermionic mean fields (3.2.1).

\item[{$\rho_0$, $\rho$ :}] the (free) spectral density,
in or out of thermal equilibrium (2.1.1, 2.1.3, 2.3.2).

\item[{${}^*\rho_{\mu\nu}$, ${}^*\rho_L$, ${}^*\rho_T$ :}]
the gluon spectral density in the HTL approximation (4.4.3, B.1.3).

\item[{$S$, ${}^*S$ :}] the fermion propagator: in general
(3.2.3) and in the HTL approximation (4.3.1, B.2.3).

\item[{$\Sigma$, $\Sigma^>$, $\Sigma^<$, $\Sigma_R$,
$\Sigma_A$ :}] self-energies for scalars (2.1.3, 2.2.3),
quarks (3.3.1), or gluons (7.1); the associated Wigner functions 
are denoted as $\Sigma^<(k,X)$, etc. (2.3.2, 7.1).

\item[{$\sigma^{\mu i}$, $\sigma$ :}] electromagnetic (1.4, 3.1)
and colour (7.3.3) conductivities.

\item[{${\rm T}$, ${\rm T}_\tau$, $\tilde {\rm T}$,  ${\rm T}_C$ :}]
symbols for operator ordering, 
in real- or complex-time (2.1, 2.2, 2.2.2).

\item[{$U(x,y|A)$, $U(x,y)$} :] Parallel transporters, or Wilson lines
(1.3, 3.1, 3.3.2, 4.1.1).

\item[{$U(t,t_0)$,  $U_j(t_2, t_1)$, $U_j(z,t_0)$ :}]
evolution operators, in real- or complex-time (2.1,
2.2.1, 2.2.2).

\item[{ $W(x, {\bf v})$, $W_a^\mu(x, {\bf v})$,
$W_a(x, {\bf v})$ :}] 
reduced density matrices for charge (1.2) or colour 
 (4.1.2, 7.2) oscillations of the hard particles.

\item[{$\omega_n = 2n \pi  T$, or $\omega_n = (2n+1) \pi  T$ with
integer $n$ :}]
 Matsubara frequencies for bosons or fermions (2.1.1, Appendix B).

\item[{$Z$, $Z_{cl}$  :}]  the thermal partition function: quantum (2.1)
and classical  (2.1.4, 2.2.4).

\item[{$Z[j]$, $Z_{cl}[J]$ :}]
 the generating functional of thermal correlations:
 quantum (2.2.2, 3.2.1) and  classical (4.4.3).

\item[{$\tilde Z[j,A]:$}] the generating functional in the background field
gauge (3.2.1).

\item[{$\zeta^a$ and $\bar\zeta^a\,$:}] the anticommuting  ``ghost'' fields
(3.2.1).

\end{list}

\parindent 30pt

\newpage
\setcounter{equation}{0}
\section{Diagrammatic calculations of hard thermal loops}

In this appendix, we present a few explicit calculations of Feynman
diagrams in the imaginary time formalism. In particular, we obtain in
this fashion the gluon and fermion self-energies in the hard thermal loop
(HTL) approximation.  Unless otherwise stated, all calculations are 
for massless QCD at zero chemical potential. Ultraviolet divergences are
regulated by dimensional
 continuation ($4 \to d=4-2\epsilon$),
but we keep the fermions as four-component objects,
${\rm Tr} (\gamma_{\mu}\gamma_{\nu})=4 g_{\mu\nu}$.
We   denote a generic four-momentum as
 $k^\mu = (k_0,{\bf k})$, $k_0 = i\omega_n
= i n\pi T$, with $n$ even (odd) for bosonic (fermionic) fields.
The scalar product is defined with the Minkowski metric,
so that, for instance, $k^2=k^2_0-{\bf k}^2=-\omega_n^2-{\bf k}^2$.
The  measure of loop integrals is denoted by the following condensed notation:
\beq\label{mesure}
\int[{\rm d}k]\equiv T\sum_{n, even}\int ({\rm d} {\bf k})\, ,
\qquad\,\,\,\,
 \int\{{\rm d}k\}\equiv  T\sum_{n, odd}\int ({\rm d} {\bf k})\, ,
\eeq
where
\beq 
\int ({\rm d} {\bf k})
\equiv \int\frac{{\rm d}^{d-1}k}{(2\pi)^{d-1}}.\nonumber \eeq
For a free scalar particle with mass $m$,
the Matsubara propagator is given by eq.~(\ref{D0F}), namely
(in this appendix, we prefer to denote 
this propagator as $\Delta$, rather than $G_0$) :
\beq\label{delta}
\Delta(k) \,=\,\frac{1}{\omega_n^2+{\bf  k}^2
+m^2}\,=\, \frac{ -1}{k^2 - m^2}
\,,\qquad k_0=i\omega_n= i\,2n\pi T\,,\eeq
while for a massive fermion we have:
\beq\label{fermion}
S_0(k)\,=\,{({\slashchar k}+m) }\,\tilde\Delta(k)\,,\qquad
\tilde\Delta(k)\equiv \frac{ -1}{k^2-m^2}\,,\,\,\,\,\,
 k_0 =i\omega_n=i(2n+1)\pi T\,,\eeq
with ${\slashchar k}\equiv \gamma^\mu k_\mu =
i\omega_n\gamma^0-{\bf  k}\cdot{\bfgamma}$.
Note that the only difference between the functions $\Delta(k)$
and $\tilde\Delta(k)$  lies in  the odd or even character
of the corresponding Matsubara frequencies.
The gluon propagator in the covariant gauge, with a gauge-fixing term 
$(\del^\mu A_\mu^a)^2/2\lambda$, is:
\beq\label{photon}
G^0_{\mu\nu}(k)
= -\,g_{\mu\nu}\,\Delta(k)\,+\,(\lambda -1)
k_\mu k_\nu\,\Delta^2(k)\,,\eeq
with $\Delta(k)$ given by eq.~(\ref{delta}) with $m=0$. 
The Coulomb gauge propagator will be also used.

\begin{figure}
\protect \epsfxsize=12cm{\centerline{\epsfbox{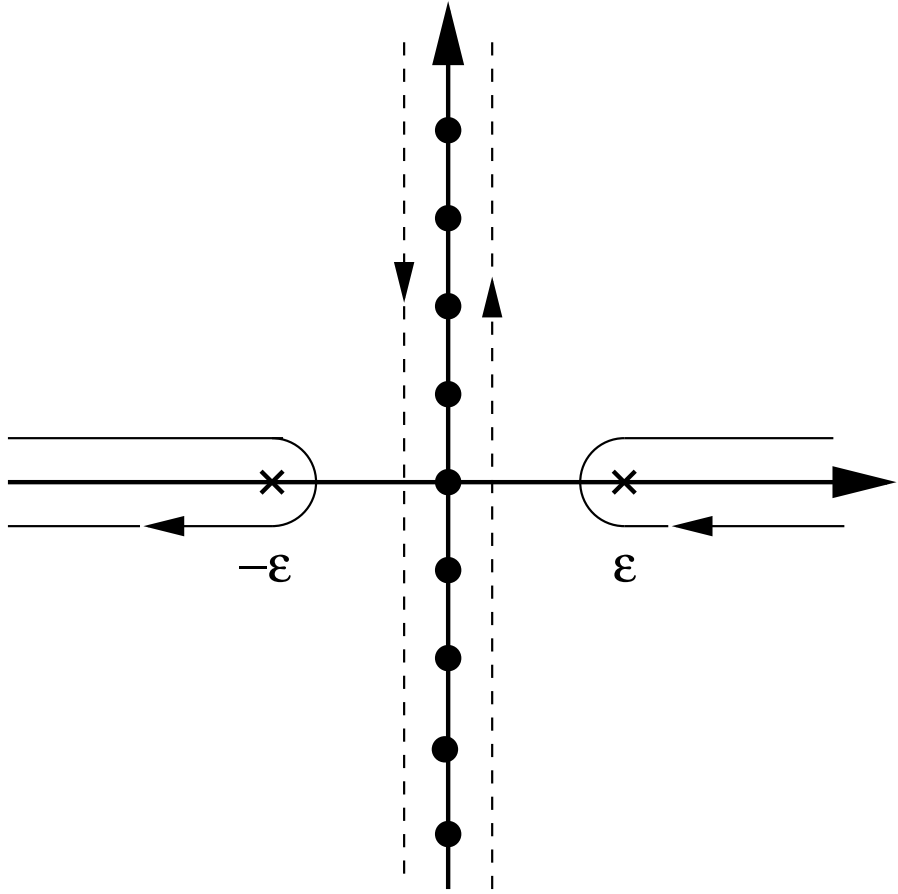}}}
	 \caption{Integration contour for  eqs.~(\ref{SUM2})--(\ref{SUM20}).
Dashed line: original contour used to reproduce the Matsubara
sum. Solid line:
deformed contour used to calculate the integral. The dots
on the imaginary axis correspond to Matsubara frequencies
$\omega_n$, the crosses on the real axis to $\pm \varepsilon_k$.}
\label{sumCont}
\end{figure}

When computing Feynman graphs, we have to calculate 
 sums over the internal Matsubara frequencies.
These may be done by appropriate contour integration, or by introducing a
mixed Fourier representation of the propagators, as we shall explain
shortly.  As an example of the use of contour integration, let us compute
the  imaginary-time propagator
$\Delta(\tau)$  by performing the following 
frequency sum (cf. eq.~(\ref{SUM})):
\beq\label{SUM2}
\Delta(\tau, {\bf k})\,=\,T\sum_n
{\rm e}^{-i\omega_n\tau}\Delta(i\omega_n, {\bf k}),
\eeq
where $\omega_n= 2n\pi T$.
For $0<\tau <\beta$, we have $\Delta(\tau)=\Delta^>(\tau)$, and we can 
replace the sum in eq.~(\ref{SUM2}) by:
\beq \label{SUM20}
\Delta^>(\tau,{\bf k})=-\oint {{\rm d}\omega\over 2\pi
i}\frac{{\rm e}^{-\omega\tau}}{{\rm e}^{-\beta\omega}-1}
\Delta(\omega,{\bf k})
,\eeq
where $\Delta(\omega,{\bf k})=(\varepsilon_k^2-\omega^2)^{-1}$,
  $\varepsilon_k = \sqrt {{\bf  k}^2 +m^2}$, and
 the integration contour is
indicated in fig.~\ref{sumCont}.
Note that the choice of the function 
is such that one can close the
contour without getting contribution from grand circles at
infinity. The integration is then trivial and yields:
\beq\label{delta>}
\Delta^>(\tau, {\bf k})
=\frac{1}{2\varepsilon_k}\left\{
(1+N_k)\,{\rm e}^{-\varepsilon_k\tau}+N_k\,{\rm e}^{\varepsilon_k\tau}\right\},
\eeq
where $N_k\equiv N(\varepsilon_k)=1/({\rm e}^{\beta \varepsilon_k}-1)$.
For $\tau<0$ one could  use instead
\beq
\Delta^<(\tau)=\oint {{\rm d}\omega\over 2\pi
i}\frac{{\rm e}^{-\omega\tau}}{{\rm e}^{\beta\omega}-1}\Delta(\omega),
\eeq
which gives:
\beq\label{Delta<(tau)}
\Delta^<(\tau, {\bf k})=
\frac{1}{2\varepsilon_k}\left\{
N_k \,{\rm e}^{-\varepsilon_k\tau}+(1+N_k)\,
 {\rm e}^{\varepsilon_k\tau}\right\}.\eeq
By putting together the previous results, we derive an expression
for $\Delta(\tau, {\bf k})$ valid for $-\beta\le \tau \le \beta$:
\beq\label{DtauS}
\Delta(\tau, {\bf k})&=&\int_{-\infty}^{+\infty}\frac{{\rm d}k_0}{2\pi}\,
{\rm e}^{-k_0\tau}\rho_0(k)\,\Bigl(\theta(\tau)+N(k_0)\Bigr)\,,\eeq
where $\rho_0(k)$  is the  spectral density for a free particle, eq.~(\ref{rho0}). 
For fermions, we obtain similarly
\beq\label{DtauF}
\tilde\Delta(\tau, {\bf  k})
&=&\int_{-\infty}^{+\infty}\frac{{\rm d}k_0}{2\pi}\,
{\rm e}^{-k_0\tau}\rho_0(k)\,\Bigl(\theta(\tau) - n(k_0)\Bigr)\,.\eeq
where $n(k_0)=1/({\rm e}^{\beta k_0}+1)$. By using the simple identities
\beq
{\rm e}^{\beta k_0} N(k_0)\,=\,1+N(k_0)\,,\qquad
{\rm e}^{\beta k_0} n(k_0)\,=\,1-n(k_0)\,, 
\eeq
one can easily verify that the functions  $\Delta(\tau)$
and $\tilde\Delta(\tau)$ obey the expected (anti)periodicity
conditions for $-\beta\le \tau \le \beta$. For instance, for
$0 < \tau \le \beta$ :
\beq 
\Delta(\tau-\beta, {\bf k})\,=\,\Delta(\tau, {\bf k}),\qquad
\tilde\Delta(\tau-\beta, {\bf k})\,=\,-\, \tilde\Delta(\tau, {\bf k}).
\eeq

The corresponding representations for the quark and for the gluon
propagators are then (cf. eqs.~(\ref{fermion}) and (\ref{photon})):
\beq\label{FMix}
S_0(\tau, {\bf  k})&=&\int_{-\infty}^{+\infty}\frac{{\rm d}k_0}{2\pi}\,
{\rm e}^{-k_0\tau}\rho_0(k)\,({\slashchar k}+m)\,\Bigl(\theta(\tau)-n(k_0)\Bigr)\,,
 \nonumber \\
G^0_{\mu\nu}(\tau, {\bf  k})
&=&\int_{-\infty}^{+\infty}\frac{{\rm d}k_0}{2\pi}\,
{\rm e}^{-k_0\tau}\rho_{\mu\nu}(k)\,\Bigl(\theta(\tau)+N(k_0)\Bigr)\,,\eeq
where
\beq\label{a4}
\rho_{\mu\nu}(k)=\rho_{\mu\nu}^F(k)+\left(1-{\lambda}\right)
\rho_{\mu\nu}^\lambda(k),
\eeq
and  $\rho_{\mu\nu}^F(k)=-g_{\mu\nu}2\pi\epsilon(k_0)\delta(k^2)$ 
is the gluon spectral density in Feynman gauge ($\lambda =1$),
while $\rho_{\mu\nu}^\lambda(k) \equiv - 2\pi \epsilon(k_0)\,
k_\mu k_\nu \,\delta ' (k^2)\,$ [$\delta ' (k^2)$ is the
derivative of $\delta (k^2)$ {\wr} $k^2$].
In the Feynman gauge, $G^0_{\mu\nu}(\tau, {\bf  k})$ reduces to 
$-g_{\mu\nu}\Delta (\tau, {\bf  k})$.

The mixed Fourier representation of the propagators given by
eqs.~(\ref{DtauS})--(\ref{FMix})  can be used to facilitate  the 
evaluation of Matsubara sums in Feynman diagrams. To this aim,  we write,
for instance,
\beq\label{DeltanS}
\Delta(i\omega_n, {\bf  k})& =& \int_0^\beta {\rm d}\tau\,{\rm e}^{i\omega_n\tau}\,
\Delta^>(\tau, {\bf  k})\,,\eeq
and perform the sums over Matsubara frequencies by using (with integer $l$) :
\beq\label{summing}
{T}\sum_n {\rm e}^{i\omega_n\tau}\,=\,\sum_l (\pm)^l
\delta(\tau-l\beta)\,,
\eeq 
where the plus (minus) sign corresponds to even (odd) Matsubara
frequencies. Explicit examples of the procedure will be given below. As a
trivial application, we use  eqs.~(\ref{DeltanS}),
(\ref{summing}) and (\ref{delta>})  to obtain
(henceforth, we set $m=0$, and therefore $\varepsilon_k=k$):
\beq
{T}\sum_n \frac{1}{\omega_n^2+{\bf  k}^2}\,=\,\Delta^>(\tau=0, {\bf  k})
\,=\,\frac{1+2N_k}{2k},\eeq
 so that:
\beq\label{TP}
\int [{\rm d}k]\,\Delta(k)&=&
 \int\frac{{\rm d}^{d-1}k}{(2\pi)^{d-1}}\,\frac{1+2N_k}{2k}
\nonumber\\
&=& \int\frac{{\rm d}^{3}k}{(2\pi)^{3}}\,\frac{N_k}{k}\,=\,
\frac{T^2}{12}\,.\eeq
In going from the first to the second line of eq.~(\ref{TP}),
we have used the fact that, under dimensional regularization,
\beq \int\frac{{\rm d}^{d-1}k}{(2\pi)^{d-1}}\,\frac{1}{2k}\,=\,0,\eeq
and we have set $d=4$ in the evaluation of the temperature-dependent
piece, which is UV finite. (If some other UV regularization scheme
is being used --- e.g., an upper cut-off $\Lambda$ --- then the 
zero-temperature piece in eq.~(\ref{TP}) gives a non-vanishing
contribution, which is quadratically divergent as $\Lambda \to \infty$.
This divergence can be removed by renormalization at $T=0$,
and the final result is the same as above. See also the discussion
 in Sect.  2.3.3, after eq.~(\ref{LAMBDA}).)
The integral (\ref{TP}) appears, for instance, in the calculation
of the tadpoles in fig. 2 and in fig.~\ref{figgluon}.c below, 
or in the evaluation of the gauge field fluctuations in
Sect. B.1.4. In an entirely similar way, we obtain:
\beq\label{TPF}
\int \{ {\rm d}k\}\,\tilde\Delta(k)\,=\,
 \int\frac{{\rm d}^{d-1}k}{(2\pi)^{d-1}}\,\frac{1-2n_k}{2k}
\,=\,-\, \int\frac{{\rm d}^{3}k}{(2\pi)^{3}}\,\frac{n_k}{k}\,=\,
-\,\frac{T^2}{24}\,.\eeq

Finally, let us consider the Fourier transform of the 2-point function
$\Delta^<(t,{\bf k})$ obtained from eq.~(\ref{Delta<(tau)}) by analytic
continuation to real time. Let us denote by $\Delta_T^<(t,{\bf k})$ the
finite temperature contribution. We have:
\beq
\Delta_T^<(t,{\bf k})=\frac{N_k}{k}\cos(kt),
\eeq
where we have taken $\varepsilon_k=k$. The Fourier transform of this
expression is easily obtained in the form:
\beq
\Delta_T^<(t,{\bf x})=\frac{1}{4\pi^2 x}\sum_{n\geq
1}\left(\frac{x+t}{(\beta n)^2+(x+t)^2}+\frac{x-t}{(\beta
n)^2+(x-t)^2}\right).
\eeq
By using the summation formula:
\beq
\sum_{n\geq 0}\frac{1}{n^2\beta^2 +y^2}=\frac{1}{2y^2}
+\frac{\pi}{2\beta y}\coth\left(\frac{\pi y}{\beta}\right),
\eeq
one gets:
\beq\label{deltaT1}
\Delta_T^<(t,{\bf x})=\frac{T}{8\pi x}\left( h(\pi Tx_+)+h(\pi
Tx_-)\right),
\eeq
with $x_\pm \equiv x\pm t$, and:
\beq
h(u)\equiv \coth(u)-\frac{1}{u}.
\eeq
For $u\ll 1$, $h(u)\approx u/3$, so eq.~(\ref{deltaT1}) yields
\beq
\Delta_T^<(t=0, x=0)\,=\,\frac{T^2}{12}\,,\eeq
in agreement with eq.~(\ref{TP}). 
For $u\gg 1$, $h(u)= 1-1/u+{\cal O}(e^{-u})$, so at large $x$
the equal-time ($t=0$) two-point function decreases as:
\beq\label{largex}
\frac{\Delta_T^<(t=0,x)}{\Delta_T^<(t=0, x=0)}\,\simeq\,
\frac{3}{\pi}\,\frac{1}{xT}\quad{\rm for}\quad x\gg 1/\pi T\,.\eeq
This slow decrease ($\sim 1/x$) is due to the static ($\omega_n=0$)
Matsubara mode. To clearly see this, let us rederive
eq.~(\ref{largex}) starting with eq.~(\ref{SUM}) (with $\tau=0$)
where the contributions of the various Matsubara modes are
explicitly separated. We have:
\beq
\Delta(0, {\bf x})&=&T\sum_n
\int\frac{{\rm d}^{3}k}{(2\pi)^{3}}\,\frac{
{\rm e}^{i{\bf k}\cdot{\bf x}}}
{\omega_n^2+{\bf  k}^2}\,=\,\frac{T}{4\pi x}\,\sum_n
{\rm e}^{-|\omega_n|x}\nonumber\\
&=&\frac{T}{4\pi x}\,\biggl\{
1+2\sum_{n\geq 1} {\rm e}^{-\omega_n x}\biggr\}
\,=\,\frac{T}{4\pi x}\,\coth(\pi Tx),
\eeq
where the first term  within the braces
in the second line, $\sim 1/x$, is the contribution of the
static mode.

\subsection{The soft gluon polarization tensor}

To one-loop order, the gluon polarization tensor
$\Pi_{\mu\nu}^{ab}(p)$ is given by the four  diagrams in fig.
\ref{figgluon}, which we shall evaluate  in the hard thermal
loop approximation, valid when
 the external gluon line carries
 energy and momentum  of order $gT$.  Since the colour
structure of this tensor is trivial,
 $\Pi_{\mu\nu}^{ab}=\delta^{ab}\Pi_{\mu\nu}$, we shall omit colour
indices in what follows.
For complementarity with the analysis in Sect. 3, where
the Vlasov equations which determine the HTL $\Pi_{\mu\nu}$
have been constructed in the Coulomb gauge, here we shall
rather work in the Feynman gauge (i.e., the covariant gauge
with $\lambda =1$; cf. eq.~(\ref{photon})). As already emphasized
in the main text, the final result for the HTL will be gauge-independent.

\subsubsection{The quark loop}

\begin{figure}
\protect \epsfxsize=15cm{\centerline{\epsfbox{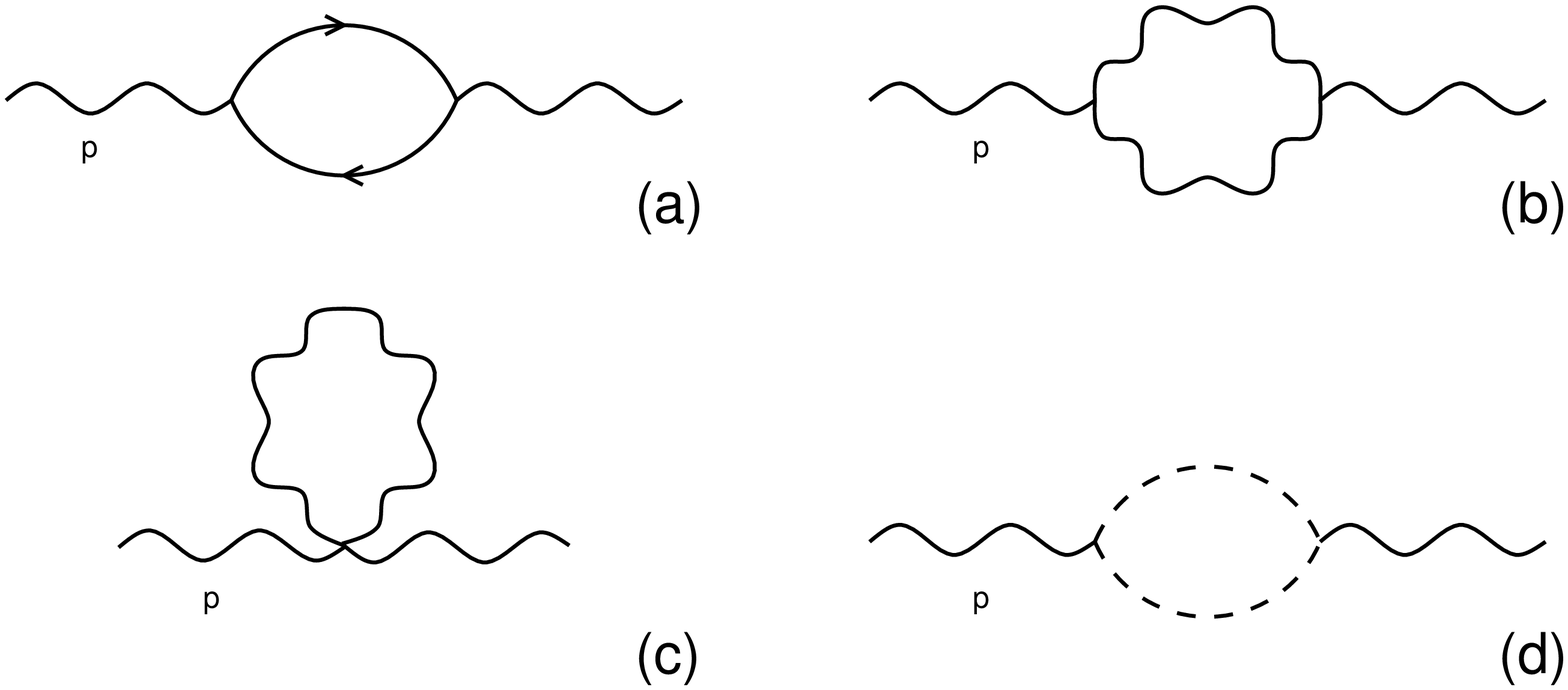}}}
	 \caption{The one-loop gluon polarization tensor}
\label{figgluon}
\end{figure}
Consider first the contribution 
of the quark loop in fig.  \ref{figgluon}.a, which we denote
by $\Pi_{\mu\nu}^{(a)}$.  A straightforward application of
the Feynman rules gives, for $N_f$ quark flavors,
\beq\label{Pia1}
\Pi_{\mu\nu}^{(a)}(i\omega_n, {\bf  p})= \frac{g^2 N_f}{2}
\int \{{\rm d}k\}\,{\rm Tr}\,\biggl\{\gamma_\mu\,S_0
(i\omega_r, {\bf  k})\,
\gamma_\nu\,S_0(i\omega_r-i\omega_n, {\bf  k}-{\bf  p})\biggr\},\eeq
with $p_0=i\omega_n=i\,2n\pi T$ and $k_0=i\omega_r=i\,(2r+1)\pi T$,
with integers $n$ and $r$. 
The trace in eq.~(\ref{Pia1}) refers to spin variables only (the colour trace has been 
already evaluated and it is responsible for  the factor $1/2$).
It is easily verified that
$\Pi_{\mu\nu}^{(a)}(p)$ is transverse, $p^\mu \Pi_{\mu\nu}^{(a)}(p)=0$,
so that it has only two independent components (see below).

In the calculation, the external energy $p_0=i\omega_n$
 is purely imaginary and discrete to start with. 
To implement the property that $p_0$ is soft, an analytic continuation
 to real energy is necessary. This becomes possible after performing
 the sum over the internal  Matsubara frequency $i\omega_r$. 
To do so, we use the mixed Fourier 
representations of the two quark propagators in eq.~(\ref{Pia1})
together with the identity (\ref{summing}), and obtain:
\beq\label{Pia11}
\Pi_{\mu\nu}^{(a)}(i\omega_n, {\bf  p})=-\,\frac{g^2 N_f}{2}
\int ({\rm d}{\bf k})\int_0^\beta {\rm d}\tau\,{\rm e}^{-i\omega_n \tau}
{\rm Tr}\,\Bigl\{\gamma_\mu\,S_0
(\beta-\tau, {\bf  k})\,
\gamma_\nu\,S_0(\tau, {\bf  k}-{\bf  p})\Bigr\}.\eeq
We then use the integral representation (\ref{FMix}) for the two propagators
in eq.~(\ref{Pia11}), and denote by $k_0$ and $q_0$
 the respective energy variables. The $\tau$-integral
 is then easily evaluated:
\beq\label{inttau}
\int_0^\beta {\rm d}\tau\,{\rm e}^{-i\omega_n\tau}
\,{\rm e}^{-k_0(\beta-\tau)}\,{\rm e}^{-q_0\tau}=
\,\frac{{\rm e}^{-\beta(i\omega_n+q_0)}-{\rm e}^{-\beta k_0}}{k_0-q_0-i\omega_n}=
\,\frac{{\rm e}^{-\beta q_0}-{\rm e}^{-\beta k_0}}{k_0-q_0-i\omega_n}\,,\eeq
where we used the fact that ${\rm exp}(-i\beta\omega_n) =1$.
By also using the identity:
\beq
\Bigl(1 - n(k_0)\Bigr)\Bigl(1 - n(q_0)\Bigr)
({\rm e}^{-\beta q_0}-{\rm e}^{-\beta k_0})\,=\,n(q_0)-n(k_0),\eeq
we finally get:
\beq\label{Pia12}\Pi^{(a)}_{\mu\nu}
(i\omega_n, {\bf  p})= \frac{g^2 N_f}{2}\,\int ({\rm d}{\bf k})
\int_{-\infty}^\infty
\frac{{\rm d}k_0}{2\pi}\,\int_{-\infty}^\infty\frac{{\rm d}q_0}{2\pi}\,\rho_0(k)\,
\rho_0(q) \qquad\nonumber\\\,\times {\rm
Tr}\,\Bigl(\gamma_\mu\,{\slashchar {k}}\,
\gamma_\nu\,{\slashchar {q}}\Bigr)
\frac{n(k_0)-n(q_0) }{k_0-q_0-i\omega_n}\,,\eeq
with the  notations  $k^\mu=(k_0, {\bf  k})$, $q^\mu=(q_0, {\bf  q})$
and ${\bf  q}\equiv {\bf  k}-{\bf  p}$. 

The expression (\ref{Pia12}) can now be  continued 
in the complex energy plane by simply replacing $i\omega_n
\to p_0$ with $p_0$ off the real axis:
\beq\label{Pia2}
\Pi^{(a)}_{\mu\nu}(p_0, {\bf  p})= 2g^2 N_f\int ({\rm d}{\bf k})
\int_{-\infty}^\infty
\frac{{\rm d}k_0}{2\pi}\,\int_{-\infty}^\infty\frac{{\rm d}q_0}{2\pi}\,\rho_0(k)\,
\rho_0(q)\qquad\nonumber\\ \times
\Bigl[k_\mu q_\nu +q_\mu k_\nu-g_{\mu\nu}(k\cdot q)\Bigr]\,
\frac{n(k_0)-n(q_0)}{k_0-q_0-p_0}\,.\eeq
In going from eq.~(\ref{Pia12}) to eq.~(\ref{Pia2}), we have also
performed the spin trace.

 Let us now focus on the spatial components of the
polarization tensor (\ref{Pia2}). After integration
 over $k_0$ and $q_0$, one obtains:
\beq\label{Pia3}\lefteqn{ 
\Pi^{(a)}_{ij}(p_0, {\bf  p})= \frac{g^2 N_f}{2}\int  ({\rm d}{\bf k})
\,\frac{1}{\varepsilon_k \varepsilon_q}} \nonumber\\
 & &\Biggl\{\Bigl( k_i q_j + q_i k_j +\delta_{ij}
(\varepsilon_k \varepsilon_q - {\bf k \cdot q})\Bigr)\left(
\frac{n(\varepsilon_k)-n(\varepsilon_q)}{\varepsilon_k - \varepsilon_q + p_0}
\,+\,\frac{n(\varepsilon_k)-n(\varepsilon_q)}{\varepsilon_k - \varepsilon_q -p_0}
\right) \nonumber\\ & &
+\,\Bigl( k_i q_j + q_i k_j -\delta_{ij}
(\varepsilon_k \varepsilon_q + {\bf k \cdot q})\Bigr)\left(
\frac{ 1- n(\varepsilon_k)-n(\varepsilon_q)}{\varepsilon_k + \varepsilon_q + p_0}
\,+\,\frac{1-n(\varepsilon_k)-n(\varepsilon_q)}{\varepsilon_k + \varepsilon_q - p_0}
\right)\Biggr\}\,.\nonumber \\
& & \eeq
As $T \to 0$, the statistical factors
vanish, and 
\beq\label{Pia0}
\Pi_{T=0}^{ij}(p)= \frac{g^2 N_f}{2}\int 
({\rm d}{\bf k})\frac{
 k^i q^j + q^i k^j -\delta^{ij}
(\varepsilon_k \varepsilon_q + {\bf k \cdot q})}
{\varepsilon_k \varepsilon_q}\left(
\frac{ 1}{\varepsilon_k + \varepsilon_q + p_0}
+\frac{1}{\varepsilon_k + \varepsilon_q - p_0}
\right).\nonumber \\ \eeq
In four space-time dimensions, this expression develops ultraviolet
divergences, which are eliminated by the gluon wave-function
 renormalization \cite{IZ}.
The thermal contribution $\Pi_{T}\equiv \Pi-\Pi_{T=0}$
has no ultraviolet divergences since the statistical factors $n(\varepsilon_k)$
are exponentially decreasing for $k\gg T$. Thus, in evaluating
$\Pi_T$,  we can set $d=4$.

To isolate the HTL in eq.~(\ref{Pia3}),
we set $p_0\to \omega+i\eta$, with  real $\omega$ and $\eta\to 0^+$
(retarded boundary conditions), 
 and assume that both $\omega$ and $p\equiv |{\bf p}|$ are of the order $gT$.
 By definition, the hard thermal loop 
 is the leading piece in the expansion of $\Pi_{\mu\nu}(p)$
 in powers of $g$, including
the assumed $g$-dependence of the external four-momentum 
\cite{BP90}. It can be verified
 that the HTL in eq.~(\ref{Pia3}) arises
 entirely from the  integration over {\it hard} loop momenta $k\sim T$.
The contribution of the soft momenta $k\simle gT$ is suppressed
(here, by a factor $g^2$)  
because of the smallness of the associated phase space.
Furthermore, the contribution of the very high momenta, $k\gg T$, is
 exponentially suppressed by the thermal occupation numbers.
 This last argument does not apply to the vacuum piece, eq.~(\ref{Pia0});
 however, after renormalization, the finite contribution of $\Pi_{T=0}(p)$ 
 is of the order $g^2 p^2$, which for $p\sim gT$ is down by a factor of $g^2$
as compared to the HTL (we anticipate that the latter is 
${\cal O}(g^2 T^2)$).

Let us evaluate now the contribution of the hard loop momenta $k\sim T$.
Since $p \sim gT \ll k $, we can write:
\beq
\varepsilon_q\equiv |{\bf  k}-{\bf  p}|\simeq k-{\bf  v}\cdot {\bf  p}\,\eeq
which allows us  to simplify the energy denominators
in eq.~(\ref{Pia3}) as follows:
\beq\label{approx}
\varepsilon_k+\varepsilon_q\pm\omega\,\simeq\, 2k,\qquad
\varepsilon_k-\varepsilon_q\pm\omega\,\simeq\,
 {\bf  v}\cdot {\bf  p}\pm\omega\,.\eeq
In these equations,  ${\bf  v}\equiv {\bf  k}/k$ denotes the velocity of
the hard  particle, $|{\bf  v}|=1$.

Note that we have two types of energy denominators:
Those involving the {\it difference} of the two internal quark
energies, which are {\it soft\,}  $(\varepsilon_k-\varepsilon_q \simeq\,
 {\bf  v}\cdot {\bf  p}\sim gT)$, and those involving the {\it sum} of the
respective energies, which are {\it hard\,} 
$(\varepsilon_k+\varepsilon_q\simeq 2k \sim T)$. 
The soft denominators, whose form is reminiscent of the
Bloch-Nordsieck approximation described in section 6.3, are associated
with the scattering of  the soft gluon  on the hard thermal quarks.
Because of the Pauli principle, such processes occur with the following
statistical weight:
\beq\label{PS} n(\varepsilon_k)[1-n(\varepsilon_q)]-
[1-n(\varepsilon_k)] n(\varepsilon_q)=
n(\varepsilon_k)-n(\varepsilon_q)\,\simeq\,{\bf  v}\cdot{\bf  p}
\,\frac{{\rm d}n}{{\rm d}k}\,,\eeq
that is, they are suppressed by one power of $g$ at soft momenta $p \sim gT$.
Because of this suppression, they contribute to the same order
as the terms involving hard denominators, associated
 to vacuum-like processes where the soft gluon field
turns into a virtual quark-antiquark pair. These are accompanied by
statistical factors like \beq
[1-n(\varepsilon_k)][1-n(\varepsilon_q)]-n(\varepsilon_k)n(\varepsilon_q)
=1- n(\varepsilon_k)- n(\varepsilon_q) \,\simeq\,1-2n(k)\,.\eeq
Note however that, to the order of interest, the hard denominators
are independent of the external energy and momentum, so that they give
only a constant contribution to $\Pi_{\mu\nu}$. After
performing the following simplifications:
\beq\label{approx1}
k_i q_j + q_i k_j +\delta_{ij}
(\varepsilon_k \varepsilon_q - {\bf k \cdot q}) &\simeq& 2k_i k_j\,,\nonumber\\
k_i q_j + q_i k_j -\delta_{ij}
(\varepsilon_k \varepsilon_q + {\bf k \cdot q}) &\simeq& 2(k_i k_j - \delta_{ij} k^2)\,,\eeq
we obtain:
\beq\label{Piahtl}
 \Pi^{(a)}_{ij}(\omega, {\bf  p})&\approx & -
 2{g^2 N_f}\int({\rm d}{\bf k}) \left\{\frac{k_i k_j}{k^2}\,
\frac{{\rm d}n}{{\rm d}k}\,\frac{{\bf  v}\cdot{\bf  p}}
{\omega - {\bf  v}\cdot{\bf  p}}+
\frac{k_i k_j - \delta_{ij} k^2}{k^2}\,\frac{n(k)}{k}\right\}\\\nonumber
&=&\frac{g^2 T^2 N_f}{6}\int\frac{{\rm d}\Omega}{4\pi}\,\frac{\omega\,v_iv_j}
{\omega - {\bf  v}\cdot{\bf  p}}\,,\eeq
where the second line follows from the first one by using:
\beq \int_0^\infty {\rm d}k\,k\,n(k)= -\frac{1}{2}\,\int_0^\infty {\rm d}
k\,k^2\,\frac{{\rm d}n}{{\rm d}k}\,=\,\frac{\pi^2 T^2}{12}.\eeq
The angular integral  $\int {\rm d}\Omega$ runs over all the
directions of the unit vector ${\bf  v}$.

The other space-time components of $\Pi_{\mu\nu}^{(a)}$ can be computed in a similar
fashion. The complete result to leading order in $g$ reads
\beq\label{DPia} \Pi_{\mu\nu}^{(a)}(\omega, {\bf  p}) \approx
\frac{g^2 T^2 N_f}{6}
\left \{-\delta^0_\mu\delta^0_\nu \,+\,\omega \int\frac{{\rm d}\Omega}{4\pi}
\frac{v_\mu\, v_\nu} {\omega - {\bf  v}\cdot {\bf  p}
+i\eta}\right\}\,.\eeq
This coincides with the corresponding result of the kinetic theory (namely,
the quark piece of eq.~(\ref{Pi})).

\subsubsection{The ghost and gluon loops}

Consider now the other pieces of the one-loop polarization tensor,
as given by the diagrams with one gluon or ghost loop in
Figs. \ref{figgluon}.b -- d. 
The tadpole diagram \ref{figgluon}.c is easily evaluated as
(in Feynman's gauge) :
\beq\label{Pic1}
\Pi_{\mu\nu}^{(c)}\,=\,-\,g_{\mu\nu}\,g^2 (d-1) N \int [{\rm d}k]
\,\Delta (k)
\,.\eeq
It is momentum-independent. 
The factor $N$ (number of colours) arises from the colour
trace, while the factor $(d-1)$ originates from the
 four-gluon vertex. 

The contribution of the gluon loop in fig. \ref{figgluon}.b is:
\beq\label{Pib1}
\Pi_{\mu\nu}^{(b)}(p)= \frac{g^2 N}{2}
\int [{\rm d}k]\,\Gamma_{\sigma\mu\lambda}^0 (-p+k, p, -k)\,
G_0^{\lambda \rho}(k)\,
\\\nonumber \Gamma_{\rho\nu\eta}^0 (-k, p, -p+k)\,G_0^{\eta\sigma}
(p-k)\,,\eeq
where $\Gamma^0_{\mu\nu\rho}$
is the bare three-gluon vertex, 
\beq 
\Gamma^0_{\mu\nu\rho}(p,q,k)\,=\,
g_{\mu\nu}(p-q)_\rho\,+\,g_{\nu\rho}(q-k)_\mu\,+\,g_{\mu\rho}
(k-p)_\nu\,.\eeq
Instead of performing the Matsubara sum directly in eq.~(\ref{Pib1}),
which would be complicated by the energy dependence of the vertices,
we follow Ref. \cite{BP90}  and make first some of the simplifications which
are allowed as long as we are interested only in the {\htl}.
 Since, by assumption, the external momentum $p$ is soft,
while the integral over  $k$ is dominated by hard internal momenta,
 the terms linear in $p$ in the three-gluon vertex can be neglected
next to those linear in $k$:
\beq \label{approx2}
\Gamma^0_{\sigma\mu\lambda} (-p+k, p, -k)\,\simeq\,
\Gamma^0_{\sigma\mu\lambda} (k, 0, -k)\,=\,\Gamma_{\alpha\mu\sigma\lambda}
\,k^{\alpha}\,,\eeq
with the notation
\beq 
\label{Gamma}
\Gamma_{\mu\nu\rho\lambda}\equiv\,2g_{\mu\nu}g_{\rho\lambda}-g_{\mu\rho}
g_{\nu\lambda}-g_{\mu\lambda}g_{\nu\rho}.\eeq
(The reader might doubt of the validity
of eq.~(\ref{approx2}) at this stage
since, strictly speaking,
the external energy is generally hard to start with, $p_0=i 2n\pi T$.
Note, however, that   $p_0$ is   a dummy
variable in the summation over the internal Matsubara frequencies
$k_0$. After
the analytic continuation, $p_0\to \omega$ becomes soft, with $\omega \sim
gT$, while $k_0$ remains hard, $k_0\sim T$.) Since 
\beq
\Gamma_{\alpha\mu\sigma\lambda}\,\Gamma_{\beta\nu\rho\eta}
g^{\lambda\rho}g^{\eta\sigma}\,=\, 
4(d-2)g_{\alpha\mu} g_{\beta\nu}\,+\,2(g_{\alpha\beta} g_{\mu\nu}
\,+\,g_{\alpha\nu} g_{\beta\nu})\,,\eeq
and $G^0_{\mu\nu}(k)= -g_{\mu\nu}\Delta (k)$ in  Feynman's gauge,
we deduce that
\beq\label{Pib2}
\Pi_{\mu\nu}^{(b)}(p) \approx \,-\,\frac{g^2 N}{2}
\int [{\rm d}k]\, \Bigl((4d-6)k_\mu k_\nu + 2g_{\mu\nu} k^2
\Bigr)\Delta(k)\Delta(p-k)\,.\eeq
After performing  similar manipulations on the ghost loop in fig. \ref{figgluon}.d, we obtain (in covariant gauges, the ghost propagator
coincides with the scalar propagator $\Delta(k)$) :
\beq\label{Pid1}
\Pi_{\mu\nu}^{(d)}(p) \approx {g^2 N}
\int [{\rm d}k]\,k_\mu k_\nu \,\Delta(k)\Delta(p-k)\,.\eeq

By adding together eqs.~(\ref{Pic1}), (\ref{Pib2}) and (\ref{Pid1}),
we end up with the following expression for the
one-gluon-loop polarization tensor:
\beq\label{Pig}
\Pi_{\mu\nu}^{(g)}(p) \approx -\,(d-2){g^2 N}
\int [{\rm d}k]\,\left\{2\,k_\mu k_\nu\,\Delta(k)\Delta(p-k)
\,+\,g_{\mu\nu}\,\Delta(k)\right \},\eeq 
where the upperscript $g$ stays for ``gluons''.
Although this has been derived here in Feynman's gauge, 
the final result (\ref{Pig}) is actually gauge-fixing independent
\cite{Klimov81,Weldon82a,BP90}, as also shown by the kinetic theory
\cite{qcd}. The factor $d-2=2$ in  eq.~(\ref{Pig}) shows that 
only the two physical, transverse, degrees of freedom of the hard
gluons are involved in the {\htl}. The contributions of the unphysical degrees
of freedom (longitudinal gluons and ghosts) mutually cancel in the sum
of eqs.~(\ref{Pic1}), (\ref{Pib2}) and (\ref{Pid1}).

At this point, it is interesting to consider also the
contribution of the quark loop in fig.~\ref{figgluon}.a that
 would have been obtained if, before summing over the Matsubara
frequencies, we had  implemented the same kinematical approximations as
above. For the quark loop in eq.~(\ref{Pia1}), this amounts to the
replacement
\beq
{\rm Tr}\,(\gamma_\mu\,{\slashchar {k}}\,
\gamma_\nu\,({\slashchar {k}}-{\slashchar {p}}))\,\simeq\,4(2k_\mu k_\nu
- g_{\mu\nu}k^2),\eeq
and one gets (we recall that $S(k)=\slashchar{k}\tilde \Delta(k)$)
\beq\label{Pias}
\Pi_{\mu\nu}^{(a)}(p) \approx 2{g^2 N_f}
\int \{{\rm d}k\}\,\left\{2\,k_\mu k_\nu\,\tilde\Delta(k)\tilde\Delta(k-p)
\,+\,g_{\mu\nu}\,\tilde\Delta(k)\right \}.\eeq 
The similitude between eqs.~(\ref{Pias}) and (\ref{Pig}) suggests
that the HTL content of eq.~(\ref{Pig})
can be obtained by following the same steps as for the quark
loop in Sect. B.1.1. The resulting expression is identical to
 eq.~(\ref{DPia}), with the factor
 $g^2 T^2 N_f/6$ replaced by $g^2 T^2 N/3$.
Thus, the complete {\htl}
for the soft gluon self-energy reads  \cite{Klimov81,Weldon82a}
\beq\label{DPi}\Pi_{\mu\nu}(\omega, {\bf  p})\,=\,m_D^2
\left \{-\delta^0_\mu\delta^0_\nu \,+\,\omega \int\frac{{\rm d}\Omega}{4\pi}
\frac{v_\mu\, v_\nu} {\omega - {\bf  v}\cdot {\bf  p}
+i\eta}\right\}\,,\eeq
with the Debye mass in eq.~(\ref{omegap}).
Eq.~(\ref{DPi}) coincides with the expression (\ref{Pi}) derived from kinetic
theory.

\begin{figure}
\protect \epsfysize=2.5cm{\centerline{\epsfbox{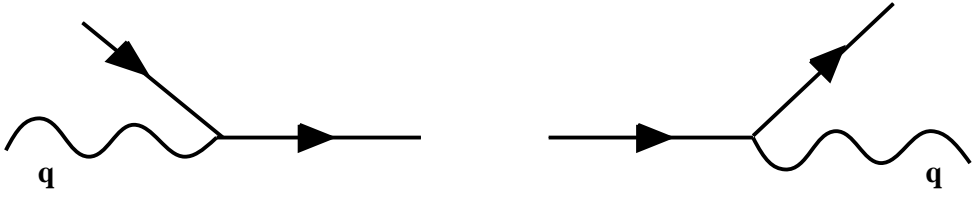}}}
	 \caption{Illustration of the Landau damping mechanism:
 a space-like photon (or gluon) can be absorbed or emitted
 by an on-shell thermal fermion.}
\label{landau}
\end{figure}

The small imaginary part $i\eta$ in the denominator of
eq.~(\ref{DPi}) implements the retarded boundary conditions. 
This is relevant  only for {\it space-like}
($\omega^2<{\bf p}^2$) external momenta, since it is only for such
momenta that the energy denominator $\omega - {\bf  v}\cdot {\bf  p}$
 can vanish. In that case, the 
polarization tensor (\ref{DPi}) develops an imaginary part,
\beq\label{ImPi2}
{\rm Im}\, \Pi_{\mu\nu}(\omega, {\bf  p})\,=\,-\,\pi m_D^2\,\omega\int 
\frac {{\rm d}\Omega}{4\pi}\,{ v_{\mu}v_{\nu}}
\,\delta (\omega - {\bf  v}\cdot{\bf  p})\,,\eeq 
which describes the absorption or the emission of a 
 space-like gluon, with four-momentum $p^\mu=(\omega,{\bf p})$,
by a hard particle (quark or gluon) from the thermal  bath
(see, fig.~\ref{landau} for an example).
According to the eq.~(\ref{ImPi2}),
${\rm Im}\, \Pi_{\mu\nu}(\omega, {\bf  p})$ vanishes
linearly in the static limit, $\omega \to 0$.
This may be easily understood by inspection of the thermal phase
space available for the processes in  fig.~\ref{landau}. This is
proportional to:
\beq n_1(1-n_2)-(1-n_1)n_2\,=\,n_1 - n_2\,,\eeq
where $n_1$ and $n_2$ are the statistical factors for the two
 thermal fermions, with energies $\epsilon_1$ 
and $\epsilon_2=\epsilon_1 + \omega$, respectively.
As $\omega\to 0$, we may write $n_1 - n_2 \simeq -\omega
({\rm d}n/{\rm d} \epsilon_1)$, which vanishes linearly
with $\omega$.

\subsubsection{The HTL gluon propagator}

In order to construct the propagator of the  soft gluon
in the HTL approximation, we have to invert the equation:
\beq\label{D0*}
{}^*\!G_{\mu\nu}^{-1}\equiv G^{-1}_{0\,\mu\nu}+ \Pi_{\mu\nu},\eeq
where $ \Pi_{\mu\nu}$ is the polarization tensor of eq.~(\ref{DPi}),
which is transverse:
\beq \label{trans}
p^\mu \Pi_{\mu\nu}(p)\,=\,0\,.\eeq
Actually, this property holds for the whole one-loop contribution
to $\Pi_{\mu\nu}$ \cite{GPY81,Weldon82a}, but,
in contrast to what happens at zero temperature,
it is generally {\it not} satisfied
(except in ghost-free gauges, like axial gauges)
beyond the one-loop approximation \cite{KK85,Schulz95}.

In order to invert eq.~(\ref{D0*}), we need to fix the gauge.
The physical interpretation is more transparent in the Coulomb gauge,
 where the only non-trivial components of ${}^*\!G_{\mu\nu}$
 are the electric (or longitudinal)
one, ${}^*\!G_{00}(p_0, {\bf p})\equiv {}^*\!\Delta_L(p_0,p)$,
 and the magnetic (or transverse) one,
${}^*\!G_{ij}(p_0, {\bf p})=(\delta_{ij}-\hat p_i\hat p_j){}^*\!\Delta_T(p_0,p)$.
At tree-level, we have $\Delta_L(p_0,p)=-1/p^2$, corresponding to the
instantaneous Coulomb interaction,  and $\Delta_T(p_0,p)=-1/(p_0^2-p^2)$,
whose poles at $p_0=\pm p$ are associated with
the massless transverse gluon excitations.

 Being transverse, the polarization tensor (\ref{DPi})
 is determined by  only two scalar functions, which we choose to be  its 
 longitudinal ($L$) and transverse ($T$) components
with respect to ${\bf p}$. Specifically, we write:
 \beq\label{Pitens}
\Pi_{00}(p_0,{\bf p})= -\Pi_L(p_0,{p}),\qquad
\Pi_{0i}(p_0,{\bf p})= -\,\frac{p_0 p_i}{p^2}\,
 \Pi_L(p_0,{p}),\\ \nonumber
\Pi_{ij}(p_0,{\bf p})=
(\delta_{ij}-\hat p_i\hat p_j) \Pi_T(p_0,{p})
- \hat p_i\hat p_j \,\frac{p_0^2}{p^2}\,
 \Pi_L(p_0,{p})\,,\eeq
where $p=|{\bf p}|$ and $\hat p_i = p_i/p$. These definitions are
appropriate for the Coulomb gauge. 
 The explicit forms of the scalar functions
$\Pi_{L,\,T}(p_0,p)$ follow easily from eqs.~(\ref{DPi}) and (\ref{Pitens}):
\beq\label{plpt}
\Pi_L(p_0, p)\,=\,m_D^2\,\Bigl(1\,-\,
Q({p_0}/{p})\Bigr)\,,\\ \nonumber
\Pi_T(p_0, p)\,=\,\frac{m_D^2}{2}\,\frac{p_0^2}{p^2}\,\left (
1\,-\,\frac{p_0^2 - p^2}{p_0^2}\, Q({p_0}/{p})\right )\,,\eeq
with  \beq\label{Q}
Q(x)\equiv \frac{x}{2}\,\ln\frac{x+1}{x-1}\,.\eeq
For real energy $p_0= \omega+i\eta$ and space-like
momenta ($|\omega|<p$), the function $Q(\omega/p)$
has a non-vanishing imaginary part:
\beq
{\rm Im}\,Q(\omega/p)\,=\,
-\,\frac{\pi\omega}{2 p}\,\theta (p-|\omega|).\eeq
Correspondingly, the polarization
functions $\Pi_L$ and $\Pi_T$ acquire imaginary parts
which describe the Landau damping of soft space-like gluons.

At high temperature, and in the {\htl} approximation, we have then:
 \beq\label{effd}
{}^*\!\Delta_L(p_0,p)\,=\,\frac{- 1}{p^2 +\Pi_L(p_0,p)},\qquad
{}^*\!\Delta_T(p_0,p)\,=\,\frac{-1}{p_0^2-p^2 -\Pi_T(p_0,p)}\,.\eeq
These functions can be given the following spectral representations
\beq\label{Dspec}
{}^*\!\Delta_T(\omega, {p})&=&\int_{-\infty}^{\infty}\frac{{\rm d}p_0}{2\pi}
\,\frac{{}^*\!\rho_T(p_0, p)}{p_0-\omega}\,,\nonumber\\
{}^*\!\Delta_L(\omega, p)&=&-\,\frac{1}{p^2}
\,+\int_{-\infty}^{\infty}\frac{{\rm d}p_0}{2\pi}
\,\frac{{}^*\!\rho_L(p_0, p)}{p_0-\omega}\,,\eeq
where ${}^*\!\rho_{L}$ and ${}^*\!\rho_{T}$ are the 
corresponding spectral densities,
\beq\label{rhos}
{}^*\!\rho_{L,T}(p_0,p) = 2{\rm Im}\,{}^*\!\Delta_{L,T}
(p_0+i\eta ,p)\,.\eeq
Note the subtraction performed in the spectral representation
of ${}^*\!\Delta_L(\omega, p)$: this is necessary since
${}^*\!\Delta_L(\omega,p)\to -1/p^2$ as $|\omega|\to \infty$.
By taking $\omega\to 0$ in eqs.~(\ref{Dspec}), and using
${}^*\!\Delta_L(0,p)=-1/(p^{2}+m_{D}^{2})$ and 
${}^*\!\Delta_T(0,p)=1/p^{2}$, one obtains the
following ``sum rules'':
\beq\label{SRULES0}
\int \frac{{\rm d}p_{0}}{2\pi p_{0}}\,\,
{}^*\!\rho_L(p_0, p)&=&
\frac{1}{p^{2}}-\frac{1}{p^{2}+m_{D}^{2}}\,,\nonumber\\
\int \frac{{\rm d}p_{0}}{2\pi p_{0}}\,\,{}^*\!\rho_T(p_0, p)&=&
\frac{1}{p^2}\,.
\eeq
Besides, the transverse density ${}^*\!\rho_T$ satisfies
the usual sum-rule (\ref{SRrho}) :
\beq\label{SRrhoT}
\int \frac{{\rm d}p_{0}}{2\pi}\,p_{0}\,
{}^*\!\rho_T(p_0, p)&=& 1.\eeq

The spectral functions  ${}^*\!\rho_L$ and ${}^*\!\rho_T$
have the following structure (with $s=L$ or $T$):
\beq\label{rhos0}
{}^*\!\rho_{s}(p_0,p)&\equiv& 2\,{\rm Im}\,{}^*\!\Delta_{s}(p_0+i\eta ,p)
\nonumber\\
&=& 2\pi\epsilon(p_0)\,z_s(p)\,\delta(p_0^2 -\omega_s^2(p))
 +\beta_s (p_0,p) \theta (p^2- p_0^2),\eeq
where, in the second line, the $\delta$-function 
corresponds to the (time-like) poles
of the resummed propagator ${}^*\!\Delta_s(p_0,p)$,  with energy
$p_0=\pm {\omega_s(p)}$ and residue $ z_s(p)$ (cf. Sect. 4.3.1),
while the function $\beta_s (p_0,p)$, with support at space-like
momenta, corresponds to Landau damping (cf. Sect. 4.3.3).
Specifically, the mass-shell residues are defined by :
\beq
{}^*\!\Delta_{s}(p_0,p)\,\approx\,\frac{-z_s(p)}{p_0^2-
\omega_s^2(p)}\qquad {\rm for\,\,\,\,} p_0^2\approx\omega_s^2(p)\,,
\eeq 
which, together with the pole condition 
${}^*\!\Delta_{s}^{-1}(p_0=\pm {\omega_s},p) =0$ and
eqs.~(\ref{plpt}) for $\Pi_T$ and $\Pi_L$,
implies the following expressions for $z_T$ and $z_L$ :
\beq
z_T^{-1}(p)\,=\,\frac{2\omega_T^2(\omega_T^2-p^2)}
{m_D^2\omega_T^2-(\omega_T^2-p^2)^2}\,,\qquad 
z_L^{-1}(p)\,=\,\frac{2\omega_L^2(\omega_L^2-p^2)}
{p^2(m_D^2 +p^2-\omega_L^2)}\,.\eeq 
It can be verified on the above formulae that the residues
$z_s(p)$ are positive functions \cite{emt}, with the following
limiting behaviour: At small $p$, $p\ll \omega_{pl}$
(with $\omega_{pl}\equiv m_D/{\sqrt 3}$ the frequency
of the longwavelength plasma oscillations; cf. Sect. 4.3.1),
\beq
z_T(p)\,\approx\,1\,-\,\frac{p^2}{5\omega_{pl}^2}\,,
\qquad z_L(p)\,\approx\,\frac{\omega_{pl}^2}{p^2}\,,\eeq
while for large momenta, $p\gg \omega_{pl}$,
\beq
z_T(p)\,\approx\,1\,-\,\frac{3\omega_{pl}^2}{4p^2}\left(\ln
\frac{8p^2}{3\omega_{pl}^2}-3\right),\qquad z_L(p)\,\approx\,
\frac{8p^2}{3\omega_{pl}^2}\,
\exp\left\{ -\frac{2p^2 }{3\omega_{pl}^2}-2\right\}.
\eeq 
Thus, as mentioned in Sect. 4.3.1, the longitudinal mode
is exponentially suppressed at $p\gg gT$.

The functions
$\beta_L$ and $\beta_T$ are explicitly given by
(cf. eq.~(\ref{plpt})) :
\beq\label{RHOLT}
\beta_{L}(p_0,p)&=&\pi m_D^2\,\frac{p_0}{p}\,\Big
|{}^*\!\Delta_L(p_0,p)\Big|^2,\nonumber\\
\beta_{T}(p_0,p)&=&\pi m_D^2\,\frac{p_0(p^2-p_0^2)}{2p^3}\,
\Big|{}^*\!\Delta_T(p_0,p)\Big|^2,\eeq 
and satisfy the following sum-rules, which follow by
inserting the decomposition (\ref{rhos0}) in
eqs.~(\ref{SRULES0}) and (\ref{SRrhoT}) :
\beq\label{SRULESbeta}
\int \frac{{\rm d}p_{0}}{2\pi p_{0}}\,
\beta_L(p_0, p)&=&
\frac{1}{p^{2}}-\frac{1}{p^{2}+m_{D}^{2}}-
\frac{z_L(p)}{\omega_L^2}\,
,\nonumber\\
\int \frac{{\rm d}p_{0}}{2\pi p_{0}}\,\beta_T(p_0, p)&=&
\frac{1}{p^2}-\frac{z_T(p)}{\omega_T^2}\,,\nonumber\\
\int \frac{{\rm d}p_{0}}{2\pi}\,p_{0}\,
\beta_T(p_0, p)&=& 1-z_T(p).\eeq

\subsubsection{The gauge field fluctuations}

We can use the spectral functions that we have just obtained to 
calculate the magnitude of the gauge field fluctuations. These are 
given by
\beq\label{Afluct}
\langle A^{2}\rangle\,=\, \int\frac{{\rm d}^{4}k}{(2\pi)^4}
\,N(k_{0})\,\rho(k)
\eeq
where $\rho(k)$ is one of the spectral functions (\ref{rhos}). 

The dominant fluctuations in the plasma have momenta $k\sim T$, and can be
estimated  using the free spectral density 
$\rho_0(k)=2\pi\epsilon(k_{0})\delta (k^2)$. One then finds (ignoring the
vacuum contribution, see eq.~(\ref{TP}):
\beq
\langle A^{2}\rangle_T\,=\, \int \frac{{\rm d}^{3}k}{(2\pi)^3}\,
\frac{N(\varepsilon_{k})}{\varepsilon_{k}}\,\sim \,T^{2}.
\eeq

The longwavelength fluctuations can be evaluated using the classical 
field approximation whereby $N(k_{0})\approx T/k_{0}$. One then obtains
\beq
\langle A^{2}\rangle_{soft}\approx T\int {\rm d}^{3}k\,\int \frac{{\rm 
d}k_{0}}{2\pi k_{0}}\rho(k),
\eeq
where the momentum integral is to be limited to soft momenta $k\ll T$. The
$k_0$ integral can be calculated with the help of the ``sum rules''
(\ref{SRULES0}).
In the case of longitudinal (electric) fields with typical momenta 
$k\sim gT$, one obtains
\beq
\langle A^{2}\rangle_{gT}\approx T m_{D}^{2}\int  \frac{{\rm 
d}^{3}k}{k^{2}(k^{2}+m_{D}^{2})}\sim Tm_{D}\sim gT^{2}.
\eeq
Transverse (magnetic) fields are not screened at small frequencies 
in the HTL approximation. Therefore:
\beq
\int \frac{{\rm d}k_{0}}{2\pi 
k_{0}}\,{}^*\!\rho_{T}(k_{0},k)\,=\,\frac{1}{k^{2}}.
\eeq
Then the typical fluctuations at the scale $k\sim \mu$ are
\beq
\langle A^{2}\rangle_{\mu}\approx T \int^\mu  \frac{{\rm 
d}^{3}k}{k^{2}}\sim \mu T.
\eeq
In particular, for $\mu\sim g^2T$, the magnetic fluctuations $\langle
A^{2}\rangle_{g^2T}\sim g^2T^2$ become non perturbative: $g\sqrt{\langle
A^{2}\rangle_{\mu}}\sim \mu$ for $\mu\sim g^2T$.
Fluctuations with longer wavelengths $\lambda \gg 1/g^2T$
are expected to be damped by the non-perturbative
magnetic screening at the scale $g^2T$.

\begin{figure}
\protect \epsfysize=20.cm{\centerline{\epsfbox{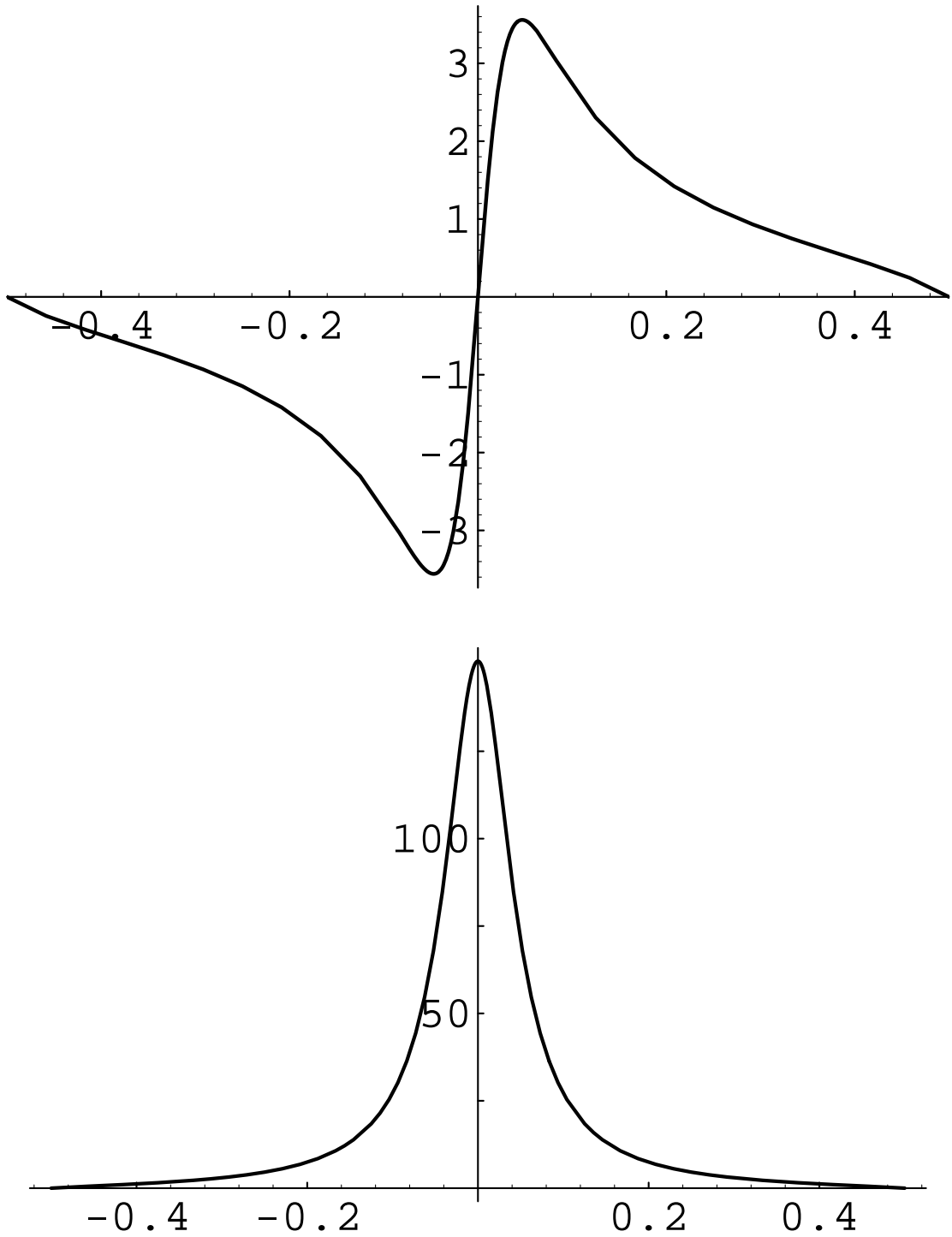}}}
	 \caption{The functions $\beta_T(q_0,q)$ and $\beta_T(q_0,q)/q_0$ 
in terms of $q_0/\omega_{pl}$ 
for $q=0.5\, \omega_{pl}$.  All the quantities are made adimensional
by multiplying them by appropriate powers of $\omega_{pl}$.}
\label{rho2}
\end{figure}

Note that the ultrasoft magnetic fluctuations $\langle
A^{2}\rangle_{g^2T}$ carry very soft 
frequencies $k_0\simle g^4T$. Indeed, for $k\sim g^2T$, the
contribution of the magnetic modes to the integral in
eq.~(\ref{Afluct}) is saturated by the Landau damping piece
$\beta_T$ of the spectral density ${}^*\!\rho_{T}$,
as it can be easily verified on the
second equation (\ref{SRULESbeta}). This gives
(cf. eqs.~(\ref{RHOLT}) and (\ref{deltaT0})) :
\beq\label{betat0}
\frac{1}{k_0}\,\beta_T(k_0\ll k)\,\simeq\, 
\frac{\pi}{2}\, \frac{m_D^2 \, k}
{k^6\,+\, (\pi m_D^2 k_0/4)^2}\,.\eeq
For $k\sim g^2T$, this is strongly peaked at small 
frequencies $k_0\simle k^3/m_D^2 \simeq g^4T$, as illustrated in
fig.~\ref{rho2}.

\subsection{The soft fermion propagator}

The one-loop fermion self-energy is displayed in fig.~\ref{figfermion},
and is evaluated as:
\begin{figure}
\protect \epsfysize=2.cm{\centerline{\epsfbox{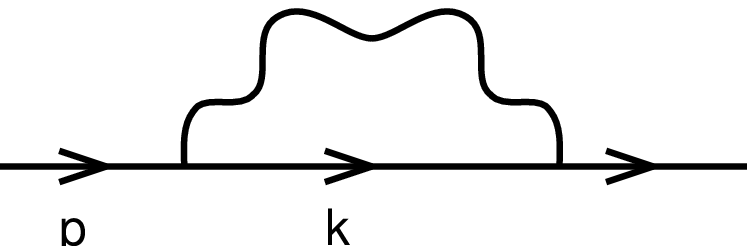}}}
	 \caption{The one-loop quark self-energy}
\label{figfermion}
\end{figure}
\beq\label{Sig1}
\Sigma(p)=-\,g^2 C_{\rm f}\,\int [{\rm d}q] 
\,\gamma^\mu\,S_0(p-q)\,\gamma^\nu
\,G_{\mu\nu}(q)\,,\eeq
where $p^0=i\omega_n = i (2n +1)\pi  T$, $q^0= i\omega_m= 
i \,2m\pi T$,  with integers $n$ and
 $m$, and the  gluon propagator $G_{\mu\nu}$ is here taken, 
for convenience, in the Coulomb gauge.
In view of further applications, we shall compute eq.~(\ref{Sig1})
with a {resummed} gluon propagator of the general form (\ref{Dspec}).

\subsubsection{The one-loop self-energy}

After performing the Matsubara sum and the continuation to complex values 
of the external energy $p_0$, we obtain the analytic 
one-loop self-energy in the form:
\beq\label{S123}
\Sigma(p)=\Sigma_C({\bf p})+\Sigma_L(p)+\Sigma_T(p),\eeq
where $\Sigma_C$ denotes the 
contribution of the instantaneous Coulomb interaction, as arising from
the term $-1/p^2$ in the second line of eq.~(\ref{Dspec}):
\beq\label{SC}
\Sigma_C({\bf  p})=g^2 C_{\rm f} \int ({\rm d}{\bf k})\,\frac{1}{({\bf p-k})^2}\,
\int_{-\infty}^{+\infty}\frac{{\rm d}k_0}{2\pi}\,\rho_0(k)\Bigl(1-n(k_0)\Bigr)
\gamma^0 {\slashchar k}\gamma^0\,,\eeq
while $\Sigma_L(p)$ and $\Sigma_T(p)$ are respectively the contributions
of the longitudinal and transverse gluons:
\beq\label{SL}
\Sigma_L(p)=-g^2 C_{\rm f} \int ({\rm d}{\bf q})
\int_{-\infty}^{+\infty}\frac{{\rm d}k_0}{2\pi}
\int_{-\infty}^{+\infty}\frac{{\rm d}q_0}{2\pi}\,
\rho_0(k)\rho_L(q)\, \gamma^0 {\slashchar k}\gamma^0\,
\frac{1+N(q_0)-n(k_0)} {k_0+q_0-p_0}\,,\nonumber \\ \eeq
and 
\beq\label{ST}\lefteqn{
\Sigma_T(p)=-g^2 C_{\rm f} \int ({\rm d}{\bf q})
\int_{-\infty}^{+\infty}\frac{{\rm d}k_0}{2\pi}
\int_{-\infty}^{+\infty}\frac{{\rm d}q_0}{2\pi}\,
\rho_0(k)\rho_T(q)}\qquad\qquad \nonumber\\
 & & \times (\delta_{ij}-\hat q_i \hat q_j) \,\gamma^i
 {\slashchar k}\gamma^j\,
\,\frac{1+N(q_0)-n(k_0)} {k_0+q_0-p_0}\,.\eeq
In these equations, $k^\mu=(k_0,{\bf k})$ (with ${\bf k=p-q}$) is the 
four-momentum carried by the internal quark line. 

Consider now the strict one-loop calculation, i.e., the one involving the
tree-level  gluon propagator.  For a bare gluon, $\rho_L=0$,
and therefore $\Sigma_L=0$ as well. Furthermore, $\rho_T=\rho_0$, so that
eq.~(\ref{ST}) becomes (after performing the Dirac trace):
\beq\label{ST1}
\Sigma_T(p_0, {\bf p})&=&-2\,g^2 C_{\rm f} \int ({\rm d}{\bf q})
\int_{-\infty}^{+\infty}\frac{{\rm d}k_0}{2\pi}
\int_{-\infty}^{+\infty}\frac{{\rm d}q_0}{2\pi}\,
\rho_0(k)\rho_T(q)\, \nonumber\\
&{}&\qquad\qquad\Bigl( k_0\gamma_0 - (\bfgamma \cdot \hat {\bf q})
({\bf k}\cdot\hat{\bf q})\Bigr)
\frac{1+N(q_0)-n(k_0)} {k_0+q_0-p_0}\,.\eeq
 The energy integrations can be easily performed with the result:
\beq\label{Sig5}\lefteqn{ 
\Sigma_T(p_0, {\bf  p}) =-\,2\, g^2 C_{\rm f}
\int ({\rm d}{\bf q})\,\frac{1}{2\varepsilon_k\,2\varepsilon_q} }
\nonumber\\ &{}&\,\,\Biggl\{
\Bigl(\varepsilon_k\gamma_0 - (\bfgamma \cdot \hat {\bf q})
({\bf k}\cdot\hat{\bf q})\Bigr)
\left(\frac{1-n(\varepsilon_k)+N(\varepsilon_q)}
 {\varepsilon_k+\varepsilon_q-p_0}\,+\,\frac{n(\varepsilon_k)+N(\varepsilon_q)}
 {\varepsilon_k-\varepsilon_q-p_0}\right)\nonumber\\&{}&-
\Bigl(\varepsilon_k\gamma_0 +(\bfgamma \cdot \hat {\bf q})
({\bf k}\cdot\hat{\bf q})\Bigr)
\left(\frac{n(\varepsilon_k)+N(\varepsilon_q)}
 {\varepsilon_k-\varepsilon_q+p_0}+
\frac{1-n(\varepsilon_k)+N(\varepsilon_q)}
 {\varepsilon_k+\varepsilon_q+p_0}\right)\Biggr\},\nonumber \\\eeq
where $\varepsilon_q\equiv |{\bf  q}|$ and $\varepsilon_k\equiv |{\bf  k}|
=|{\bf  p}-{\bf  q}|$.
The zero-temperature piece of this expression contains ultraviolet divergences
 which can be removed by renormalization \cite{IZ}.
The finite-temperature piece is UV finite, because of the thermal
factors, and we can set $d=4$ in its evaluation.

\subsubsection{The hard thermal loop}

To isolate the HTL in the previous expressions, we consider  soft external 
energy and momentum, $\omega \sim p \sim gT \ll T$,
and keep only the leading order contribution.
Once again, this contribution arises by integration over
 {\it hard} loop momenta, $q\sim T$, and it can be isolated
by performing  the same kinematical approximations as 
in the previous subsections.  It can then  be verified that the
Coulomb piece (\ref{SC}) contains no HTL (i.e., no contribution
proportional to $T^2$). Thus, only the physical,
transverse, gluons contribute to the HTL,
in agreement with the results from kinetic theory (cf. Sect. 3.5.1).

Consider eq.~(\ref{Sig5}) for $p_0= \omega+i\eta$, with $\omega$  real.
For $p\sim gT \ll q\sim T$, we can write:
$\varepsilon_k \simeq q-{\bf  v}\cdot {\bf  p}\,$
with ${\bf v}\equiv \hat{\bf q}\,$, and, to leading order
 in $g$, we can even replace ${\bf k}\simeq
-\,{\bf q}$ and $\varepsilon_k \simeq q$ everywhere except in the
energy denominators. For instance, we shall write
\beq
\varepsilon_k\gamma_0 + (\bfgamma \cdot \hat {\bf q})
({\bf k}\cdot\hat{\bf q})\,\simeq \,q\, (\gamma_0 - {\bf v\cdot \bfgamma})
\,=\, q \,{\slashchar{v}}\,,\eeq
where $v^\mu\equiv (1, {\bf v})$.
As in the previous discussion of the gluon HTL, we encounter both
soft and hard energy denominators,
\beq\label{approx3}
\varepsilon_k-\varepsilon_q\pm\omega\,\simeq\,
 {\bf  v}\cdot {\bf  p}\pm\omega\,,\qquad\,\,\,\,
\varepsilon_k+\varepsilon_q\pm\omega\,\simeq\, 2k\,,
\eeq
but in the present case, only those terms involving soft
denominators survive to leading order. 
Indeed, the corresponding numerators in eq.~(\ref{Sig5})
involve the sum of
statistical factors $n(\varepsilon_k)+N(\varepsilon_q)
\approx n(q)+N(q)$ which, unlike the difference 
$n(\varepsilon_k)-n(\varepsilon_q)$ in eq.~(\ref{PS}), 
is not suppressed when the external momentum $p$ is soft.
We thus get:
\beq\label{Sig6}
\Sigma(\omega, {\bf  p})&\approx& 2\, g^2 C_{\rm f} 
\int\frac{{\rm d}^3 q}{(2\pi)^3}\,\frac{n(q)+N(q)}{2q\,2q}\,
\biggl(\frac{\gamma_0 - {\bf v\cdot \bfgamma}}
{\omega-{\bf  v}\cdot {\bf  p}}\,+\,
\frac{\gamma_0 + {\bf v\cdot \bfgamma}}
{\omega+ {\bf  v}\cdot {\bf  p}}\biggr)\nonumber\\
&=& \frac{g^2 C_{\rm f}}{(2\pi)^3}\int_0^\infty {\rm d}q\,q\,(n(q)+N(q))
\,\int {\rm d}\Omega\,
\frac{{\slashchar v}}{\omega-{\bf  v}\cdot {\bf  p}},\eeq
 up to terms of higher order in $g$. The radial integral
is evaluated as:
\beq \int_0^\infty {\rm d}q\,q\,(n(q)+N(q))\,=\,\frac{\pi^2 T^2}{12}\,+\,
\frac{\pi^2 T^2}{6}\,=\,\frac{\pi^2 T^2}{4},\eeq
so that, finally \cite{Klimov81,Weldon82b},
\beq\label{Sigma2}
\Sigma(\omega, {\bf  p})&\approx& \omega_0^2 \int\frac{{\rm d}\Omega}{4\pi}\,
\frac{\slashchar v}{\omega-{\bf  v}\cdot {\bf  p}+i\eta},\eeq
where $\omega_0$  is the fermionic plasma frequency given in 
eq.~(\ref{omega0}). We thus recover the quark HTL (\ref{Sigma})
derived from kinetic theory. Eq.~(\ref{Sigma2}) has been
obtained here in the Coulomb gauge, but it is gauge-fixing
independent, as shown in Refs.
\cite{Klimov81,Weldon82b,BP90,qcd} .

 After  performing the angular integration in eq.~(\ref{Sigma2}),
   the final result for the quark HTL may be rewritten as:
 \beq\label{Sigma1}
\Sigma(\omega, {\bf  p})\,=\,a(\omega, p)\,\gamma^0\,+\,b(\omega, p)
{\hat{\bf  p}}\cdot{\bfgamma},\eeq
where:
\beq\label{S-ab}
a(\omega, p)\equiv\frac{\omega_0^2}{\omega}\,Q(\omega/p),\qquad
b(\omega, p)\equiv -\frac{\omega_0^2}{p}\,\Bigl[Q(\omega/p)\,-\,1\Bigr],\eeq
and the function $Q(x)$ is defined in eq.~(\ref{Q}).
For real energy ($p_0= \omega+i\eta$) and space-like
momenta ($|\omega|<p$), the functions in eq.~(\ref{S-ab})
develop imaginary parts which describe the Landau damping of
the soft fermion field.

\subsubsection{The  propagator of the  soft quark}

The propagator of the  soft quark in the {\htl} approximation
is obtained by inverting the Dyson-Schwinger equation:
\beq\label{SDS}
{}^*\!S^{-1}(p)=S_0^{-1}(p)+\Sigma(p).\eeq
To this aim, it is useful to observe that both the tree-level propagator $S_0(p)$ and the
self-energy $ \Sigma(p)$ are chirally symmetric (e.g., $\left\{
\gamma^5, \Sigma(p)\right\} =0$), so that they can be decomposed into
simultaneous eigenstates of chirality ($\gamma_5$) and helicity
($\bfsigma\cdot {\bf p}$). Specifically, we write:
\beq\label{S000}
S_0(\omega, {\bf p})=-\frac {\omega\gamma_0 -{\bf p}\cdot {\bfgamma}}
{\omega^2-p^2}=\,
\frac{-1}{\omega- p}\, h_+(\hat {\bf p})
+ \frac{-1}{\omega+ p}\, h_-(\hat {\bf p}),\eeq
where  $h_\pm(\hat {\bf p}) = (\gamma^0\mp \hat {\bf p}\cdot \bfgamma)/2\,$
and ${\hat{\bf  p}}\equiv {\bf  p}/p\,$, and similarly:
 \beq\label{Sigma1pol}
\Sigma(\omega, {\bf  p})\,=\,h_-(\hat {\bf p})\Sigma_+(\omega,p)\,-\,
h_+(\hat {\bf p})\Sigma_-(\omega,p),\eeq
with (cf. eq.~(\ref{Sigma1}))
\beq\label{PIpm}
\Sigma_\pm(\omega,p)\,\equiv\,\pm \,\frac{1}{2}\,{\rm tr}\,\Bigl(
h_\pm(\hat {\bf p}) \Sigma(\omega, {\bf p})\Bigr)\,=\,
\pm \,a(\omega, p) + b(\omega, p).\eeq
Note that the matrices $\Lambda_{\pm}(\hat {\bf p})\equiv
h_\pm(\hat {\bf p})\gamma^0$ project 
onto  spinors  whose chirality is equal ($h_+$) or opposite $(h_-)$ 
to the  helicity. (To see this, it is useful to
recall the identity
$\sigma^i=\gamma_5 \gamma^0 \gamma^i$, where $\sigma^i\equiv
-\epsilon_{ijk}\gamma^j\gamma^k/2i$,
so that $\gamma_5\Lambda_\pm=(\gamma_5\pm \bfsigma\cdot \hat
p)/2$.)
 
By using the above representations, 
one can easily invert eq.~(\ref{SDS}) to get the full propagator:
\beq\label{S*exp}
{}^*\!S(\omega, {\bf p})={}^*\!\Delta_+(\omega,p) h_+(\hat {\bf p})
+{}^*\!\Delta_-(\omega,p) h_-(\hat {\bf p}),\eeq
where
\beq\label{DPM}
{}^*\!\Delta_\pm(\omega,p)
 &=&\frac{-1}{\omega\mp (p + \Sigma_\pm(\omega,p))}\,.\eeq
The poles of the effective propagator ${}^*\!S(\omega,{\bf p})$ determine
the mass-shell condition  for soft fermionic excitations, 
to leading order in $g$  (cf. Sect. 4.3.1).

\subsubsection{The one-loop damping rate}

To leading order in $g$, the (hard) fermion damping rate 
can be readily extracted from the previous formulae in this section.
As explained in Sect. 6.3, this requires the resummation
of the internal gluon line, which is soft in the kinematical
regime of interest. By using eqs.~(\ref{SL}) and (\ref{ST})
with the gluon spectral densities in the HTL approximation 
(cf. eqs.~(\ref{rhos})--(\ref{RHOLT})), we obtain:
\beq\label{g1}
\gamma &\equiv&  -\,\frac{1}{4p}\,{\rm tr}\left(
{\slashchar p}\, {\rm Im}\,{}^*\!\Sigma_R(p_0+i\eta,{\bf p})\right)\Big |_
{p_0=p}\nonumber\\
&=&\pi
g^2 C_{\rm f} \int ({\rm d}{\bf q})
\int_{-\infty}^{+\infty}\frac{{\rm d}k_0}{2\pi}
\int_{-\infty}^{+\infty}\frac{{\rm d}q_0}{2\pi}
\delta(k_0+q_0-p)\,
\Bigl[1+N(q_0)-n(k_0)\Bigr]\nonumber\\
&\mbox{}&\qquad\times\rho_0(k)\biggl\{2\Bigl[k_0 - ({\bf v}\cdot \hat{\bf
q}) ({\bf k}\cdot \hat{\bf q})\Bigr]\,{}^*\!\rho_T(q)\,+\,
\Bigl[ k_0 + ({\bf v}\cdot {\bf k})\Bigr] \,{}^*\!\rho_L(q)\biggr\},\eeq
where $k^\mu=(k_0, {\bf k})$, ${\bf k=p-q}$ and  ${\bf v}=\hat {\bf p}$.
After performing the integral over $k_0$, and using the inequality
$q\ll p$ to do a few kinematical approximations, e.g.:
\beq\label{kinapp}
\epsilon_{p-q} \,\simeq\, p-{\bf v}\cdot {\bf q} &=& p-q\cos \theta,
\nonumber\\
1+N(q_0)-n(\epsilon_{p-q})&\simeq& N(q_0)\,\simeq\, T/q_0\,,\eeq
we can bring eq.~(\ref{g1}) into the form:
\beq\label{g4}
\gamma \simeq \pi g^2C_{\rm f}\int \frac{{\rm d}^4q}{(2\pi)^4}
 \,\frac{T}{q_0}\, \delta(q_0 -q\cos \theta)
\,\left({}^*\!\rho_L(q)+(1-\cos^2\theta)
{}^*\!\rho_T(q)\right).\eeq
The energy-conserving $\delta$-function singles out
the  space-like  pieces
$\beta_L$ and $\beta_T$ of the gluon spectral functions
(cf. eq.~(\ref{rhos0})) :
\beq\label{g5}
\gamma  \simeq \alpha TC_{\rm f}  \,\int_0^{\infty}{\rm d}q \,q\,
 \int_{-q}^{q}\frac{{\rm d}q_0}{ 2\pi q_0}\, 
\left\{\beta_L(q_0,q)+\left(1-\frac{q_0^2}{q^2}\right)
\beta_T(q_0,q)\right\}.\eeq
By also using eq.~(\ref{RHOLT}), one can see that this 
expression for $\gamma$ is identical to that obtained 
 for the interaction rate in the Born approximation,
eq.~(\ref{G2L}).

\end{document}